\documentclass[11pt,english]{article}
\usepackage[a4paper, tmargin=0.7in,bmargin=0.8in,lmargin=0.9in,rmargin=0.9in]{geometry}
\usepackage[skip= 0.5\baselineskip plus 2pt, indent=\baselineskip]{parskip}
\usepackage{setspace} 
\usepackage[utf8]{inputenc}
\usepackage[T1]{fontenc} 
\usepackage{appendix} 
\usepackage{import}
\usepackage{graphicx}
\usepackage{wrapfig}
\usepackage[backend=bibtex, sorting=none, bibstyle=ieee, style=numeric-comp, giveninits]{biblatex}
\bibliography{QdSEntropyReferences}
\AtEveryBibitem{\clearfield{abstract}}\AtEveryBibitem{\clearlist{abstract}}
\AtEveryBibitem{\clearfield{issn}}\AtEveryBibitem{\clearlist{issn}}
\AtEveryBibitem{\clearfield{language}}\AtEveryBibitem{\clearlist{language}}
\AtEveryBibitem{\clearfield{keywords}}\AtEveryBibitem{\clearlist{keywords}}
\AtEveryBibitem{\clearfield{note}}\AtEveryBibitem{\clearlist{note}}
\AtEveryBibitem{\clearfield{url}}\AtEveryBibitem{\clearlist{url}}
\AtEveryBibitem{%
\ifentrytype{online}{}{\clearfield{urlyear}\clearfield{urlmonth}\clearfield{urlday}}}
\usepackage[font=small,labelfont=bf]{caption}
\usepackage{subcaption}
\usepackage[explicit]{titlesec} 
\usepackage{babel}
\usepackage{csquotes}
\usepackage{mathtools}
\usepackage{amsthm}
\usepackage{amssymb}
\usepackage{dsfont}
\usepackage{mathrsfs} 
\usepackage[bbgreekl]{mathbbol}
\usepackage{esint}
\usepackage{cancel}
\usepackage{float}
\usepackage{booktabs,array,arydshln}
\usepackage[inline]{enumitem}
\usepackage{xargs}
\usepackage{xcolor}
\definecolor{darkgreen}{rgb}{0,0.5,0}
\definecolor{darkblue}{rgb}{0,0,0.6}
\definecolor{darkorchid}{rgb}{0.6, 0.2, 0.8}
\usepackage[colorlinks=true,citecolor=darkgreen,linkcolor=darkblue,urlcolor=darkorchid,pdfauthor={Bhavya Bandaru},pdftitle={Quantum de Sitter Entropy and Sphere Partition Functions: A-Hypergeometric Approach to Higher Loop Corrections}]{hyperref}
\usepackage{cleveref}
\numberwithin{equation}{section}
\numberwithin{figure}{section}
\numberwithin{table}{section} 
\creflabelformat{figure}{(#2#1#3)}

\theoremstyle{definition}
\newtheorem{exmp}{Example}[section] 
\newtheorem*{excont}{Example \continuation}
\newcommand{\continuation}{??}
\newenvironment{continueexample}[1]
 {\renewcommand{\continuation}{\ref{#1}}\excont[continued]}
 {\endexcont}

\usepackage{pict2e,picture}
\usepackage{scalerel,stackengine}

\usepackage{tikz}
\usepackage[compat=1.1.0]{tikz-feynman}
\usetikzlibrary{tikzmark, shapes.geometric, arrows,fit,calc}
\newcommand{\drawdot}[2][black]{\fill[#1] (#2) circle (0.05cm)}
\def\centerarc[#1](#2)(#3:#4:#5);%
{\draw[#1]([shift=(#3:#5)]#2) arc (#3:#4:#5);}

\def\drawcenterarrow[#1](#2,#3)(#4,#5);%
{\draw[#1,->] (#2,#3) -- (#2/2 + #4/2, #3/2 + #5/2); \draw[#1] (#2/2 + #4/2, #3/2 + #5/2) -- (#4,#5); }

\DeclareMathOperator{\csch}{csch}

\DeclareMathOperator\Vol{Vol} 
\DeclareMathOperator\Real{Re}
\DeclareMathOperator\im{i}
\def\transpose#1{{#1}^{\!{\scriptscriptstyle{\rm T}}}}
\def\geodissq{\mathtt w} \def\geodis{w} \def\xdotY{\sigma}
 \def\bdel{{\bar\Delta}} \def\del{{\Delta}}
\def\CC{{\mathds C}} \def\DD{{\mathbb D}} \def\RR{{\mathds R}} \def\ZZ{{\mathds Z}} \def\NN{{\mathds N}}
\def\mI{{\mathcal I}} \def\mA{{\mathcal A}} \def\mD{{\mathcal D}}
\newcommand{\de}{\,\text{d}} \newcommand{\De}{\,\text{D}}
\newcommand{\prop}{\; \propto \;}
\newcommand{\embed}[1]{\mathbb{#1}}
\def\volR{{\rm Vol} \, \mathds R^*}
\def\Xint#1{\mathchoice
   {\XXint\displaystyle\textstyle{#1}}%
   {\XXint\textstyle\scriptstyle{#1}}%
   {\XXint\scriptstyle\scriptscriptstyle{#1}}%
   {\XXint\scriptscriptstyle\scriptscriptstyle{#1}}%
   \!\int}
\def\XintAdd#1#2#3{\mathchoice
   {\XXint\displaystyle\textstyle{#1}}%
   {\XXint\textstyle\scriptstyle{#1}}%
   {\XXint\scriptstyle\scriptscriptstyle{#1}}%
   {\XXint\scriptscriptstyle\scriptscriptstyle{#1}}%
   \!\int^{#2}_{#3}}
\def\XXint#1#2#3{{\setbox0=\hbox{$#1{#2#3}{\int}$}
     \vcenter{\hbox{$#2#3$}}\kern-.5\wd0}}
\def\dashint{\Xint-}
\def\circint{\ointctrclockwise}

\makeatletter
\newcommand*\bdot{\,\mathpalette\bigcdot@{.6}}
\newcommand*\bigcdot@[2]{\mathbin{\vcenter{\hbox{\scalebox{#2}{$\m@th#1\bullet$}}}}}
\makeatother

\makeatletter
\DeclareRobustCommand{\embednab}{{\mathpalette\em@Nabla\relax}}
\newcommand{\em@Nabla}[2]{%
  \begingroup
  \sbox\z@{$\m@th#1\nabla$}%
  \dimendef\Dht=6 \dimendef\Dwd=8
  \setlength{\Dwd}{\wd\z@}%
  \setlength{\Dht}{\ht\z@}%
  \begin{picture}(\Dwd,\Dht)
  \put(0,0){$\m@th#1\nabla$}
  \put(.6\Dwd,.2\Dht){\line(-1,2){.36\Dht}}
  \end{picture}%
  \endgroup
}
\makeatother

\makeatletter
\DeclareRobustCommand{\embedDelta}{{\mathpalette\em@Delta\relax}}
\newcommand{\em@Delta}[2]{%
  \begingroup
  \sbox\z@{$\m@th#1\Delta$}%
  \dimendef\Dht=6 \dimendef\Dwd=8
  \setlength{\Dwd}{\wd\z@}%
  \setlength{\Dht}{\ht\z@}%
  \begin{picture}(\Dwd,\Dht)
  \put(0,0){$\m@th#1\Delta$}
  \put(0.65\Dwd,.05\Dht){\line(-0.5,1){0.34\Dht}}
  \end{picture}%
  \endgroup
}
\makeatother

\begin{document}
\emergencystretch 3em

\begin{center} 
{\Large  Quantum de Sitter Entropy and Sphere Partition Functions: \\[0.25\baselineskip] $\mathcal A$-Hypergeometric  Approach to Higher Loop Corrections} \\[\baselineskip]
Bhavya Bandaru \\
\it{\footnotesize Department of Physics, Columbia University}\\[2\baselineskip]
{\bf Abstract}
\end{center}
\par In order to find quantum corrections to the de Sitter entropy, a new approach to higher loop Feynman integral computations on the sphere is presented. 
Arbitrary scalar Feynman integrals on a spherical background are brought into the generalized Euler integral ($\mathcal A$-hypergeometric series/GKZ system) form by expressing the massive scalar propagator as a quotient of a bivariate radial Mellin transform of the massless scalar propagator in one higher dimensional Euclidean flat space.
This formulation is expanded to include massive and massless vector fields by construction of similar embedding space propagators. Vector Feynman integrals are shown to be sums over generalized Euler integrals formed of underlying scalar Feynman integrals.
Granting existence of general spin embedding space propagators, the same is shown to be true for general spin Feynman integrals. 
\clearpage

\begin{singlespacing}
  \tableofcontents
  \vfill
\end{singlespacing}

\section{Introduction}\label{ch:Introduction}
\begin{figure}[htbp]
\centering
\includegraphics[width=\textwidth]{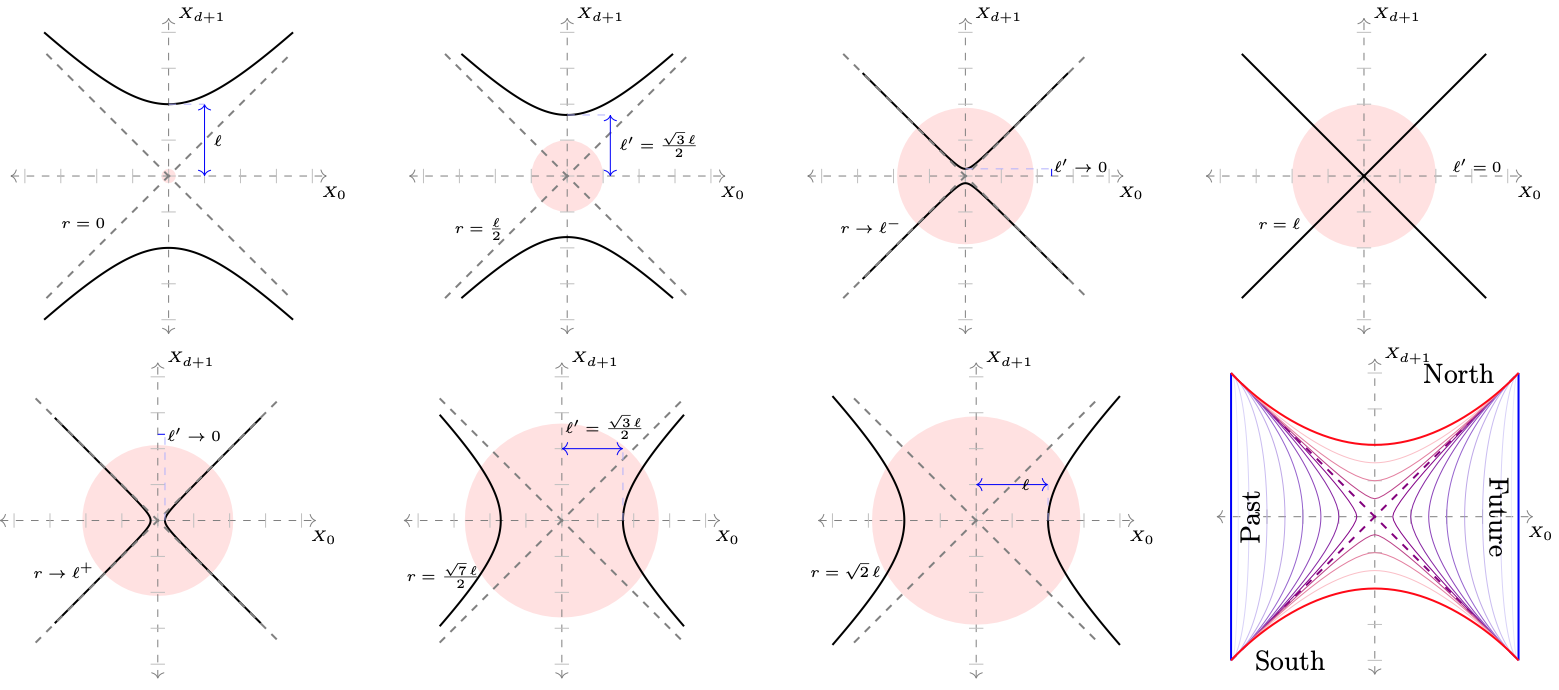}
\caption{North and South Static Patches of De Sitter}\label{fig:dSstaticcoordinates}
\begin{flushleft}\singlespacing \vspace{-\baselineskip} 
The red circles are representative of $r \, \hat{\Omega}_{d-1}$ in \cref{eq:dSstaticcoordinates}. The dashed lines are asymptotes depicting the horizon at $r = \ell$.  
\end{flushleft}
\end{figure}

Observations of distant cosmological objects (light \cite{Bennett_2013} and matter \cite{SupernovaSearchTeam:1998fmf,Schmidt_1998,Perlmutter_2000}) have all but categorically confirmed that this universe is undergoing an accelerated expansion.
A positive cosmological constant $\Lambda > 0$ in the Einstein-Hilbert action implies such an accelerated expansion:
\begin{equation}
\begin{aligned}
  \mathsf{S}_{\rm grav} & = \frac  {1} {8\pi\, \mathsf G}  \,\int \de^{d+1}x \, \sqrt {- \mathsf g} \, \big(\tfrac 1 2 \, \mathsf R - \Lambda\big).
\end{aligned}
\end{equation}
The first order variation of $\mathsf S_{\rm grav}$ with respect to a small metric perturbation results in Einstein's field equations in vacuum: $\delta \mathsf S_{\rm grav} = 0 \implies \mathsf R_\mu{}_\nu{} - \mathsf g_\mu{}_\nu{} \, (\tfrac 1 2 \, \mathsf R - \Lambda) = 0$.
With a constant positive curvature of $ R =  \tfrac{2  \, (d+1)} {(d-1)} \, \Lambda$ and Ricci scalar $R_\mu{}_\nu{}  = \frac {2 }{(d-1)} \, \Lambda\,  g_\mu{}_\nu{}$ that is proportional to the metric tensor $g_\mu{}_\nu{}$, $(d+1)$-dimensional de Sitter space dS${}_{(d+1)}$ is the maximally symmetric vacuum solution to Einstein's equations.
dS${}_{(d+1)}$ can be embedded in Lorentzian flat space of one higher dimension $\RR^{1,(d+1)}$, with a mostly positive metric $\eta = (-1, \, + 1, \cdots + 1)$, as a hyperboloid sheet that gets parameterised in flat coordinates $(X_0, \, \vec X)$ by $- X_0^2 + \vec X^2 = \ell^2$, where $\vec X \in \RR^{(d+1)}$ and $\ell$ is a length scale that is related to the cosmological constant by
\begin{equation}\label{eq:ellandLambdarelation}
\begin{aligned}
  \ell^2 =  \frac 1 {\Lambda} \, \frac {d \, (d-1)}{2}.
\end{aligned}
\end{equation}
A natural parameterisation of the flat coordinates covering the entire dS hyperboloid is $(X_0, \, \vec X) = \ell \, (\sinh t, \, \cosh t \, \hat\Omega_{d})$, where $\hat\Omega_d$ refers to coordinates on the $d$-sphere.
The metric in these coordinates is time dependent: $\de s^2 = \ell^2 \, (- \de t^2 + \cosh^2 t \, \de \hat\Omega_{d}^2)$, where $\de \hat\Omega_d^2$ is the distance measure on $S^d$.
Null, real and imaginary distances between two points imply light-like, space-like and time-like separations respectively.
Representing dS in so-called static coordinates, characterised by a time independent metric (i.e. time translations leave the metric invariant):
\begin{equation}\label{eq:dSstaticcoordinates}
\begin{aligned}
  (X_0 , \, X_1, \cdots X_d, \, X_{d+1}) & = (\sqrt{\ell^2 - r^2} \,  \sinh t , \, r \, \hat\Omega_{d-1}, \, \pm \sqrt{\ell^2 - r^2} \,  \cosh t), && r \le \ell
  \\  \de s^2 & = - \left(1 - \frac {r^2}{\ell^2}\right) \, \ell^2 \, \de t^2 + \frac {1} {(1 - \frac {r^2} {\ell^2})} \, \de r^2 +r^2 \, \de \Omega_{d-1}^2,
\end{aligned}
\end{equation}
splits the space-time into causally disconnected North and South patches, as depicted in \cref{fig:dSstaticcoordinates}, enclosed by an event horizon, a $(d-1)$-sphere of radius $r = \ell$ and area $A_{\rm hor} =  \frac {2 \, \pi^{\frac d 2}} {\Gamma(\frac d 2)} \, \ell^{d-1}$.
Static coordinates, however, do not cover the entirety of dS, requiring the patches labelled Past and Future to instead be parameterised as
\begin{equation}
\begin{aligned}
  (X_0 , \, X_1, \cdots X_d, \, X_{d+1}) & = (\pm \sqrt{r^2 - \ell^2} \,  \cosh t , \, r \, \hat\Omega_{d-1}, \, \sqrt{r^2 - \ell^2} \,  \sinh t), && r > \ell.
\end{aligned}
\end{equation}
\noindent
There is perhaps more to be said about what de Sitter space doesn't have or allow than it does.
Since it isn't asymptotically flat, there is no scattering matrix.
It doesn't have a boundary and so cannot be described in terms of boundary conditions/correlators, unlike Anti-de Sitter space.
There is also no conserved positive energy: Picking a global notion of time on dS makes it so that it flows in the `right' direction, i.e. from the past to future, only in the southern static patch.
So although the time-like Killing vector of the southern static patch's coordinates can be used to define a Hamiltonian, it would not maintain its positivity if extended to the entire space.
Finally, even inserting an observer to simply look at the space isn't as trivial as it may sound because it will influence the metric itself and physical details of the observer, which are totally extrinsic to the properties of dS, become relevant.
Some useful reviews discussing the nature of de Sitter space in detail are \cite{Spradlin:2001pw,Witten:2001kn,Anninos:2012qw,Galante:2023uyf}.
Fortunately, there is at least one unambiguously defined non-trivial calculable quantity in empty de Sitter space, its entropy.
The macroscopic dS entropy $\mathcal S$ is given by the logarithm of the Euclidean path integral \cite{Gibbons:1976ue}
\begin{equation}
\begin{aligned}
  \mathcal S & = \log  \mathcal Z, 
\end{aligned}
\end{equation}
defined on the $(d+1)$-sphere, resulting from Wick rotation of the time-like coordinate $t \rightarrow - \im \, \tau$ (also see \cite{Hartle:1983ai}) of the southern static patch of dS${}_{(d+1)}$:
\begin{equation}\label{eq:SeuclideandS}
\begin{aligned}
\begin{array}{lll}
  \displaystyle\mathsf S_{\rm Lor} = \mathsf S_{\rm grav} + \cdots & \displaystyle\rightarrow & \displaystyle\mathsf S_{\rm Eucl}  = \frac  {1} {8\pi\, \mathsf G}  \,\int \de^{d+1}x \, \sqrt {\mathsf g} \, \big(\Lambda - \tfrac 1 2 \, \mathsf R\big) + \cdots, \\ 
  \displaystyle\int \De\mathsf g \, \cdots \, e^{i \mathsf S_{\rm Lor}[\mathsf g, \, \cdots]} & \displaystyle\rightarrow & \mathcal Z = \displaystyle\int \De\mathsf g \, \cdots \, e^{- \mathsf S_{\rm Eucl}[\mathsf g, \, \cdots]} ,
\end{array} 
\end{aligned}
\end{equation}
where $\cdots$ is used to indicate possible matter fields.
Expansion around the saddle point of the partition function $\mathcal Z$, corresponding to the $(d+1)$-dimensional round sphere of radius $\ell$ as given in \cref{eq:ellandLambdarelation}, leads to the tree-level result \cite{Gibbons:1977mu,Hawking:1978jz}: 
\begin{equation}
\begin{aligned}
   \mathcal S_{\rm tree} &  = \frac  {A_{\rm hor}} {4 \, G}. 
\end{aligned}
\end{equation}
The tree-level entropy $\mathcal S_{\rm tree}$ doesn't provide any additional characterisation of dS, only supplying a dimensionless coupling constant to the low energy effective field theory of gravity \cite{Donoghue:1993eb,Donoghue:2012zc}, far from showcasing the expected non-triviality of $\mathcal S$.

\subsection{Higher loop corrections to de Sitter entropy: {\it why?} }
Before searching for {\it the} microscopic model of quantum gravity in dS, it is important to define what {\it a} microscopic model of quantum gravity in dS is, namely what features is it expected to show, how to recognise it if one does come across it, what is the duck test?
\par 
The dS entropy $\mathcal S$ on its own is just a number and doesn't convey much beyond setting a scale.
If it is to be compared with something, it would have to look within itself.
Less philosophically, non-local quantum corrections to the entropy, i.e. those that cannot be absorbed into local field redefinitions, contain important information regarding the space itself, with each higher order correction translating information regarding the geometry of the space into numbers.
\par
As discussed in detail in \cite{Anninos:2020hfj}, these non-local quantum corrections to $\mathcal S$, represented as a series expansion in terms of $\mathcal S_{\rm tree}$, are the features being sought in a microscopic model.
A simple depiction of this idea can be found at $1$-loop order of pure $3$D ($d=2$) gravity with the horizon fluctuating around $S^{1}$ of radius $\ell$:
\begin{equation}\label{eq:1loop3Dentropy}
\begin{aligned}
  \mathcal S^{(0)} & = \frac {\pi} {4 \, \mathsf G}( 3 \, \ell - \Lambda \, \ell^3) \xrightarrow{\partial_\ell \mathcal S^{(0)} = 0} \frac {2 \pi } {4 \, \mathsf G} \, \ell, \quad \mathcal S^{(1)} & = \underbracket{-\frac {9 \pi} {2 \, \epsilon} \, \ell}_{\text{divergent}}  + \underbracket{5 \,\log (-2\pi\im) - 3 \,\log \left(\frac {2 \pi } {4 \, \mathsf G} \, \ell \right) }_{\text{finite}},
\end{aligned}
\end{equation}
where $\epsilon$ is some UV cutoff and the finite part is notably cutoff independent.
In terms of the Ricci scalar $\mathsf R$, traceless Ricci tensor $\mathsf Q_\mu{}_\nu{} = \mathsf  R_\mu{}_\nu{} - \frac 1 {d+1} \, \mathsf  R \, \mathsf g_\mu{}_\nu{}$ and Weyl tensor $\mathsf W_\mu{}_\nu{}_\rho{}_\sigma$, $ \{\mathsf R, \, \mathsf Q, \, \mathsf W\} \prop \ell^{-2}$, the most general diffeomorphism invariant form of a shift to the action $\mathsf S_{\rm Eucl}$ in \cref{eq:SeuclideandS} is
\begin{equation}\label{eq:generalcountertermandmoredactionS3}
\begin{aligned}
  \mathsf S_{\rm c} & = \int \sqrt{\mathsf g} \, \left( c_{\Lambda} \, \Lambda  - c_{\mathsf R} \, \mathsf R - \ell_{c}^2 \, \left( c_{\mathsf R^2} \, \mathsf R^2 + c_{\mathsf Q^2} \,\mathsf  Q^2 + c_{\mathsf W^2} \, \mathsf W^2\right) + \mathcal O(\ell^{-6}) \right),
\end{aligned}
\end{equation}
where $\ell_c$ is a length scale and $c$ are dimensionless constants. 
On the round sphere, both $\mathsf Q$ and $\mathsf W$ vanish, and $\mathsf R\Big|_{S^{d+1}} = \frac {d \, (d+1)} {\ell^2}$. 
$\mathsf S_{\rm c}$ serves to parameterise all possible curvature corrections, counter-terms, and local metric field redefinitions of the form $\mathsf  g_\mu{}_\nu{} \rightarrow \mathsf g_\mu{}_\nu{} + \ell_{c}^2 \, (c_0 \, \Lambda \, \mathsf  g_\mu{}_\nu{} + c_1 \, \mathsf R \, \mathsf  g_\mu{}_\nu{} + c_2 \, \mathsf Q_\mu{}_\nu{})$.
The contribution of $\mathsf S_{c}$ at $1$-loop order to the entropy is
\begin{equation}\label{eq:generalcountertermandmoredentropyat1loopS3}
 \begin{aligned}
  \mathcal S_{c} & = - 2\pi^2  \, c_{\Lambda} \, \Lambda \, \ell^{3}  + 12\pi^2 \, c_{\mathsf R} \, \ell + 2\pi^2 \,\sum_{n=2}^{\infty} c_{\mathsf R^{n}}  \, 6^n \, \ell^{2n-2}_{c} \, \ell^{3-2n},
\end{aligned}
\end{equation}
with terms $\propto \, \{\ell^3, \, \ell, \, \ell^{-1}, \, \ell^{-3} \cdots\}$ i.e. only odd powers of $\ell$. 
Setting $c_{\Lambda} = 0, \; c_{\mathsf R} =  \frac {3} {8 \pi \, \epsilon}$ ensures that the renormalized values of $\Lambda, \, G$ remain the same as at tree-level.
The finite terms of $\mathcal S^{(1)}$ in \cref{eq:1loop3Dentropy} remain unchanged by local operations and constitute invariant data of the quantum theory.
An accurate microscopic model of pure 3D gravity is expected to replicate these terms. 
\par 
This idea applies to theories with matter content too. For example, inclusion of a massive scalar field with some curvature coupling to the above setup:
\begin{equation}
 \begin{aligned}
  \mathsf S_{\phi} & = \frac 1 2 \,\int \sqrt{\mathsf g} \,  \phi \, \left( - \nabla^2 + m^2 + \frac {1-\eta} {6} \, \mathsf R \right) \, \phi, \quad \nu^2 \equiv m^2 \, \ell^2 -  \eta  
 \end{aligned}
\end{equation} 
supplies an additional contribution to the entropy at $1$-loop order equalling:
\begin{equation}\label{eq:1loopcorrectionforscalaronS3}
\begin{aligned}
  \mathcal S^{(1)}_{\phi} & = \underbracket{\frac {\pi} {2 \,\epsilon^3} \,\ell^3 - \frac {\pi} {4 \,\epsilon}  \,\nu^2\,\ell}_{\text{divergent}}  + \underbracket{{\frac {\pi }{6}\,\nu^3 + \frac {\nu^2 \,\log(1-e^{-2 \pi\,\nu})} {2} - \frac {\nu \,\text{Li}_{2}(e^{-2 \pi\,\nu})} {2 \pi} - \frac {\text{Li}_{3}(e^{-2 \pi\,\nu})} {(2 \pi)^2}}}_{\text{finite}}, 
\end{aligned}
\end{equation}
where $\text{Li}_n$ are polylogarithms. 
The renormalisation condition $\lim\limits_{\ell \rightarrow \infty} \partial_\ell \mathcal S^{(1)} = 0$ sets the counter-terms in \cref{eq:generalcountertermandmoredactionS3} to
\begin{equation}
\begin{aligned}
  \mathsf S_{c} & \rightarrow \int \sqrt{\mathsf g} \, \left( \frac {1} {4 \pi \,\epsilon^3}  - \frac {m^2} {8  \pi\,\epsilon}  + \frac {m^3} {12 \pi} - \left(\frac {3} {8  \pi\, \epsilon} - \frac {\eta} {48  \pi\,\epsilon} + \frac {m \, \eta} {48 \pi}\right) \, \mathsf R\right) + \cdots.
\end{aligned}
\end{equation}
Thus, entropy at $1$-loop order $\mathcal S^{(1)}$ represented as an expansion in orders of $\ell$:
\begin{equation}
\begin{aligned}
  \mathcal S^{(1)} & = \underbracket{5 \,\log (-2\pi\im) - 3 \,\log \left(\frac {2 \pi } {4 \, \mathsf G} \, \ell \right)}_{\text{gravity : \cref{eq:1loop3Dentropy}}} + \underbracket{\frac {\pi \, \eta }{4} \, m \, \ell - \frac {\pi}{6} \, m^3 \, \ell^3  + \cdots}_{\ell^{\text{odd}}}
  \\  {\scriptstyle \ell^{\text{even}}} & \left[\begin{aligned}
& + \frac {\pi}{4} \, \sqrt\eta \,\cot(\pi \, \sqrt\eta ) \, m^2 \, \ell^2 + \frac {\pi^2}{16}\, \left( \csc^2(\pi\,\sqrt{\eta}) - \frac {\cot(\pi\,\sqrt{\eta})}{\pi \,\sqrt{\eta}} \right) \, m^4 \, \ell^4 \\ 
& - \frac {\text{Li}_3(e^{- 2  \pi \im\, \sqrt{\eta}})} {4 \pi^2} - \frac {\im \sqrt{\eta} \, \text{Li}_2(e^{- 2  \pi \im\, \sqrt{\eta}})}{2\pi} - \frac {\eta \, \log(1 - e^{- 2  \pi \im\, \sqrt{\eta}})}{2} + \frac {\im \, \eta^{\frac 3 2}}{6} + \cdots
\end{aligned} \right.
\end{aligned}
\end{equation}
has some counter-term dependence appearing as terms in odd powers of $\ell$ but the terms in even powers of $\ell$ are invariant non-local quantum corrections.
\par
As a major step towards generating such model constraining data, character integral and closed form expressions of quantum corrections at $1$-loop order for arbitrary field content on spherical and (A)dS backgrounds are presented in \cite{Anninos:2020hfj}. 
Their group theoretic approach is, however, not generalisable to account for non-trivial interactions. 
It is at this stage that developing a viable approach to higher loop Feynman integral computations becomes necessary. 
Some works advancing the search for quantum corrections to dS entropy with this aim are \cite{Anninos:2021ene,Muhlmann:2021clm,Anninos:2022hqo,Muhlmann:2022duj,Anninos:2023exn}. 

\subsection{Higher loop corrections to de Sitter entropy: {\it how?} } \label{sec:Higherloopcorrectionshow}
A new formulation of Feynman integrals on the Euclideanisation of $(d+1)$-dimensional de Sitter, i.e. the $(d+1)$-sphere, that is well suited for analytic and numerical integration, is presented in this work.
The propagator of a massive scalar field on $S^{d+1}$ (of radius $\ell \equiv 1$) is \cite{Candelas:1975,Dowker:1976,Bunch:1978yq}
\begin{equation}
\begin{aligned}
  G(\hat X, \hat Y) = \frac {\Gamma(\Delta) \,\Gamma(d - \Delta)} {(4 \pi)^{\frac{d+1}{2}} } \, {}_2F_1(\Delta, \, d - \Delta; \, \tfrac {d+1} {2}; \, \tfrac {1 + \cos \theta} {2}), \quad \cos \theta = \hat X \cdot \hat Y,
\end{aligned}
\end{equation}
where the mass $m$ and mass parameter $\Delta$ are related by $ m^2 = \Delta\, (d - \Delta) = \frac {d^2}{4} + \nu^2$.
Using this propagator, the integral describing the so-called $3$-melon Feynman diagram:
\vspace{-0.5\baselineskip}
\begin{figure}[H]
  \centering
\begin{tikzpicture}
\draw (0,0) circle (1);
\draw (-1,0) -- (1,0);
\drawdot{-1,0}; 
\drawdot{1,0}; 
\node at (1,0)[anchor=west]{$\hat X$};
\node at (-1,0)[anchor=east]{$\hat Y$};
\node at (0,1)[anchor=south]{\small$P_1$}; 
\node at (0.6,0.7) [anchor=south west]{\small$\Delta$}; 
\node at (0.6,-0.7) [anchor=north west]{\small$\Delta$}; 
\node at (0.5,0.1) [anchor=north]{\small$\Delta$}; 
\node at (0,-0.1)[anchor=south]{\small$P_2$}; 
\node at (0,-1)[anchor=north]{\small$P_3$}; 
\node at (-2,0)[anchor=east]{$\mathcal I^{[\Delta]}_{2,3} = $};
\end{tikzpicture}
\caption{$3$-Melon Feynman Diagram with mass parameter $\Delta$}\label{fig:intro3-MelonFeynmanDiagram}
\end{figure}
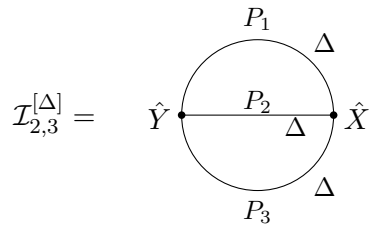\vspace{-\baselineskip}\noindent
consisting of $3$ massive scalar propagators of the same mass parameter $\Delta$ connecting $2$ internal vertices $(\hat X, \, \hat Y)$ is
\begin{equation}
\begin{aligned}
  \mathcal I^{[\Delta]}_{2,3} & = \int \de \Omega_{\hat X} \, \de \Omega_{\hat Y} \, \big(G(\hat X \cdot \hat Y) \big)^3
  \\ & = \Omega_{d+1} \, \Omega_{d} \, \Big(\frac {\Gamma(\Delta) \,\Gamma(d - \Delta)} {(4 \pi)^{\frac{d+1}{2}} }\Big)^3 \int_{0}^{\pi} \de \theta \, \sin^{d} \theta \, \Big({}_2F_1(\Delta, \, d - \Delta; \, \tfrac {d+1} {2}; \, \tfrac {1 + \cos \theta} {2}) \Big)^3.
\end{aligned}
\end{equation}
Since hypergeometric functions aren't particularly amenable to analytical integration, apart from some special cases of $\Delta$ or $d$ in which ${}_2F_1$ reduces to simpler functions, there is no obvious way to simplify this expression further.
\par
In flat space, Feynman integrals featuring massive scalar propagators (when written in terms of Bessel K functions $ \prop (\frac {m}{|X - Y|})^{\frac{d-1}{2}} \, K_{\frac{d-1}{2}}(m \, |X - Y|)$) may look similarly daunting in position space but they are quickly remedied by using the momentum space representation of the massive scalar propagator: $\int \frac {\de^{d+1}P} {(2 \pi)^{d+1}}\frac {e^{- \im \, P \, (X - Y)}}{(P^2 + m^2)}$, which allows better analytic treatment of Feynman integrals, especially when considering them in general dimensions.
\par
Unlike flat space, there is no global momentum space representation of massive scalar propagators on the sphere. 
Shifting to angular momentum space, there is an eigenfunction expansion of this propagator:
\begin{equation}
\begin{aligned}
  G( \hat X, \,  \hat Y) =  \sum_{Y \in \Upsilon} \frac {1} {L\, (L + d) + \Delta\, (d - \Delta)} \, Y_{L,m}^*( \hat X) \, Y_{L,m}( \hat Y), 
\end{aligned}
\end{equation}
where $Y_{L,m}$ are $(d+1)$-dimensional scalar spherical harmonics labelled by $(L,m)$, $L\, (L + d)$ is their eigenvalue, and the sum is over their entire orthonormal basis $\Upsilon$ \cite{Camporesi:1990wm}. 
However, using such an expansion to rewrite the $3$-melon integral results in sums over coefficients stemming from integrals over multiple spherical harmonics, i.e. Wigner $3$-j and general SO$(d+2)$ $3$-j symbols, and, if considering Feynman diagrams with more propagators, higher numbered-$j$ symbols. 
Not only are such sums `harder' than momentum space integrals, they aren't suited for dimensional regularisation.
\par 
Further, the integral measure itself grows more complicated with an increasing number of internal points in the Feynman diagram.
Though, this can be somewhat sidestepped by clever reparameterisations like shifting to stereographic coordinates, the simplicity of the Gaussian integrals over internal points of their flat space counterparts remains unmatched. 
\par 
Completely avoiding the aforementioned problems, the main result is a set of rules (\cref{ch:dSFeynmanIntegrals}) that allow Feynman integrals on the sphere to be read off from the Feynman diagram itself, directly resulting in an integral form which has an algorithmic solution.
Applying these rules to, for example, the $3$-melon diagram in \cref{fig:intro3-MelonFeynmanDiagram}, its ``incidence matrix'' is
\begin{equation}
\begin{aligned}
  L_{2,3} & =   \left(\begin{array}{c|c c}
               & \hat X  & \hat Y \\ 
              \hline
              P_1 & \lambda_1  & - \mu_1 \\ 
               P_2 & \lambda_2  & - \mu_2 \\ 
               P_3 &  \lambda_3  & - \mu_3 \\ 
            \end{array} \right),  
\end{aligned}
\end{equation}
where each pair of parameters $\{\lambda, \, \mu\}$ are associated with a propagator $P$ connecting $\hat X$ and $\hat Y$.
Then the Feynman integral is given by
\begin{equation}
\begin{aligned}
  \mathcal I^{[\Delta]}_{2,3} & \prop \int_0^\infty \frac {\de \lambda_1} {\lambda_1} \,\frac {\de \lambda_2} {\lambda_2} \,\frac {\de \lambda_3} {\lambda_3} \,\frac {\de \mu_1} {\mu_1} \, \frac {\de \mu_2} {\mu_2} \, \frac {\de \mu_3} {\mu_3} \, \frac {(\lambda_1 \, \lambda_2 \, \lambda_3)^{d - \Delta} \, (\mu_1 \, \mu_2 \, \mu_3)^{\Delta}} {\big(f_{2,3}\big)^{\frac {d+2}{2}}},
\end{aligned}
\end{equation}
where $f_{2,3}$ is a polynomial in $(\lambda, \, \mu)$ defined as
\begin{equation}\label{eq:3melonf23inintro}
\begin{aligned}
  f_{2,3} \vcentcolon\!&= \det (\mathds 1_2 + \transpose{L}_{2,3} \, L_{2,3} )  = \det     \left( \begin{smallmatrix}
        1 + \lambda_1^2 + \lambda_2^2 + \lambda_3^2   & - (\lambda_1 \, \mu_1 + \lambda_2 \, \mu_2 + \lambda_3 \, \mu_3) \\ - (\lambda_1 \, \mu_1 + \lambda_2 \, \mu_2 + \lambda_3 \, \mu_3) & 1 + \mu_1^2 + \mu_2^2 + \mu_3^2 
      \end{smallmatrix} \right)
      \\ & = 1 + \sum_{i=1}^{3} (\lambda_i^2 + \mu_i^2) + (\lambda_1 \, \mu_2 - \lambda_2 \, \mu_1)^2 + (\lambda_2 \, \mu_3 - \lambda_3 \, \mu_2)^2 + (\lambda_3 \, \mu_1 - \lambda_1 \, \mu_3)^2.
\end{aligned}
\end{equation}
Following the same pattern, the $n$-melon Feynman integral is
\begin{equation}
\begin{aligned}
  \mathcal I^{[\Delta]}_{2,n} & \prop \int_0^\infty \; \frac 1 {(f_{2,n})^{\frac {d+2}{2}}} \; \prod_{i=1}^{n} \frac {\de \lambda_i} {\lambda_i} \,\frac {\de \mu_i} {\mu_i} \, \lambda_i^{d-\Delta} \, \mu_i^{\Delta} \; ,
  \\ f_{2,n} & = 1 + \sum_{i=1}^{n} (\lambda_i^2 + \mu_i^2) + \tfrac 1 2 \sum_{i,j = 1}^{n} (\lambda_i \, \mu_j - \lambda_j \, \mu_i)^2.
\end{aligned}
\end{equation}
In a similar vein, arbitrary scalar Feynman diagrams can be directly converted into parametric integrals taking the form:
\begin{equation}
\begin{aligned}
  \mathcal I_{F} & \prop \int_{0}^{\infty} \; \Big(\det (\mathds 1 + \transpose{L} \, L)\Big)^{-\frac {d+2}{2}} \; \prod_{i = 1}^{n_P} \frac {\de \lambda_i} {\lambda_i} \,\frac {\de \mu_i} {\mu_i} \, \lambda_i^{d - \Delta_i} \, \mu_i^{\Delta_i} , 
\end{aligned}
\end{equation}
where the incidence matrix, $L$, can be read off from the Feynman diagram. 
These integrands are symmetric under the exchange $(\lambda \leftrightarrow \mu)$. 
Some examples of incidence matrices are given in \cref{tb:AdjacencymatricesofsomeFeynmanDiagrams}, where the mass parameters $\Delta$ label the momenta $P$.
\begin{table}[htbp]
\caption{Incidence matrices of some Feynman Diagrams}\label{tb:AdjacencymatricesofsomeFeynmanDiagrams}
\begin{center}
\begin{tabular}{| c | c |}
\hline 
Feynman Diagram & Incidence matrix $L$\\
\hline
\rule{0pt}{2em} $p$-melon : \raisebox{-2em}{\begin{tikzpicture}
\draw (0,0) circle (0.5);
\drawdot{0.5, 0};
\drawdot{-0.5, 0};
\draw [black,thick,dotted] plot [smooth, tension=1.5] coordinates{ (-0.5, 0) (0,0.3) (0.5, 0)};
\draw [black,thick,dotted] (-0.5, 0)  --  (0.5, 0);
\draw [black,thick,dotted] plot [smooth, tension=1.5] coordinates{ (-0.5, 0) (0,-0.3) (0.5, 0)};
\node at (-0.5,0)[anchor=east]{\tiny$X_1$};
\node at (0.5,0)[anchor=west]{\tiny$X_2$};
\node at (0,0.5)[anchor=south]{\tiny$\Delta_1$}; 
\node at (0,-0.5)[anchor=north]{\tiny$\Delta_p$};
\end{tikzpicture}}  & $   \left( \begin{smallmatrix}
      \lambda_1 & -\mu_1 \\
      \lambda_2 & -\mu_2\\
       \cdots & \cdots \\
       \lambda_p & -\mu_p \\
    \end{smallmatrix} \right)$ \\[2ex] 
  \hline
\rule{0pt}{3em}pacman : \raisebox{-2em}{\begin{tikzpicture}
\draw (0,0) circle (0.5);
\drawdot{0.3, 0.4};
\drawdot{0.3, -0.4};
\drawdot{-0.5, 0};
\draw (-0.5, 0) -- (0.3, 0.4);
\draw (-0.5, 0) -- (0.3, -0.4);
\node at (0,0.5)[anchor=south]{\tiny$\Delta_1$};
\node at (0.2,0.4)[anchor=north]{\tiny$\Delta_2$};
\node at (0.2,-0.4)[anchor=south]{\tiny$\Delta_3$};
\node at (0,-0.5)[anchor=north]{\tiny$\Delta_4$};
\node at (0.5,0)[anchor=west]{\tiny$\Delta_5$};
\node at (-0.5,0)[anchor=east]{\tiny$X_1$};
\node at (0.3, 0.4)[anchor=west]{\tiny$X_2$};
\node at (0.3, -0.4)[anchor=west]{\tiny$X_3$};
\end{tikzpicture}} & $    \left( \begin{smallmatrix}
       \lambda_1 & - \mu_1 &0 \\
       \lambda_2 & - \mu_2 & 0 \\
       \lambda_3 & 0 &  - \mu_3 \\
       \lambda_4 & 0 &  - \mu_4 \\
       0 & \lambda_5 & - \mu_5 \\
    \end{smallmatrix} \right)$ \\[4ex] 
    \hline
\rule{0pt}{3em}pillbox : \raisebox{-2em}{\begin{tikzpicture}
\draw (0,0) -- (0.75,0) -- (0.75,1) -- (0,1) -- cycle;
\drawdot{0, 0};\node at (0,0)[anchor=north]{\tiny$X_1$};
\node at (0.4,0.1)[anchor=north]{\tiny$\Delta_1$};
\drawdot{0.75, 0};\node at (0.75, 0)[anchor=north]{\tiny$X_2$};
\node at (0.65, 0.5)[anchor=west]{\tiny$\Delta_2$};
\node at (1.1, 0.5)[anchor=west]{\tiny$\Delta_6$};
\drawdot{0.75,1};\node at (0.75,1)[anchor=south]{\tiny$X_3$};
\node at (0.4,0.9)[anchor=south]{\tiny$\Delta_3$};
\drawdot{0, 1};\node at (0, 1)[anchor=south]{\tiny$X_4$};
\node at (0.1, 0.5)[anchor=east]{\tiny$\Delta_4$};
\node at (- 0.4, 0.5)[anchor=east]{\tiny$\Delta_5$};
\centerarc[](0.75,0.5)(-90:90:0.5cm);
\centerarc[](0,0.5)(90:270:0.5cm);
\end{tikzpicture}} & $    \left( \begin{smallmatrix}
    \lambda_1 & -\mu_1 & 0 & 0 \\ 
    0 & \lambda_2 & - \mu_2 & 0 \\
    0 & 0 & \lambda_3 & - \mu_3\\
    - \mu_4 & 0 & 0 & \lambda_4\\
    \lambda_5 & 0 & 0 & - \mu_5\\
    0 & - \mu_6 & \lambda_6 & 0 \\
    \end{smallmatrix} \right)$\\[4ex] 
    \hline
\rule{0pt}{3em}peace : \raisebox{-2em}{\begin{tikzpicture}
\draw (0,0) circle (0.5);
\draw (0.4,0.3) -- (0,0) -- (0,-0.5);
\draw (-0.4,0.3) -- (0,0);
\node at (0,0.5)[anchor=south]{\tiny$\Delta_4$};
\node at (-0.3,-0.3)[anchor=north east]{\tiny$\Delta_5$};
\node at (0.3,-0.3)[anchor=north west]{\tiny$\Delta_6$};
\drawdot{0, 0};\node at (0.1,0.1)[anchor=north east]{\tiny$X_0$};
\drawdot{0.4, 0.3};\node at (0.4, 0.3)[anchor=west]{\tiny$X_1$};
\node at (0.2, 0.05)[anchor=south]{\tiny$\Delta_1$};
\node at (-0.2, 0.05)[anchor=south]{\tiny$\Delta_2$};
\drawdot{-0.4, 0.3};\node at (-0.4, 0.3)[anchor=east]{\tiny$X_2$};
\drawdot{0,-0.5};\node at (0,-0.5)[anchor=north]{\tiny$X_3$};
\node at (-0.1,-0.2)[anchor=west]{\tiny$\Delta_3$};
\end{tikzpicture}} & $    \left( \begin{smallmatrix}
      \lambda_1 & - \mu_1 & 0 & 0 \\
      \lambda_2 & 0& - \mu_2 & 0 \\
      \lambda_3 & 0 & 0 & - \mu_3 \\ 
      0 & \lambda_4 & - \mu_4 & 0 \\ 
      0 & 0& \lambda_5    & - \mu_5 \\ 
      0 & - \mu_6& 0 & \lambda_6 \\
    \end{smallmatrix} \right)$\\[4ex] 
    \hline
\end{tabular}
\end{center}
\end{table}\noindent
It is also possible to include external legs in Feynman diagrams with minimal modifications to this formulation (\cref{sec:GeneralisedCorrelationFunctions}).
The basic stucture of the Feynman integral remains the same with additional polynomials (usually $1$ but possibly more for reducible diagrams) in $(\lambda, \, \mu)$ appearing in the denominator of the integrand. 
\par
This simplification to sphere scalar Feynman integral representations hinges upon the use of a ``momentum-space''-like representation of the massive scalar propagator on the sphere (\cref{sec:scalars}):
\begin{equation}
\begin{aligned}
  G(\hat X, \, \hat Y) & \prop \int_{0}^{\infty} \frac {\de \mu} {\mu} \, \mu^{\Delta}  \int \frac {\de^{d+2}P}{(2 \pi)^{d+2}}\frac {e^{- \im \, P\,(\hat X - \mu \,\hat Y)}} {P^2},
\end{aligned}
\end{equation}
where the momentum\footnote{There is no direction to this momentum, however, some direction may be appropriated to it for notational convenience. There is also no momentum conservation at the vertices.} $P$ is integrated over Euclidean {\it flat} space $\RR^{d+2}$ that serves as the embedding space for $S^{d+1}$ and $\hat X, \, \hat Y$ are embedding space coordinates of points on $S^{d+1}$. 
This formulation can be extended to higher spin Feynman integrals too (\cref{sec:HigherSpinFeynmanIntegralsSetup}), by using higher spin propagator expressions similar to the ``momentum-space''-like scalar propagator expression. 
To this end, massive and massless vector propagator expressions compatible with this formulation have been found (\cref{sec:VectorFields}). 
\par
As may have been noticed, there is a consistent pattern to these Feynman integral representations. 
Namely, they all appear to be multivariate generalisations of Mellin transformations, the univariate version of which, for some function $f(s)$, is 
\begin{equation}
\begin{aligned}
  (\mathcal M \circ f)(\Delta) & = \int_0^{\infty} \frac {\de s}{s} \, s^{\Delta} \, f(s).
\end{aligned}
\end{equation}
In the case at hand, the analogue of the function $f$ always takes the form of a polynomial raised to some arbitrary exponent. These types of integrals are a subset of so-called generalized Euler integrals. 
More rigorously, the structure of a generalized Euler integral always follows the pattern:
\begin{equation}
  \mathcal F_{[\alpha, \, \beta]}(z)  = \int_{x \in \sigma} \, \frac {\de x_1} {x_1} \cdots \frac {\de x_n} {x_n} \, \frac {x_1^{\beta_1} \cdots x_n^{\beta_n}} {(z_1 \, m_1 + z_2 \, m_2 + \cdots + z_N \, m_N)^{\alpha}}, \quad m_i = x_1^{s_{1,i}} \cdots  x_n^{s_{n,i}},
\end{equation}
where $m_i$ are monomials in the variables $\{x_1, \, \cdots x_n\}$ with non-negative integer exponents $s_i\in \ZZ_{\ge 0}$, the integration contour of $x$ avoids the singularities of the integrand, $\{\alpha, \, \beta\}$ are generic parameters, and the integral $\mathcal F_{[\alpha, \, \beta]}$ is a function of the coefficients $z$ in the polynomial. Interpreting the sphere Feynman integrals to be in this form, only certain special values of $\alpha, \, \beta, \, z$ are actually physically relevant, making this generalisation appear excessive. 
\par
However, there is a huge benefit to it. 
Apart from a natural consequence of this description being manifest dimensional regularisation, these integrals show many scaling symmetries, by virtue of which they satisfy certain PDEs known as Gel'fand-Kapranov-Zelevinsky (GKZ) systems, \cite{GELFAND1990255}. 
If and only if these integrals are taken to be in their completely generalized form, the solution space to the aforementioned PDEs exactly equals the integrals (i.e. if the full generalisation isn't considered, the solutions to the PDEs will be a superset of the functions describing the integral). 
The solutions to these systems of equations can be found algorithmically and take the form of multivariate hypergeometric series, called \emph{$\mathcal A$-hypergeometric functions}, \cite{GELFAND1990255}. 
Restricting these series solutions to the relevant physical parameters solves the Feynman integral. 

\subsection{Overview}
The main results of \cref{sec:scalars,sec:VectorFields} are the propagators on $S^{d+1}$, \cref{eq:scalarpropunfixed,eq:vectorprop,eq:masslessvectorpropagatormainformula}, expressed as quotients of momentum space representations of the Euclidean flat space massless propagators in one higher dimensional embedding space $\RR^{d+2}$. 
The construction of massive scalar propagators, starting from simplistic observation $\RR^{d+2} / \RR_+ = S^{d+1}$, is discussed in \cref{sec:scalars}. 
In a similar vein but with more care devoted to the details of the embedding space representation of the Mellin transformed basis of fields, described in \cref{app:CoordinateSystems,sec:MellinTransformedBasisofFields}, the vector propagators are built in \cref{sec:VectorFields} by gauge fixing the higher dimensional massless vector field in different ways (\cref{sec:EmbeddingSpaceRepresentationof(GaugeFixed)MasslessVectorAction}) to get to the massive (\cref{sec:MassiveVectorPropagator}) and massless (\cref{sec:MasslessVectorPropagator}) cases. 
\par
\Cref{ch:dSFeynmanIntegrals} is devoted to the construction of sphere Feynman integrals in embedding space. 
\Cref{sec:EmbeddingSpaceFormulationofFeynmanIntegralsonSd+1Scalar} contains the set of rules to convert any scalar Feynman diagram into its generalized Euler integral form. 
In \cref{sec:HigherSpinFeynmanIntegralsSetup}, it is shown that higher spin integrals can also be turned into generalized Euler integrals, as long as the higher spin propagators featured in the Feynman diagrams (both internal and external) can be represented as parameter derivatives of the scalar propagator in embedding space i.e. \cref{eq:operatorwithscalarprop}.  
The construction of a ``master'' generalized Euler integral form of any Feynman diagram is described. 
Based on observations made in \cref{eq:ModeIntVerificationofScalarPropagatorinEmbeddingSpaceFormalism,eq:ModeIntVerificationofVectorPropagatorinEmbeddingSpaceFormalism} and some experimental explorations into higher spin propagator constructions,\footnote{It is trivial to write a multitude of embedding space forms of higher spin propagators but the gauge conditions being portrayed would be obscure. In the language of \cite{Allen:1986tt}, only the transverse part of the propagator can be established to be correct. For actual computational purposes, such propagators would be incomplete.} it is hypothesized that propagators of arbitrary spin can be represented in this embedding space form. 
Granting this hypothesis, the master integral form encodes information of all possible Feynman diagrams with the same graphical structure/incidence matrix but different spin propagators. 
As described in \cref{sec:GeneralisedCorrelationFunctions}, all (scalar) correlation functions can be laconically written as \cref{eq:generalizedcorrelationfunctionaseulerintegral}, with the same caveats regarding extentions to higher spin applying. 
\par
Generalized Euler integral constructions of scalar Feynman integrals upto $3$-loops are given in \cref{sec:ExplicitConstructionofScalarFeynmanIntegrals}, with some explicitly worked out results, e.g.
\begin{equation}
\begin{aligned}
\begin{tikzpicture}[baseline=0ex]
 \drawdot{-0.5,0}; \drawdot{0.3,0.4}; \drawdot{0.3,-0.4}; \node at (0,-0.5)[anchor=north]{\tiny$\Delta_2$}; \node at (0,0.5)[anchor=south]{\tiny$\Delta_1$};
 \node at (0.5,0)[anchor=west]{\tiny$\Delta_3$};
 \node at (-0.5,0)[anchor=east]{\tiny$X_2$};
 \node at (0.3,0.4)[anchor=west]{\tiny$X_1$};
 \node at (0.3,-0.4)[anchor=west]{\tiny$X_{3}$};
 \draw (0,0) circle (0.5);
 \end{tikzpicture} & = \tfrac {\Gamma(1-d)}{d \, \pi} \Big( \tfrac {(\sin (\pi \, \Delta_1) + \sin (\pi \, \bar\Delta_1))\, \Gamma(\Delta_1) \,\Gamma(\bar \Delta_1) } {(\nu_2^2 - \nu_1^2)(\nu_3^2 - \nu_1^2)} + \tfrac {(\sin (\pi \, \Delta_2) + \sin (\pi \, \bar\Delta_2))\, \Gamma(\Delta_2) \,\Gamma(\bar \Delta_2) } {(\nu_1^2 - \nu_2^2)(\nu_3^2 - \nu_2^2)} 
 \\ & \phantom{= \tfrac {\Gamma(1-d)}{d \, \pi} \Big(}+ \tfrac {(\sin (\pi \, \Delta_3) + \sin (\pi \, \bar\Delta_3))\, \Gamma(\Delta_3) \,\Gamma(\bar \Delta_3) } {(\nu_1^2 - \nu_3^2)(\nu_2^2 - \nu_3^2)} \Big).
\end{aligned}
\end{equation}
Representative examples are used to illustrate the relation of vector Feynman integrals to underlying scalar Feynman integrals in \cref{sec:vectorFeynmanIntegrals}. 
\Cref{ch:discussion} highlights the next few concrete steps and some unaswered questions.
\par
Though possibly superfluous to some readers, details of the coordinate conventions used to describe the sphere $S^{d+1}$ and its embedding space $\RR^\DD$ have been noted in \cref{app:CoordinateSystems}. 
Further, \cref{sec:MellinTransformedBasisofFields} is crucial cross-reference for details on the Mellin transformation used to convert (scale invariant) fields in $\RR^{d+2}$ to fields on $S^{d+1}$ and back, in keeping with these coordinate conventions. 
\Cref{ch:Eulerintegrals} provides a cursory review of the theory of generalized Euler integrals, GKZ systems of partial differential equations that annihilate them, and their solutions, $\mathcal A$-hypergeometric functions. 
The GKZ ideal is motivated and introduced in \cref{sec:AHypergeometricSystem}. 
A basic explanation of ideals in Weyl algebras, specifically holonomic ideals (which GKZ ideals are), their general properties and approaches to building their solutions is given in \cref{sec:IdealstoSolutions}. 
The focus then shifts to the properties and constructions of $\mathcal A$-hypergeometric solutions to GKZ systems in \cref{sec:SolutionsofAHypergeometricSystems}.

\vfill

\section{Scalar propagators on the sphere}\label{sec:scalars}
Feynman diagrams associated with scalar fields on Minkowski space $\RR^{(\DD-1,1)}$ or Euclidean space $\RR^{\DD}$ have integral representations, related by Wick rotation, that can be brought into the form of generalized Euler integrals, hence having $\mathcal A$-hypergeometric series representations. 
A Feynman integral consisting of $n_P$ scalar propagators and $n_l$ loops takes the form:
\begin{equation}
\begin{aligned}
  \mathcal F^{n_l}_{n_{P}} = \int^{\RR^{(\DD-1,1)}}\limits_{P_{\rm int}} \, \prod_{i = 1}^{n_{P}} \, \frac 1 {(P_i^2 - m_i^2)^{\eta_i}} \quad \xrightarrow{P^{\DD} \rightarrow - \im P^{\DD}} \quad \int^{\RR^{\DD}}\limits_{P_{\rm int}} \, \prod_{i = 1}^{n_{P}} \, \frac 1 {(P_i^2 + m_i^2)^{\eta_i}},
\end{aligned}
\end{equation} 
where the product is over all momentum space propagators, the integral is over internal undetermined momenta $P_{\rm int}$,  and the parameter $\eta$ allows further generalisation. 
Using Schwinger parameterisation, with limits over the sum and product suppressed for notational brevity: 
\begin{equation}\label{eq:flatfeynmanintschwinger}
\begin{aligned}
  \mathcal F^{n_l}_{n_{P}} & = \int^{\RR^{\DD}}\limits_{P_{\rm int}} \, \int^{*}_{z} \, \prod \, \frac {z_i^{\eta_i}} {\Gamma(\eta_i)} \, e^{- \sum z_i\, (P_i^2 + m_i^2)},
\end{aligned}
\end{equation}
the integral over $P_{\rm int}$ can be represented as a tractable Gaussian integral, 
\begin{equation}\label{eq:LeePomeranskyrepresentationstillgaussian}
\begin{aligned}
  \mathcal F^{n_l}_{n_{P}} & =\vcentcolon \int^{\RR^{\DD}}\limits_{P_{\rm int}} \, \int^{*}_{z} \, \prod \, \frac {z_i^{\eta_i}} {\Gamma(\eta_i)} \, e^{- (\transpose{P_{\rm int}} \, U \, P_{\rm int} + \transpose{P_{\rm int}} \, W + \transpose{W} \, P_{\rm int} + \mathcal P)} 
  \\ & = \int^{*}_{z} \, \prod \, \frac {z_i^{\eta_i}} {\Gamma(\eta_i)} \, \frac {e^{- (\mathcal P - \transpose{W} \, U^{-1} \, W)} } {(\det U)^{\frac {\DD} 2}},
\end{aligned}
\end{equation}
where $\mathcal P$ and matrices $U$, $W$ are functions of the external momenta $P_{\rm ext}$, masses $m$ and Schwinger parameters $z$.
$\det U \equiv \mathcal U$ and $\det U \,(\mathcal P - \transpose{W} \, U^{-1} \, W) \equiv \mathcal F$ are called the first and second Symanzik polynomials and are homogenous in $z$ of orders $n_l$ and $n_l+1$ respectively. 
When convergent, such integrals can be written in the Lee-Pomeransky representation \cite{Lee:2013hzt} as:
\begin{equation}\label{eq:LeePomeranskyrepresentation}
\begin{aligned}
  \mathcal F^{n_l}_{n_{P}}  & = \frac {\Gamma(\sum \eta_i - n_l \frac {\DD} 2)} {\prod \Gamma(\eta_i)} \,\int^{*}_{z} \, \prod \, z_i^{\eta_i} \, \delta(1 - \raisebox{1pt}{$\scriptstyle\sum$} \, z_i) \, \frac {\mathcal F^{n_l \,\frac {\DD} 2 - \sum \eta_i}} {{\mathcal U}^{(n_l+1)\,\frac {\DD} 2 - \sum \eta_i} } 
  \\ & = \frac {\Gamma(\frac {\DD} 2)} {\Gamma((n_l+1) \,\frac {\DD} 2 - \sum \eta_i) \,\prod \Gamma(\eta_i)} \,\int^{*}_{z} \,\frac { \prod z_i^{\eta_i} } {{(\mathcal U + \mathcal F)}^{\frac {\DD} 2}},
\end{aligned}
\end{equation}
taking the form of an Euler integral, as described in \cref{eq:toocutetobetrue2}. 
General tensorial flat space Feynman diagrams can be reduced to a sum over scalar integrals \cite{Davydychev:1991va}, which can be explicitly confirmed by taking the form of general higher spin flat space propagators into account \cite{Lindwasser:2023zwo}. 
Thus, parametric representations of flat space Feynman integrals are indeed a subset of generalized Euler integrals, and solutions of $\mathcal A$-hypergeometric systems \cite{Kashiwara1976OnAC,Kashiwara:1977nf}. 
Some recent works discussing the status of flat space feynman integrals are \cite{Mizera:2022+7,Klausen:2021yrt}.
\par
The propagator of a scalar field on $S^{d+1}$ of mass $m^2 = \Delta\,(d - \Delta)$ is known to be 
\begin{equation}\label{eq:scalarprophypergeometricform}
\begin{aligned}
  G(\hat X, \, \hat Y) & =  \frac {\Gamma(\Delta) \,\Gamma(d - \Delta)} {(4 \pi)^{\frac{d+1}{2}} \,\Gamma(\frac {d+1} {2})} \, {}_2 F_1 (\Delta, \, d - \Delta; \, \tfrac {d+1} {2}; \, \tfrac {1 + \hat X \cdot \hat Y} {2}),
\end{aligned}
\end{equation}
some of its earliest sources being \cite{Candelas:1975,Dowker:1976,Bunch:1978yq}, that find it as a solution to the differential equation satisfied by it or construct it explicitly as a sum over eigenmodes satisfying the scalar wave equation. 
The choice of solution spaces in either procedure fixes the vacuum state.
In constrast to flat space, computations of higher loop Feynman integrals using \cref{eq:scalarprophypergeometricform} have two major stumbling blocks: \begin{enumerate*}[label=(\roman*)]
	\item integrals of the Gauss hypergeometric function ${}_2 F_1$ are in general not analytically conducive (though \cref{eq:scalarprophypergeometricform} does reduce to simpler functions on odd dimensional spheres, \cref{tb:oddSprop}, and at specific mass parameters in the complementary series on even dimensional spheres, \cref{tb:evenSprop}), and
	\item when considering more than two points on the sphere, the distance measures between these points become non-trivial. 
\end{enumerate*}
However, taking inspiration from the simplicity of the Gaussian integral form in \cref{eq:flatfeynmanintschwinger}, it becomes worthwhile to explore alternate avenues to eventually make higher loop computations on the sphere algorithmically possible, just like their flat space counterparts \cite{nasrollahpoursamami2016periods,Bitoun:2017nre,delaCruz:2019skx,Klausen:2019hrg}. 
Recognising the $(d+1)$-sphere as a quotient of $\RR^{\DD}$ ($\DD \equiv d + 2$) by $\RR_+$ and building upon this observation serves as a detour around the aforementioned issues, as will be illustrated in the following, by the construction and use of the embedding space representation of the massive scalar on $S^{d+1}$ that will, by design, produce sphere Feynman integrals in the generalized Euler integral form.
\par
A quick refresher of the massive scalar propagator on the sphere is given in \cref{sec:MassiveScalarPropagatorinPositionSpace}, before laying the groundwork for embedding space representations of sphere propagators, essentially formed by recovering the scalar Laplacian on $S^{d+1}$ from the Laplacian on $\RR^{\DD}$ as a quotient (\cref{sec:EmbeddingSpaceRepMassiveScalarPropagator}). 
Along with verifying the propagator (\cref{eq:ModeIntVerificationofScalarPropagatorinEmbeddingSpaceFormalism}), its various possible position space forms are noted (\cref{sec:EquivalentFormsoftheScalarPropagator}). 
Their limiting behaviors imply they satisfy different conditions on the vacuum state \cite{Burges:1984qm,Allen:1985ux}.

\subsection{Massive scalar propagator in position space}\label{sec:MassiveScalarPropagatorinPositionSpace}
The action representing a massive scalar field $\Phi$ on a $(d+1)$-sphere is
\begin{equation}\label{eq:actionofmassivescalarfieldonthesphere}
  S^{[0]} = \int^{S^{d+1}} \bar\Phi \, \big( - \nabla^2 + m^2 \big) \, \Phi
\end{equation}
where $\bar \Phi = \Phi, \; \Phi^*$ for real and complex fields respectively. 
The propagator of $\Phi$, $G(\hat X,  \,\hat Y) \equiv \big( - \nabla^2 + m^2 \big)^{-1}$, is defined such that it exhibits the property
\begin{equation}\label{eq:greensfunctiondefinitionsphere}
  \big( - \nabla^2 + m^2 \big) \, G(\hat X, \,\hat Y) = \delta(\hat X - \hat Y) \implies \Phi(\hat Y) = \int_{\hat X} G(\hat X, \,\hat Y) \,\big( - \nabla^2 + m^2 \big) \, \Phi(\hat X).
\end{equation}
The construction of the position space $\delta$ function and the relation between the Euclidean and dS propagators is reviewed in \cite{Dowker:1976}. 
Given a complete orthonormal basis of eigenvectors $\Phi_{\omega}$, labelled by $\omega$, with eigenvalues $\lambda_\omega$ of the scalar Laplacian, $- \nabla^2$, 
\begin{equation}
\begin{aligned}
  & \exists c_{\omega} \; | \; \Phi(\hat X) = \sum_{\omega} \,c_{\omega}  \,\Phi_{\omega}, && - \nabla^2 \, \Phi_{\omega} (\hat X) = \lambda_{\omega}  \,\Phi_{\omega} (\hat X), && \int_{\hat X} \bar \Phi_{\omega} (\hat X) \,\Phi_{\omega'} (\hat X) = \delta_{\omega, \,\omega'},
\end{aligned}
\end{equation}
the propagator is
\begin{equation}\label{eq:propandmodegiveinverseeigenvalue}
\begin{aligned}
  G(\hat X, \,\hat Y) = \sum_{\omega} \,\frac {\bar\Phi_{\omega} (\hat X)  \, \Phi_{\omega}(\hat Y)} {\lambda_{\omega} + m^2} \implies \int_{\hat X, \, \hat Y} \Phi_{\omega}(\hat X) \, G(\hat X, \,\hat Y) \, \bar\Phi_{\omega'}(\hat Y) = \frac 1 {\lambda_\omega + m^2} \delta_{\omega,  \,\omega'}.
\end{aligned}
\end{equation} 
Spin-$s$ symmetric transverse traceless eigentensors of the Laplacian on $S^{d+1}$ have eigenvalues labelled by $n \ge 0$, $\lambda_n + s = (n+s) \,(n+d+s)$ \cite{Camporesi:1994ga}. 
As such, it is useful to define scalar mass parameters $\Delta, \, \bar \Delta$, such that $\lambda_n + m^2 = (n+\Delta)  \,(n + \bdel)$,
\begin{equation}\label{eq:scalarmassparameterdef}
\begin{aligned}
  & m^2 = \vcentcolon \Delta  \,\bdel, && \Delta \vcentcolon= \tfrac {d} {2} + \im \nu, && \bdel \vcentcolon= (d - \Delta) = \tfrac {d} {2} - \im \nu, && \nu = \sqrt{m^2 - \tfrac {d^2} {4}},
\end{aligned}
\end{equation}
where $\bdel = \del^*$ for $\nu \in \RR$. 
The scalar propagator $G(\hat X, \hat Y)$ satisfies the equations of motion when $\hat X \neq \hat Y$. 
Since there are no unique points on the sphere and it is rotationally invariant, $G(\hat X,\hat Y)$ is purely a function of the geodesic distance $\theta$:
\begin{equation}
\begin{aligned}
  & \xdotY \equiv  \hat X \cdot \hat Y, && \theta \equiv \cos^{-1} \xdotY, && \geodissq \equiv \geodis^2 \equiv \tfrac {1 + \xdotY} {2} = \cos^2 \tfrac \theta 2.
\end{aligned}
\end{equation}
The scalar Laplacian on $S^{d+1}$ of unit radius in terms of the geodesic distance is
\begin{equation}
\begin{aligned}
  - \nabla^2 & = - \frac 1 {\sin^d \theta} \,\partial_\theta \,\sin^d \theta \,\partial_\theta = - \partial^2_\theta - d \, \cot \theta \, \partial_\theta 
 = - (1 - \geodissq) \, \geodissq \,\partial^2_\geodissq - (d+1) \, (\tfrac{1} {2 }  - \geodissq ) \, \partial_\geodissq.
\end{aligned}
\end{equation}
A standard procedure to construct two-point functions is to solve the equations of motion, \cref{eq:greensfunctiondefinitionsphere}, away from $\theta = 0$ \cite{Allen:1985ux,Allen:1985wd},
\begin{equation}
\begin{aligned}
  &\Big( (1 - \geodissq) \, \geodissq \,\partial^2_\geodissq + (d+1) \,\big(\tfrac{1} {2 }  - \geodissq \big) \,\partial_\geodissq - \Delta\,(d-\Delta) \Big) \, G(\geodissq) = 0,
\end{aligned}
\end{equation}
which is noted to remain unchanged when all $\geodissq$ are changed to $\geodissq' = 1 - \geodissq$ (corresponding to $\theta \rightarrow \pi - \theta$). 
Thus, the solution space to this ODE is
\begin{equation}\label{eq:solutionspaceofscalarODEusual}
 \begin{aligned}
  G(\geodissq) \prop & \,{}_2F_1 (\Delta, \, d - \Delta\, ; \, \tfrac{d + 1} {2 } \,; \, \geodissq ) \,\oplus \,\geodissq^{\frac{1 - d} {2 }} \, {}_2F_1 ( \tfrac{1} {2} + \im \nu \, , \, \tfrac{1} {2} - \im \nu \, ; \, 1 - \tfrac{d - 1} {2 } \, ; \, \geodissq ) 
  \\ & \oplus \,{}_2F_1 (\Delta, \, d - \Delta\, ; \, \tfrac{d + 1} {2 } \,; \, \geodissq' ) \,\oplus \,\geodissq'{}^{\frac{1 - d} {2 }} \, {}_2F_1 ( \tfrac{1} {2} + \im \nu \, , \, \tfrac{1} {2} - \im \nu \, ; \, 1 - \tfrac{d - 1} {2 } \, ; \, \geodissq'  ).
\end{aligned}
\end{equation}
Only the first solution is continuous at $\geodissq = 0, \; \theta = \pi$ corresponding to the Euclidean, or upon analytic continuation, the Bunch-Davies vacuum state, \cite{Bunch:1978yq}. 
In the $\ell \rightarrow \infty$ limit of $S^{d+1}$, the space appears flat, just like it does in the $\theta \rightarrow 0$ limit. 
The normalisation of the chosen solution can hence be set by comparing the flat space limit to the flat space scalar propagator. 
The embedding space formalism, presented next, will naturally produce the proper normalisation, inheriting the correct factors from the higher dimensional flat space.

\subsection{Embedding space formalism of the massive scalar propagator}\label{sec:EmbeddingSpaceRepMassiveScalarPropagator}
The action of a massless scalar field $\embed \Phi(X)$ in $\RR^\DD$ written in flat and spherical coordinates $(t, \, \hat X)$ is
\begin{equation}
\begin{aligned}
  \embed S^{[0]} & = \int_{X} \partial^{I}\embed {\bar \Phi} \, \partial_{I}\embed \Phi  = \int_{X} \embed {\bar \Phi} \, (- \partial^2) \,\embed \Phi 
  = \int_{X} \embed {\bar \Phi} \, e^{- 2 t} \,\Big( - (d + \partial_{t}) \,\partial_{t} - g^\mu{}^\nu{} \, \partial_{\mu} \, \partial_{\nu} \Big) \,\embed \Phi 
\end{aligned}
\end{equation}
with the Green's function
\begin{equation}
\begin{aligned}
  \embed G(X,\, Y) & \equiv \embed G(e^t \,\hat X, \, e^{s}\, \hat Y) = \frac 1 {4 \pi^{\DD} } \int_{P} \frac {e^{- 2 \,\im \,P\,(X - Y)}} {P^2} = \frac 1 {4 \pi^{\DD} } \int_{P} \frac {e^{- 2 \,\im\, P\,(e^t \,\hat X - e^{s}\, \hat Y)}} {P^2} 
  \\ & = \frac {\Gamma(\frac {d}{2})} {4 \pi^{\DD} } \, \frac 1  {|X - Y|^{d}}
\end{aligned}
\end{equation}
by definition satisfying
\begin{equation}\label{eq:greensfunctiondefinition}
\begin{aligned}
  \embed \Phi(Y) & = \int_{X} \embed G(X,\,Y) \,(- \partial^2) \,\embed \Phi(X)
  \\ 
  \embed \Phi(e^{s} \,\hat Y) & = \int_{t} e^{d \,t} \int_{\hat X} \embed G(X,\,Y) \, \Big( - (d + \partial_{t}) \,\partial_{t} - g^\mu{}^\nu{} \, \partial_{\mu} \, \partial_{\nu} \Big) \,\embed \Phi(e^t\, \hat X). 
\end{aligned}
\end{equation}
By representing the scalar fields $\embed \Phi $ in terms of their radial Mellin transforms $\Phi_{[\Delta]}$ in \cref{eq:greensfunctiondefinition}
\begin{equation}\label{eq:greensfunctiondefinitionMellinised}
\begin{aligned}
  \circint_{\Delta} e^{- \Delta\, s} \,\Phi_{[\Delta]}(\hat Y) & = \circint_{\Delta} \int_{\hat X} \int_{t} e^{(d - \Delta)\, t} \,\embed G(X,\,Y) \, \Big( - \nabla^2 + (d - \Delta)\,\Delta \Big) \,\Phi_{[\Delta]}(\hat X)
\end{aligned}
\end{equation}
the massive scalar Laplacian on $S^{d+1}$, given in \cref{eq:actionofmassivescalarfieldonthesphere} with mass as defined in \cref{eq:scalarmassparameterdef}, can be recognised. 
A couple of simple variable substitutions, namely $t \rightarrow t + s$, $P \rightarrow P \, e^{- s}$, brings this into a suggestive form that distributes over $\Delta$,
\begin{equation}\label{eq:scalarpropderivationstep1}
\begin{aligned}
  \circint_{\Delta} e^{- \Delta\, s} \,\underbracket{\Phi_{[\Delta]}(\hat Y)} & = \circint_{\Delta} e^{- \Delta\, s} \overbracket{\int_{\hat X} \int_{t} e^{(d - \Delta) \,t} \,\embed G(e^t \,\hat X, \,\hat Y) \, \Big( - \nabla^2 + (d - \Delta)\,\Delta \Big)\Phi_{[\Delta]}(\hat X)}
\end{aligned}
\end{equation}
with the bracketed terms completely independent of $s$ or alternately $|Y|$, resulting in
\begin{equation}\label{eq:scalarpropderivationalmostcomplete}
\begin{aligned}
  \Phi_{[\Delta]}(\hat Y) & = \int_{\hat X} \int_{t} e^{(d - \Delta) \,t} \,\embed G(e^t \,\hat X, \,\hat Y) \, \Big( - \nabla^2 + (d - \Delta)\,\Delta \Big) \,\Phi_{[\Delta]}(\hat X).
\end{aligned}
\end{equation}
Comparing \cref{eq:scalarpropderivationalmostcomplete} to \cref{eq:greensfunctiondefinitionsphere}, the massive scalar propagator on the $d+1$-sphere is found to be
\begin{equation}
\begin{aligned}
  G(\hat X,\, \hat Y) & = \int_{t} e^{(d - \Delta) \,t} \, \embed G(e^t \,\hat X, \,\hat Y) = \int^*_{\lambda} \lambda^{\bar\Delta}\,\embed G(\lambda \, \hat X, \,\hat Y) = \int^*_{\mu} \mu^{\Delta}\,\embed G(\hat X, \,\mu \, \hat Y)
\end{aligned}
\end{equation}
or equivalently,
\begin{equation}
\begin{aligned}
  G(\hat X, \,\hat Y) & = \int^*_{\lambda} |\lambda \,X|^{\bar\Delta} \,|Y|^\Delta \, \embed G(\lambda \, X, \,Y) = \int^*_{\mu} |X|^{\bar\Delta}\, |\mu \,Y|^\Delta \, \embed G(X, \,\mu \, Y) 
  \\ & = \int^*_{q} q^{i \nu} \,|X|^{\bar\Delta} \,|Y|^\Delta \, \embed G(\tfrac 1 {\sqrt q} \,X, \,\sqrt q \, Y).
\end{aligned}
\end{equation}
It can be represented in a more symmetric form, 
\begin{equation}\label{eq:scalarpropunfixed}
\begin{aligned}
  G_{[\Delta]}(\hat X, \, \hat Y) \vcentcolon 
   =  & \, \frac 1 {\Vol \RR^*} \,\frac 1 {4 \pi^{\DD} } \,\int_{\lambda, \,\mu}^* \, |\lambda \,X|^{\bar\Delta} \, |\mu \,Y|^{\Delta}  \int_{P} \frac {e^{- 2 \im \, P\,(\lambda \,X - \mu \,Y)}} {P^2}
   \\ =  & \, \frac 1 {\Vol \RR^*} \,\frac 1 {4 \pi^{\DD} } \int_{\lambda, \,\mu, \, \tau}^* \, |\lambda \,X|^{\bar\Delta} \, |\mu \,Y|^{\Delta}  \, \tau \,\int_{P} e^{- \tau \,P^2 - 2 \im \, P\,(\lambda \,X - \mu \,Y)}
\end{aligned}
\end{equation}
where the division by $\Vol \RR^*$ indicates the prescription to fix the scaling redundancy $\hat {\rm T}$ in the integration variables when the mass parameters satisfy $\Delta + \bdel = d$,
\begin{equation}\label{eq:scalarpropscaleinvariances}
\begin{aligned}
  \hat {\rm T} \, : \, G_{[\Delta]}(\hat X, \, \hat Y) \rightarrow G_{[\Delta]}(\hat X, \, \hat Y) \, : \, \{P, \, \tau, \, \lambda, \, \mu\} & \mapsto  \{\rho^{-1}\, P, \, \rho^2 \tau, \, \rho \, \lambda, \, \rho\, \mu\}, \quad \rho \in \RR_+.
\end{aligned}
\end{equation}
For the sake of notational brevity, this prescription will be denoted by $\displaystyle\Xint{\times}$. 
It is emphasised that although the domain of definition of the propagator in \cref{eq:scalarpropunfixed} has been expanded to the entirety of the embedding space $\RR^\DD$, it is still {\it independent} of $|X|$, \, $|Y|$. 
Generalisation to a wider class of propagators $G_{[\Delta]}^{\eta}(\hat X, \, \hat Y)$ parameterised by $\eta$ ($\Delta + \bdel + 2 \,\eta = \DD$)
\begin{equation}\label{eq:scalarpropunfixedgeneral}
\begin{aligned}
  & G_{[\Delta]}^{\eta}(\hat X, \, \hat Y) & = \frac 1 {\Vol \RR^*} \frac { 1  } {4^{\eta} \pi^{\DD}} \int^{*}_{\lambda, \mu} \int_{P} |\lambda \, X|^{\bar \Delta} \, |\mu \, Y|^{\Delta} \,\frac {e^{-2 \im\, P \,(\lambda X - \mu Y)}} {P^{2 \,\eta}},
\end{aligned}
\end{equation}
now exhibiting the expanded scale invariance $\hat{\rm T}^{\eta} \, : \, G_{[\Delta]}^{\eta}(\hat X, \, \hat Y) \rightarrow G_{[\Delta]}^{\eta}(\hat X, \, \hat Y)$,
\begin{equation}
\begin{aligned}
  \hat{\rm T}^{\eta} \, : \{P, \, X, \, Y, \, \tau, \, \lambda, \, \mu\} & \mapsto  \{\rho_1^{-1}\, P, \, \rho_2^{-1}\, X, \, \rho_3^{-1}\, Y, \, \rho_1^2 \tau, \, \rho_1 \, \rho_2 \, \lambda, \, \rho_1 \, \rho_3 \, \mu\},
\end{aligned}
\end{equation}
where $\{\rho_1, \, \rho_2, \, \rho_3 \}  \in \RR_+$, allows the integral to be interpreted as a generalized Euler integral with {\it generic} parameters by assuming the dimension $\DD$ can take any value in $\CC$, i.e. by evaluating the Feynman integrals containing this propagator in dimensional regularisation. 
Further, $G^{\eta}(\hat X, \, \hat Y)$ can be used to define the extension of $\delta(\hat X - \hat Y)$ on $S^{d+1}$ to the embedding space, $\RR^\DD$, as 
\begin{equation}\label{eq:embeddingspacescalardeltafunction}
\begin{aligned}
  \delta(\hat X - \hat Y) & = \lim_{\eta \rightarrow 0} G^{\eta} (\hat X, \, \hat Y) 
  \\ & = \lim_{\eta \rightarrow 0} \frac { 1  } {4^{\eta} \pi^{\DD}} \Xint{\times}^{*}_{\lambda, \mu, \tau} \int_{P} |\lambda \, X|^{\DD - \Delta} \, |\mu \, Y|^{\Delta} \, \frac {\tau^{\eta}} { \Gamma(\eta)} \, e^{- \tau\, P^2 -2 \im\, P \,(\lambda \,X - \mu \,Y)},
\end{aligned}
\end{equation}
for any value of $\Delta$. 
For the sake of future notational convenience, $\mathcal F \circ G_{[\Delta]}^\eta(\hat X, \, \hat Y)$ is defined as
\begin{equation}\label{eq:operatorwithscalarprop}
\begin{aligned}
  \mathcal F \circ G_{[\Delta]}^\eta(\hat X, \, \hat Y) & \vcentcolon=  \frac { 1  } {4^{\eta} \pi^{\DD}} \Xint{\times}^{*}_{\lambda, \mu} \int_{P} |\lambda \, X|^{\bar \Delta} \, |\mu \, Y|^{\Delta} \, \mathcal F \, \frac {e^{-2 \im\, P \,(\lambda \,X - \mu \,Y)}} {P^{2 \,\eta}}. 
\end{aligned}
\end{equation}
When $|\mathcal F|$ is polynomial in $\{\lambda \, |P| \, |X|, \, \mu \, |P| \, |Y|, \, \tau \, |P|^{2}\}$, $\mathcal F \circ G_{[\Delta]}^\eta(\hat X, \, \hat Y) $ is $\hat{\rm T}^{\eta}$ invariant. 
For contour integration involving extension of $\lambda, \, \mu$ to $\CC$: note that it is possible to extend the integration range of the parameters from $\RR_+$ to $\RR$ by a simple change of variables, related by $G\big|_{\RR} = 4 \, (-1)^{\del + \bdel + 1} \, \sin (\pi \bdel) \, \sin (\pi \del) \, G\big|_{\RR_+}$. 
In addition, considering a gauge fixed form of this integral for illustration, say \cref{eq:scproptau1gauge}, $G$ falls off at large $|\lambda|, \, |\mu|$.

\subsection{Equivalent forms of the scalar propagator}\label{sec:EquivalentFormsoftheScalarPropagator}
The family of de Sitter invariant vacuum states, usually presented as a superposition of solutions, \cref{eq:solutionspaceofscalarODEusual}, can instead be captured by the embedding space propagator formulation, in different ``gauges'' used to fix the scale invariance. 
The scale invariance $\hat {\rm T}$ in \cref{eq:scalarpropscaleinvariances} can be fixed by any of the following (inexhaustive) list of conditions (including the relevant Faddeev-Popov determinant):
\begin{equation}
\begin{aligned}
  & ({\rm i}) \; |P| \, \delta(|P| - 1), \quad  ({\rm ii}) \; 2 \, \tau \, \delta(\tau - 1), \quad ({\rm iii}) \; \lambda \, \delta(\lambda - 1) \cong |n + 1| \, \lambda\, \mu^n \, \delta(\lambda \, \mu^n - 1),
\end{aligned}
\end{equation}
where $n \neq - 1$. 
As it turns out, (ii) and (iii) are equivalent conditions related by trivial variable rescalings.
The different conditions can be used to study the behaviour of the propagator as an expansion around different values of $\xdotY = \cos \theta$ or equivalently $\geodissq = \cos^2 \frac \theta 2$. 
\begin{enumerate}[left=-12pt,label=(\roman*)]
	\item Restricting the momentum integral to $S^{d+1}$, yields a form of $G(\hat X, \,\hat Y)$
\begin{equation}
\begin{aligned}
  G(\hat X,\,  \hat Y) & = \frac 1 {4 \pi^{\DD}} \int_{\hat P} \int^{*}_{\lambda}\, \lambda^{\bar \Delta} \, e^{-2 \im \lambda \,\hat P \cdot \hat X }  \int^{*}_{\mu} \mu^{\Delta} \, e^{2 i\, \mu \,\hat P \cdot \hat Y} 
  \\ & = \frac {1} {4 \pi^{\DD}} \int_{\hat P} \frac{\Gamma(\bar \Delta) \,\Gamma(\Delta)}{ ( 2 \im \hat X \cdot \hat P )^{\bar \Delta} \, ( - 2 \im \hat P \cdot \hat Y )^{\Delta} }
\end{aligned}
\end{equation}
resembling products of bulk to boundary (and back to bulk) propagators in AdS, \cite{Penedones:2016voo}.
\begin{figure}[htbp]
  \centering
  \scalebox{0.7}{\begin{tikzpicture}
  \fill[gray!5!white, opacity = 0.3] (0,0) circle (2);
  \draw[gray,dashed,<->] (-2.2,0) -- (2.2,0);
  \draw[gray,dashed,<->] (0,-2.2) -- (0,2.2);
  \foreach \x in {-2,-1.5,-1,-0.5,0,0.5,1,1.5,2} \draw[thin, lightgray] (\x,-0.1) -- (\x,0.1);
  \foreach \x in {-2,-1.5,-1,-0.5,0,0.5,1,1.5,2} \draw[thin, lightgray] (-0.1,\x) -- (0.1,\x);
  \draw[dotted] plot [smooth, tension=1.25] coordinates{ (-2,0) (0,-0.5) (2,0)};
  \draw (0,0) circle (2);
  \draw[dotted] plot [smooth, tension=1.25] coordinates{ (-2,0) (0,0.5) (2,0)};
  \draw plot [smooth, tension=1.25] coordinates{ (-0.375,-0.0625) (0,-0.13) (0.3,0)};
  \draw[dashed] (-1.5,-0.25) -- (0,0);
  \node at (0,0) [anchor = north east]{\small$\theta$};
  \fill (2,0) circle (0.05cm);
  \node at (2,0)  [anchor = north west]{\small$\hat X$};
  \fill (-1.5,-0.25) circle (0.05cm);
  \node at (-1.5,-0.25)  [anchor = north]{\small$\hat Y$};
  \fill (0.6,1.1) circle (0.05cm);
  \node at (0.6,1.1)   [anchor = south]{\small$\hat P$};
  \draw[->,thick] plot [smooth, tension=1.25] coordinates{  (-1.5,-0.25) (-0.6,0.6) (0.6,1.1)};
  \draw[->,thick] plot [smooth, tension=1.25] coordinates{ (0.6,1.1) (1.5,0.8) (2,0) };
  \draw[dashed] (2,0) -- (0,0) -- (0.6,1.1);
  \draw[gray,dashed,<->] (-2.2-8,0) -- (2.2-8,0);
  \draw[gray,dashed,<->] (0-8,-2.2) -- (0-8,2.2);
  \foreach \x in {-2,-1.5,-1,-0.5,0,0.5,1,1.5,2} \draw[thin, lightgray] (\x-8,-0.1) -- (\x-8,0.1);
  \foreach \x in {-2,-1.5,-1,-0.5,0,0.5,1,1.5,2} \draw[thin, lightgray] (-0.1-8,\x) -- (0.1-8,\x);
  \fill (2-8,0) circle (0.05cm);
  \node at (2-8,0)  [anchor = north west]{\small$\hat X$};
  \fill (-1.5-8,-0.25) circle (0.05cm);
  \node at (-1.5-8,-0.25)  [anchor = north]{\small$\hat Y$};
  \draw (-10,2) -- (-6,2); 
  \node at (-6,2)  [anchor = west]{\small$P$};
  \draw[->,thick] (-1.5-8,-0.25) -- (0.6-8,2); \draw[->,thick] (0.6-8,2) -- (2-8,0);
  \draw[<->,dotted,thick] (-5,0) -- (-3,0);
  \end{tikzpicture}}
  \caption{dS Scalar Propagator as an analogue of AdS Bulk to Boundary Propagators}\label{fig:AnalogueofAdSpropagatorform}
\end{figure}
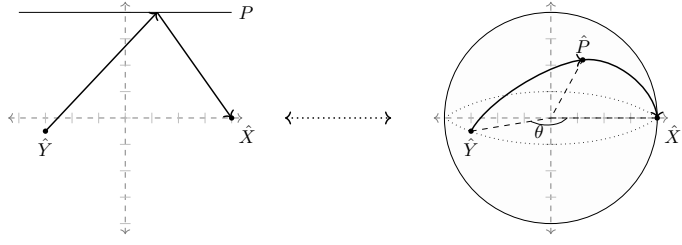\vspace{-0.5\baselineskip}\noindent
	\item Applying the condition $\tau = 1$ 
\begin{equation}\label{eq:scproptau1gauge}
\begin{aligned}
  G(\hat X,\,  \hat Y) & = \frac {1} { 2 \pi^{\frac {\DD} {2}} } \int^*_{\lambda, \mu} \lambda^{\bar \Delta} \, \mu^{\Delta} \,e^{- |\lambda \, \hat X - \mu \, \hat Y|^2} 
\end{aligned}
\end{equation}
allows the integral to be evaluated as an expansion around $\xdotY = -1$, i.e. $\geodissq = 0$,
\begin{equation}
\begin{aligned}
  G(\hat X, \, \hat Y) & = \frac {\Gamma(\frac{\bar \Delta + \Delta} {2})} { 4 \pi^{\frac {\DD} {2}} } \dashint_r^{*} \frac {r^{\bar \Delta} \, (1-r)^{\Delta}} {(1 - 4 \, r \, (1-r) \, \geodissq)^{\frac {\bar \Delta + \Delta} {2}}}, \quad \dashint_r^{*} \equiv \int_0^1 \frac {\de r} {r \, (1-r)}
\end{aligned}
\end{equation}
by expanding the integrand as a series in increasing powers of $\geodissq, \, r$. 
The series remains convergent for $|4 \, r \, (1-r) \, \geodissq| < 1$, which is the case for $\geodissq, \, r \in (0,1)$:
\begin{equation}
\begin{aligned}
  G(\hat X, \,\hat Y) & = \sum_{n = 0}^{\infty} \frac {\Gamma(\frac {\bar \Delta + \Delta} {2} + n)} { 4 \pi^{\frac {\DD} {2}} } \,\frac {4^{n} \, \geodissq^{n}} { n!} \, \dashint_r^{*} r^{\bar \Delta + n} \, (1-r)^{\Delta + n}.
\end{aligned}
\end{equation}
The integral over $r$ is in the Feynman parameterisation form and is easily evaluated to
\begin{equation}
\begin{aligned}
  G(\hat X, \, \hat Y) & = \sum_{n = 0}^{\infty} \frac {\Gamma(\frac {\bar \Delta + \Delta} {2} + n)} { 4 \pi^{\frac {\DD} {2}} } \, \frac {\Gamma(\bar \Delta + n) \, \Gamma(\Delta + n)} { \Gamma(\bar \Delta + \Delta + 2 n) } \,\frac {4^{n} \, \geodissq^{n}} { n!} 
  \\ & = \frac {1} { 2^{\bar \Delta + \Delta + 1}\pi^{\frac {\DD - 1} {2}} } \sum_{n = 0}^{\infty}  \frac {\Gamma(\bar \Delta + n) \, \Gamma(\Delta + n)} { \Gamma(\frac {\bar \Delta + \Delta + 1} {2} + n) } \,\frac {\geodissq^{n}} { n!},
\end{aligned}
\end{equation}
which is the series representation of 
\begin{equation}
  G(\hat X,\, \hat Y) =  \frac {\Gamma(\bar \Delta) \, \Gamma(\Delta)} { 2^{\bar \Delta + \Delta + 1} \, \pi^{\frac {\DD - 1} {2}} \, \Gamma(\frac {\bar \Delta + \Delta  + 1} {2})} \, {}_2F_1(\bar \Delta, \,\Delta, \, \tfrac {\bar \Delta + \Delta  + 1} {2}, \, \geodissq),
\end{equation}
matching \cref{eq:scalarprophypergeometricform} when the $\bdel + \del = d$. 
Another approach to proceeding from \cref{eq:scproptau1gauge} onwards is to rescale the variable $\lambda \rightarrow \lambda \, \mu $, the resulting form of which is equivalent to that stemming from the third condition $\delta(\mu - 1)$, \cref{eq:scproplam1gauge}.

	\item It is useful to illustrate that $G(\hat X, \, \hat Y)$ resulting from $\delta(\lambda - 1)$: 
\begin{equation}\label{eq:scproplam1gauge}
\begin{aligned}
  G(\hat X, \, \hat Y) & = \frac {\Gamma(\frac {\bar \Delta + \Delta} {2})} { 4 \pi^{\frac {\DD} {2}} } \int^*_{\mu} \frac {\mu^{\Delta}} {(1 + \mu^2 - 2 \, \mu \, \xdotY)^{\frac {\bar \Delta + \Delta} {2}}} 
\end{aligned}
\end{equation}
is indeed equivalent to \cref{eq:scalarprophypergeometricform} by considering its expansion around $\xdotY = -1$ once again. This can be done by representing the integral as
\begin{equation}\label{eq:scproplam1gaugePlus1}
\begin{aligned}
  G(\hat X, \, \hat Y) & = \frac {\Gamma(\frac {\bar \Delta + \Delta} {2})} { 4 \pi^{\frac {\DD} {2}} } \int^*_{\mu} \,\frac {\mu^{\Delta}} {((1 + \mu)^2 - 4 \, \mu \, \geodissq)^{\frac {\bar \Delta + \Delta} {2}}} 
  \\ & = \frac {\Gamma(\frac {\bar \Delta + \Delta} {2})} { 2 \pi^{\frac {\DD} {2}} } \,\int^*_{\mu} \,\frac {\mu^{2 \, \Delta}} {\big((1 - 2 \, \mu \, \geodis + \mu^2 )\,  (1 + 2 \, \mu \, \geodis + \mu^2 ) \big)^{\frac {\bar \Delta + \Delta} {2}}}
\end{aligned}
\end{equation} 
and evaluating it by using the techniques reviewed in \cref{ch:Eulerintegrals}. 
The $\mathcal A$-hypergeometric system is defined by
\begin{equation}\label{eq:scproplam1gaugePlus2}
\begin{aligned}
  \mathcal A & = \left(\begin{smallmatrix}
z_{0,1} & z_{1,1} & z_{2,1} \\ \midrule \\  1 & 1 & 1 \\  0 & 0 & 0 \\  0 & 1 & 2 \\ 
\end{smallmatrix} \middle| \begin{smallmatrix}
z_{0,2} & z_{1,2} & z_{2,2} \\ \midrule \\  0 & 0 & 0 \\ 1 & 1 & 1 \\ 0 & 1 & 2 
\end{smallmatrix} \right) , && \gamma = \left( \begin{smallmatrix}
\frac {\bdel + \del} {2} \\ \frac {\bdel + \del} {2} \\ 2 \Delta
\end{smallmatrix} \right), 
\\ \mathcal J & = \langle \partial_{\bdot,2}^2 - \partial_{\bdot,1} \,\partial_{\bdot,1}\; ; \; \partial_{1,\bdot} \, \partial_{2,\star} - \partial_{1,\star} \, \partial_{2,\bdot} \rangle, && z = \{1, \, - 2 \geodis, \, 1, \, 1, \, 2 \geodis, \, 1\}
\end{aligned}
\end{equation}
where $\bdot, \, \star$ are placeholders referring to the different coefficients in the first or second polynomials. 
Some possible bases of the kernel, $\mathcal K$, are
\begin{equation}\label{eq:scproplam1gaugePlus3}
\begin{aligned}
  \transpose{\mathcal K} & = \left( \begin{smallmatrix}
t_1 & t_2 & - (t_1 + t_2) & - (t_1 + t_3) & 2 t_3 - t_2 & t_1 + t_2 - t_3 \\ 
\midrule
t_1 + t_2 - t_3 & 2 t_3 - t_2 &  - (t_1 + t_3) &  - (t_1 + t_2) & t_2 &  t_1
\end{smallmatrix} \right).
\end{aligned}
\end{equation}
Using a weight vector belonging to the class $\{1, \, 1,\, 0, \, 0, \, 1, \, 1\}$, results in the root system $\mathcal R$:
\begin{equation}\label{eq:scproplam1gaugePlus4}
\begin{aligned}
  \mathcal R& = \left( \begin{smallmatrix}
0 & 0 & -\frac {\bdel+\del} {2} \\ 
0 & 0 & -\frac {\bdel+\del} {2} \\
\frac {\del - \bdel} {2}& 0 & -\del \\
\frac {\del - \bdel - 1} {2} & 1 & -\del - \frac 1 2\\
\end{smallmatrix} \middle|  \begin{smallmatrix}
- \bdel & 0 & \frac {\bdel - \del} {2}\\
-\bdel - \frac 1 2  & 1 & \frac {\bdel - \del - 1} {2}\\
-\frac {\bdel+\del} {2} & 0 & 0 \\
-\frac {\bdel+\del} {2}& 0 & 0 
\end{smallmatrix} \right). 
\end{aligned}
\end{equation}
The second and fourth series are odd in $\geodis$ and so expectedly vanish in the relevant physical limit. 
The normalisation, $\mathcal N$, of the series will be
\begin{equation}
\begin{aligned}
  \mathcal N & =    \left\{  \tfrac { \Gamma(\bdel) \,\Gamma(\frac {\del - \bdel} {2})   } {2 \,\Gamma(\frac {\bdel + \del} {2})} , \;\; - \tfrac { \Gamma(\bdel + \frac 1 2) \,\Gamma(\frac {\del - \bdel + 1} {2})   } {2 \,\Gamma(\frac {\bdel + \del} {2})} , \;\; \tfrac { \Gamma(\del) \,\Gamma(\frac {\bdel - \del} {2})   } {2 \,\Gamma(\frac {\bdel + \del} {2})} , \;\; - \tfrac { \Gamma(\del + \frac 1 2)\, \Gamma(\frac {\bdel - \del + 1} {2})  } {2 \,\Gamma(\frac {\bdel + \del} {2})}       \right\}.
\end{aligned}
\end{equation}
The first and third series differ only by a $\Delta$-dependent factor, and together sum precisely to \cref{eq:scalarprophypergeometricform}. 
The Euler transformation of ${}_2F_1$, \cref{eq:scpropEulertransform}, gives an expansion around $\geodissq = 1$.
The benefit of using the integral representation of $G(\hat X, \, \hat Y) $ is that it naturally encodes analytic continuation within it. For example, considering a weight vector $\cong \{1, \, 0,\, 1, \,1, \, 0,\, 1\}$ results in an expansion in negative powers of $\geodissq$, i.e. an expansion around $\geodissq \rightarrow \infty$:
\begin{equation}
\begin{aligned}
  G(\hat X, \, \hat Y) & = \frac { \big( (-1)^{\frac {d+1}{2}} \, \csch (\pi \nu) \, \sin (\pi \, \Delta) - 1 \big) }  { 4 \pi^{\frac {d} {2}} \, \Gamma(1 + \im \nu) \, \sin (\pi (\frac{d}{2} + 2 \im \nu) )}  \, \frac {\Gamma(\Delta)}{(2\geodissq)^{\Delta}} \, {}_2F_1(\Delta, \, \tfrac {1} {2} + \im \nu, \, 1 + 2 \im \nu, \, \tfrac 1 {\geodissq})
  \\ &  + (\Delta \leftrightarrow \bar\Delta),
\end{aligned}
\end{equation}
satisfying the condition that $G$ fall off with increasing $\geodissq$, as is required for Anti-de Sitter space. 
Alternately, exploring the behaviour of \cref{eq:scproplam1gauge} around the limits $\xdotY = 0$ and $\xdotY \rightarrow \infty$ gives:
\begin{equation}\label{eq:scalarproparoundxdoty0}
\begin{aligned}
  G(\hat X, \, \hat Y) & = \tfrac {\Gamma(\frac{\Delta}{2}) \,\Gamma(\frac{\bar\Delta}{2})} {8 \pi^{\frac{d+2}{2}}} \,{}_2F_1(\tfrac {\bar \Delta} {2}, \,\tfrac {\Delta} {2}, \,\tfrac {1} {2}, \,\xdotY^{2}) + \xdotY \,\tfrac {\Gamma(\frac{\Delta+1}{2}) \,\Gamma(\frac{\bar\Delta+1}{2})} {4 \pi^{\frac{d+2}{2}}} \, {}_2F_1(\tfrac {\bar \Delta+1} {2}, \,\tfrac {\Delta+1} {2}, \,\tfrac {3} {2}, \,\xdotY^{2})
\end{aligned}
\end{equation}
and
\begin{equation}
\begin{aligned}
  G(\hat X, \, \hat Y) & = \tfrac {\Gamma(\bar \Delta) \,\Gamma(\frac {\Delta - \bar \Delta} {2}) } {4 \pi^{\frac{d+2}{2}}} \,{}_2F_1(\tfrac{\Delta}{2}, \,\tfrac{\Delta+1}{2}, \,1 + \tfrac {\Delta - \bar \Delta } {2},\, \tfrac 1 {\xdotY^2})  + (\Delta \leftrightarrow \bar\Delta).
\end{aligned}
\end{equation}
\end{enumerate}

\clearpage
\section{Vector propagators on the sphere}\label{sec:VectorFields}
The ideas used to build the ``momentum-space''-like massive scalar propagator representation in \cref{sec:EmbeddingSpaceRepMassiveScalarPropagator} can be expanded to derive embedding space representations of massive and massless vector propagators on $S^{d+1}$. 
These expressions, given in \cref{eq:vectorprop,eq:masslessvectorpropagatormainformula}, have the same general structure as the scalar propagator, \cref{eq:scalarpropunfixed}, and similarly result in sphere Feynman integrals in the generalized Euler integral form. 

\Cref{sec:PositionSpaceConsiderationsofVectorPropagators} reviews some background about the properties of propagators of the massive and massless (photon) vector (gauge fixed) actions on $S^{d+1}$, including the standard approaches to constructing them, i.e. by explicitly summing over eigenmodes of the Laplacian and/or solving the ODEs defining it away from the coincident point limit, first presented in \cite{Allen:1985wd}. 
Unfortunately, the two-point (Wightman) functions in \cite{Allen:1985wd,Frob:2013qsa}, aren't able to serve as `propagators' on the sphere. 
This can be verified by explictly integrating them against test functions in simple cases (say finding the eigenvalue of the first Killing vector on $S^3$), resulting in the expected realisation that they don't account for some $\delta$-function at $\theta = 0$. 
Further, as is expected, the graviton two-point (Wightman) functions in \cite{Allen:1986tt,Turyn:1988af} also can't be used as graviton propagators. 

This is followed by the action of the massless vector in embedding space and gauges that allow its proper reduction to the massive and massless vector Laplacians on $S^{d+1}$ (\cref{sec:EmbeddingSpaceRepresentationof(GaugeFixed)MasslessVectorAction}), subsequently deriving the corresponding propagators (\cref{sec:MassiveVectorPropagator,sec:MasslessVectorPropagator}). 
The special case of a massless vector on $S^{3}$ is considered separately (\cref{sec:masslessvectoronS3}). 
When compared to the aforementioned literature on this topic, it is found that the vector propagators expectedly do differ by what can be described as a $\delta$-function/longitudinal piece but match away from the coincident point limit (\cref{sec:positionspaceformofvectorprops}).

\subsection{Position space considerations of vector propagators}\label{sec:PositionSpaceConsiderationsofVectorPropagators}
The action representing a massive vector field $A$ on a $d+1$-sphere is
\begin{equation}\label{eq:massivevectoraction}
\begin{aligned}
  S^{[1]} & = \tfrac 1 2 \int^{S^{d+1}}  \bar A^\mu{} \,\big(- g_\mu{}_\nu{}  \, \nabla^2 + \nabla_\nu{}  \, \nabla_\mu{} + g_\mu{}_\nu{} \, m^2\big) \,A^\nu{}
  \\ &  = \tfrac 1 2 \int^{S^{d+1}} \bar A^\mu{}\, \big(- g_\mu{}_\nu{}  \, \nabla^2 + \nabla_\mu{}  \, \nabla_\nu{} + g_\mu{}_\nu{} \, (m^2 + d) \big)\, A^\nu{},
\end{aligned}
\end{equation}
where $ \bar A =  A, \,  A^*$ for real and complex fields respectively. 
When massless, the gauge symmetry $A^\mu{} \sim A^\mu{} + \nabla^\mu{} \Psi$ for any scalar field $\Psi$, is fixed in what is commonly known as the $R_\xi$ gauge (Landau gauge: $\xi \rightarrow 0$, Feynman gauge: $\xi = 1$) by $\tfrac 1 {2 \, \xi} (\nabla \cdot \bar A) \, (\nabla \cdot A)$, resulting in the gauge fixed action:
\begin{equation}\label{eq:masslessvectoractioninRgauge}
\begin{aligned}
  S^{[1]}_{\rm GF} & = \tfrac 1 2  \int^{S^{d+1}}  \bar A^\mu{}  \,\big(- g_\mu{}_\nu{}  \,\nabla^2 + (1 - \tfrac 1 {\xi}  ) \,\nabla_\mu{} \, \nabla_\nu{} + d \big) \, A^\nu{}
  \\ & = \tfrac 1 2  \int^{S^{d+1}}  \bar A^\mu{}  \,\big(- g_\mu{}_\nu{}  \,\nabla^2 + (1 - \tfrac 1 {\xi}  ) \,\nabla_\nu{} \, \nabla_\mu{} + \tfrac d \xi \big) \, A^\nu{}.
\end{aligned}
\end{equation}
The propagator $G^{\nu \nu'}(\hat X, \, \hat Y)$ is defined such that it satisfies
\begin{equation}\label{eq:vectorodesetup}
\begin{aligned}
  & K_\mu{}_\nu{} \, G^{\nu \nu'}(\hat X,  \,\hat Y) = \delta_\mu{}^{\nu'}{}  \,\delta(\hat X - \hat Y), \quad K_\mu{}_\nu{} = (- \nabla^2 + m^2 + d) \, g_\mu{}_\nu{} + \nabla_\mu  \,\nabla_\nu,
  \\ & A^{\mu}{}(\hat X) = \int_{\hat Y} G^{\mu}{}^{\mu'}{} (\hat X, \, \hat Y)\, K_{\mu'}{}_{\nu'}{} \, A^{\nu'}{} (\hat Y),
\end{aligned}
\end{equation}
where $\partial_{\mu}, \; \partial_{\mu'}$ form the bases of the tangent spaces at $\hat X, \; \hat Y$ respectively. 
Thus, $\delta_\mu{}^{\nu'}{}$ should be interpreted to be
\begin{equation}
  \delta_\mu{}^{\nu'}{} = \Lambda_\mu{}^{\mu'} \, \delta_{\mu'}^{\nu'}{} = \Lambda^{\nu'}{}_{\nu}{} \, \delta_{\mu}^{\nu}{}, 
\end{equation}
where $\delta_{\mu}^{\nu}{}, \; \delta_{\mu'}^{\nu'}{}$ are the usual Kronecker delta functions, and $\Lambda_\mu{}^{\mu'}$ and $\Lambda^{\nu'}{}_{\nu}{}$ parallel transport the basis vectors $\partial_{\mu} \in T_{\hat X}(S^{d+1})$ at $\hat X$ to the tangent space at $\hat Y$, $T_{\hat Y}(S^{d+1})$, along some path $\gamma_{\hat X, \hat Y}$ and vice versa. 
$\Lambda^\mu{}_{\mu'}$ is obviously path-dependent, and so the choice of $\gamma_{\hat X, \hat Y}$ needs to be canonicalised in order to properly define the parallel propagator, $\Lambda_\mu{}^{\mu'}$. 
One such choice of $\gamma_{\hat X, \hat Y}$ is the {\it signed} geodesic $\theta$, with the associated unit normalised tangent vectors and parallel propagator along the geodesic (depicted in \cref{fig:CanonicalTangentSpaces}) given in \cite{Allen:1985wd}:
\begin{equation}\label{eq:positionspaceunitvectors}
\begin{aligned}
  &\theta_\mu{} = \nabla_{\hat X^\mu}{} \theta, && \nabla_{\hat X^\mu}{} \theta_{\nu} = \cot \theta  \, (g_\mu{}_\nu{} - \theta_\mu{} \, \theta_\nu{}), && \theta_{\nu'} =  - \nabla_{\hat Y^{\nu'}}{} \theta,  && \Lambda_\mu{}^{\nu'}{} = - \sin \theta  \, \nabla_{\hat X^\mu}{} \theta^{\nu'} - \theta_\mu{} \, \theta^{\nu'}.
\end{aligned}
\end{equation}
These unit vectors may, alternately, be defined as: 
\begin{equation}\label{eq:positionspaceunitvectors2}
\begin{aligned}
  & \sigma_\mu{} = \nabla_{\hat X^\mu}{} \cos \theta = - \sin \theta \, \theta_\mu{}, \quad \sigma_{\nu'} =  \nabla_{\hat Y^{\nu'}}{} \cos \theta =  - \sin \theta \, \theta_{\nu'}, 
  \\ & \sigma_\mu{}_{\nu'}{} = \nabla_{\hat X^\mu}{} \nabla_{\hat Y^{\nu'}}{} \cos \theta = \Lambda_{\mu\nu'} + 2 \, \sin^2 (\tfrac {\theta}{2}) \, \theta_{\mu} \, \theta_{\nu'}.
\end{aligned}
\end{equation}
Thus, the only two tensorial objects the vector propagator may depend upon are: $\{\theta_\mu{} \, \theta_{\nu'}{}, \; \Lambda_\mu{}_{\nu'}{} \}$ or equivalently $\{\sigma_\mu{} \, \sigma_{\nu'}, \; \sigma_\mu{}_{\nu'}{}\}$.
\begin{figure}[htbp]
  \centering
\begin{tikzpicture}
\fill[gray!5!white, opacity = 0.3] (0,0) circle (2);
\fill[gray!30!white, opacity = 0.3,rotate = 30] (1.2,-0.1) -- (0.7,-0.25) -- (0.85,-0.75) -- (1.35,-0.6)-- cycle;
\fill[gray!30!white, opacity = 0.3,rotate = 30] (-1.85,-0.45) -- (-1.35,-0.6) -- (-1.15,-0.05) -- (-1.65,0.1) -- cycle;
\draw[gray,dashed,<->] (-2.2,0) -- (2.2,0);
\draw[gray,dashed,<->] (0,-2.2) -- (0,2.2);
\foreach \x in {-2,-1.5,-1,-0.5,0,0.5,1,1.5,2} \draw[thin, lightgray] (\x,-0.1) -- (\x,0.1);
\foreach \x in {-2,-1.5,-1,-0.5,0,0.5,1,1.5,2} \draw[thin, lightgray] (-0.1,\x) -- (0.1,\x);
\draw[dotted,rotate = 30] plot [smooth, tension=1.25] coordinates{ (-2,0) (0,-0.5) (2,0)};
\draw (0,0) circle (2);
\draw[rotate = 30,dotted] plot [smooth, tension=1.25] coordinates{ (-2,0) (0,0.5) (2,0)};
\draw[rotate = 30] plot [smooth, tension=1.25] coordinates{ (-0.375,-0.0625) (0,-0.13) (0.25,-0.1)};
\draw[dashed,rotate = 30] (-1.5,-0.25) -- (0,0) -- (1,-0.4);
\node at (0,0) [anchor = south east]{\small$\theta$};
\fill[rotate = 30] (1,-0.4) circle (0.05cm);
\draw[->,thick,rotate = 30] (1,-0.4) -- (0.5,-0.55);
\draw[dashed,->,thick,rotate = 30] (1,-0.4) -- (0.85,0.25);
\draw[rotate = 30] (1.2,-0.1) -- (0.7,-0.25) -- (0.85,-0.75) -- (1.35,-0.6)-- cycle;
\fill[rotate = 30] (-1.5,-0.25) circle (0.05cm);
\draw[->,thick,rotate = 30] (-1.5,-0.25) -- (-1,-0.45);
\draw[dashed,->,thick,rotate = 30] (-1.5,-0.25) -- (-1.25,0.315);
\draw[rotate = 30] (-1.85,-0.45) -- (-1.35,-0.6) -- (-1.15,-0.05) -- (-1.65,0.1) -- cycle;
\draw[blue,->,thick,opacity = 0.7,rotate = 30] plot [smooth, tension=1.25] coordinates{ (-1.5,-0.25)(-0.3,-0.5) (1,-0.4) };
\end{tikzpicture}
\quad 
\begin{tikzpicture}
\fill[gray!5!white, opacity = 0.3] (0,0) circle (2);
\fill[gray!30!white, opacity = 0.3,rotate=125] (1.2,-0.1) -- (0.7,-0.25) -- (0.85,-0.75) -- (1.35,-0.6)-- cycle;
\fill[gray!30!white, opacity = 0.3] (1.2,-0.1) -- (0.7,-0.25) -- (0.85,-0.75) -- (1.35,-0.6)-- cycle;
\fill[gray!30!white, opacity = 0.3,rotate=25] (-1.85,-0.45) -- (-1.35,-0.6) -- (-1.15,-0.05) -- (-1.65,0.1) -- cycle;
\draw[gray,dashed,<->] (-2.2,0) -- (2.2,0);
\draw[gray,dashed,<->] (0,-2.2) -- (0,2.2);
\foreach \x in {-2,-1.5,-1,-0.5,0,0.5,1,1.5,2} \draw[thin, lightgray] (\x,-0.1) -- (\x,0.1);
\foreach \x in {-2,-1.5,-1,-0.5,0,0.5,1,1.5,2} \draw[thin, lightgray] (-0.1,\x) -- (0.1,\x);
\draw (0,0) circle (2);
\draw[dotted] (0,0) -- (1,-0.4);
\fill (1,-0.4) circle (0.05cm);
\draw[->,thick] (1,-0.4) -- (0.5,-0.55);
\draw[dashed,->,thick] (1,-0.4) -- (0.85,0.25);
\draw (1.2,-0.1) -- (0.7,-0.25) -- (0.85,-0.75) -- (1.35,-0.6)-- cycle;
\draw[dotted,rotate=125] (0,0) -- (1,-0.4);
\fill[rotate=125] (1,-0.4) circle (0.05cm);
\draw[->,thick,rotate=125] (1,-0.4) -- (0.5,-0.55);
\draw[dashed,->,thick,rotate=125] (1,-0.4) -- (0.85,0.25);
\draw[rotate=125] (1.2,-0.1) -- (0.7,-0.25) -- (0.85,-0.75) -- (1.35,-0.6)-- cycle;
\draw[dotted,rotate=25] (0,0) -- (-1.5,-0.25);
\fill[rotate=25] (-1.5,-0.25) circle (0.05cm);
\draw[->,thick,rotate=25] (-1.5,-0.25) -- (-1,-0.45);
\draw[dashed,->,thick,rotate=25] (-1.5,-0.25) -- (-1.25,0.315);
\draw[rotate=25] (-1.85,-0.45) -- (-1.35,-0.6) -- (-1.15,-0.05) -- (-1.65,0.1) -- cycle;
\end{tikzpicture}
  \caption{"Canonical" choice of Tangent Spaces on the Sphere wrt Geodesics}\label{fig:CanonicalTangentSpaces}
  \begin{flushleft}\singlespacing \vspace{-0.5\baselineskip} 
  In the first image, normalised unit vectors are tangent to the geodesic. The parallel propagator carries the tangent space at one point to the other along the geodesic in the direction of these unit vectors. However, if such a geodesic was absent/not considered like in the second image, all arbitrary choices of tangent spaces would be indistinguishable.
  \end{flushleft}
\end{figure}
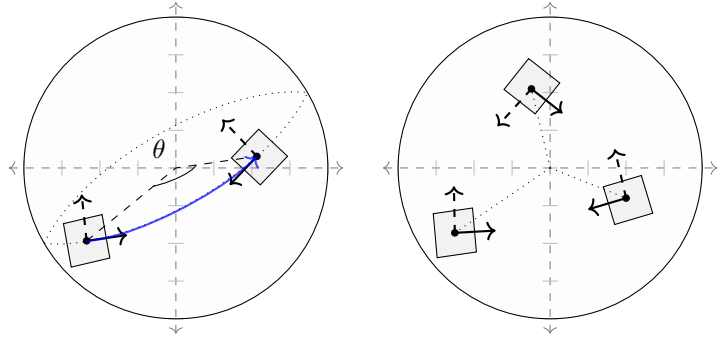\noindent
An orthonormal basis of eigenmodes of $A^\mu{}$ consists of transverse spherical vector harmonics satisfying $\nabla \cdot A = 0$ and longitudinal modes, which are normalised covariant derivations of scalar spherical harmonics. 
The longitudinal modes form a basis of the kernel of the massless vector Laplacian. 
The mass parameter $\Delta$ is 
\begin{equation}\label{eq:massivevectormassdef}
\begin{aligned}
  m^2 & = \vcentcolon (\Delta - 1)\,(\bar\Delta - 1), && \bdel = d - \del,   && \Delta \underset{m = 0}{=} 1
  \\ m^2 & = (\tfrac {d}{2} - 1)^2 + \nu^2, && m^2 + d = \Delta \, \bar\Delta + 1, && \nu \underset{m = 0}{=} \im \tfrac {d-2}{2},
\end{aligned}
\end{equation}
defined as such because the eigenvalue of $n$\textsuperscript{th} transverse mode of the massive Laplacian operator equals $(n+\Delta + 1) \, (n + \bar\Delta + 1)$. 
Given a complete orthonormal basis of transverse $A^\nu_{\omega}$ and longitudinal $\Phi^\nu_{\omega'} = \nabla^\nu{} \Phi_{\omega'}$ eigenvectors, labelled by $\omega, \, \omega'$, the propagator $G^{\nu \nu'}(\hat X, \, \hat Y)$ for a massive field can be written as a sum over transverse and longitudinal components:
\begin{equation}\label{eq:sumdefofvecpropmassive}
\begin{aligned}
  G^{\nu \nu'}(\hat X, \, \hat Y) & = \sum_{\omega} \frac {A^\nu_{\omega}(\hat X) \, \bar A^{\nu'}_{\omega}(\hat Y)} {\lambda_\omega + d + m^2} + \sum_{\omega'} \frac {\nabla^\nu{} \Phi_{\omega'} (\hat X) \, \nabla^{\nu'}{}\bar \Phi_{\omega'}(\hat Y)} {m^2}
  \\ & = \sum_{\omega} \frac {A^\nu_{\omega}(\hat X) \, \bar A^{\nu'}_{\omega}(\hat Y)} {\lambda_\omega + d + m^2} + \frac 1 {m^2} \,\nabla_{\hat X}^\nu{} \, \nabla_{\hat Y}^{\nu'}{} \, \lim_{\epsilon \rightarrow 0} G_{[\epsilon] \setminus 0}(\hat X, \, \hat Y),
\end{aligned}
\end{equation}
where $\nabla_{\hat X / \hat Y}$ is the covariant derivative at $\hat X/ \hat Y$ and the operators commute with each other when $\hat X \neq \hat Y$, $\Omega_{d+1}$ is the volume of $S^{d+1}$ and
\begin{equation}\label{eq:excludingzeromodefromlongitudinalvectorprop}
\begin{aligned}
  G_{[\Delta]  \setminus 0}(\hat X, \, \hat Y) & \vcentcolon = G_{[\Delta]}(\hat X, \, \hat Y) - \frac 1 {\Delta \, (d - \Delta) \, \Omega_{d+1}},
\end{aligned}
\end{equation}
in order to exclude the zero mode of $\Phi_{\omega'}$ from the sum because the corresponding longitudinal mode is absent. 
In the $R_\xi$ gauge (\cref{eq:masslessvectoractioninRgauge}), the propagator $G_{[1]}{}^{\nu \nu'}{}(\hat X, \, \hat Y)$ of a massless vector field is similarly written as a sum over transverse and longitudinal pieces:
\begin{equation}\label{eq:sumdefofvecpropmassless}
\begin{aligned}
  G_{[1]}{}^{\nu \nu'}(\hat X, \, \hat Y) & = \sum_{\omega} \frac {A^\nu_{\omega}(\hat X) \, \bar A^{\nu'}_{\omega}(\hat Y)} {\lambda_\omega + d} + \xi \, \sum_{w'} \frac {\nabla^\nu{} \Phi_{\omega'} (\hat X) \, \nabla^{\nu'}{}\bar \Phi_{\omega'}(\hat Y)} {\lambda_{\omega'}}
  \\ & = \sum_{\omega} \frac {A^\nu_{\omega}(\hat X) \, \bar A^{\nu'}_{\omega}(\hat Y)} {\lambda_\omega + d}  + \xi \,  \nabla_{\hat X}^\nu{} \, \nabla_{\hat Y}^{\nu'}{} \, \lim_{\epsilon, \, \epsilon' \rightarrow 0} \int_{\hat S}  G_{[\epsilon]  \setminus 0}(\hat X, \, \hat S)  \, G_{[\epsilon']  \setminus 0}(\hat S, \, \hat Y). 
\end{aligned}
\end{equation}
A useful result in this context is given in \cref{eq:Propagatorwithvertexinsertion} corresponding to the Feynman diagram \cref{fig:feynmandiagramofPropagatorwithvertexinsertion}. 
The transverse and longitudinal components of $G^{\nu \nu'}$ are independent and satisfy the ODE \cref{eq:vectorodesetup} away from $\theta = 0$ independently. 
Thus, the transverse components $G_{T}^{\nu \nu'}(\theta \neq 0)$ given by the sums in \cref{eq:sumdefofvecpropmassive,eq:sumdefofvecpropmassless}, that are as yet undetermined, can be found as the solution to the aforementioned ODE additionally satisfying the conditions $\nabla_\nu{} G_{T}^{\nu \nu'} = \nabla_{\nu'}{} G_{T}^{\nu \nu'} = 0$. 
\par
These definitions are, however, not enough to properly characterise the coincident point limit of $G^{\nu \nu'}$, in particular the trace of this object, making the formulation lacking for loop computations. 
This can be verified by explicit computation of some simple test cases, e.g. the inner product of the massive vector propagator on an odd-dimensional sphere with the $n = 1$ transverse modes, i.e. killing vectors, denoted by $\psi^{(1)}$: $\psi^{(1),(a)}_\nu = (\delta^a_\nu)$ and $\psi^{(1),(ab)}_\nu = (x^a{} \, \delta^b_\nu - x^b{} \, \delta^a_\nu)$ modulo normalisation, should be 
\begin{equation}
\begin{aligned}
  \left\langle\psi^{(1)}_\nu{}(x) \, G^\nu{}^{\nu'}(x,y) \, \psi^{(1)}_{\nu'}(y) \right\rangle \overset{?}{=} \frac {\frac{(d+1)(d+2)}{2}} {(\bdel+ 1)(\del + 1)},
\end{aligned}
\end{equation}
but for $d = 2, \, 4$, it deviates from the expected value by $- \frac {2}{\nu^2}, \, - \frac {3}{1 + \nu^2}$ respectively. 
The embedding space description of $G^{\nu \nu'}$ doesn't have this drawback, as has been explicitly verified for general transverse and longitudinal modes in \cref{eq:ModeIntVerificationofVectorPropagatorinEmbeddingSpaceFormalism}.

\subsection{Radial quotient of the massless vector action in embedding space}\label{sec:EmbeddingSpaceRepresentationof(GaugeFixed)MasslessVectorAction}
The action of a massless vector field $\embed A_I{}$ in $\RR^\DD$:
\begin{equation}\label{eq:flatspacemasslessvectoractionflatcoord}
\begin{aligned}
  \embed S^{[1]} & = \tfrac 1 4 \int_{X} \embed {\bar F}^I{}^J{}  \,\embed {F}_I{}_J{} = \tfrac 1 2  \int_{X} \embed {\bar A}^I{}  \,\Big( -  \partial^2 \, \delta_I{}_J{} +  \partial_J  \,\partial_I  \Big) \, \embed {A}^J{}
\end{aligned}
\end{equation}
exhibits a gauge symmetry $\delta\embed A_M = \partial_M \embed \Psi$, $\delta\embed {\bar A}_M = \partial_M \embed {\bar \Psi}$, parameterised by some scalar field $\embed \Psi,  \; \embed {\bar\Psi}$, that leaves the field strength $\embed F, \, \embed {\bar F}$ and action $\embed S^{[1]}$ unchanged. 
In the $R_\xi$ gauge, with gauge fixing action $\tfrac 1 {2 \, \xi}  \, (\embednab \cdot \embed {\bar A})  \, (\embednab \cdot \embed {A})$, the propagator $\embed G_I{}_J{}(X, \,Y)$ of a massless vector field is
\begin{equation}\label{eq:flatspaceRxigaugepropagatorofmasslessvector}
\begin{aligned}
  \embed G_I{}_J{}(X, \,Y) & = \int_{P} \Big( \delta_I{}_J{} - (1 - \xi)  \,\tfrac {P_I  \,P_J} {P^2} \Big) \, \frac {e^{- \im P \,(X-Y)}} {P^2}.
\end{aligned}
\end{equation}
The $2$-point function of the field strength $\embed F$ remains independent of the gauge parameter $\xi$:
\begin{equation}\label{eq:2pointfunctionofflatspacevectorfieldstrength}
\begin{aligned}
  & \left\langle\embed F_{IJ}{}(X) \, \embed F_{I'J'}{}(Y) \right\rangle = \int_{P} 4 \, P_[{}_I{} \, \delta_J{}_]{}_[{}_{J'}{} \, P_{I'}{_]{}} \, \frac {e^{- \im P \,(X-Y)}} {P^2}. 
\end{aligned}
\end{equation}
In spherical coordinates, with field strength $\embed F$ defined as
\begin{equation}\label{eq:fieldstrengthinsphericalcoordinates}
\begin{aligned}
  \embed {F}_\mu{}_\nu{} & = \embednab_\mu{} \embed A_\nu{} - \embednab_\nu{} \embed A_\mu{}, && \embed {F}_\mu{} \vcentcolon = \embed {F}_\mu{}_t{} = \embednab_\mu{} \embed A_t{} - \embednab_t{} \embed A_\mu{},
\end{aligned}
\end{equation}
the same action is given by
\begin{equation}\label{eq:flatspacemasslessvectoractionsphericalcoord}
\begin{aligned}
  \embed S^{[1]} & = \tfrac 1 2 \int_{X} \embed {\bar F}^\mu{}  \,\embed {F}_\mu{} + \tfrac 1 2 \embed {\bar F}^\mu{}^\nu{}  \,\embed {F}_\mu{}_\nu{} = \tfrac 1 2 \int_{X} \embed g^\mu{}^\nu{}  \,\embed g^t{}^t{} \, \embed {\bar F}_\mu{}  \,\embed {F}_\nu{} + \tfrac 1 2 \,\embed g^\mu{}^\sigma{}  \,\embed g^\nu{}^\rho{} \, \embed {\bar F}_\mu{}_\nu{}  \,\embed {F}_\sigma{}_\rho{} 
  \\ & = \tfrac 1 2 \int_{X} - \embed g_t{}_t{}  \,\embed {\bar A}^t{} \, \embednab^\rho{}  \,\embednab_\rho{}  \,\embed {A}^t{} + \embed {\bar A}^\mu{}  \,\embednab_t  \,\embednab_\mu{}  \,\embed {A}^t{} + \embed {\bar A}^t{}  \,\embednab_\nu{} \,\embednab_t  \,\embed {A}^\nu{} + \embed {\bar A}^\mu{}  \,\embed K_\mu{}_\nu{}  \,\embed {A}^\nu{} 
  \\ & = \tfrac 1 2 \int_t e^{\DD t} \int_{\hat X} - \embed {\bar A}^t{}  \,\nabla^2  \,\embed {A}^t{} + \embed {\bar A}^\mu{}   \,(\partial_t + d) \, \nabla_\mu  \,\embed {A}^t{} + \embed {\bar A}^t{}  \,(\partial_t + 2)  \,\nabla_\nu{}  \,\embed {A}^\nu{} + \embed {\bar A}^\mu{}  \,K_\mu{}_\nu{}  \,\embed {A}^\nu{} 
\end{aligned}
\end{equation}
where the spherical components of the Laplacian operators $\embed K_\mu{}_\nu{}, \, K_\mu{}_\nu{}$ are
\begin{equation}
\begin{aligned}
  \embed K_\mu{}_\nu{} & = - \embed g_\mu{}_\nu{}  \,(\embednab^t  \,\embednab_t + \embednab^\rho  \,\embednab_\rho) + \embednab_\nu  \,\embednab_\mu
  \\ K_\mu{}_\nu{} & = - g_\mu{}_\nu{}  \,\nabla^2 + \nabla_\nu  \,\nabla_\mu - g_\mu{}_\nu{} \, (\partial_t^2 + (d + 2)  \,\partial_t + 2  \,d).
\end{aligned}
\end{equation}
The gauge symmetry in these coordinates is expectedly generated by the transformations: 
\begin{equation}\label{eq:flatspacemasslessvectorgaugesymmetrysphericalcoord}
\begin{aligned}
  & \delta\embed A^\mu = \embednab^\mu  \embed \Psi, \quad  \delta\embed {A}^t{} = e^{- 2 t}  \,\partial_t  \,\embed \Psi = \embednab^t \embed \Psi, \quad  \delta\embed {\bar A}^\mu = \embednab^\mu \embed {\bar \Psi}, \quad  \delta\embed {\bar A}^t{} = e^{- 2 t} \, \partial_t \embed {\bar \Psi} = \embednab^t \embed {\bar \Psi}.
\end{aligned}
\end{equation}
Representing $\embed A^\nu{}, \, \embed A^t{}$ in terms of their radial Mellin transformed fields $A_{[\Delta]}^\nu{}, \, \chi_{[\Delta]}$, as defined in \cref{sec:ReextensiontoEmbeddingSpace}, the action in \cref{eq:flatspacemasslessvectoractionsphericalcoord} becomes
\begin{equation}
\begin{aligned}
  \embed S^{[1]} & 
  = \tfrac 1 2 \circint_{\Delta} \int_{t} e^{(\bdel + 1) t} \int_{\hat X} g_\mu{}_\nu{}  \,\Big(\nabla^\mu{}  \,\embed {\bar A}^t{} + (\bdel - 1)  \,\embed {\bar A}^\mu{}\Big) \, \Big( \nabla^\nu{} \, \chi_{[\Delta]} + (\Delta - 1)  \,{A}_{[\Delta]}^\nu{}\Big)
  \\ & \phantom{\tfrac 1 2 \circint_{\Delta} \int_{t} e^{(\bdel + 1) t} \int_{\hat X}} + \embed {\bar A}^\mu{}  \,\underbracket{\Big(- g_\mu{}_\nu{}  \,\nabla^2 + \nabla_\nu  \,\nabla_\mu\Big)}  \,{A}_{[\Delta]}^\nu{}
  \\ & = \tfrac 1 2 \circint_{\Delta} \int_{t} e^{(\bdel + 1) t} \int_{\hat X} \embed {\bar A}^t{}  \,( -\nabla^2 )  \,\chi_{[\Delta]} - (\Delta - 1)  \,\embed {\bar A}^t{}  \,\nabla_\nu{}  \,{A}_{[\Delta]}^\nu{} + (\bdel - 1)  \,\embed {\bar A}^\mu{}  \, \nabla_\mu  \,\chi_{[\Delta]}
  \\ & \phantom{\tfrac 1 2 \circint_{\Delta} \int_{t} e^{(\bdel + 1) t} \int_{\hat X}} + \embed {\bar A}^\mu{}  \,\underbracket{\Big(- g_\mu{}_\nu{}  \,\nabla^2 + \nabla_\nu  \,\nabla_\mu + g_\mu{}_\nu{} \, (\Delta - 1) \,(\bdel - 1)\Big)} \, {A}_{[\Delta]}^\nu{},
\end{aligned}
\end{equation}
where the bracketed terms are the massless and massive vector Laplacians on $S^{d+1}$ respectively, given in \cref{eq:massivevectoraction} with mass as defined in \cref{eq:massivevectormassdef}. 
By interpreting the remaining integral over $t$ as a Mellin transform of $\embed {\bar A}$, $\embed S^{[1]}$ can be represented as
\begin{equation}\label{eq:mellintransformedversionofmasslessactioninflatspace}
\begin{aligned}
  \embed S^{[1]} & 
  = \tfrac 1 2 \circint_{\Delta} \int_{\hat X}  \,g_\mu{}_\nu{} \, \Big(\nabla^\mu{}  \,{\bar \chi}_{[\bdel]} + (\bdel - 1)  \,{\bar A}_{[\bdel]}^\mu{}\Big)  \,\Big( \nabla^\nu{}  \,\chi_{[\Delta]} + (\Delta - 1)  \,{A}_{[\Delta]}^\nu{}\Big)
  \\ & \phantom{\tfrac 1 2 \circint_{\Delta} \int_{\hat X}} + {\bar A}_{[\bdel]}^\mu{}  \,\Big(- g_\mu{}_\nu{}  \,\nabla^2 + \nabla_\nu  \,\nabla_\mu\Big)  \,{A}_{[\Delta]}^\nu{}
\end{aligned}
\end{equation}
with the gauge symmetry in these new coordinates $(\Delta, \hat X)$ being
\begin{equation}\label{eq:flatspacemasslessvectorgaugesymmetries2}
\begin{aligned}
  & \delta A^\nu_{[\Delta]} = \nabla^\nu \Psi_{[\Delta]}, && \delta \chi_{[\Delta]}  = - (\Delta - 1)  \,\Psi_{[\Delta]}, && \delta \bar A^\mu_{[\bar \Delta]} = \nabla^\mu \bar \Psi_{[\bar\Delta]}, &&\delta \bar\chi_{[\bar \Delta]}  =  - (\bar \Delta - 1)  \,\bar \Psi_{[\bar \Delta]}.
\end{aligned}
\end{equation}

\subsubsection{BRST gauge fixing}\label{sec:vectorsBRSTgaugefixing}
A general form of the gauge fixing action is
\begin{equation}
\begin{aligned}
  \embed S^{[1]}_{\rm GF} & = \tfrac 1 {2} \int \Big(\bar \beta_1 \, \embednab_\mu \embed {\bar A}^\mu + \bar \beta_2 \, \embednab_t \embed {\bar A}^t\Big) \Big( \beta_1 \, \embednab_\mu \embed A^\mu + \beta_2 \, \embednab_t \embed A^t \Big) + \embed S^{[1]}_{\rm ghost}
  \\ \embed S^{[1]}_{\rm ghost} & = - \int \embed c^\dagger \Big( \bar \beta_1 \, \embednab_\mu \embednab^\mu \embed {\bar c} + \bar \beta_2 \, \embednab_t \embednab^t \embed {\bar c} \Big) + \embed {\bar c}^\dagger \Big( \beta_1 \, \embednab_\mu \embednab^\mu  \embed c + \beta_2\,  \embednab_t \embednab^t \embed c \Big),
\end{aligned}
\end{equation}
where $\embed c, \embed {\bar c}$ and $\embed c^\dagger, \embed {\bar c}^\dagger$ are ghost and anti-ghost fields. Upon integrating by parts,
\begin{equation}
\begin{aligned}
  \embed S^{[1]}_{\rm GF} & = - \tfrac {1} {2} \int \embed {\bar A}^t {K}'{}_t{}_t{} \embed {A}^t + \embed {\bar A}^t {K}'{}_t{}_\nu{} \embed {A}^\nu  +  \embed {\bar A}^\mu {K}'{}_\mu{}_t{} \embed {A}^t  + \embed {\bar A}^\mu {K}'{}_\mu{}_\nu{} \embed {A}^\nu + \embed S^{[1]}_{\rm ghost},
\end{aligned}
\end{equation}
where
\begin{equation}
\begin{aligned}
  & {K}'{}_t{}_t{} =  \bar \beta_2 \, \beta_1 \, (d+1) \, \partial_t  + \bar \beta_2 \, \beta_2 \,\partial_t \,(\partial_t + 1), \quad {K}'{}_\mu{}_\nu{}  = \bar \beta_1 \, \beta_1 \, \nabla_\mu \nabla_\nu  +  \bar \beta_1 \, \big(\beta_1 -  \beta_2 \big)\, g_\mu{}_\nu{} \, \partial_t, 
  \\ & {K}'{}_t{}_\nu{} = \bar \beta_2 \, \beta_1 \, \partial_t  \, \nabla_\nu , \quad {K}'{}_\mu{}_t{} = \big( \bar \beta_1 \, \beta_1 \,(d+2) + \bar \beta_1 \, \beta_2 \, \partial_t  \big) \, \nabla_\mu.
\end{aligned}
\end{equation}
Once again representing the fields $\embed {\bar A}, \, \embed A$ in terms of their radially Mellin transformed basis, the gauge fixing action in $(\Delta, \, \hat X)$ coordinates is
\begin{equation}
\begin{aligned}
  \embed S^{[1]}_{\rm GF} & = \tfrac {1} {2} \circint_{\Delta} \int_{\hat X} \bar \beta_2 \, (\Delta + 1) \, \big(\beta_1 \, (d+1) - \beta_2 \, \Delta \big) \, \bar \chi_{[\bar\Delta]} \, \chi_{[\Delta]}
  \\ & \phantom{\tfrac {1} {\xi} \circint_{\Delta} \int_{\hat X} } + \bar \beta_2 \, \beta_1 \, (\Delta + 1)  \, \bar \chi_{[\bar\Delta]} \, \nabla_\nu A_{[\Delta]}^\nu + \bar \beta_1 \, \big( \beta_1 \,(d+2) - \beta_2 \,  (\Delta + 1)   \big) \, \nabla_\mu {\bar A}_{[\Delta]}^\mu \, \chi_{[\Delta]}
  \\ &  \phantom{\tfrac {1} {\xi} \circint_{\Delta} \int_{\hat X} } - \bar \beta_1 \, \beta_1 \, {\bar A}_{[\Delta]}^\mu \, \nabla_\mu \,\nabla_\nu A_{[\Delta]}^\nu + \bar \beta_1 \, \big(\beta_1 -  \beta_2 \big) \, (\Delta + 1)\, g_\mu{}_\nu{} \, {\bar A}_{[\Delta]}^\mu \,  A_{[\Delta]}^\nu.
\end{aligned}
\end{equation}
Deciding to leave the mass term of $\bar A, \, A$ unchanged in the total gauge fixed action constrains $\embed S^{[1]}_{\rm GF}$ to either of the following:
\begin{equation}
\begin{aligned}
  \embed S^{[1]}_{\rm GF}\Big|_{\beta_1 = \beta_2} & = \tfrac {\beta_1} {2} \circint_{\Delta} \int_{\hat X} \bar \beta_2 \,  (\Delta + 1) \, (\bdel+1) \, \bar \chi_{[\bar\Delta]} \, \chi_{[\Delta]} - \bar \beta_1 \, {\bar A}_{[\Delta]}^\mu \, \nabla_\mu \,\nabla_\nu A_{[\Delta]}^\nu 
  \\ & \phantom{\tfrac {1} {\xi} \circint_{\Delta} \int_{\hat X} } + \bar \beta_2 \, (\Delta + 1)  \, \bar \chi_{[\bar\Delta]} \, \nabla_\nu A_{[\Delta]}^\nu + \bar \beta_1 \, (\bar\Delta + 1) \, \nabla_\mu {\bar A}_{[\Delta]}^\mu \, \chi_{[\Delta]}
  \\ \embed S^{[1]}_{\rm GF}\Big|_{\bar \beta_1 = 0} & = \tfrac {\bar \beta_2} {2} \circint_{\Delta} \int_{\hat X} (\Delta + 1)  \, \bar \chi_{[\bar\Delta]}\, \Big( \big(\beta_1 \, (d+1) - \beta_2 \, \Delta \big) \, \chi_{[\Delta]} + \beta_1  \, \nabla_\nu A_{[\Delta]}^\nu \Big).
\end{aligned}
\end{equation}
Upon requiring the cross terms $\chi$-$\bar A$ and $\bar \chi$-$A$ to vanish in the total gauge fixed action, $\embed S^{[1]}_{\rm GF}$ is further constrained to:
\begin{equation}\label{eq:generalcrosstermcancellinggaugefixingactionforvectors}
\begin{aligned}
  \embed S^{[1]}_{\rm GF}\Big|_{R_\xi} & = \tfrac {1} {2} \circint_{\Delta} \int_{\hat X}   (\Delta - 1) \, (\bdel+1) \, \bar \chi_{[\bar\Delta]} \, \chi_{[\Delta]} - \tfrac {(\bar\Delta - 1) } {(\bar\Delta + 1)} \, {\bar A}_{[\Delta]}^\mu \, \nabla_\mu \,\nabla_\nu A_{[\Delta]}^\nu
  \\ & \phantom{\tfrac {1} {\xi} \circint_{\Delta} \int_{\hat X} } + (\Delta - 1)  \, \bar \chi_{[\bar\Delta]} \, \nabla_\nu A_{[\Delta]}^\nu + (\bar\Delta - 1) \, \nabla_\mu {\bar A}_{[\Delta]}^\mu \, \chi_{[\Delta]}.
\end{aligned}
\end{equation}
The total gauge fixed action, $\embed S^{[1]} + \embed S^{[1]}_{\rm GF}\Big|_{R_\xi} $, represented in $(\Delta, \, \hat X)$ coordinates at $\Delta = 1$ comes to resemble the massless vector action on $S^{d+1}$ in the $R_\xi$ gauge, \cref{eq:masslessvectoractioninRgauge}, specifically at $\xi = \tfrac {d}{d - 2}$.

\subsubsection{Reduction to massive field}\label{sec:massivevecpropproofjustification}
Evidently, a gauge choice of \cref{eq:mellintransformedversionofmasslessactioninflatspace} in which the scalar fields $\bar \chi, \, \chi$ (and when $\Delta, \, \bar\Delta \neq 1$, equivalently $\embed {\bar A}_{t}, \, \embed A_{t}$) vanish, the gauge fixed action will take the form of \cref{eq:massivevectoraction}, representing a massive vector field on $S^{d+1}$. 
However, the radial gauge ($\embed A_t = 0$) does not completely fix the gauge symmetry of \cref{eq:flatspacemasslessvectoractionsphericalcoord}, instead reducing it to $\embed A_\mu \sim \embed A_\mu + \embednab_\mu \embed \Psi(\hat X)$, requiring some additional condition to completely fix this gauge symmetry, the most natural being $\embednab^\mu{} \embed A_\mu = 0$. 
Regardless of the choice of this condition, the gauge invariant quantity $\embed F_\mu$ will remain the same,
\begin{equation}
\begin{aligned}
  \embed F_\mu^{\rm rad} &  = - \partial_t \embed A_\mu \rightarrow - \partial_t \,\big(\embed A_\mu + \embednab_\mu \embed \Psi(\hat X)\big) = - \partial_t \embed A_\mu 
  = \circint_{\Delta} (\Delta-1) \, e^{-(\Delta-1) \, t} \, A_{[\Delta]}{}_\mu{} 
\end{aligned}
\end{equation}
consequently allowing its evolution to be studied in terms of a propagator in any gauge. 
Changing back to flat coordinates for convenience,
\begin{equation}\label{eq:FandSformassivevectorinradialgaugedefinition}
\begin{aligned}
  \embed F_I^{\rm rad} & = \embed{\hat e}^{\mu}_I \, \embed F_\mu^{\rm rad} = \circint_{\Delta} e^{-\Delta \, t} \, F_{[\Delta]}{}_I{}, \quad F_{[\Delta]}{}_I{} \vcentcolon= {\hat e}^{\mu}_I \, (\Delta-1) \, A_{[\Delta]}{}_\mu{} =  (\Delta-1) \, A_{[\Delta]}{}_I{}, 
  \\ \embed S^{[1]}_{\rm rad} & = \tfrac 1 2 \int_{X} \embed {\bar F}^\mu{}  \,\embed {F}_\mu{} + \cdots = \tfrac 1 2 \int_{t} \int_{\hat X} e^{d \, t} \, g^\mu{}^\nu{} \, \embed {\bar F}_\mu{}  \,\embed {F}_\nu{} + \cdots = \tfrac 1 2 \int_{X} \embed {\bar F}^{\rm rad}_I{} \, e^{- 2 \, t} \, \delta^I{}^J{}  \,\embed {F}^{\rm rad}_J{} + \cdots,
\end{aligned}
\end{equation}
its associated $2$-point function, found using \cref{eq:2pointfunctionofflatspacevectorfieldstrength}, is
\begin{equation}
\begin{aligned}
  \embed G^{\rm rad}_{II'}(X, \, Y) = \left\langle\embed F^{\rm rad}_{I}(X) \, \embed F^{\rm rad}_{I'}{}(Y) \right\rangle = \int_{P} 4 \, X^J \, P_[{}_I{} \, \delta_J{}_]{}_[{}_{J'}{}  \, P_{I'}{_]{}} \, Y^{J'} \, \frac {e^{- \im P \,(X-Y)}} {P^2}. 
\end{aligned}
\end{equation}
Thus, in the same vein as \cref{eq:greensfunctiondefinition,eq:greensfunctiondefinitionMellinised,eq:scalarpropderivationstep1,eq:scalarpropderivationalmostcomplete} of the scalar case, the propagator equation 
\begin{equation}
\begin{aligned}
  \embed F^{\rm rad}_{I'}(Y) & = \int_{X} \embed G^{\rm rad}_{I'I} \, e^{- 2 \, t} \, \delta^I{}^J{}  \, \embed F^{\rm rad}_{J}(X) 
\end{aligned}
\end{equation}
represented in terms of Mellin transformed fields
\begin{equation}
\begin{aligned}
  \circint_{\Delta} e^{- \Delta \, s} \,  F_{[\Delta]}{}_{I'}{}(\hat Y) & 
  = \circint_{\Delta} e^{-\Delta \, s} \int_{\hat X} \, F_{[\Delta]}^I(\hat X) 
  \\ & \phantom{\circint_{\Delta} e^{-\Delta \, s} \int_{\hat X}}\times \int_{t} e^{(d - \Delta + 1) \, t} \int_{P} 4 \, \hat X^J \, P_[{}_I{} \, \delta_J{}_]{}_[{}_{J'}{}  \, P_{I'}{_]{}} \, \hat Y^{J'} \, \frac {e^{- \im P \, (e^t \hat X - \hat Y)}} {P^2} 
\end{aligned}
\end{equation}
distributes over $\Delta$ into $|Y| = e^s$ independent parts to give 
\begin{equation}\label{eq:propofFofvectoronsphere}
\begin{aligned}
  \left\langle F_{[\Delta]}{}_{I}{}(\hat X) \,  F_{[\Delta]}{}_{I'}{}(\hat Y) \right\rangle & = \int_{t} e^{(d - \Delta + 1) \, t} \int_{P} 4 \, \hat X^J \, P_[{}_I{} \, \delta_J{}_]{}_[{}_{J'}{}  \, P_{I'}{_]{}} \, \hat Y^{J'} \, \frac {e^{- \im P \, (e^t \hat X - \hat Y)}} {P^2},
\end{aligned}
\end{equation}
and, upon making the identification \cref{eq:FandSformassivevectorinradialgaugedefinition}, the propagator of a massive vector field on $S^{d+1}$ as described by \cref{eq:mellintransformedversionofmasslessactioninflatspace} in the radial gauge (further detailed \cref{sec:MassiveVectorPropagator}).

\subsubsection{Reduction to massless field}
A massless vector field $A$ on $S^{d+1}$ is represented by \cref{eq:mellintransformedversionofmasslessactioninflatspace} in a gauge which results in the decoupling of $\bar A, \, A$ and $\bar \chi, \, \chi$. 
However, at $\Delta = 1$, i.e. when the corresponding vector field $A_{[\Delta = 1]}$ on $S^{d+1}$ is massless, neither is the gauge condition $\chi = 0$ proper nor is the identification \cref{eq:FandSformassivevectorinradialgaugedefinition} valid. 
Alternately, applying the tangentiality condition on the field $\embed {A}$, $X \cdot \embed A = 0$, decouples the radial $\embed {A}_t$ and spherical $\embed {A}_\mu$ components of the vector field by enforcing $\embed {A}_t = 0$ identically on the projected field,
\begin{equation}
\begin{aligned}
  & \embed {A}_I^{\rm tan} =  \embed {\hat t}^J_I \, \embed {A}_J,
\end{aligned}
\end{equation}
where $\embed {\hat t}$ is the tangential projection operator, as defined in \cref{eq:tangentialprojectionoperator}, and $\embed F_\mu$ takes the form:
\begin{equation}\label{eq:FandSformasslessvectorintangentialgaugedefinition}
\begin{aligned}
  & \embed F_\mu\Big|_{\embed A = \embed {A}^{\rm tan}} = \embednab_\mu \embed A_t - \embednab_t \embed A_\mu \Big|_{\embed A_t = 0} = - \embednab_t \embed A_\mu = - \partial_t \embed A_\mu + \embed A_\mu = \circint_{\Delta} \Delta \, e^{-(\Delta-1) \, t} \, A_{[\Delta]}{}_\mu{},
\end{aligned}
\end{equation}
where the slice of $\embed F_\mu$ at $\Delta = 1$ is the radial average. 
Proceeding similarly to \cref{sec:massivevecpropproofjustification}, the relevant $2$-point function of $\embed F_\mu$ in flat coordinates, constructed in terms of \cref{eq:flatspaceRxigaugepropagatorofmasslessvector}, is 
\begin{equation}
\begin{aligned}
  \embed G^{\rm rad, \, tan}_{II'}(X, \, Y)  & = \left\langle\embed F^{\rm rad}_{I}(X) \, \embed F^{\rm tan}_{I'}{}(Y) \right\rangle = X^J \, ( \delta^K_I \partial_{X^J} - \delta^K_J \partial_{X^{I}}) \,Y^{J'} \, \partial_{Y^{J'}} \, \embed G_K{}_{K'}{}\,\embed {\hat t}^{K'}_{I'}
  \\ & = \int_{P} \Big(|PX| \, |P Y| \, \delta_I{}_{K'}{} - |P Y| \,P_I{}\, X_{K'}{} \Big) \,\Big(\delta^{K'}_{I'} - \tfrac {Y^{K'}\, Y_{I'}}{|Y|^2}\Big) \, \frac {e^{- \im P \,(X-Y)}} {P^2}
\end{aligned}
\end{equation}
and its reduction to the sphere at $\Delta = 1$ is
\begin{equation}\label{eq:Fdelta1Fdelta12pointfunction}
\begin{aligned}
  \left\langle F_{[\Delta = 1]}{}_{I}{}(\hat X) \,  F_{[\Delta = 1]}{}_{I'}{}(\hat Y) \right\rangle & = \int_{t} e^{d \, t} \int_{P} \Big(P_R{}\, \hat X^R \, \delta_I{}_{K'}{} - P_I{}\, \hat X_{K'}{} \Big) \,P_{R'} \, \hat Y^{R'} 
  \\ & \phantom{= \int_{t} e^{d \, t} \int_{P} } \times \Big(\delta^{K'}_{I'} - \hat Y^{K'}\, \hat Y_{I'}\Big) \, \frac {e^{- \im P \, (e^t \hat X - \hat Y)}} {P^2}.
\end{aligned}
\end{equation}
The identifications \cref{eq:FandSformassivevectorinradialgaugedefinition,eq:FandSformasslessvectorintangentialgaugedefinition} give the correspondence of \cref{eq:Fdelta1Fdelta12pointfunction} to the propagator of a massless vector field on $S^{d+1}$ in the $R_{\xi = \frac {d} {d-2}}$ gauge, where the effective gauge parameter $\xi$ is found by comparing the masslike term $\embed {\bar F}^\mu \, \embed {F}_\mu \Big|_{\Delta = 1} \sim (d - 2) \, \bar A^\mu{} \, A_\mu{}$ in this radial gauge-tangential projection hybrid setup to that in \cref{eq:masslessvectoractioninRgauge}, i.e. $\tfrac {d}{\xi}  \, \bar A^\mu{} \, A_\mu{}$, matching \cref{eq:generalcrosstermcancellinggaugefixingactionforvectors} (further detailed in \cref{sec:MasslessVectorPropagator}).

\subsection{Massive vector propagator}\label{sec:MassiveVectorPropagator}
The propagator of the massive vector field with mass parameter as given in \cref{eq:massivevectormassdef}, as originating from the $2$-point function of the field strength in $(\Delta, \, \hat X)$ coordinates, \cref{eq:propofFofvectoronsphere}, is
\begin{equation}\label{eq:vectorprop}
  \begin{aligned}
    & G_I{}_{I'}(\hat X, \, \hat Y) = \mathcal G_{II'} \circ G_{[\Delta]}(\hat X, \, \hat Y), \qquad G_\nu{}_{\nu'}{}(\hat X, \, \hat Y) = \hat {e}_\nu^I \, \hat {e}_{\nu'}^{I'}  \, G_I{}_{I'}(\hat X, \, \hat Y)
  \\ & \mathcal G_{II'} = \frac {4 \, \lambda \, \mu} {(\bar \Delta - 1)\, (\Delta - 1)} \, \hat X^J \, P_[{}_I{} \, \delta_J{}_]{}_[{}_{J'}{}  \, P_{I'}{_]{}} \, \hat Y^{J'} 
  \\ & \phantom{\mathcal G_{II'} } = \frac {4 \, \lambda \, \mu} {(\bar \Delta - 1)\, (\Delta - 1)} \, \Big( |P X|\,|P Y|\, \delta_{I}{}_{I'}{} - |P X|\, Y_{I}{} \, P_{I'}{}  - P_{I}{} \, X_{I'}{} \, |P Y|   + |X Y|\, P_{I}{} \, P_{I'}{} \Big),
  \end{aligned}
\end{equation}
following the notation in \cref{eq:operatorwithscalarprop}. 
The operator $\mathcal G_{II'}$ is manifestly bilocally tangential $(X^I \, \mathcal G_{II'} = \mathcal G_{II'} \, Y^{I'}  = 0)$ and transverse $(\partial^{I}\mathcal G_{II'} = \partial^{I'} \mathcal G_{II'} = 0)$. 
Its scale invariance, as required by \cref{eq:operatorwithscalarprop}, can be made more manifest in a variety of ways:
\begin{equation}
\begin{aligned}
  \mathcal G_{II'} & = \frac 1 {(\Delta-1)(\bar \Delta - 1) }  \Big( \theta_{\lambda} \, \theta_{\mu} \, \delta_{II'} - \theta_{\lambda} \, Y_{I} \, \partial_{Y^{I'}} - \theta_{\mu} \, X_{I'}  \, \partial_{X^I} + |X Y| \, \partial_{X^I} \, \partial_{Y^{I'}} \Big)
  \\ & = \frac {\bar \Delta \, \Delta} {(\bar \Delta - 1)\, (\Delta - 1)} \, \Big(\delta_{I}{}_{I'}{} - \tfrac {1} {|PY|}\, Y_{I}{} \, P_{I'}{}  - \tfrac {1} {|PX|}\, P_{I}{} \, X_{I'}{}   + \tfrac {1} {|PX| \, |PY|}\, |X Y|\, P_{I}{} \, P_{I'}{} \Big),
\end{aligned}
\end{equation}
where \cref{eq:negativepowerofPX} was used. 
Although $\mathcal G_{II'}$ is defined over the entire embedding space, it is dependent on {\it only} $(\hat X, \, \hat Y)$, i.e. $\mathcal G_{II'} (X,\,Y) = \mathcal G_{II'} (\hat X,\,\hat Y)$. 
Purely in terms of position space operators, it is 
\begin{equation}\label{eq:positionspaceoperatorformassivevectorprop}
\begin{aligned}
  \mathcal G_{II'} & = \frac {1} {(\bar \Delta - 1)\, (\Delta - 1)} \, \hat X^J \, \partial_[{}_{\hat X^I}{} \, \delta_J{}_]{}_[{}_{J'}{}  \, \partial_{\hat Y^{I'}}{_]{}} \, \hat Y^{J'}.
\end{aligned}
\end{equation}
However, since the derivatives in the above do not commute with the rest of the expression, in practice it is more useful to consider an equivalent formulation based on \cref{eq:feynmanintegraloperatorreduction}:
\begin{equation}\label{eq:massivevectorpropformthatgetsused}
\begin{aligned}
  \mathcal G_{II'} & = \frac 1 {(\bar \Delta - 1) \, (\Delta - 1)} \Big( \bar \Delta \, \Delta \, \delta_{I}{}_{I'}{}  + 2 \im \, \bar \Delta\, \mu \, Y_I{} \, P_{I'}{} - 2 \im \, \Delta  \, \lambda \, P_I{} \, X_{I'}{} 
  \\ & \phantom{= \frac 1 {(\bar \Delta - 1) \, (\Delta - 1)} \Big( } + \tfrac {2 \im \, \lambda \, \mu} {\sqrt{\alpha_X \, \alpha_Y}} \,P_I{} \, P_{I'}{} \,  \partial_\beta\, e^{- 2 \im \, \sqrt{\alpha_X \, \alpha_Y} \, \beta \, |X Y| } \Big|_{\beta = 0} \Big).
\end{aligned}
\end{equation}

\subsection{Massless vector propagator}\label{sec:MasslessVectorPropagator}
The propagator of the massless vector field, i.e. with mass parameter $\Delta = 1$, in the $R_\xi$ gauge,  $\xi =  \tfrac {d} {d-2}$, given in \cref{eq:masslessvectoractioninRgauge}, as implied by \cref{eq:Fdelta1Fdelta12pointfunction}, is
\begin{equation}\label{eq:masslessvectorpropagatormainformula}
\begin{aligned}
  G^{(o)}_{II'}(\hat X, \, \hat Y) & = \mathcal G^{(o)}_{II'} \circ G_{[\Delta = 1]}(\hat X, \, \hat Y), \qquad G^{(o)}_{\nu\nu'}{}(\hat X, \, \hat Y) = \hat {e}_\nu^I \, \hat {e}_{\nu'}^{I'}  \, G^{(o)}_{II'}(\hat X, \, \hat Y)
  \\ \mathcal G^{(o)}_{II'} & = \frac {4 \,\lambda \,\mu} {d-2} \, |PY| \, \Big( |PX| \,\delta_I{}_{I'} - P_I{} \, X_{I'}  -  \tfrac {|PX|}{|Y|^2} \, Y_{I}\, Y_{I'} + \tfrac {|XY| }{|Y|^2} \, P_I{} \, Y_{I'}\Big). 
\end{aligned}
\end{equation}
The last $2$ terms of $\mathcal G^{(o)}_{II'}$ will be irrelevant in proper formulations of closed Feynman integrals containing this propagator because they will, by default, be paired with tangential objects. 
Its tangentiality is obvious: $X^I \, \mathcal G^{(o)}_{II'} = \mathcal G^{(o)}_{II'}\, Y^{I'} = 0$. 
Some reformulation akin to the massive case confirms its scale invariance too:
\begin{equation}
\begin{aligned}
  \mathcal G^{(o)}_{II'} & = \frac {d-1} {d-2}  \, \Big(  \delta_I{}_{I'}{}  - \frac {P_I{} \, X_{I'}{}} {|PX|}  - \frac {Y_{I} \, Y_{I'}}{|Y|^2} +  \frac {|XY| }{|Y|^2 |PX|} \, P_I{} \, Y_{I'}\Big).
\end{aligned}
\end{equation}
Though less pleasing to the eye, it is more practical to use in Feynman integrals as:
\begin{equation}\label{eq:masslessvectorpropformthatgetsused}
\begin{aligned}
  \mathcal G^{(o)}_{II'} & = \frac {1} {d-2}  \, \Big( (d-1) \, \delta_I{}_{I'}{}  - 2 \im \, \lambda   \, P_I{} \, X_{I'}{} 
  \\ & \phantom{ = \frac {1} {d-2}  \, \Big( } \underbracket{- (d-1) \,\frac {Y_{I} \, Y_{I'}}{|Y|^2} - \frac {\lambda}{\sqrt{\alpha_X \, \alpha_Y} } \, \frac {P_I{} \, Y_{I'}} {|Y|^2} \, \partial_\beta\, e^{- 2 \im \, \sqrt{\alpha_X \, \alpha_Y} \, \beta \, |X Y| } \Big|_{\beta = 0}}_{\text{vanishing in closed loops}} \Big).
\end{aligned}
\end{equation}

\subsubsection{\texorpdfstring{$S^3$}{3-Sphere}}\label{sec:masslessvectoronS3}
For $d=2$, i.e. on $S^3$, the massless propagator is constructed in the $\xi \rightarrow \infty$ limit of the $R_\xi$ gauge in embedding space:
\begin{equation}\label{eq:masslessvectorpropagatord2}
\begin{aligned}
  \mathcal G^{(o)}_{II'} & =  \lim_{\xi \rightarrow 0} \Big( \delta^{K}_{I} - \frac {X^{K} \, X_{I}}{|X|^2} \Big) \, \Big( \delta_{KK'} + \frac {P_K \, P_{K'}} {\xi \, P^2} \Big) \, \Big( \delta^{K'}_{I'} - \frac {Y^{K'} \, Y_{I'}}{|Y|^2} \Big).
\end{aligned}
\end{equation}

\subsection{Position space forms of vector propagators}\label{sec:positionspaceformofvectorprops}
The position space form of the massive vector propagator, $G_{\nu\nu'}{}(\hat X, \, \hat Y)$ in \cref{eq:vectorprop}, can be found by applying the operator form of $\mathcal G_{II'}$ in \cref{eq:positionspaceoperatorformassivevectorprop} on the scalar propagator \cref{eq:scalarprophypergeometricform}. 
It is useful to recall that derivatives of the hypergeometric function ${}_2F_1$ simply induce shifts to its parameters, written concisely by using Pochhammer symbols \cref{eq:pochhammersymboldef} as $\partial^{n}_x \, {}_2F_1(a, \, b,\, c,\, x) =  \frac {a^{(n)} \, b^{(n)}} {c^{(n)}} \, {}_2F_1(a+n,\, b+n,\, c+n,\, x)$.
Working through the details, the position space {\it propagator}, in terms of the geodesic distance $\theta = \cos^{-1} (\xdotY)$, the unit vectors defined along it \cref{eq:positionspaceunitvectors,eq:positionspaceunitvectors2}, and the massive scalar propagator $G_{[\Delta]}$ (where the effective scalar mass $m_{\rm sc}^2 = \del \, \bdel = m_{\rm vec}^2 + d - 1$), is found to be 
\begin{equation}
\begin{aligned}
  G_{\mu\nu'}(\xdotY) & = \underbracket{\tfrac {1} {m^2} \, \Big( (\Lambda_{\mu\nu'} + \theta_\mu{} \, \theta_{\nu'}) \, (\sin^2 \theta \, \partial_{\xdotY} - \xdotY) + \theta_\mu{} \, \theta_{\nu'} \Big) \, \partial_{\xdotY} G_{[\Delta]}}_{\text{matching Wightman function of \cite{Allen:1985wd} away from }\theta = 0}  
  \\ & + \tfrac {1} {m^2} \, \sigma_\mu{}_{\nu'}{} \, \underbracket{(-\nabla^2 + \bdel \, \del) \, G_{[\Delta]}}_{\rm \cref{eq:greensfunctiondefinitionsphere}},
\end{aligned}
\end{equation}
with the second line showing the expected position space $\delta$-function. 
The explicit form of the scalar propagator was not used and so the different gauges of \cref{sec:EquivalentFormsoftheScalarPropagator} can be directly translated into this context too. 
In the same way, the massless vector propagator, \cref{eq:masslessvectorpropagatormainformula}, in position space is found to be
\begin{equation}
   G^{(o)}_{\mu \nu'} = \left( \frac {d-1} {d-2}\, \sigma_{\mu \nu'} + \frac {1} {d-2} \, \sigma_\mu \, \sigma_{\nu'} \, \partial_\sigma \right) \, G_{[\Delta = 1]}.
\end{equation}
On $S^3$, \cref{eq:masslessvectorpropagatord2} takes the rather formal form: 
\begin{equation}
  G^{(o)}_{\mu \nu'} = \sigma_\mu{}_{\nu'}{} \, G_{[\Delta = 1]} + \lim\limits_{\del, \bdel \rightarrow 0} \hat\delta_\mu{}_{\nu'}, \quad \text{where} \; \hat\delta_\mu{}_{\nu'} \prop (\sigma_\mu{}_{\nu'}{} \, \partial_\sigma + \sigma_\mu{}\, \sigma_{\nu'}{} \, \partial_\sigma^2 ) \, G_{[\Delta]}.
\end{equation}

\section{Feynman integrals on the sphere as generalized Euler integrals}\label{ch:dSFeynmanIntegrals}
Starting from as early as \cite{Christensen:1979iy}, all the way up to \cite{Senatore:2009cf,Marolf:2010zp,Marolf:2011sh,Polyakov:2012uc,Giombi:2015haa,Fei:2015oha}, practically culminating in \cite{Anninos:2020hfj}, which gives closed form expressions for $1$-loop integrals of arbitrary field content, higher than $1$-loop integrals on a de Sitter background have been recognised to be particularly challenging, with some recent results at $2$-loops being \cite{Kamenshchik:2021tjh,Cacciatori:2024zrv}. 
On the other hand, approaches to higher loop Feynman integral computations in flat space have flowed towards the idea of solving differential equations to do integrals, \cite{Chetyrkin:1981qh,Tkachov:1981wb,Harlander:1998dq,Gehrmann:1999as,Broadhurst:1998rz,Broadhurst:1991fi,Kotikov:1991hm,Caffo:1998du,Kotikov:1991pm,Baikov:1996iu,Laporta:2000dsw,Czakon:2004bu,Vermaseren:2005qc,Baikov:2016tgj,Herzog:2017ohr,Luthe:2017ttg,Tkachov:1996wh}, eventually, as predicted by \cite{Kashiwara1976OnAC}, converging into the Euler integral-GKZ ideal-$\mathcal A$-hypergeometric language, \cite{1977387,Bogner:2007mn,Kalmykov:2008ofy,KALMYKOV2012103,nasrollahpoursamami2016periods},  with increasingly higher loop computations (both analytic and numerical) being explored using this technology, \cite{Bitoun:2017nre,Klausen:2019hrg,delaCruz:2019skx,Kalmykov:2020cqz,Klausen:2021yrt,Mizera:2021icv,Fevola:2023fzn}.
This is greatly supported by the Lee-Pomeransky representation of Feynman integrals, \cite{Lee:2013hzt}, of which the formulation suggested in this work is reminiscent. 
\par
\Cref{sec:EmbeddingSpaceFormulationofFeynmanIntegralsonSd+1Scalar} explains the construction of scalar (loop) integrals and provides a systematic approach to converting them into generalized Euler integrals. 
Each integral is described by a single polynomial (the denominator in the integrand of \cref{eq:gaugefixednormalisedofscalarfeynmanint}, which is factorisable if the diagram is reducible), in $2 \, n_P$ variables ($n_P$ being the total number of propagators). 
Specifically, it is the determinant of a matrix, \cref{eq:Uparsetmatrixform}, encoding the graphical structure of the Feynman diagram, making its origin similar to the first Symanzik polynomial in \cref{eq:LeePomeranskyrepresentationstillgaussian}.
\par
Though currently only valid for spin-$1$ fields, \cref{sec:HigherSpinFeynmanIntegralsSetup} describes how to modify the formulation with a perturbing matrix, \cref{eq:higherspinUparset}, to account for terms observed in higher spin computations. 
This results in a ``master'' integral form which, granting that higher spin propagators can also be brought into the scalar-based embedding space form, \cref{eq:operatorwithscalarprop}, can be used to find all spin Feynman integrals with the same incidence matrix.
\Cref{sec:GeneralisedCorrelationFunctions} extends the previous setup to generalized correlation functions by accounting for possible external legs in the Feynman diagrams. 
The additional component appearing in \cref{eq:generalizedcorrelationfunctionaseulerintegral} becomes the analogue of the second Symanzik polynomial in \cref{eq:LeePomeranskyrepresentationstillgaussian}.
\par
The proposed procedure is implemented in \cref{sec:ExplicitConstructionofScalarFeynmanIntegrals} for all scalar diagrams upto $3$-loops and some vector diagrams that are representative of their relation to scalar integrals. 
In particular, vector integrals are found to be sums over corresponding scalar integrals.

\subsection{Scalar Feynman integrals}\label{sec:EmbeddingSpaceFormulationofFeynmanIntegralsonSd+1Scalar}
A general scalar Feynman integral on $S^{d+1}$ with $n_P$ propagators and $n_V$ internal vertices ($n_F = n_P + n_V$), takes the form 
\begin{equation}
\begin{aligned}
  \mathcal I_{F} & = \int^{S^{d+1}}_{\hat X} {\int^{\RR^\DD}_{P}} \, G_{F} (\hat X) = \Xint{\times}^{\RR^\DD}_{X} \frac {\de^{\DD} X}{|X|^{\DD}} \, \int^{\RR^\DD}_{P} \, G_{F} (X),
\end{aligned}
\end{equation}
where $G_{F}$ is the set/product\footnote{The multi-index notation is used wherever applicable. So here $G_{F}$ is also used to denote $\prod G_i$ for all $G_i \in G_F$.} of propagators forming the Feynman diagram. 
When each propagator in $G_{F}$ is represented in terms of \cref{eq:scalarpropunfixed}, upon Schwinger parameterisation, $\mathcal I_{F}$ can be consistently simplified to a gaussian integral over $W = \{P, \, X\}$: \def\parset{\varsigma}
\begin{equation}\label{eq:generalformofscalarFeynmanintegral}
\begin{aligned}
  \mathcal I_{F} & = \frac {\mathcal N_{F}} {\Vol \mathcal G} \,\int^*_{\parset} {\int^{\RR^\DD}_{W}} \, \lambda^{\bar\Delta} \, \mu^{\Delta} \, \tau^{\eta} \, \alpha^{\Delta_{X}} \, e^{- \transpose{W} \, U(\parset)  \, W},
\end{aligned}
\end{equation}
where $\mathcal G = (\RR^*)^{n_F}$ is the group of scale transformations, the set of integration variables is $\parset = \{\lambda_{1, \, \cdots \, n_P}, \, \mu_{1, \, \cdots\,n_P}, \, \tau_{1, \, \cdots\,n_P}, \, \alpha_{1, \, \cdots \, n_V}\}$, $\eta$ is a parameter generalising the scalar propagator, given in \cref{eq:scalarpropunfixedgeneral}, by default set to $1$, $\Delta$ are the mass parameters, $\Delta_X$ are the weights associated with internal vertices $X$, $\mathcal N_{F}$ is the normalisation constant, and $U(\parset)$ is a symmetric $n_F \times n_F$ matrix, referred to as the weighted incidence matrix, encoding the Feynman diagram $F$. $U(\parset)$ is constructed according to the following rules:
\begin{enumerate} 
  \item $U_{i, \, i} = \tau_i$ for $i \in \{1, \cdots n_P\}$ and $U_{n_P + j, \, n_P + j}{} = \alpha_i$ for $j \in \{1, \cdots n_V\}$.
  \item If propagator $P_i$ originates from vertex $X_j$ ($X_j \rightarrow P_i$), then $U_{i, \,n_P + j} = U_{n_P + j, \, i} = i \, \lambda_i$.
  \item If propagator $P_i$ terminates at vertex $X_j$ ($ X_j \leftarrow P_i$), then $U_{i, \,n_P + j} = U_{n_P + j, \, i} = - i \, \mu_i$.
  \item All other elements of $U$ are $0$.
\end{enumerate}
Given its symmetric form, it may better visualised as
\begin{equation}\label{eq:Uparsetmatrixform}
\begin{aligned}
  U(\parset) & = \left(\begin{array}{c|c}
              \tau_{n_P} & \im L(\lambda, \, \mu)  \\ 
              \hline
              \im \transpose{L}(\lambda, \, \mu)  & \alpha_{n_V}  \\ 
            \end{array} \right),
\end{aligned}
\end{equation}
where $\tau_{n_P}$ and $\alpha_{n_V}$ are $(n_P \times n_P)$ and $( n_V \times n_V)$ weighted diagonal matrices, and $L(\lambda, \, \mu)$, referred to as the incidence matrix, is a $(n_P \times n_V)$ matrix representing the propagator and vertex connections, constructed as follows:
\begin{enumerate} 
  \item For every $X_j \rightarrow P_i$, $L_i{}_j{} = \lambda_i$.
  \item For every $ X_j \leftarrow P_i$, $L_i{}_j{} = \mu_i$.
  \item All other elements of $L$ are $0$.
\end{enumerate}
The vertex weights are defined as
\begin{equation}
\begin{aligned}
  \Delta_{X_j} & = \tfrac 1 2 \big(\DD - \sum_{i \, | \, X_j \rightarrow P_i} \bar \Delta_i - \sum_{i \, | \, X_j \leftarrow P_i} \Delta_i \big).
\end{aligned}
\end{equation}
The normalisation constant is
\begin{equation}
\begin{aligned}
  \mathcal N_{F} & = \frac {1} {4^{\eta} \, \pi^{n_P \, \DD}} \, \frac 1 {\Gamma(\eta)} \, \frac 1 {\Gamma(\Delta_X)}, \quad \mathcal N_{F}\Big|_{\eta = 1} = \frac {1} {4^{\eta} \, \pi^{n_P \, \DD}} \, \frac 1 {\Gamma(\Delta_X)}.
\end{aligned}
\end{equation}
Upon fixing the scaling redundancy $\mathcal G$ by setting all $\tau = \alpha = 1$ 
\begin{equation}
\begin{aligned}
  \mathcal I_{F} & =  2^{n_F} \, {\mathcal N}_{F} \, \int^*_{\parset} {\int\limits^{\RR^\DD}_{W}} \, \lambda^{\bar\Delta} \, \mu^{\Delta} \, e^{- \transpose{W} \, \bar U(\lambda, \, \mu)  \, W}, && \bar U(\lambda, \, \mu) = U(\parset) \big|_{\tau = \alpha = 1}
\end{aligned}
\end{equation}
and integrating over the momentum and position spaces $W$, the integral becomes
\begin{equation}\label{eq:FeynmanintegralforScalarsFormula}
\begin{aligned}
  \mathcal I_{F} & = \bar{\mathcal N}_{F} \int^*_{\lambda, \, \mu} \, \frac {\lambda^{\bar\Delta} \, \mu^{\Delta} } { (\det \bar U)^{\frac {\DD}{2}} } = \frac {\bar{\mathcal N}_{F}} {\Gamma(\frac {\DD}{2})} \int^*_{\lambda, \, \mu} \, \lambda^{\bar\Delta} \, \mu^{\Delta} \, z^{\frac {\DD}{2}} \, e^{- z \, \det \bar U(\lambda, \, \mu)},  
\end{aligned}
\end{equation}
where unit-weighted $\bar U(\lambda, \, \mu)$ is now a sparse matrix with unit diagonal and
\begin{equation}\label{eq:gaugefixednormalisedofscalarfeynmanint}
\begin{aligned}
  \bar{\mathcal N}_{F} = (2 \, \pi^{\frac \DD 2})^{n_F} \, {\mathcal N}_{F}, \quad \bar{\mathcal N}_{F}\Big|_{\eta = 1} = (4 \pi^{\DD})^{\frac{n_V - n_P}{2}} \, \frac 1 {\Gamma(\Delta_X)}.
\end{aligned}
\end{equation}
Presented in this form, it is clear that all scalar Feynman integrals on the sphere have generalized Euler integral representations \cref{eq:toocutetobetrue2} in terms of at most $2 \, n_P$ integration variables, and hence correspond to $\mathcal A$-hypergeometric functions. 
The complexity of these representations are significantly greater than their flat space counterparts, \cref{eq:LeePomeranskyrepresentation}, which have at most $n_P$ integration variables.

\subsection{Higher spin Feynman integrals}\label{sec:HigherSpinFeynmanIntegralsSetup}
A general Feynman integral consisting of higher spin propagators, that can be represented as $\mathcal F \circ G^{[\Delta]}_\eta$ defined in \cref{eq:operatorwithscalarprop}, similarly takes the form
\begin{equation}\label{eq:generalfeynmanintwithfsigma}
\begin{aligned}
  \mathcal I_{F} & = \frac {\mathcal N_{F}} {\Vol \mathcal G} \int^*_{\parset} {\int\limits^{\RR^\DD}_{W}} \, \lambda^{\bar\Delta} \, \mu^{\Delta} \, \tau^{\eta} \, \alpha^{\Delta_{X}} \, f(\parset, \, W) \, e^{- \transpose{W} \, U(\parset)  \, W},
\end{aligned}
\end{equation}
where $f(\parset, \, W)$ is polynomial in
\begin{equation}
\begin{aligned}
  & \lambda_i \, \underbracket{|P_i \, X_j|}_{X_j \rightarrow P_i}, && \mu_i\,  \underbracket{|P_i \, X_j|}_{X_j \leftarrow P_i}, && \sqrt{\alpha_{i} \, \alpha_{j}} \, |X_i \, X_j| \ni \frac {\lambda_i \, \mu_i}{\tau_i} \, \underbracket{|X_j{} \,X_k{}|}_{X_j \rightarrow P_i \rightarrow X_k} 
  \\ &  \alpha_i \, |X_i|^2, && \tau_i \, |P_i|^2, && \sqrt{\tau_{i} \, \tau_{j}} \, |P_i \, P_j|,
\end{aligned}
\end{equation}
and products thereof, and hence invariant under scale transformations generated by $\mathcal G$. 
$X^{2}$ and $P^2$ can simply be absorbed into the Schwinger parameterisation by $\alpha, \, \tau$, eventually leading to simple shifts in the parameters appearing in the normalisation constant:
\begin{equation}
\begin{aligned}
  & X^{2 \, \Delta'_{X}} \implies \Delta_X \rightarrow \Delta_X - \Delta'_{X}, && P^{2 \, \eta'} \implies \eta \rightarrow \eta - \eta'.
\end{aligned}
\end{equation}
Since the inner products of position and momentum vectors belong to the set of elements in $\{\transpose{W} \, W\}$, a generating function $\transpose{W} \, {U}'(\parset, \, \parset') \, W$, where $\parset' = \{\lambda', \, \mu', \, \beta\}$, can be used to perturb the Gaussian integral $e^{- \transpose{W} \, U(\parset) \, W}$, allowing $f(\parset, \, W)$ to be replaced by an operator $f[\partial_{\parset'}]$ that is polynomial in $\langle \partial_{\lambda'}, \, \partial_{\mu'}, \, \partial_{\beta} \rangle$ such that
\begin{equation}
\begin{aligned}
  f(\parset, \, W) & = f[\partial_{\parset'}] \, e^{- \transpose{W} \, {U}'(\parset, \, \parset') \, W} \Big|_{\parset' = 0}.
\end{aligned}
\end{equation} 
One such construction of $U'(\parset, \, \parset')$ and $f[\partial_{\parset'}]$ is as follows:
\begin{equation}\label{eq:fsigmatofUprescription}
\begin{aligned}
  \begin{array}{|c|c|c|c|}
    \hline 
    F & f(W) & f[\partial_{\parset'}] & U'(\parset, \, \parset') \\ 
    \hline
    X_j \rightarrow P_i & |P_i \, X_j|^{\bar s_{i}} & (-2 \im \, \lambda_i)^{- \bar s_{i}} \,\partial^{\bar s_{i}}_{\lambda_i'} & U'_{i, \, n_P + j} = U'_{n_P + j, \, i} = \im \,\lambda_i \, \lambda_i'\\
    \hline
    X_j \leftarrow P_i  & |P_i \, X_j|^{s_{i}} & (2 \im \, \mu_i)^{- s_{i}} \,\partial^{s_{i}}_{\mu_i'} & U'_{i, \, n_P + j} = U'_{n_P + j, \, i} = - \im \,\mu_i \, \mu_i'\\
    \hline 
    X_j{}, \; X_k{} & |X_j{} X_k{}|^{\omega} & (- 2 \im \, \sqrt {\alpha_j \, \alpha_k})^{- \omega} \, \partial^\omega_\beta   &  U'_{n_P + j, \, n_P + k} = U'_{n_P + k, \, n_P + j} = - \im \, \sqrt {\alpha_j \, \alpha_k} \, \beta\\
     \hline
     P_j{}, \; P_k{} & |P_j{} P_k{}|^{\omega'} & (- 2 \im \, \sqrt {\tau_j \, \tau_k})^{- \omega'} \, \partial^{\omega'}_{\beta}   &  U'_{j, \,k} = U'_{k, \,j} = - \im \, \sqrt {\tau_j \, \tau_k} \, \beta\\
     \hline
  \end{array}\,.
\end{aligned}
\end{equation}
If there exists a propagator $P_i$ from $X_j$ to $X_k$ i.e. $X_j \rightarrow P_i \rightarrow X_k$:
\begin{equation}
\begin{aligned}
& |X_j{} X_k{}|^\omega \rightarrow (\tfrac {\tau_i} {- 2 \im \, \lambda_i \, \mu_i})^{\omega} \, \partial^\omega_\beta \, e^{- \transpose{W} \, {U}' \, W} \Big|_{U' = 0}, && U'_{n_P + j, \, n_P + k} = U'_{n_P + k, \, n_P + j} = - \im \, \tfrac {\lambda_i \, \mu_i}{\tau_i} \, \beta
\end{aligned}
\end{equation}
can also be used but note that this operator doesn't commute with the remaining prescription in \cref{eq:fsigmatofUprescription}. 
Further, a couple of trivial changes of variables allows $f[\partial_{\lambda'}, \, \partial_{\mu'}]$ to be simplified to
\begin{equation}
\begin{aligned}
  \partial^{\bar s_{i}}_{\lambda_i'} & \rightarrow (-1)^{\bar s_i} \, \frac {\Gamma(\bar \Delta_i)}{ \Gamma(\bar \Delta_i - \bar s_{i})}, && \partial^{ s_{i}}_{\mu_i'} \rightarrow (-1)^{s_i} \, \frac {\Gamma(\Delta_i)}{ \Gamma(\Delta_i - s_{i})},
\end{aligned}
\end{equation}
leaving only some $f[\partial_\beta]$ behind. 
Thus, given a general monomial
\begin{equation}\label{eq:generalmonomialforfeynmanintegralwithspin}
\begin{aligned}
  f(W) = X^{2 \, \Delta'_{X}} \, P^{2 \, \eta'} \, |P\, X|^{\bar s} \, |P \, X'|^{s} \, |X \, X'|^{\omega} \, |P \, P'|^{\omega'},
\end{aligned}
\end{equation}
the Feynman integral
\begin{equation}
\begin{aligned}
  \mathcal I_{F} & = \frac {\mathcal N_{F}} {\Vol \mathcal G} \int^*_{\parset} {\int\limits^{\RR^\DD}_{W}} \, \lambda^{\bar\Delta} \, \mu^{\Delta} \, \tau^{\eta} \, \alpha^{\Delta_{X}} \, f(W) \, e^{- \transpose{W} \, U(\parset)  \, W}
\end{aligned}
\end{equation}
can be rewritten as
\begin{equation}\label{eq:feynmanintegraloperatorreduction}
\begin{aligned}
  \mathcal I_{F} & = \bar {\mathcal N}_{F}(f)  \int^*_{\lambda, \, \mu, \, z} \lambda^{\bar\Delta - \bar s} \, \mu^{\Delta - s} \, z^{\frac {\DD}{2}}  \, f_{\omega + \omega'}(\lambda, \, \mu) \, e^{- z \, \det \bar U(\lambda, \, \mu)}
  \\ \bar {\mathcal N}_{F}(f) & = \frac{\bar {\mathcal N}_{F}} {\Gamma(\frac {\DD}{2})} \, \frac{(-1)^{s + \omega + \omega'}} {(2 i)^{\bar s + s + \omega + \omega'}} \, \frac {\Gamma(\eta)} {\Gamma(\eta - \eta')} \, \frac {\Gamma(\Delta_X)} {\Gamma(\Delta_X - \Delta'_X)} \, \frac {\Gamma(\bar \Delta)}{ \Gamma(\bar \Delta - \bar s)} \, \, \frac {\Gamma(\Delta)}{ \Gamma(\Delta - s)},  
\end{aligned}
\end{equation}
where $f_{\omega + \omega'}(\lambda, \, \mu)$ is a polynomial,
\begin{equation}
\begin{aligned}
  f_{\omega + \omega'}(\lambda, \, \mu) & = e^{z \, \det (\bar U(\lambda, \, \mu) + \bar U'(\beta))} \, \partial^{\omega+\omega'}_{\beta} \, e^{- z \, \det (\bar U(\lambda, \, \mu) + \bar U'(\beta))}, \quad \bar U'(\beta) & = U'(\beta)\Big|_{\tau = \alpha = 1}.
\end{aligned}
\end{equation}
In particular, $f_\omega$ is spanned by the zeroth to $\omega$\textsuperscript{th} Hermite polynomials in $(\lambda, \, \mu)$. 
Thus, \cref{eq:generalfeynmanintwithfsigma} is a sum over scalar Feynman integrals with generic parameters, and hence also a generalized Euler integral. 
Interestingly enough this type of formulation also allows negative orders of $|PX|$ ($X \rightarrow P$) to be considered:
\begin{equation}\label{eq:negativepowerofPX}
\begin{aligned}
  & \lambda \circ G 
  = \tfrac {\lambda} {- 2 \im \lambda\, |PX|} \partial_{\lambda'} \circ G \, e^{- \transpose{W} \, U'(\lambda') \, W} \Big|_{\lambda' = 0}
  = \tfrac {\bar \Delta} {2 \im \, |PX|} \circ G \implies \tfrac 1 {|PX|} \circ G = \tfrac {2 \im  } {\bar \Delta} \,\lambda \circ G
\end{aligned}
\end{equation}
and similarly if $X \leftarrow P$, $\tfrac 1 {|PX|} \circ G = \tfrac {- 2 \im  } {\Delta} \, \mu \circ G$, implying that \cref{eq:feynmanintegraloperatorreduction} is true for $\bar s, \, s \in \ZZ$, not just $\NN$. 
Thus, the reduced form of $\bar U'$, the perturbation incidence matrix, is sparse and effectively takes the form 
\begin{equation}\label{eq:higherspinUparset}
\begin{aligned}
  \bar U'(\parset, \, \parset') & = \im \left(\begin{array}{c|c}
               L_P'(\beta) &  L'(\lambda\lambda', \, \mu \mu')  \\ 
              \hline
              \transpose{L'}(\lambda\lambda', \, \mu \mu')  & L_V'(\beta) \\ 
            \end{array} \right) \rightarrow \bar U'(\beta) = \im \left(\begin{array}{c|c}
              L_P'(\beta) & 0  \\ 
              \hline
              0  & L_V'(\beta) \\ 
            \end{array} \right),
\end{aligned}
\end{equation}
where $L_V', \, L_P'$ are $n_V \times n_V, \, n_P \times n_P$ symmetric matrices with non-zero entries $L'_i{}_j{} = L'_j{}_i{} = \beta_{ij}$ corresponding to the terms $|X_i \, X_j|$ and $P_i \, P_j$ in $f(W)$, as defined in \cref{eq:generalmonomialforfeynmanintegralwithspin}. 
\par
Thus, the perturbation incidence matrix $\bar U'(\beta)$ with a maximal $L'_V, \, L'_P$, i.e. with no non-zero entries, forms a gaussian ``master'' integral, $\mathcal I_F(\beta)$. $\mathcal I_F(\beta)$, once solved in complete generality of the defining parameters $\{\DD, \, \Delta, \, \bar \Delta\}$, can be used to find all Feynman integrals with the same incidence matrices $L(\lambda, \, \mu)$ as its derivatives with respect to the perturbative parameters $\beta$ at $\beta = 0$.

\subsection{Generalized correlation functions}\label{sec:GeneralisedCorrelationFunctions}
A Feynman integral representing a generalized $n$-point scalar correlation function $\mathcal J_F$ on $S^{d+1}$, with $n_P$ internal propagators ($P$), $n_Y$ internal vertices ($Y$), and $n$ external legs of the propagators ($Q$) and external vertices ($\hat X$), can also be brought into the Euler integral format. 
Using the now familiar embedding space formulation of the scalar propagator, $\mathcal J_F(\hat X)$, can be written as
\begin{equation}
\begin{aligned}
  \mathcal J_F(\hat X) & = \frac {1} {\Vol (\RR^*)^{n + n_Y}} \int_{Q} \int_{Y} \frac {\de^{\DD} Y} {|Y|^{\DD}}\int^*_{\lambda, \, \mu} \lambda^{\bar\Delta_Q}_{Q} \, |\mu_{Q} Y|^{\Delta_Q} \, \frac {e^{- 2 \im \, Q \,(\lambda_Q \, \hat X - \mu_Q \, Y)}} {4 \pi^{\DD} \, Q^2} \, G_P(Y),
\end{aligned}
\end{equation}
which upon integrating out the external propagator momenta $Q$ (where $\eta_Q$ has been tacitly assumed to be $1$ but can be easily reinstated if needed), 
\begin{equation}
\begin{aligned}
  \mathcal J_F(\hat X) & = \frac {(2 \pi^{\frac {\DD}{2}})^{- n}} {\Vol (\RR^*)^{n_Y}}\int^*_{\lambda, \, \mu} \lambda^{\bar\Delta_Q}_{Q} \, \mu_{Q}^{\Delta_Q}  e^{- \lambda_Q^2}\int_{Y} \frac {\de^{\DD} Y} {|Y|^{\DD}}\,|Y|^{\Delta_Q} \, e^{- (\mu^2_Q \, Y^2 - 2 \, \lambda_Q \, \mu_Q \, Y \cdot \hat X)} \, G_P(Y)
\end{aligned}
\end{equation}
comes to resemble \cref{eq:generalformofscalarFeynmanintegral} when represented in its Schwinger parameteric form:
\begin{equation}
\begin{aligned}
  \mathcal J_F(\hat X) & = \frac {\mathcal N_F} {\Vol (\RR^*)^{n_Y + n_P}}\int^*_{\lambda, \, \mu, \tau_P, \alpha_P} \! \! \! \! \! \! \! \! \! \! \! \! \! \! \! \! \lambda^{\bar\Delta} \, \mu^{\Delta}  e^{- \lambda_Q^2}\int_{W} \tau_P^{\eta_P} \, \alpha_Y^{\Delta_Y} \, e^{- (\transpose{W} \, U \, W - \transpose{V} \, W - \transpose{W} \, V)}
  \\ \mathcal N_F &  = \frac {1} {(2 \pi^{\frac {\DD}{2}})^{(n + 2 \, n_P) }  \, \Gamma(\eta_P) \, \Gamma(\Delta_Y)},
\end{aligned}
\end{equation}
where the weighted incidence matrix $U$ is appropriately modified to represent the gaussian integral 
\begin{equation}
\begin{aligned}
  U & = \left(\begin{array}{c|c}
              \tau_{P} & \im L(\lambda_P, \, \mu_P)  \\ 
              \hline
              \im \transpose{L}(\lambda_P, \, \mu_P)  & \alpha_{Y} + \mu_Q^2 \\ 
            \end{array} \right)
\end{aligned}
\end{equation}
and $V$ is a vector with the components $\{ \underbracket{0 \cdots 0}_{n_P}, \, \underbracket{\lambda_Q \, \mu_Q \, \hat X}_{n_Y}\}$. 
Fixing all scaling symmetries and integrating over $W$, leads to
\begin{equation}\label{eq:generalizedcorrelationfunctionaseulerintegral}
\begin{aligned}
  \mathcal J_F(\hat X) & = \mathcal {\bar N}_F\int^*_{\lambda, \, \mu} \lambda^{\bar\Delta} \, \mu^{\Delta} \, e^{- \lambda_Q^2 } \, \frac { e^{\transpose{V} \, \bar U^{- 1} \, V}} {(\det \bar U)^{\frac {\DD}{2}}} , \quad \bar U = \left(\begin{array}{c|c}
              1 & \im L(\lambda_P, \, \mu_P)  \\ 
              \hline
              \im \transpose{L}(\lambda_P, \, \mu_P)  & 1 + \mu_Q^2 \\ 
            \end{array} \right)
  \\ \mathcal {\bar N}_F &  = \frac {1} {(2 \pi^{\frac {\DD}{2}})^{(n + n_P - n_Y) }  \, \Gamma(\eta_P) \, \Gamma(\Delta_Y)},
\end{aligned}
\end{equation}
from which one of the parameters in $(\lambda_Q, \, \mu_Q)$ can always be trivially integrated out, finally resulting in an Euler integral representation of the generalized $n$-point scalar correlation function, $\mathcal J_F(\hat X)$, in terms of $(2 \, n_P + 2 \, n_Q - 1)$ integration variables. 
\par
This setup can be extended to include higher spin fields in the same vein as \cref{sec:HigherSpinFeynmanIntegralsSetup}, with some obvious additions to the prescription as needed, e.g.
\begin{equation}
\begin{aligned}
  & Q_I \rightarrow \tfrac 1 {- 2 \im \, \lambda_Q} \partial_{\hat X^I}, \quad |\transpose{V}_i \cdot W| \rightarrow \tfrac 1 {2} \partial_{v} \, e^{v \, (\transpose V \, W + \transpose W \, V)}\Big|_{v = 0}.
\end{aligned}
\end{equation}
This brings all generalized correlation functions into the Euler integral format.

\subsection{Explicit constructions of scalar Feynman integrals}\label{sec:ExplicitConstructionofScalarFeynmanIntegrals}
\begin{figure}[H]
\centering
\begin{subfigure}[t]{0.25\textwidth}
\centering
\begin{tikzpicture}
\drawdot{0,0}; \node at (0,0)[anchor=north]{\tiny$X$};
\drawdot{2,0}; \node at (2,0)[anchor=north]{\tiny$Y$};
\draw [->] (0,0) -- (1,0);
\draw (1,0) -- (2,0);
\node at (1,0)[anchor=south]{\tiny$\Delta$};
\end{tikzpicture}
\caption{Propagator}\label{fig:feynmandiagramofPropagator}
\end{subfigure}
\hfill 
\begin{subfigure}[t]{0.33\textwidth}
\centering
\begin{tikzpicture}
\drawdot{0,0}; \node at (0,0)[anchor=north]{\tiny$X_1$};
\drawdot{1,0}; \node at (1,0)[anchor=north]{\tiny$Y$};
\drawdot{2,0}; \node at (2,0)[anchor=north]{\tiny$X_2$};
\draw [->] (0,0) -- (0.5,0); \draw (0.5,0) -- (1,0); \node at (0.5,0)[anchor=south]{\tiny$\Delta_1$};
\draw (1,0) -- (1.5,0); \draw [->] (2,0) -- (1.5,0); \node at (1.5,0)[anchor=south]{\tiny$\Delta_2$};
\end{tikzpicture}
\caption{Propagator with vertex insertion}\label{fig:feynmandiagramofPropagatorwithvertexinsertion}
\end{subfigure}
\hfill
\begin{subfigure}[t]{0.38\textwidth}
\centering
\begin{tikzpicture}
\drawdot{0,0}; \node at (0,0)[anchor=north]{\tiny$X_1$};
\drawdot{1,0}; \node at (1,0)[anchor=north]{\tiny$Y_1$};
\drawdot{2,0}; \node at (2,0)[anchor=north]{\tiny$Y_n$};
\drawdot{3,0}; \node at (3,0)[anchor=north]{\tiny$X_2$};
\draw[densely dotted] (1,0) -- (2,0);
\draw [->] (0,0) -- (0.5,0); \draw (0.5,0) -- (1,0); \node at (0.5,0)[anchor=south]{\tiny$\Delta_1$};
\draw (2,0) -- (2.5,0); \draw [->] (3,0) -- (2.5,0); \node at (2.5,0)[anchor=south]{\tiny$\Delta_2$};
\node at (1.5,0)[anchor=south]{\tiny$\Delta$};
\end{tikzpicture}
\caption{Propagator with $n$ vertex insertions}\label{fig:feynmandiagramofPropagatorwithnvertexinsertions}
\end{subfigure}
\\ 
\begin{subfigure}[t]{0.25\textwidth}
\centering
\begin{tikzpicture}
\drawdot{0,-0.5}; \node at (0,-0.5)[anchor=north]{\tiny$X$};
\node at (0,0.5)[anchor=north]{\tiny$\Delta$};
\draw (0,0) circle (0.5);
\node at (-0.5,-0.03)[anchor=east]{$\mathcal I_{[\Delta]} = $};
\end{tikzpicture}
\caption{$1$-loop}\label{fig:feynmandiagramof1loop}
\end{subfigure}
\hfill 
\begin{subfigure}[t]{0.33\textwidth}
\centering
\begin{tikzpicture}
\drawdot{0,-0.5}; \drawdot{0,0.5}; \node at (0,-0.5)[anchor=north]{\tiny$X$}; \node at (0,0.5)[anchor=south]{\tiny$Y$};
\node at (0.5,0)[anchor=west]{\tiny$\Delta_1$};
\node at (-0.5,0)[anchor=east]{\tiny$\Delta_2$};
\draw (0,0) circle (0.5);
\end{tikzpicture}
\caption{$1$-loop with $1$ vertex insertion}\label{fig:feynmandiagramof1loopwith1vertexinsertion}
\end{subfigure}
\hfill 
\begin{subfigure}[t]{0.38\textwidth}
\centering
\begin{tikzpicture}
\drawdot{-0.5,0}; \drawdot{0.3,0.4}; \drawdot{0.3,-0.4}; \node at (0,-0.5)[anchor=north]{\tiny$\Delta_2$}; \node at (0,0.5)[anchor=south]{\tiny$\Delta_1$};
\node at (0.5,0)[anchor=west]{\tiny$\Delta$};
\node at (-0.5,0)[anchor=east]{\tiny$X_2$};
\node at (0.3,0.4)[anchor=west]{\tiny$X_1$};
\node at (0.3,-0.4)[anchor=west]{\tiny$X_{n+1}$};
\centerarc[](0,0)(57:303:0.5cm);
\centerarc[densely dotted](0,0)(-57:57:0.5cm);
\end{tikzpicture}
\caption{$1$-loop with $n$ vertex insertions}\label{fig:feynmandiagramof1loopwithnvertexinsertions}
\end{subfigure}
\caption{$2$-point function and related Feynman diagrams}\label{fig:2pointfunctionandrelatedFeynmandiagrams}
\label{fig:2pointfunctionandrelatedfeynmandiagrams}
\end{figure} \vspace{-\baselineskip}
\subsubsection{Coincident point limit and \texorpdfstring{$1$}{1}-loop character integral}
The coincident point limit of the scalar propagator, i.e. geodesic distance $\theta = 0$ or alternately $\xdotY = 1$ in say \cref{eq:scproplam1gauge}, is expectedly divergent:
\begin{equation}
\begin{aligned}
  G_{[\Delta]} \equiv G_{[\Delta]}(\xdotY)\Big|_{\xdotY = 1} & = \frac {\Gamma(\frac {d} {2})} { 4 \pi^{\frac {\DD} {2}} } \int^*_{\mu} \frac {\mu^{\Delta}} {(1 + \mu^2 - 2 \, \mu )^{\frac {d} {2}}} = \frac {\Gamma(\frac {d} {2})} { 4 \pi^{\frac {\DD} {2}} } \int^*_{\mu} \frac {\mu^{\Delta}} {|1 - \mu|^{d}}, 
\end{aligned}
\end{equation}
appearing in the integral as a pole of the integrand at $\mu = 1$. The $1$-loop integral, \cref{fig:feynmandiagramof1loop}, is simply
\begin{equation}
 \begin{aligned}
   \mathcal I_{[\Delta]} = \Omega_{d+1} \times G_{[\Delta]} = \frac {1} {d}  \int^*_{\mu} \frac {\mu^{\Delta}} {|1 - \mu|^{d}}.
 \end{aligned}
\end{equation}
Though in this case needlessly convoluted, strictly following the rules given in \cref{sec:EmbeddingSpaceFormulationofFeynmanIntegralsonSd+1Scalar} results in the same:
\begin{equation}\label{eq:scalar1loopcharacterintegralform}
\begin{aligned}
  \mathcal I_{[\Delta]} & =  \frac 1 {(\Vol \RR^*)^2} \,\frac 1 {4 \pi^{\DD} } \,\int_{\lambda, \,\mu}^* \, \lambda^{\bar\Delta} \, \mu^{\Delta}  \int_{P, X} \frac {e^{- 2 \im \,P\,X\,(\lambda - \mu)}} {P^2 \, X^2}
  \\ & =  \int_{\lambda, \,\mu}^* \, \frac {\lambda^{\bar\Delta} \, \mu^{\Delta} } {(1 + (\lambda - \mu)^2)^{\frac {\DD} {2}}}  =  \frac 1 {d} \int_{\mu}^* \, \frac {\mu^{\Delta} } { (1 + \mu^2 - 2 \, \mu)^{\frac d 2}}  =  \frac 1 {d} \int_{\mu}^* \, \frac {\mu^{\Delta} } { |1 - \mu|^{d}}
  \\ & =  \frac 1 {d} \int_{0}^1  \, \frac {\de \mu} {\mu}\, \frac {\mu^{\Delta} + \mu^{\bar \Delta} } { |1 - \mu|^{d}},
\end{aligned}
\end{equation}
recreating the character integral formulation of scalar $1$-loop integrals from \cite{Anninos:2020hfj}, wherein they are explicitly evaluated, \cref{eq:1loopcorrectionforscalaronS3} being one such result. Explicitly using the limiting value of $\lim\limits_{\xdotY \rightarrow 1} G(\xdotY) = G_{[\Delta]}$, i.e. assuming $d<1$:
\begin{equation}\label{eq:oneloopcharacterintegralevalutated}
\begin{aligned}
  \mathcal I_{[\Delta]} = \Omega_{d+1} \times \lim_{\sigma \rightarrow 1} G_{[\del]}(\sigma) & = \frac {2 \pi^{\frac {d+2} {2}}}{\Gamma(\frac {d+2} {2})} \times \frac {\Gamma(\Delta) \,\Gamma(\bar \Delta)} {(4 \pi)^{\frac{d+1}{2}} \,\Gamma(\frac {d+1} {2})} \times  \frac {\Gamma(\frac {d+1} {2}) \, \Gamma(\frac {d+1} {2} - d)} {\Gamma(\frac {d+1} {2} - \Delta) \, \Gamma(\frac {d+1} {2} - \bar \Delta)} 
  \\ & = \Gamma(\Delta) \,\Gamma(\bar \Delta) \times \frac {2^{- d} \sqrt\pi}{\Gamma(1 + \frac {d} {2}) \, \Gamma(\frac {-d} {2})} \times  \frac {\Gamma(\frac {1 - d} {2}) \, \Gamma(\frac {-d} {2})} {\Gamma(\frac {1} {2} - \im \nu) \, \Gamma(\frac {1} {2} + \im \nu)} 
  \\ & = 2 \, \sin (\tfrac {\pi \, d}{2}) \, \cosh (\pi \, \nu) \, \Gamma(\Delta) \,\Gamma(\bar \Delta) \, \frac {\Gamma(1-d)}{d \, \pi}
  \\ & = (\sin (\pi \, \Delta) + \sin (\pi \, \bar\Delta))\, \Gamma(\Delta) \,\Gamma(\bar \Delta) \, \frac {\Gamma(1-d)}{d \, \pi}. 
\end{aligned}
\end{equation}
Note that \cref{eq:oneloopcharacterintegralevalutated} is an important result that will keep reappearing. 

\subsubsection{\texorpdfstring{$1$}{1}-Melon : \texorpdfstring{$2$}{2}-point function integrated}
The simplest possible scalar Feynman integral is the integral of the $2$-point function over all space, a $1$-melon so to say:
\begin{equation}
\begin{aligned}
  \mathcal I_{2,1} & = \frac 1 {4 \pi^{\DD} } \,\XintAdd{\times}{\RR^\DD}{X,Y} \, \Xint{\times}_{\lambda, \,\mu}^* \, |\lambda \,X|^{\bar\Delta} \, |\mu \,Y|^{\Delta}  \int_{P}^{\RR^\DD} \frac {e^{- 2 \im \, P\,(\lambda \,X - \mu \,Y)}} {P^2}
  \\ & = \frac {2 \pi^{\frac \DD 2}} {\Gamma(\frac {\DD - \Delta} {2})  \,\Gamma(\frac {\DD - \bar \Delta} {2})} \int^*_{\lambda, \mu} \frac {\lambda^{\bar \Delta} \, \mu^{\Delta}} {(1 + \lambda^2 + \mu^2)^{\frac \DD 2}} = \frac {2 \pi^{\frac \DD 2}} {\Gamma(\frac \DD 2)} \frac {\Gamma(\frac \del 2)  \,\Gamma(\frac \bdel 2)} {4 \, \Gamma(\frac {\DD - \Delta} {2}) \, \Gamma(\frac {\DD - \bar \Delta} {2}) } = \frac {\Omega_{d+1}} {\del  \,\bdel}.
\end{aligned}
\end{equation}

\subsubsection{\texorpdfstring{$2$}{2}-Melon}\label{sec:2melonfeynmanint}
The $2$-melon Feynman diagram, \cref{fig:feynmandiagramof1loopwith1vertexinsertion}, with $2$ propagators and $2$ internal vertices forming a $1$-loop diagram, is encoded in the $4 \times 4$ unit-weighted incidence matrix:
\begin{equation}
\begin{aligned}
  \bar U_{2,2} & = \left( \begin{smallmatrix}
        1 & 0 & \im  \lambda_1 & - i \mu_1 \\ 0 & 1 & \im  \lambda_2 & - \im  \mu_2 \\  \im  \lambda_1  & \im  \lambda_2 & 1 & 0 \\ - \im  \mu_1  & - \im  \mu_2 & 0 & 1 \\ 
      \end{smallmatrix} \right).
\end{aligned}
\end{equation}
Using \cref{eq:FeynmanintegralforScalarsFormula}, the corresponding Feynman integral is 
\begin{equation}\label{eq:2melonfeynmaneulerintegralform}
\begin{aligned}
  \mathcal I_{2,2} & 
      = \mathcal N_{2,2} \int^*_{\lambda, \, \mu} \, \frac {\lambda_1^{\bar\Delta_1} \, \mu_1^{\Delta_1} \, \lambda_2^{\bar\Delta_2} \, \mu_2^{\Delta_2}} { (1 + \lambda_1^2 + \lambda_2^2 + \mu_1^2 + \mu_2^2 + \lambda_1^2 \, \mu_2^2 + \lambda_2^2 \, \mu_1^2 - 2 \, \lambda_1 \, \lambda_2 \, \mu_1 \, \mu_2)^{\frac {\DD}{2}} },
      \\ \mathcal N_{2,2}  & = \frac 1 {\Gamma(\frac {\DD - \del_1- \del_2}{2}) \, \Gamma(\frac {\DD - \bdel_1- \bdel_2}{2})} .
\end{aligned}
\end{equation}
\Cref{exmp:runningexample} solves the generalized Euler integral version of $\mathcal I_{2,2}$ in detail. At the physically relevant parameters, $\mathcal I_{2,2}$ is found to be proportional to \cref{eq:2Pexamplefinalresultforarbitraryparameters}. 
In particular, it is
\begin{equation}\label{eq:2melonfeynmanintfinalresult}
\begin{aligned}
   \mathcal I_{2,2} & 
   = \frac {\Gamma (1-d) \left( (\sin \pi\Delta_2 + \sin \pi\bar\Delta_2) \Gamma (\Delta_2 ) \Gamma ( \bar\Delta_2 )  - (\sin \pi\Delta_1 + \sin \pi\bar\Delta_1) \Gamma (\Delta_1) \Gamma (\bar\Delta_1 ) \right)}{  d \, \pi \, (\nu_1^2-\nu_2^2) },
\end{aligned}
\end{equation}
or upon recognizing \cref{eq:oneloopcharacterintegralevalutated}
\begin{equation}\label{eq:2melonfeynmanintfinalresult2}
\begin{aligned}
  \mathcal I_{2,2} & = \frac {\mathcal I_{[\Delta_2]} - \mathcal I_{[\Delta_1]}} {(\nu_1^2-\nu_2^2)} .
\end{aligned}
\end{equation}

\subsubsection{Propagator with \texorpdfstring{$1$}{1} vertex insertion}\label{eq:Propagatorwithvertexinsertion}
The correlation function, $G_{[\del_{(1,2)}]} (\hat X_1, \, \hat X_2)$, depicted in \cref{fig:feynmandiagramofPropagatorwithvertexinsertion}, is given by the integral:
\begin{equation}\label{eq:propwith1vertexinsertioninitialintegral}
\begin{aligned}
G_{[\del_{(1,2)}]}(\xdotY_{12})  & = \frac 1 {(4 \pi^{\DD})^2} \XintAdd{\times}{*}{\lambda, \mu} \lambda_1^{\bar \Delta_1} \, \mu_1^{\Delta_1} \lambda_2^{\bar \Delta_2} \, \mu_2^{\Delta_2} \int_{P}^{\RR^{\DD}} \Xint{\times}_{Y}^{\RR^{\DD}} \frac {e^{-2 \im \,  P_1 \, (\lambda_1 \, \hat X_1 - \mu_1 \, Y) -2 \im\,  P_2\,  (\lambda_2 \, \hat X_2 - \mu_2 \, Y)}} {P_1^{2}  \, P_2^{2} \, Y^{\DD - \embedDelta}}, 
\end{aligned}
\end{equation}
where $\embedDelta = \del_1 + \del_2, \; \bar{\embedDelta} = \bdel_1 + \bdel_2$. 
Integrating over the internal vertex $Y$ and momenta $P$, it can be brought into the euler integral form:
\begin{equation}
\begin{aligned}
G_{[\del_{(1,2)}]}(\xdotY_{12})  & = \frac {1} {2 \pi^{\frac {\DD} {2}} \, \Gamma(\frac {\DD - \embedDelta} {2})} \int^*_{\lambda, \mu}  \frac {\lambda_1^{\bar \Delta_1} \, \mu_1^{\Delta_1} \lambda_2^{\bar \Delta_2} \, \mu_2^{\Delta_2}} {(1 + \mu_1^2 + \mu_2^2)^{\frac {\DD - \bar{\embedDelta}} 2} } e^{-(\lambda_1^2 + \lambda_2^2 + \lambda_1^2 \, \mu_2^2 + \lambda_2^2 \, \mu_1^2 - 2\, \xdotY_{12} \, \lambda_1 \, \lambda_2 \, \mu_1\,  \mu_2 )}
\\ & = \frac {\Gamma(\frac {\bar{\embedDelta}} {2})} {4 \pi^{\frac {\DD} {2}} \,\Gamma(\frac {\DD - \embedDelta} {2})} \int^*_{\lambda, \mu}  \frac {\mu_1^{\Delta_1}\,\mu_2^{\Delta_2}\, \lambda_1^{\bar \Delta_1} \, (1 + \mu_1^2 + \mu_2^2)^{- \frac {\DD - \bar{\embedDelta}} 2}} { (1 + \mu_1^2 + \lambda_1^2 + \lambda_1^2\, \mu_2^2  - 2\,\xdotY_{12}\, \lambda_1 \,\mu_1 \,\mu_2)^{\frac {\bar{\embedDelta}} {2}}} ,
\end{aligned}
\end{equation}
encoded in the $\mathcal A_{12}$ matrix and $\gamma_{12}$ vector:
\begin{equation}
\begin{aligned}
  \mA_{12} & =  \left( \begin{smallmatrix}
      1 & 1 & 1 & 0 & 0 & 0 & 0 & 0 \\
      0 & 0 & 0 & 1 & 1 & 1 & 1 & 1 \\
      0 & 2 & 0 & 0 & 2 & 0 & 0 & 1\\
      0 & 0 & 2 & 0 & 0 & 0 & 2 & 1\\
      0 & 0 & 0 & 0 & 0 & 2 & 2 & 1
    \end{smallmatrix} \right), \quad \gamma_{12} = \{\tfrac {\DD - \bar{\embedDelta}} {2}, \, \tfrac {\bar{\embedDelta}} {2}, \, \Delta_1, \, \Delta_2, \, \bar \Delta_1\}.
\end{aligned}
\end{equation}
Picking a weight in the class of $\{0,1,1,0,0,1,0,1\}$ to form the indicial ideal and find the roots (see \cref{app:Propwith1vertexExtraDetails} for details), the log-free series solution as an expansion around $\xdotY_{12} = 0$ is found to be:
\begin{equation}
\begin{aligned}
  G_{[\del_{(1,2)}]}(\xdotY_{12})  & = \frac {1} {2 \pi^{\frac {\DD} {2}} } \Bigg(\frac {\Gamma(\frac {\bar\Delta_2}{2}) \, \Gamma(\frac {\Delta_2}{2}) \,{}_2F_1(\frac {\bar\Delta_2}{2}, \, \frac {\Delta_2}{2}, \, \frac 1 2, \, \xdotY_{12}^2) - [2 \rightarrow 1]} {4 \, (d - \Delta_1 - \Delta_2) \, (\Delta_1 - \Delta_2)} 
  \\ & + \xdotY \, \frac {\Gamma(\frac {\bar\Delta_2 + 1}{2}) \, \Gamma(\frac {\Delta_2+ 1}{2}) \,{}_2F_1(\frac {\bar\Delta_2+ 1}{2}, \, \frac {\Delta_2+ 1}{2}, \, \frac 3 2, \, \xdotY_{12}^2) - [2 \rightarrow 1]} {2 \, (d - \Delta_1 - \Delta_2) \, (\Delta_1 - \Delta_2)} \Bigg),
\end{aligned}
\end{equation}
which,  upon comparison to \cref{eq:scalarproparoundxdoty0}, can be seen to precisely match the expected result:
\begin{equation}\label{eq:propwith1vertexinsertionfinalresult}
\begin{aligned}
  G_{[\del_{(1,2)}]}(\xdotY_{12})  & = \frac {G_{[\Delta_1]}(\xdotY_{12}) } {(\nu_2^2 - \nu_1^2)}  + \frac {G_{[\Delta_2]}(\xdotY_{12})} {(\nu_1^2 - \nu_2^2)}.
\end{aligned}
\end{equation}
This result is related to the $2$-melon result from \cref{sec:2melonfeynmanint}, \cref{eq:2melonfeynmanintfinalresult,eq:2melonfeynmanintfinalresult2}, by the identification $\hat X_2 = \hat X_1$ and integration over this now internal vertex, i.e. $\mathcal I_{2,2} = \Omega_{d+1} \times \lim\limits_{\xdotY_{12} \rightarrow 1} G_{[\del_{(1,2)}]}(\xdotY_{12})$.

\subsubsection{Propagator and \texorpdfstring{$1$}{1} loop with \texorpdfstring{$n$}{n} vertex insertions}\label{eq:Propagatorwithmultiplevertexinsertions}
The integral corresponding to \cref{fig:feynmandiagramofPropagatorwithnvertexinsertions} at $n=2$, i.e. consisting of $2$ internal points, $2$ external points, and $3$ propagators, is
\begin{equation}
\begin{aligned}
  G_{[\del_{(1,2,3)}]}(\xdotY_{12}) & = \frac 1 {(4 \pi^{\DD})^3} \Xint{\times}_{\lambda, \mu}\Xint{\times}_{Y} \int_{P}\frac { \lambda_1^{\bar\Delta_1} \, |\mu_1 \,Y_1|^{\Delta_1} |\lambda_2 \, Y_1|^{\bar\Delta_2} \, |\mu_2 \,Y_2|^{\Delta_2} \, |\lambda_3 \, Y_2|^{\bdel_3} \, \mu_3^{\del_3}} {P_1^2 \, P_2^2 \, P_3^2}
  \\ & \phantom{\frac 1 {(4 \pi^{\DD})^3} \Xint{\times}_{\lambda, \mu}\Xint{\times}_{Y} \int_{P}} \times e^{- 2 \im P_1 \,(\lambda_1 \, \hat X_1 - \mu_1 \, Y_1) - 2 \im P_2 \,(\lambda_2 \, Y_1 - \mu_2 \, Y_2) - 2 \im P_3 \,(\lambda_3 \, Y_2 - \mu_3 \, \hat X_2)},
\end{aligned}
\end{equation}
where the argument of $G_{[\del_{(\cdots)}]}$ is suppressed when it is $(\hat X_1, \, \hat X_2) \equiv \xdotY_{12}$. Within this integral, \cref{eq:propwith1vertexinsertioninitialintegral} can be recognised by putting $\hat Y_2$ back on the sphere:
\begin{equation}
\begin{aligned}
  G_{[\del_{(1,2,3)}]} (\xdotY_{12}) & = \frac 1 {4 \pi^{\DD}} \Xint{\times}_{\lambda_3, \mu_3}\int_{\hat Y_2} \int_{P_3} \frac {\lambda_3^{\bdel_3} \, \mu_3^{\del_3}}{P_3^2} e^{ - 2 \im P_3 \,(\lambda_3 \, \hat Y_2 - \mu_3 \, \hat X_2)} \, G_{[\del_{(1,2)}]} (\hat X_1, \, \hat Y_2).
\end{aligned}
\end{equation}
Using \cref{eq:propwith1vertexinsertionfinalresult}, it can be simplified by to
\begin{equation}
\begin{aligned}
  G_{[\del_{(1,2,3)}]} (\xdotY_{12}) & = \frac {1} {(\nu_1^2 - \nu_2^2)} \,  \frac 1 {4 \pi^{\DD}} \Xint{\times}_{\lambda_3, \mu_3}\int_{\hat Y_2} \int_{P_3} \frac {\lambda_3^{\bdel_3} \, \mu_3^{\del_3}}{P_3^2} e^{ - 2 \im P_3 \,(\lambda_3 \, \hat Y_2 - \mu_3 \, \hat X_2)}
  \\ & \phantom{\frac {2 \pi^{\frac{\DD}{2}} } {(\nu_1^2 - \nu_2^2)} \,  \frac 1 {4 \pi^{\DD}} \Xint{\times}_{\lambda_3, \mu_3}\Xint{\times}_{Y_2} \int_{P_3}} \times (G_{[\del_{2}]} (\hat X_1, \, \hat Y_2) - G_{[\del_1]} (\hat X_1, \, \hat Y_2)).
\end{aligned}
\end{equation}
Upon reinstating the integral formulation of these propagators and the scale invariant measure of $Y_2$, this integral simply becomes
\begin{equation}
\begin{aligned}
  G_{[\del_{(1,2,3)}]} (\xdotY_{12}) & = \frac {1} {(\nu_1^2 - \nu_2^2)} \, \big( G_{[\del_{2,3}]}(\xdotY_{12}) - G_{[\del_{1,3}]}(\xdotY_{12}) \big).
\end{aligned}
\end{equation}
Expanding it out, it becomes:
\begin{equation}
\begin{aligned}
  G_{[\del_{(1,2,3)}]} (\xdotY_{12}) & = \frac { G_{[\del_{1}]}(\xdotY_{12}) } {(\nu_2^2 - \nu_1^2)(\nu_3^2 - \nu_1^2)} + \frac { G_{[\del_{2}]}(\xdotY_{12}) }{(\nu_1^2 - \nu_2^2)(\nu_3^2 - \nu_2^2)}  +  \frac {G_{[\del_{3}]}(\xdotY_{12}) }{(\nu_1^2 - \nu_3^2)(\nu_2^2 - \nu_3^2)}.
\end{aligned}
\end{equation}
In the same vein as the relation between \cref{eq:2melonfeynmanintfinalresult,eq:propwith1vertexinsertionfinalresult}, \cref{fig:feynmandiagramof1loopwithnvertexinsertions} at $n=3$ is
\begin{equation}
\begin{aligned}
  \mathcal I_{3,3} & = \Omega_{d+1} \times \lim_{\xdotY_{12} \rightarrow 1} G_{[\del_{(1,2,3)}]} (\xdotY_{12})
  \\ & = \frac {\mathcal I_{[\Delta_1]}} {(\nu_2^2 - \nu_1^2)(\nu_3^2 - \nu_1^2)}  + \frac {\mathcal I_{[\Delta_2]}} {(\nu_1^2 - \nu_2^2)(\nu_3^2 - \nu_2^2)}  + \frac {\mathcal I_{[\Delta_3]}} {(\nu_1^2 - \nu_3^2)(\nu_2^2 - \nu_3^2)}.
\end{aligned}
\end{equation}
This process can be inductively continued to find closed form expressions of the Feynman diagrams in \cref{fig:feynmandiagramofPropagatorwithnvertexinsertions,fig:feynmandiagramof1loopwithnvertexinsertions} for arbitrary $n$:
\begin{equation}
\begin{aligned}
  G_{[\del_{(1,\cdots n)}]} (\xdotY_{12})  & =\sum_{i=1}^{n} \frac { G_{[\del_{i}]}(\xdotY_{12})  } {\prod\limits_{j\neq i} (\nu_j^2 - \nu_i^2)}, \quad \mathcal I_{n,n}  = \sum_{i=1}^{n} \frac {\mathcal I_{[\Delta_i]} } {\prod\limits_{j\neq i} (\nu_j^2 - \nu_i^2)} .
\end{aligned}
\end{equation}
\begin{figure}[H]
\begin{subfigure}[t]{0.3\textwidth}
\centering
\begin{tikzpicture}
\drawdot{0,0}; \node at (0,0)[anchor=north]{\tiny$Y$};
\drawdot{1,0}; \node at (1,0)[anchor=west]{\tiny$X_1$};
\drawdot{-0.6,-0.7}; \node at (-0.6,-0.7)[anchor=north]{\tiny$X_2$};
\drawdot{-0.6,0.7}; \node at (-0.6,0.7)[anchor=south]{\tiny$X_3$};
\draw [->] {(1,0)} -- (0.5,0); \draw  (0.5,0) -- (0,0); \node at (0.5,0)[anchor=south]{\tiny$\Delta_1$};
\draw [->] {(-0.6,-0.7)} -- (-0.3,-0.35); \draw  (-0.3,-0.35) -- (0,0); \node at(-0.2,-0.45)[anchor=south east]{\tiny$\Delta_2$};
\draw [->] {(-0.6,0.7)} --  (-0.3,0.35); \draw  (-0.3,0.35) -- (0,0); \node at(-0.2,0.45)[anchor=north east]{\tiny$\Delta_3$};
\end{tikzpicture}
\caption{$3$-point function}
\end{subfigure}
\hfill
\begin{subfigure}[t]{0.3\textwidth}
\centering
\begin{tikzpicture}
\draw (0,0) circle (0.5);
\draw[->] (-1.2,0) -- (-0.8,0);
\draw (-0.8,0) -- (-0.5,0);
\drawdot{-0.5,0}; 
\drawdot{0.5,0}; 
\node at (-1.2,0)[anchor=east]{\tiny$X_1$};
\node at (0.5,0)[anchor=west]{\tiny$X_2$};
\node at (-0.5,0)[anchor=west]{\tiny$Y$};
\node at (0,0.5)[anchor=south]{\tiny$\Delta_2$}; 
\node at (-0.8,0)[anchor=south]{\tiny$\Delta_1$};
\node at (0,-0.5)[anchor=south]{\tiny$\Delta_3$}; 
\end{tikzpicture}
\caption{Tadpole with vertex insertion}\label{fig:feynmandiagramoftadpolewithvertex}
\end{subfigure}
\hfill
\begin{subfigure}[t]{0.3\textwidth}
\centering
\begin{tikzpicture}
\draw (0,0) circle (0.5);
\draw (-0.5,0) -- (0.5,0);
\drawdot{-0.5,0}; 
\drawdot{0.5,0}; 
\node at (0.5,0)[anchor=west]{\tiny$X$};
\node at (-0.5,0)[anchor=east]{\tiny$Y$};
\node at (0,0.5)[anchor=south]{\tiny$\Delta_1$}; 
\node at (0,0)[anchor=south]{\tiny$\Delta_2$}; 
\node at (0,-0.5)[anchor=south]{\tiny$\Delta_3$};
\end{tikzpicture}
\caption{$3$-melon}
\end{subfigure}
\\
\centering
\begin{subfigure}{0.9\textwidth}
\centering
\begin{tikzpicture}
\draw (0,0) circle (0.5);
\draw[->] (-1.2,0) -- (-0.8,0);
\draw (-0.8,0) -- (-0.5,0);
\drawdot{-0.5,0}; 
\node at (-1.2,0)[anchor=east]{\tiny$X_1$};
\node at (-0.5,0)[anchor=west]{\tiny$Y$};
\node at (-0.8,0)[anchor=south]{\tiny$\Delta_1$};
\node at (0.5,0)[anchor=west]{\tiny$\Delta_2$};
\node at (1.7,0){$\cong$};
\centerarc[](4,0)(0:180:0.5cm);
\centerarc[densely dotted](4,0)(180:360:0.5cm);
\draw[->] (2.8,0) -- (3.2,0);
\draw (3.2,0) -- (3.5,0);
\drawdot{3.5,0}; 
\drawdot{4.5,0};
\node at (2.8,0)[anchor=east]{\tiny$X_1$};
\node at (3.2,0)[anchor=south]{\tiny$\Delta_1$};
\node at (3.5,0)[anchor=west]{\tiny$Y$};
\node at (4.5,0)[anchor=west]{\tiny$Y'$};
\node at (4,0.5)[anchor=south]{\tiny$\Delta_2$};
\node at (4,-0.5)[anchor=north]{\tiny$\eta = 0$}; 
\end{tikzpicture}
\caption{Construction of tadpole with $\delta$-function propagator}\label{fig:feynmandiagramoftadpole}
\end{subfigure}
\caption{$3$-point function and related Feynman diagrams}
\label{fig:3pointfunctionandrelatedfeynmandiagrams}
\end{figure} \vspace{-\baselineskip}

\subsubsection{\texorpdfstring{$n$}{n}-Point function and \texorpdfstring{$n$}{n}-melon}
\vspace{-0.75\baselineskip} Using \cref{eq:generalizedcorrelationfunctionaseulerintegral}, a general scalar $n$-point function (corresponding to the feynman diagram \raisebox{-\baselineskip}{\scalebox{0.5}{\begin{tikzpicture}
\drawdot{0,0}; \node at (0,0)[anchor=west]{\tiny$Y$};
\node at (1,1)[anchor=south]{\tiny$X_1$}; \node at (1/2,1/2)[anchor=west]{\tiny$\Delta_1$}; \drawcenterarrow[](1,1)(0,0);
\node at (-1,1)[anchor=south]{\tiny$X_2$}; \node at (-1/2,1/2)[anchor=east]{\tiny$\Delta_2$}; \drawcenterarrow[](-1,1)(0,0);
\node at (-1,-1)[anchor=north]{\tiny$X_3$}; \node at (-1/2,-1/2)[anchor=east]{\tiny$\Delta_3$}; \drawcenterarrow[](-1,-1)(0,0);
\node at (0,-1)[anchor=north]{$\cdots$}; \drawcenterarrow[](0,-1)(0,0);
\node at (1,-1)[anchor=north]{\tiny$X_n$}; \node at (1/2,-1/2)[anchor=west]{\tiny$\Delta_n$}; \drawcenterarrow[](1,-1)(0,0);
\end{tikzpicture}}}) is given by:
\begin{equation}
\begin{aligned}
  \mathcal J_n(\hat X) & = \mathcal {\bar N}_n\int^*_{\lambda, \, \mu} \lambda^{\bar\Delta} \, \mu^{\Delta} \, e^{- \lambda^2 } \, \frac { e^{ \frac {(\sum \lambda \, \mu \, \hat X)^2} {(1 + \sum \mu^2)}}} {(1 + \sum \mu^2)^{\frac {\DD}{2}}}  = \mathcal {\bar N}_n\int^*_{\lambda, \, \mu} \lambda^{\bar\Delta} \, \mu^{\Delta} \, \frac { e^{- \lambda^2 \, (1 + \sum \mu^2) + (\sum \lambda \, \mu \, \hat X)^2}} {(1 + \sum \mu^2)^{\frac {\DD - \sum \bdel}{2}}},
\end{aligned}
\end{equation}
where
\begin{equation}
\begin{aligned}
  \mathcal {\bar N}_n = \frac {1} {(2 \pi^{\frac {\DD}{2}})^{(n - 1) } \, \Gamma(\frac {\DD - \sum \Delta} {2})}.
\end{aligned}
\end{equation}
As noted under \cref{eq:generalizedcorrelationfunctionaseulerintegral}, any one parameter $\lambda_i$ can always be integrating out in such constructions:
\begin{equation}
\begin{aligned}
  \mathcal J_n(\hat X) & = \mathcal {\bar N}_n\int^*_{\lambda, \, \mu} \frac {\lambda^{\bar\Delta} \, \mu^{\Delta} \, \delta(\lambda_i - 1)} {(1 + \sum \mu^2)^{\frac {\DD - \sum \bdel}{2}} \,  (\lambda^2 \, (1 + \sum \mu^2) - (\sum \lambda \, \mu \, \hat X)^2)^{\frac {\sum\bar\Delta}{2}} }, 
  \\ \mathcal {\bar N}_n & = \frac {\Gamma(\frac {\sum\bar\Delta}{2})} {2 \, (2 \pi^{\frac {\DD}{2}})^{(n - 1) } \, \Gamma(\frac {\DD - \sum \Delta} {2})}.
\end{aligned}
\end{equation}
The $n$-melon diagram, $\mathcal I_{2,n}$, is related to the $n$-point function, $\mathcal J_n(\hat X)$, by the identification of all $\hat X$ to the same point and integrating over this internal vertex, i.e. $\mathcal I_{2,n} = \Omega_{d+1} \times \lim\limits_{\xdotY \rightarrow 1} \mathcal J_n$. 
Represented by the incidence matrix:
\begin{equation}
\begin{aligned}
  L_1{}_i{} = \lambda_i{}, \quad L_2{}_i{} = - \mu_i, \quad U_{2,n} = \left( \begin{smallmatrix}
      \mathds 1_n &   \im \left( \begin{smallmatrix}
            \lambda_1 & - \mu_1 \\
            \cdots & \cdots \\
            \lambda_n & - \mu_n
          \end{smallmatrix} \right) \\
      \im \left( \begin{smallmatrix}
            \lambda_1 &  \cdots & \lambda_n\\
            - \mu_1 & \cdots & - \mu_n
          \end{smallmatrix} \right) & \mathds 1_2
      \end{smallmatrix} \right),
\end{aligned}
\end{equation}
using \cref{eq:FeynmanintegralforScalarsFormula}, its integral form is
\begin{equation}
\begin{aligned}
  \mathcal I_{2,n} & = \frac {1}{(4 \pi^{\DD})^{\frac{n-2}{2}} \, \Gamma(\frac {\DD - \sum \bdel}{2})\, \Gamma(\frac {\DD - \sum \del}{2})} \, \int^*_{\lambda, \mu} \frac {\lambda^{\bdel} \, \mu^{\del}}{(1 + \vec \lambda^2 + \vec\mu^2 + \frac 1 2 \sum\limits_{i,j} (\lambda_i \, \mu_j - \lambda_j \, \mu_i)^2)^{\frac {\DD}{2}}}.
\end{aligned}
\end{equation}

\subsubsection{\texorpdfstring{$3$}{3}-point function and \texorpdfstring{$3$}{3}-melon}
Specifically, the $3$-point function is
\begin{equation}
\begin{aligned}
  \mathcal J_3(\hat X) & = \frac {\Gamma(\frac {\sum\bar\Delta}{2})} {8 \pi^{\DD} \, \Gamma(\frac {\DD - \sum \Delta} {2})} \int^*_{\lambda, \, \mu} \frac {\lambda_1^{\bar\Delta_1} \,\lambda_2^{\bar\Delta_2} \, \mu_1^{\Delta_1} \, \mu_2^{\Delta_2} \, \mu_3^{\Delta_3} } {(1 + \sum \mu^2)^{\frac {\DD - \sum \bdel}{2}} \, (1 + f_{3}(\hat X))^{\frac {\sum\bar\Delta}{2}} }, 
\end{aligned}
\end{equation}
where
\begin{equation}
\begin{aligned}
  f_{3}(\hat X) & =  (1 + \lambda^2) \, (1 + \sum \mu^2) - (\lambda_1 \, \mu_1 \, \hat X_1 + \lambda_2 \, \mu_2 \, \hat X_2 + \mu_3 \, \hat X_3)^2 - 1
  \\ & =  \mu_1^2 + \mu_2^2 + \lambda_1^2 (1 + \mu_2^2 + \mu_3^2) + \lambda_2^2 (1 + \mu_1^2 + \mu_3^2)
  \\ & - 2\, (\lambda_1 \, \mu_1 \, \lambda_2 \, \mu_2 \, \xdotY_{12} + \lambda_1 \, \mu_1 \, \mu_3 \, \xdotY_{13} + \lambda_2 \, \mu_2 \, \mu_3 \, \xdotY_{23}),
\end{aligned}
\end{equation}
its irreducible form being
\begin{equation}
\begin{aligned}
  \mathcal J_3(\hat X) & =  \frac {\Gamma(\frac {\sum\bar\Delta}{2})} {16 \pi^{\DD} \, \Gamma(\frac {\DD - \sum \Delta} {2})} \int^*_{\lambda, \, \mu} \frac {\lambda_1^{\bar\Delta_1} \,\lambda_2^{\bar\Delta_2} \, \mu_1^{\Delta_1} \, \mu_2^{\Delta_2} \, \mu_3^1 } { (1 + \mu_3 \, (1 + \mu_1^2 + \mu_2^2))^{\frac {\DD - \sum \bdel}{2}} \, (1 + f'_{3}(\hat X))^{\frac {\sum\bar\Delta}{2}} }, 
\end{aligned}
\end{equation}
where
\begin{equation}
\begin{aligned}
  f'_{3}(\hat X) & =  \lambda^2_1 + \mu^2_1 - 2\,\lambda_1 \, \mu_1 \, \xdotY_{13} + \lambda^2_2 + \mu_2^2 - 2 \, \lambda_2 \, \mu_2 \, \xdotY_{23} 
  \\ & + \mu_3 \,  \Big(\lambda_1^2 + \lambda_2^2 + \lambda_1^2 \, \mu_2^2  + \lambda_2^2 \, \mu_1^2 - 2\, \lambda_1 \, \mu_1 \, \lambda_2 \, \mu_2 \, \xdotY_{12}  \Big).
\end{aligned}
\end{equation}
The GKZ system describing the generalisation of this integral has $68$ roots and as many $\mathcal A$-hypergeometric series. 
The closely related $3$-melon diagram equalling
\begin{equation}\label{eq:3meloneulerintegralform}
\begin{aligned}
  \mathcal I_{2,3} & = \frac {1}{2 \pi^{\frac \DD 2} \, \Gamma(\frac {\DD - \sum \bdel}{2})\, \Gamma(\frac {\DD - \sum \del}{2})} \, \int^*_{\lambda, \mu} \frac {\lambda^{\bdel} \, \mu^{\del}}{(f_{2,3})^{\frac {\DD}{2}}},
  \\ f_{2,3}& =  1 + \sum_{i=1}^{3} (\lambda_i^2 + \mu_i^2) + (\lambda_1 \, \mu_2 - \lambda_2 \, \mu_1)^2 + (\lambda_2 \, \mu_3 - \lambda_3 \, \mu_2)^2 + (\lambda_3 \, \mu_1 - \lambda_1 \, \mu_3)^2,
\end{aligned}
\end{equation}
has the same series complexity.

\subsubsection{Tadpole}
It is easy to evaluate \cref{fig:feynmandiagramoftadpole} and express it in terms of \cref{eq:oneloopcharacterintegralevalutated}:
\begin{equation}
\begin{aligned}
  \mathcal I_{1,2,[0]} & = G_{[\Delta_1]} \, \frac {1} {\bdel_2} \int^*_{\mu} \frac {\mu^{\del_2}} {(1 + \mu^2)^{1 + \frac {\del_2}{2}}} =  \frac {\mathcal I_{[\Delta_1]}} {\Omega_{d+1} \, \bdel_2 \, \del_2}.
\end{aligned}
\end{equation}
Using \cref{eq:propwith1vertexinsertionfinalresult}, \cref{fig:feynmandiagramoftadpolewithvertex} is:
\begin{equation}
\begin{aligned}
  \mathcal I_{1,2,[1]} & = \frac {\mathcal I_{[\Delta_1]}}{\Omega_{d+1} \, (\nu_2^2 - \nu_3^2)} \Big(\frac {1} {\bdel_3 \, \del_3} - \frac {1} {\bdel_2 \, \del_2} \Big) = \frac {\mathcal I_{[\Delta_1]}}{\Omega_{d+1} \, \bdel_2 \, \del_2 \, \bdel_3 \, \del_3},
\end{aligned}
\end{equation}
and of course, \cref{eq:Propagatorwithmultiplevertexinsertions} can be used to extend it to $n$-vertex insertions.
\begin{figure}[H]
\begin{subfigure}[t]{0.24\textwidth}
\centering
\begin{tikzpicture}
\drawdot{0,0}; \node at (0,0)[anchor=north]{\tiny$Y$};
\drawdot{0.7,0.6}; \node at (0.7,0.6)[anchor=south]{\tiny$X_4$};
\drawdot{0.7,-0.6}; \node at (0.7,-0.6)[anchor=north]{\tiny$X_1$};
\drawdot{-0.7,-0.6}; \node at (-0.7,-0.6)[anchor=north]{\tiny$X_2$};
\drawdot{-0.7,0.6}; \node at (-0.7,0.6)[anchor=south]{\tiny$X_3$};
\draw [->] {(0.7,0.6)} -- (0.35,0.3); \draw  (0.35,0.3) -- (0,0); \node at (0.35,0.3)[anchor=west]{\tiny$\Delta_4$};
\draw [->] {(0.7,-0.6)} -- (0.35,-0.3); \draw  (0.35,-0.3) -- (0,0); \node at (0.35,-0.3)[anchor=west]{\tiny$\Delta_1$};
\draw [->] {(-0.7,-0.6)} -- (-0.35,-0.3); \draw  (-0.35,-0.3) -- (0,0); \node at(-0.35,-0.3)[anchor=east]{\tiny$\Delta_2$};
\draw [->] {(-0.7,0.6)} --  (-0.35,0.3); \draw  (-0.35,0.3) -- (0,0); \node at(-0.35,0.3)[anchor=east]{\tiny$\Delta_3$};
\end{tikzpicture}
\caption{$4$-point function}
\end{subfigure}
\hfill
\begin{subfigure}[t]{0.24\textwidth}
\centering
\begin{tikzpicture}
\draw (0.5,0) circle (0.5);
\draw (-0.5,0) circle (0.5);
\drawdot{0,0}; 
\node at (0,0)[anchor=west]{\tiny$Y$};
\node at (-1,0)[anchor=west]{\tiny$\Delta_1$};
\node at (1,0)[anchor=east]{\tiny$\Delta_2$};
\end{tikzpicture}
\caption{$2$-loops}\label{fig:2connectedloops}
\end{subfigure}
\hfill
\begin{subfigure}[t]{0.24\textwidth}
\centering
\begin{tikzpicture}
\draw (0,0) circle (0.5);
\draw (-0.5,0) -- (0.5,0);
\drawdot{-0.5,0}; 
\drawdot{0.5,0}; 
\draw[->] (-1.2,0) -- (-0.8,0);
\draw (-0.8,0) -- (-0.5,0);
\node at (0.5,0)[anchor=west]{\tiny$X_3$};
\node at (-0.5,0)[anchor=south east]{\tiny$X_2$};
\node at (-1.2,0)[anchor=east]{\tiny$X_1$};
\node at (0,0.5)[anchor=south]{\tiny$\Delta_1$}; 
\node at (0,0)[anchor=south]{\tiny$\Delta_2$}; 
\node at (0,-0.5)[anchor=south]{\tiny$\Delta_3$};
\node at (-0.8,0)[anchor=north]{\tiny$\Delta_4$};
\end{tikzpicture}
\caption{$3$-melon tadpole}\label{fig:3melontadpole}
\end{subfigure}
\hfill
\begin{subfigure}[t]{0.24\textwidth}
\centering
\begin{tikzpicture}
\draw (0,0) circle (0.5);
\drawdot{0.5, 0};
\drawdot{-0.5, 0};
\draw [black] plot [smooth, tension=1.5] coordinates{ (-0.5, 0) (0,0.3) (0.5, 0)};
\draw [black] plot [smooth, tension=1.5] coordinates{ (-0.5, 0) (0,-0.3) (0.5, 0)};
\node at (-0.5,0)[anchor=east]{\tiny$X_1$};
\node at (0.5,0)[anchor=west]{\tiny$X_2$};
\node at (0,0.5)[anchor=south]{\tiny$\Delta_1$}; 
\node at (0,0.35)[anchor=north]{\tiny$\Delta_2$};
\node at (0,-0.35)[anchor=south]{\tiny$\Delta_3$};
\node at (0,-0.5)[anchor=north]{\tiny$\Delta_4$};
\end{tikzpicture}
\caption{$4$-melon}\label{fig:4melon}
\end{subfigure}
\caption{$4$-point function and related Feynman diagrams}
\label{fig:4pointfunctionandrelatedfeynmandiagrams}
\end{figure} 
\vspace{-\baselineskip}
\subsubsection{\texorpdfstring{$2$}{2} simply connected loops}
Just like the tadpole diagram, \cref{fig:2connectedloops} is also easily evaluated in terms of the $1$-loop integral:
\begin{equation}
\begin{aligned}
  \mathcal I_{1,2,[0,0]} & = \frac {\mathcal I_{[\Delta_1]} \, \mathcal I_{[\Delta_2]} } {\Omega_{d+1}} .
\end{aligned}
\end{equation}
The addition of vertices to each loop is just as simple, since the propagator expressions of \cref{fig:feynmandiagramofPropagatorwithnvertexinsertions} can be inductively found for arbitrary $n$, as shown in \cref{eq:Propagatorwithmultiplevertexinsertions}. 
So say, $n_1, \, n_2$ vertices were dropped onto the $2$ loops of this diagram, then
\begin{equation}
\begin{aligned}
  \mathcal I_{1,2,[n_1, n_2]} & = \Omega_{d+1} \times \lim_{\sigma \rightarrow 1} G_{[\Delta_{1\cdots n_1}]}(\sigma) \times G_{[\Delta_{1\cdots n_2}]}(\sigma).
\end{aligned}
\end{equation}

\subsubsection{\texorpdfstring{$3$}{3}-melon tadpole}
\Cref{fig:3melontadpole} can expressed in terms of the $3$-melon integral, $\mathcal I_{2,3}$, given in \cref{eq:3meloneulerintegralform}, as follows:
\begin{equation}
\begin{aligned}
  \mathcal I_{1,4} & = \frac {1} {4 \pi^{\DD}} \,\mathcal I_{2,3} \, \Xint{\times}^*_{\lambda_4, \mu_4} \, \int_P \frac {\lambda_4^{\bdel_4} \, \mu_4^{\del_4} \, |X_2|^{\del_4}} {P_4^2 } \, e^{-2 \im P_4 \, (\lambda_4 \, \hat X_1 - \mu_4 \, X_2)}
  \\ & = \frac {\Gamma(\frac {\DD - \del_1 - \del_2 - \del_3}{2})} {2 \pi^{\frac {\DD}{2}} \, \Gamma(\frac {\DD - \del_1 - \del_2 - \del_3 - \del_4}{2})} \,\mathcal I_{2,3}  \, \int^*_{\lambda_4, \mu_4} \lambda_4^{\bdel_4} \, \mu_4^{\del_4}\, e^{- (\lambda_4 \, \hat X_1 - \mu_4 \, X_2)^2}
  \\ & = \frac {\Gamma(\frac {\DD - \del_1 - \del_2 - \del_3}{2})\, \Gamma(\frac {\DD - \del_4}{2})} {2 \pi^{\frac {\DD}{2}} \, \Gamma(\frac {\DD - \del_1 - \del_2 - \del_3 - \del_4}{2}) \, \bdel_4 \, \del_4} \,\mathcal I_{2,3}.
\end{aligned}
\end{equation}

\subsubsection{\texorpdfstring{$3$}{3}-Loops : pacman, pillbox, peace}
The conversion of these diagrams to their Euler integral form was already described in the introduction itself \cref{sec:Higherloopcorrectionshow}. 
\begin{figure}[H]
\centering
\begin{subfigure}{0.32\textwidth}
\centering
\begin{tikzpicture}
\draw (0,0) circle (0.5);
\drawdot{0.3, 0.4};
\drawdot{0.3, -0.4};
\drawdot{-0.5, 0};
\draw (-0.5, 0) -- (0.3, 0.4);
\draw (-0.5, 0) -- (0.3, -0.4);
\node at (-0.5,0)[anchor=east]{\tiny$X_1$};
\node at (0.3, 0.4)[anchor=west]{\tiny$X_2$};
\node at (0.3, -0.4)[anchor=west]{\tiny$X_3$};
\node at (1.5,0){$\leftarrow$};
\draw (2.5, 0) -- (3.5, 0);
\draw (3,0.5) -- (3,-0.5);
\draw (3.7,0.5)-- (3.5,0) -- (3.7,-0.5);
\draw (2.3,0.5) -- (2.5,0) -- (2.3,-0.5);
\drawdot{2.5, 0};\drawdot{3, 0};\drawdot{3.5, 0};
\node at (2.5,0)[anchor=east]{\tiny$X_2$};
\node at (2.9,0.1)[anchor=north west]{\tiny$X_1$};
\node at (3.5,0)[anchor=west]{\tiny$X_3$};
\end{tikzpicture}
\caption{Pacman}
\end{subfigure} 
\hfill
\begin{subfigure}{0.32\textwidth}
\centering
\begin{tikzpicture}
\draw (0,0) -- (0.75,0) -- (0.75,1) -- (0,1) -- cycle;
\drawdot{0, 0};\node at (0,0)[anchor=north]{\tiny$X_1$};
\drawdot{0.75, 0};\node at (0.75, 0)[anchor=north]{\tiny$X_2$};
\drawdot{0.75,1};\node at (0.75,1)[anchor=south]{\tiny$X_3$};
\drawdot{0, 1};\node at (0, 1)[anchor=south]{\tiny$X_4$};
\centerarc[](0.75,0.5)(-90:90:0.5cm);
\centerarc[](0,0.5)(90:270:0.5cm);
\node at (1.7,0.5){$\leftarrow$};
\draw (2.5, 0.5) -- (4, 0.5);
\draw (3,1) -- (3,0.5);
\draw (3.5,0) -- (3.5,0.5);
\draw (4.2,1)-- (4,0.5)-- (4.2,0);
\draw (2.3,1) -- (2.5,0.5) -- (2.3,0);
\drawdot{2.5, 0.5};\drawdot{3, 0.5};\drawdot{3.5, 0.5};\drawdot{4, 0.5};
\node at (2.5,0.5)[anchor=east]{\tiny$X_1$};
\node at (3,0.5)[anchor=north]{\tiny$X_2$};
\node at (3.5,0.5)[anchor=south]{\tiny$X_3$};
\node at (4,0.5)[anchor=west]{\tiny$X_4$};
\end{tikzpicture}
\caption{Pillbox}
\end{subfigure}
\hfill
\begin{subfigure}{0.32\textwidth}
\centering
\begin{tikzpicture}
\draw (0,0) circle (0.5);
\draw (0.4,0.3) -- (0,0) -- (0,-0.5);
\draw (-0.4,0.3) -- (0,0);
\drawdot{0, 0};\node at (0.1,0.1)[anchor=north east]{\tiny$X_0$};
\drawdot{0.4, 0.3};\node at (0.4, 0.3)[anchor=west]{\tiny$X_1$};
\drawdot{-0.4, 0.3};\node at (-0.4, 0.3)[anchor=east]{\tiny$X_2$};
\drawdot{0,-0.5};\node at (0,-0.5)[anchor=north]{\tiny$X_3$};
\node at (0.8,0){$\leftarrow$};
\draw (2.2,0.3) -- (1.8,0) -- (1.8,-0.5);
\draw (1.4,0.3) -- (1.8,0);
\drawdot{1.8, 0};\node at (1.9, 0.1)[anchor=north east]{\tiny$X_0$};
\drawdot{2.2, 0.3};\node at (2.1, 0.4)[anchor=west]{\tiny$X_1$};
\drawdot{1.4, 0.3};\node at (1.5, 0.4)[anchor=east]{\tiny$X_2$};
\drawdot{1.8,-0.5};\node at (1.8,-0.5)[anchor=north]{\tiny$X_3$};
\draw [black] plot [smooth, tension=1] coordinates{ (2.2, 0.6) (2.2, 0.3) (2.5, 0.2)};
\draw [black] plot [smooth, tension=1] coordinates{ (1.4, 0.6) (1.4, 0.3) (1.1, 0.2)};
\draw [black] plot [smooth, tension=1] coordinates{ (1.5,-0.7) (1.8,-0.5) (2.1,-0.7)};
\end{tikzpicture}
\caption{Peace}
\end{subfigure}
\caption{Irreducible $3$-loop Feynman diagrams as limits of tree-level correlation functions}
\label{fig:somefeynmandiagrams}
\end{figure}
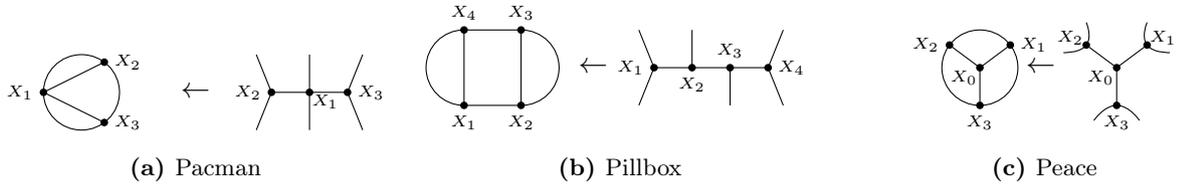 \vspace{-\baselineskip} \noindent
Directly using the incidence matrices from \cref{tb:AdjacencymatricesofsomeFeynmanDiagrams}, they are:
\begin{equation}
\begin{aligned}
  \mathcal I_{\rm pac} & =   \frac 1 {4 \pi^{\DD} \, \Gamma(\frac {\DD - \bar\Delta_1- \bar\Delta_2- \bar\Delta_3- \bar\Delta_4}{2}) \, \Gamma(\frac {\DD - \bar\Delta_5 - \Delta_1 - \Delta_2}{2}) \,  \Gamma(\frac {\DD - \del_3 - \del_4 -\del_5}{2})}
  \\ & \times \int^*_{\lambda, \, \mu} \lambda^\bdel{} \, \mu^\del{} \, \left(\det \left( \begin{smallmatrix}
        1 + \lambda_1^2 + \lambda_2^2 + \lambda_3^2 + \lambda_4^2  &  - (\lambda_1 \, \mu_1 + \lambda_2 \, \mu_2) &  - (\lambda_3 \, \mu_3 + \lambda_4 \, \mu_4) \\ 
         - (\lambda_1 \, \mu_1 + \lambda_2 \, \mu_2) & 1 + \lambda_5^2 + \mu_1^2 + \mu_2^2 & - \lambda_5 \, \mu_5 \\ 
         - (\lambda_3 \, \mu_3 + \lambda_4 \, \mu_4)  & - \lambda_5 \, \mu_5  & 1 + \mu_3^2 + \mu_4^2 + \mu_5^2
      \end{smallmatrix} \right)\right)^{- \frac {d + 2}{2}}
  \\ \mathcal I_{\rm pil} & =   \frac 1 {4 \pi^{\DD} \, \Gamma(\frac {\DD - \bdel_1 - \del_4 - \bdel_5}{2})\, \Gamma(\frac {\DD - \del_1- \bdel_2 - \del_6}{2})\, \Gamma(\frac {\DD - \del_2 - \bdel_3 - \bdel_6}{2})\, \Gamma(\frac {\DD-\del_3- \bdel_4 - \del_5}{2})}
  \\ & \times \int^*_{\lambda, \, \mu} \lambda^\bdel{} \, \mu^\del{} \, \left(\det \left( \begin{smallmatrix}
        1 + \lambda_1^2 + \lambda_5^2 + \mu_4^2 &  - \lambda_1 \, \mu_1 & 0 & - (\lambda_4 \, \mu_4 + \lambda_5 \, \mu_5) \\
          - \lambda_1 \, \mu_1 & 1 + \lambda_2^2 + \mu_1^2 + \mu_6^2 &  - (\lambda_2 \, \mu_2 + \lambda_6 \, \mu_6) & 0 \\ 
          0 & - (\lambda_2 \, \mu_2 + \lambda_6 \, \mu_6) & 1 + \lambda_3^2 + \lambda_6^2 + \mu_2^2  & - \lambda_3 \, \mu_3 \\ 
         - (\lambda_4 \, \mu_4 + \lambda_5 \, \mu_5)  & 0 & - \lambda_3 \, \mu_3  & 1 + \lambda_4^2 + \mu_3^2 + \mu_5^2
      \end{smallmatrix} \right) \right)^{- \frac {d + 2}{2}}
  \\ \mathcal I_{\rm pea} & =   \frac 1 {4 \pi^{\DD} \, \Gamma(\frac {\DD - \bdel_1 - \bdel_2- \bdel_3}{2})\, \Gamma(\frac {\DD- \del_1 - \bdel_4 - \del_6}{2})\, \Gamma(\frac {\DD - \del_2 - \del_4 - \bdel_5}{2})\, \Gamma(\frac {\DD - \del_3 - \del_5 - \bdel_6}{2})}
  \\ & \times \int^*_{\lambda, \, \mu} \lambda^\bdel{} \, \mu^\del{} \, \left(\det    \left( \begin{smallmatrix}
        1 + \lambda_1^2 + \lambda_2^2 + \lambda_3^2 & - \lambda_1 \, \mu_1 & - \lambda_2 \, \mu_2 & - \lambda_3 \, \mu_3 \\
        - \lambda_1 \, \mu_1 & 1 + \lambda_4^2 + \mu_1^2 + \mu_6^2 & - \lambda_4 \, \mu_4 & - \lambda_6 \, \mu_6 \\ 
        - \lambda_2 \, \mu_2  & - \lambda_4 \, \mu_4 & 1 + \lambda_5^2 + \mu_2^2 + \mu_4^2 & - \lambda_5 \, \mu_5 \\ 
        - \lambda_3 \, \mu_3 & - \lambda_6 \, \mu_6 & - \lambda_5 \, \mu_5  & 1 + \lambda_6^2 + \mu_3^2 + \mu_5^2
      \end{smallmatrix} \right) \right)^{- \frac {d + 2}{2}} .
\end{aligned}
\end{equation}

\subsection{Representative constructions of vector Feynman integrals}\label{sec:vectorFeynmanIntegrals}
The following examples are representative of the general procedure that relates vector Feynman integrals to purely scalar Feynman integrals.

\subsubsection{Vector \texorpdfstring{$1$}{1}-loop character integral}
The $1$-loop Feynman integral in \cref{fig:feynmandiagramof1loop}, involving the trace of the vector propagator in the coincident point limit, equals
\begin{equation}
\begin{aligned}
  \mathcal I^{(1)}_{1,1} & = \int_{\hat X} \mathcal G_{II'} \, \delta^{II'} \circ G(\hat X, \, \hat X).
\end{aligned}
\end{equation}
\paragraph{Massive:}
The trace of $\mathcal G$ in \cref{eq:vectorprop}, upon using the \cref{eq:feynmanintegraloperatorreduction} (on a part of it), reduces to
\begin{equation}
\begin{aligned}
  \mathcal G_{II'} \, \delta^{II'} & = \frac {d \, \bar \Delta \, \Delta} {(\bar \Delta - 1)\, (\Delta - 1)} + \frac {4 \, \lambda \, \mu \, X^2 \, P^2 }{ (\bar \Delta - 1)\, (\Delta - 1)}.
\end{aligned}
\end{equation}
The first term simply introduces an additional factor to the underlying $1$-loop scalar integral, $\mathcal I_{[\Delta]}$, given in \cref{eq:scalar1loopcharacterintegralform}. The second vanishes, as shown in the following:
\begin{equation}
\begin{aligned}
  \int_{\hat X} \, \frac {4 \, \lambda \, \mu \, X^2 \, P^2 }{ (\bar \Delta - 1)\, (\Delta - 1)} \circ G(\hat X, \, \hat X) & =  \int_{\hat X} \Xint{\times}^*_{\lambda, \mu} \int_{P} \frac {4 \, \lambda^{\bdel+1} \, \mu^{\del+1}}{ (\bar \Delta - 1)\, (\Delta - 1) \, (\lambda - \mu)^{\DD}} e^{- 2 \im |P \hat X|}
  \\ & = \int_{\hat X}  \Xint{\times}^*_{\lambda, \mu} \int_{P} \frac {4 \, \mu^{\del+1}}{ (\bar \Delta - 1)\, (\Delta - 1) \, (1 - \mu)^{\DD}} e^{- 2 \im |P \hat X|}
  \\ & \prop \int_{\hat X}  \int^*_{\mu} \frac {4 \, \mu^{\del+1}}{ (\bar \Delta - 1)\, (\Delta - 1) \, (1 - \mu)^{\DD}} \cancel{\int_{P} e^{- 2 \im |P \hat X|}}.
\end{aligned}
\end{equation}

\paragraph{Massless:}
The trace of $\mathcal G$ in \cref{eq:masslessvectorpropagatormainformula}, using \cref{eq:feynmanintegraloperatorreduction}, reduces to a factor:
\begin{equation}
\begin{aligned}
  \mathcal G_{II'}^{(o)}\,\delta^{II'} & = \frac {4 \,\lambda \,\mu} {(d-2)} \, |PX|^2 \,\Big( \delta_I{}_{I'} \,\delta^{II'} - 1  \Big) \cong \frac {(d+1)\, (d-1)} {(d-2)}.
\end{aligned}
\end{equation}

\subsubsection{Vector \texorpdfstring{$2$}{2}-melon}
The $2$-melon Feynman integral for vectors is:
\begin{equation}
\begin{aligned}
  \mathcal I^{(1)}_{2,2} & = \int_{\hat X, \hat Y} \mathcal G_{IK'}(\hat X, \,\hat Y) \, \mathcal G^{K'I}(\hat Y, \,\hat X) \circ G_{\Delta_1}(\hat X,\,\hat Y) \, G_{\Delta_2}(\hat Y, \, \hat X).
\end{aligned}
\end{equation}

\paragraph{Massless:}
When considering massless propagators, this reduces to a simple change in factor:
\begin{equation}
\begin{aligned}
  \mathcal I^{(1)}_{2,2} & = \frac {d \, (d -1)^2} {(d-2)^2} \, \mathcal I_{2,2},
\end{aligned}
\end{equation}
where $\mathcal I_{2,2}$ is the scalar integral from \cref{sec:2melonfeynmanint}.

\paragraph{Massive:}
Crunching through the indices and making some initial reductions, this integral splits into $10$ types of scalar Feynman integrals, all structured similar to $\mathcal I_{2,2}$ from the scalar case in \cref{eq:2melonfeynmaneulerintegralform}. 
Three of them are straight forward shifts to the values of $\Delta, \, \bdel$:
\begin{equation}
\begin{aligned}
  & \begin{array}{|c|c|c|c|c|}
  \hline
    \mathcal N & \lambda_1^{\cdots} & \mu_1^{\cdots} & \lambda_2^{\cdots} & \mu_2^{\cdots} \\
    \hline
    (d-2) \, \del_1 \, \bdel_1 \,\del_2 \, \bdel_2  &  \bdel_1 & \del_1 & \bdel_2 & \del_2 \\
    \hline
    (\bdel_1-1) \, \bdel_1 \, (\bdel_2-1) \, \bdel_2 & \bdel_1-1 & \del_1+1 & \bdel_2-1 & \del_2+1 \\
    \hline
    (\del_1-1) \, \del_1 \, (\del_2-1) \, \del_2 & \bdel_1+1 & \del_1-1 & \bdel_2+1 & \del_2-1 \\
    \hline    
  \end{array}
\end{aligned}
\end{equation}
with each scalar integral contribution taking the form:
\begin{equation}
\begin{aligned}
  \frac {\mathcal N } {(\bdel_1-1)\,(\bdel_2-1)\,(\del_1-1)\,(\del_2-1)} \,  \mathcal I_{2,2}[\del_1, \, \bdel_1, \,\del_2 \, \bdel_2].
\end{aligned}
\end{equation}
Denoting the maximal perturbation incidence matrix as $\bar U(\beta) = \{\beta_i{}_j{}\}$ in order to use the ``master'' integral form for notational convenience, the other integrals, can be written as:
\begin{equation}
\begin{aligned}
  & \frac {\mathcal N } {(\bdel_1-1)\,(\bdel_2-1)\,(\del_1-1)\,(\del_2-1)} \, \partial_{\beta}^{\omega} \,  \mathcal I_{2,2}[\del_1, \, \bdel_1, \,\del_2 \, \bdel_2][\bar U(\beta)]
\end{aligned}
\end{equation}
where
\begin{equation}
\begin{aligned}
  &\mathcal I_{2,2}[\bar U(\beta)] \vcentcolon = \bar{\mathcal N}_{F} \int^*_{\lambda, \, \mu} \, \frac {\lambda^{\bar\Delta} \, \mu^{\Delta} } { (\det (\bar U_{2,2} + \bar U(\beta)))^{\frac {\DD}{2}} },
\end{aligned}
\end{equation}
and are given by
\begin{equation}
\begin{aligned}
  & \begin{array}{|c|c|c|c|c| c |}
  \hline
    \mathcal N & \lambda_1^{\cdots} & \mu_1^{\cdots} & \lambda_2^{\cdots} & \mu_2^{\cdots} & \bar U(\beta) \\
    \hline
    4 \,\bdel_2 \,\del_1 \,(d + 1 - \bdel_1 - \del_2)  &  \bdel_1 +1 & \del_1 & \bdel_2 & \del_2 + 1 & \beta_{1,2}\\
    \hline
    4 \,\del_2 \,\bdel_1 \,(d + 1 - \del_1 - \bdel_2)  &  \bdel_1  & \del_1 + 1& \bdel_2 + 1& \del_2  & \beta_{1,2}\\
    \hline
    8 \,\bdel_1 \,\bdel_2 &  \bdel_1 & \del_1 +1 & \bdel_2 & \del_2 + 1 & \beta_{1,2}, \, \beta_{3,4}\\
    \hline
    8 \,\del_1 \,\del_2 &  \bdel_1 +1 & \del_1  & \bdel_2 + 1 & \del_2  & \beta_{1,2}, \, \beta_{3,4}\\
    \hline
    16 &  \bdel_1 +1 & \del_1 + 1 & \bdel_2 + 1 & \del_2  + 1 & \beta_{1,2}^2, \, \beta_{3,4}^2\\
    \hline
    \lim\limits_{\eta \rightarrow 0} \frac {4 \, \bdel_2 \, \del_2 }{\Gamma(\eta)} &  \bdel_1 +1 & \del_1 + 1 & \bdel_2 & \del_2  & \beta_{3,4}\\
    \hline
    \lim\limits_{\eta \rightarrow 0} \frac {4 \, \bdel_1 \, \del_1 }{\Gamma(\eta)} &  \bdel_1  & \del_1  & \bdel_2 +1& \del_2 +1 & \beta_{3,4}\\
    \hline
  \end{array},
\end{aligned}
\end{equation}
where each $\beta_i{}_j{}$ also implies $\beta_j{}_i{}$, the exponent attached to them is notational: $\beta^n$ implies $f[\partial_\beta^n]$ is used to construct the perturbing polynomial, and if not mentioned, those elements of $\bar U(\beta)$ are $0$. 
\par
Quite simply because transcribing any integral constructions more complex than this will not only be highly unsavory but also result in something entirely illegible, interested readers are invited to explore this formulation further on their own.

\section{Outlook}\label{ch:discussion}
Stemming from the overarching goal of actualizing arbitrarily higher loop Feynman integral computations of de Sitter entropy (empty and/or with arbitrary field content), the core motivation of this work was to develop a systematic way to represent and compute higher loop Feynman integrals on a spherical background, while avoiding the obstacles that appear when attempting to compute them in position space.
This involved the construction of a new class of propagator expressions in \cref{sec:scalars,sec:VectorFields}. 
Using these expressions, Feynman integrals on the sphere were lifted to embedding space in \cref{ch:dSFeynmanIntegrals}, eventually bringing integrals that were entirely intractable in position space into the domain of Generalized Euler Integrals, that afford systematic solvability, as reviewed in \cref{ch:Eulerintegrals}. 
Although, the current status far from achieves the aforementioned goal, it does suggest the next few concrete steps towards it.

\subsection{General spin sphere propagators} 
In order to vastly improve the applicability of the presented formulation, the next step is the construction of embedding space propagator expressions that remain compatible with the formulation, for gravitons, spinors and general higher spin fields.

It has been observed that when considering symmetric traceless transverse (STT) eigenmodes in \cref{eq:propagatorexpressionsverifiedtosatisfy}, any radially invariant tensor $\mathcal G$ (that remains compatible with the Feynman Integral setup in \cref{sec:HigherSpinFeynmanIntegralsSetup}) will result in either the proper mode integral eigenvalue (i.e. inverse of the Laplacian eigenvalue) or will annihilate that inner product, along the lines of \cref{eq:modeintegraloperatoreigenvalue,eq:vectorpropagatormatrixelements}. 
However, the main barrier to building higher spin propagators is ensuring that the corresponding higher dimensional massless field is gauged fixed in the precise way that replicates the massive action on the sphere as its quotient, with massless fields requiring additional gauge conditions or projections. 
As is to be expected, this difference in choice of gauge can only be detected in the propagators' behaviour with respect to the longitudinal eigenmodes.  
As is evident in \cref{sec:vectorsBRSTgaugefixing}, the current approach to picking and implementing these gauge conditions is ad hoc and needs to be systematized. 
Afterall, it is necessary to choose a ``pretty'' gauge when it comes to the graviton and higher spin fields because it vastly influences the complexity of the interaction vertices. 

Though very different from what is being sought, inspiration for this purpose was (and more still can be) found from the many works exploring propagators in de Sitter space through other approaches. 
STT eigenvectors of arbitrary spin (integer and half-integer) on a sphere, including explicit constructions thereof, with their eigenvalues and degeneracies, have been presented in \cite{Higuchi:1986wu,Camporesi:1994ga,Camporesi:1995fb}. 
An expansion in eigenfunctions of the Laplacian, the spherical representation of SO$(d+2)$ irreps, can be used to exactly build propagators on a sphere \cite{MorettePhysRev.81.848,DeWitt:1957at,Camporesi:1990wm}. 
Such a summation over vector eigenmodes has been used to find massive vector $2$-point functions in \cite{Frob:2013qsa}. 
Massive and massless vector Wightman functions have been found in \cite{Allen:1985wd} by requiring them to satisfy the equations of motion and then, based on requirements set by the vacuum state \cite{Bunch:1978yq}, regularity/``boundary'' conditions, picking out the appropriate solution, \cite{Allen:1985ux,Allen:1987tz}. This procedure was extended to graviton $2$-point functions in dS by the same authors in \cite{Allen:1986tt,Turyn:1988af}. 
Some more vector/photon position space propagator results are \cite{10.1063/1.2738361,Belokogne:2016dvd,Glavan:2022dwb}. 
Some works discussing (d)S graviton and massive spin-$2$ propagators are \cite{Antoniadis:1986sb,Gabriel:1996iy,DHoker:1999bve,Higuchi:2001uv,Garidi:2003bg,Miao:2011fc,Folkerts:2013mra,Glavan:2015ura}, many separating the propagator into transverse, vector and scalar sectors. More general discussions of dS propagators and correlation functions of cosmological observables can be found in \cite{Bros:1995js,Brunetti:2016hgw,Frob:2017gyj}. \cite{Prokopec:2022yon,Anous:2014lia,Anninos:2024fty} are connected to fermionic propagators on a dS background.
\par

\subsection{Feynman integral solution methods} 
After much searching, it appears that interpreting Feynman integrals as generalized Euler integrals and GKZ systems, and/or mathematical objects closely related to them (like Mellin-Barnes integral representations, Pfaffian Systems, Creation operators), may be the only way to consistently tackle the propagator and Feynman integral representations suggested in this work. 
\par
Apart from the aesthetical benefit of representing physically relevant quantities stemming from Feynman integrals as solutions to GKZ PDE's that impose symmetry constraints on them, there are many practical benefits to this approach, not all of which have been fully explored. 
Special cases of mass and dimension in which these systems collapse can be not only be considered in detail (for example, GKZ systems of conformally coupled scalars take simpler forms) but also reverse engineered (specific combinations of mass parameters result in sudden simplifications when the polynomials end up getting positive integer exponents, i.e. $\frac 1 {P^\alpha} \rightarrow P^{\mathbb N}$). 
\par
Making efficient use of restriction algorithms will signficantly reduce the number and complexity of series solutions to sphere Feynman integrals, which otherwise radically grow in size with increasing number of propagators. 
Of course this complexity doesn't magically vanish and instead transfers to the notoriously large complexity of Gröbner bases computations. These systems of PDEs cross pen-and-paper solvability very quickly, so larger scale computational efforts including numerical treatment are in order. 
\par
The physically relevant limits supplied by the sphere Feynman integrals always flow to a maximally singular point in the parameter space, the precise physical significance of which in the more abstract coordinates of GKZ systems is not known. Even more aspirationally, a deeper understanding of this maximally singular physically relevant point may allow direct comparison of the GKZ ideals of Feynman integrals resulting from dS entropy computations with equivalent symmetry generators of hypothesised microscopic models, foregoing the need to actually compute these integrals.

\section*{Acknowledgements}
I am extremely grateful to my advisor, Frederik Denef, without whose supervision and collaboration, this work would neither have been conceived nor progressed. I would also like to thank Dionysios Anninos for detailed comments on the draft. This work was supported in part by the U.S. Department of Energy grant DE-SC0011941.

\appendix
\section{Coordinate conventions on the sphere and its embedding space}\label{app:CoordinateSystemsandNotationalConventionsMainRef}
\subsection{Coordinate systems}\label{app:CoordinateSystems}
\subsubsection{\texorpdfstring{$(d+1)$}{(d+1)}-Sphere}\label{app:CoordinateSystemsSphere}
In order to maintain notational simplicity, the dimensions of the sphere are denoted by:
\begin{equation}
  S^{d+1} \quad : \quad d+1 \equiv D \equiv \DD - 1.
\end{equation}
The volume of $S^{d+1}$ is $\Omega_{d+1} = \tfrac {2 \pi^{\frac {d+2} {2}}} {\Gamma(\frac {d+2} {2})} $. 
General coordinates on the sphere are denoted by $w^\mu$, with the associated metric being $g_\mu{}_\nu{}$. 
Flat orthonormal coordinates, $\hat X(w)$, satisfy $|\hat X|^2 = 1$ and have the associated metric:
\begin{equation}
  \de \hat  s^2 = \delta_I{}_J{} \de \hat X^I{} \de \hat X^J{} = g_\mu{}_\nu{} \de w^\mu \de w^\nu.
\end{equation}
Tensors in flat and spherical coordinates are related by
\begin{equation}
\begin{aligned}
  & T_{I \cdots}^{\cdots} = \hat {e}^\mu_I \, T_{\mu \cdots}^{\cdots}, && \hat {e}^\mu_I \equiv \tfrac {\partial w^\mu} {\partial \hat X^I}, && T_{\cdots}^{I \cdots} = \hat {e}_\mu^I \,T_{\cdots}^{\mu \cdots}, && \hat {e}^I_\mu \equiv \tfrac {\partial \hat X^I} {\partial w^\mu}.
\end{aligned}
\end{equation}
An integral over the sphere is denoted by $\displaystyle\int_{S^{d+1}}  \equiv \int_{\hat X} \equiv \int_w \sqrt g.$
The laplacian of an $s$-indexed tensor $T$ in $\hat X(w)$ is simply $\partial_{\hat X}^2 T$.

It is an understatement to call the parameterisation of distances/triangles in curved spacetime cumbersome. 
For example, on $S^3$, the massive scalar propagator takes a very neat form: $\frac {\sinh((\pi - \theta) \, \nu)}{4 \pi \, \sinh( \pi \nu) \, \sin \theta}$. 
Given its simplicity, it becomes possible to compute the $n$-melon (i.e. $2$ internal points with $n$-propagators connecting them) Feynman integral surprisingly easily:
\begin{equation}
\begin{aligned}
  \mathcal I_{n} & = \Omega_{3} \, \Omega_{2} \, \int_{0}^{\pi } \de \theta \, \sin^{2} \theta \, \prod_{i = 1}^{n} \frac {\sinh((\pi - \theta) \, \nu_i)}{4 \pi \, \sinh( \pi \nu_i) \, \sin \theta} 
  \\ & = \lim_{d \rightarrow 2} \frac {1} {(2 \pi)^{n - 3} \prod \sinh( \pi \nu)} \, \int_{0}^{\pi} \de \theta \, \sin^{d + 1 - n} \theta \, \prod_{i = 1}^{n} (e^{\theta \, \nu_i} - e^{-\theta \, \nu_i}),
\end{aligned}
\end{equation}
which can be computed term by term in dimensional regularisation by assuming $d > n - 2$ before analytically continuing $d$ to $2$. 
Some low $n$ results, which are so easily extensible to higher $n$ that transcription is the hard part, are:
\begin{equation}
\begin{aligned}
  \mathcal I_{1} & = \frac 1 {1 + \nu^2}, \quad \mathcal I_{2} = \frac {\nu_1 \, \coth(\pi \, \nu_1) - \nu_2 \, \coth(\pi \, \nu_2)} {4 \pi \, (\nu_1^2 - \nu_2^2)}
  \\ \mathcal I_{3} & = \lim_{\epsilon \rightarrow 0} \frac {1} {48 \, \pi^2 \, \epsilon} - \frac {\ln 2 + \gamma_{\rm E}} {16 \, \pi^2} + \frac {\sinh (\pi \, \delta\nu_{12,3}) \, \big( \psi^{(0)} (\frac {1 + \im \, \delta\nu_{12,3}}{2}) + \psi^{(0)} (\frac {1 - \im \, \delta\nu_{12,3}}{2})\big)} {128 \, \pi^2 \, \sinh( \pi \nu_1)\, \sinh( \pi \nu_2)\, \sinh( \pi \nu_3)}
  \\ & + (\delta\nu_{12,3} \rightarrow \delta\nu_{23,1} ) + (\delta\nu_{12,3} \rightarrow \delta\nu_{31,2}), \quad \delta\nu_{12,3} \equiv \nu_1 + \nu_2 - \nu_3
  \\ \mathcal I_{\bar 3} & = \lim_{\epsilon \rightarrow 0} \frac {1} {48 \, \pi^2 \, \epsilon} - \frac {\ln 2 + \gamma_{\rm E}} {16 \, \pi^2} - \frac {\big(\psi^{(0)} (\frac {1 + 3\im \, \nu}{2}) +  \psi^{(0)} (\frac {1 -3 \im \, \nu}{2})\big)} {32 \, \pi^2} 
  \\ & + \frac {3  \, \big( \psi^{(0)} (\frac {1 + \im \, \nu}{2}) + \psi^{(0)} (\frac {1 - \im \, \nu}{2} ) - \psi^{(0)} (\frac {1 + 3\im \, \nu}{2}) -  \psi^{(0)} (\frac {1 -3 \im \, \nu}{2})\big)} {128 \, \pi^2 \, \sinh^2( \pi \nu)} .
\end{aligned}
\end{equation}
However, inspite of this simplified propagator, even a $1$-loop diagram (with $3$ internal vertex insertions) starts to become highly impractical:
\begin{equation}
\begin{aligned}
  I_{3,3} & = \Omega_{3}\, \Omega_{2} \, \Omega_{1} \, \int_0^{\pi} \de \theta_1 \, \sin\theta_{1} \int_{0}^{\pi} \de \theta_2 \, \sin^{2}\theta_2 \int_{0}^{2 \pi} \de \theta_3 \, \sin \theta_3 
   \\ &  \times \frac {\sinh(\theta_1 \, \nu_1)}{4 \pi \, \sinh( \pi \nu_1)} \, \frac {\sinh((\pi - \theta') \, \nu_2)}{4 \pi \, \sinh( \pi \nu_2) \, \sin \theta'} \, \frac {\sinh((\pi - \theta'') \, \nu_3)}{4 \pi \, \sinh( \pi \nu_3) \, \sin \theta''}
   \\ & \theta' = \cos^{-1} (\cos \theta_{2} \, \cos \theta_{3}), \quad \theta'' = \pi -  \cos^{-1}(\cos (\theta_{2}-\theta_{1}) \, \cos \theta_{3}).
\end{aligned}
\end{equation}
In comparison, Feynman integral computations in flat space benefit from their Gaussian structure when it comes to integrating over internal vertices, which seems to be indispensible to higher loop computations, as already noted in and around \cref{eq:flatfeynmanintschwinger}, necessitating the shift to embedding space.

\subsubsection{Embedding space \texorpdfstring{$\RR^{\DD}$}{(d+2)-dimensional Euclidean Space}}\label{app:CoordinateSystemsEmbeddingSpace}
General coordinates on Euclidean space $\RR^\DD$, serving as embedding space for $S^{d+1}$ in present context, are denoted by $x^I{}$ with metric $\embed g_I{}_J{}$. 
In flat coordinates $X$, the metric is simply $\delta_I{}_J{}$. 
Spherical coordinates $(t, w)$, $t \in \RR$, are related by $X = e^t \hat X(w)$, $\hat X = \tfrac {X} {|X|}$ and have the metric
\begin{equation}
\begin{aligned}
  \de s^2 = \delta_I{}_J{} \de X^I{} \de X^J{} =  \embed g_I{}_J{} \de x^I{} \de x^J{} = e^{2 t} (\de t^2 + \de \hat s^2) = e^{2 t} (\de t^2 + g_\mu{}_\nu{} \de w^\mu \de w^\nu). 
\end{aligned}
\end{equation}
Specifically, the metric in spherical coordinates is related to that on the embedded sphere by
\begin{equation}
  \embed g_t{}_t{} = e^{2 t}, \quad \embed g_t{}_\mu{} = 0, \quad \embed g_\mu{}_\nu{} = e^{2 t} g_\mu{}_\nu{},
\end{equation}
with the Christoffel symbols $\embed\Gamma$ of the associated Levi-Civita connection $\embednab$ given by
\begin{equation}
\begin{aligned}
  & \embed\Gamma^t{}_t{}_t{} = 1, \quad \embed\Gamma^t{}_\mu{}_\nu{} = - g_\mu{}_\nu{}, \quad \embed\Gamma^t{}_t{}_\mu{} = 0, \quad \embed\Gamma^\mu{}_t{}_t{} = 0, \quad \embed\Gamma^\mu{}_t{}_\nu{} = \delta^\mu_\nu, \quad \embed\Gamma^\mu{}_\nu{}_\rho{} = \Gamma^\mu{}_\nu{}_\rho{},
\end{aligned}
\end{equation}
where $\Gamma^\mu{}_\nu{}_\rho{}$ are Christoffel symbols of the Levi-civita connection $\nabla$ on $S^{d+1}$ and greek indices are explicitly used to indicate that they do not include $t$. 
They are related by
\begin{equation}
\begin{aligned}
  & \embednab^t \embed T = e^{- 2 t} \,\embednab_t \embed T, \quad \embednab_{\mu}\embednab_t \embed T = \embednab_t \embednab_{\mu} \embed T, \quad \embednab_t \embed T_{a_1 \cdots a_n}^{b_1 \cdots b_m} = (\partial_t + m - n) \, \embed T_{a_1 \cdots a_n}^{b_1 \cdots b_m},
  \\ & (\embednab_\mu{} - \nabla_\mu)  \,\embed T^{t'} = - g_\mu{}_\nu{}  \,\embed T^{\nu}, \quad (\embednab_\mu{} - \nabla_\mu)  \, \embed T^{\nu'} =  \delta_\mu^{\nu'} \, \embed T^t
  \\ & (\embednab_\mu{} - \nabla_\mu)  \,\embed T_{t} = - \embed T_\mu, \quad (\embednab_\mu{} - \nabla_\mu)  \, \embed T_{\nu} = g_\mu{}_\nu{} \, \embed T_{t},
\end{aligned}
\end{equation}
for some arbitrarily indexed tensor $\embed T$. 
The Riemann curvature tensor $\embed R$ remains trivial.
A integrals over $\RR^{\DD}$ and $S^{d+1}$ in embedding space coordinates are denoted by
\begin{equation}\label{eq:RRandSd+1volumeintegrals}
\begin{aligned}
  & \int^{\RR^{\DD}} \equiv \int_X \equiv \int_x \sqrt {\embed g} \equiv \int_{\hat X} \int_{t} e^{\DD t}, \quad \int^{S^{d+1}}_{\hat X} \equiv \Xint{\times}_{X} = \int^{\RR^\DD}_{X} \frac {\de^{\DD} X} {|X|^{\DD}}.
\end{aligned}
\end{equation}
When integrating over internal vertices in embedding space coordinates, it is vital to use the proper measure of $S^{d+1}$. 
It reproduces the correct volume in the ``$\alpha = 1$ gauge'':
\begin{equation}
\begin{aligned}
  \Xint{\times}^{\RR^\DD} \frac {\de^{\DD} X} {|X|^{\DD}} = \frac {1} {\Vol \RR^*} \int^*_{\alpha} \int_{X} \frac {\alpha^{\frac {\DD} {2}} \, e^{- \alpha \, X^2} } {\Gamma(\frac {\DD} {2})} = \int_{X} \frac {2 \, e^{-X^2} } {\Gamma(\frac {\DD} {2})} = \frac {2 \pi^{\frac {\DD} {2}}} {\Gamma(\frac {\DD} {2})} = \Omega_{d+1}.
\end{aligned}
\end{equation}
Tensors in flat and spherical coordinates are related by
\begin{equation}
\begin{aligned} 
  & \embed T_{\cdots}^{I \cdots} = X^I \,\embed T_{\cdots}^{t \cdots} + \hat {\embed e}_\mu^I \,\embed T_{\cdots}^{\mu \cdots}, && \embed T_{\cdots}^{t \cdots} = \frac {X_J} {|X|^2} \,\embed T_{\cdots}^{J \cdots}, && \embed T_{\cdots}^{\mu \cdots} = \hat {\embed e}^\mu_I  \, \hat {\embed t}^I_J \, \embed T_{\cdots}^{J \cdots}, 
  \\ & \embed T_{I \cdots}^{\cdots} = \frac {X_I} {|X|^2} \,\embed T_{t \cdots}^{\cdots} + \hat {\embed e}^\mu_I \, \embed T_{\mu \cdots}^{\cdots}, && \embed T_{t \cdots}^{\cdots} = X^J \,\embed T_{J \cdots}^{\cdots}, && \embed T_{\mu \cdots}^{\cdots} =  \hat {\embed e}_\mu^I \, \hat {\embed t}_I^J \, \embed T_{J \cdots}^{\cdots} ,
\end{aligned}
\end{equation}
where
\begin{equation}
\begin{aligned}
  \hat {\embed e}_\mu^I \equiv \frac{\partial X^I} {\partial w^\mu{}} = |X| \, \hat {e}_\mu^I{} , \quad \hat {\embed e}^\mu_I \equiv \frac{\partial w^\mu{}}{\partial X^I} = \frac 1 {|X|} \, \hat {e}^\mu_I, \quad \hat {\embed e}_\mu^I{}  \, \hat {\embed e}^\mu_J{} = \delta^I_J{}, \quad \hat {\embed e}_\mu^I{} \, \hat {\embed e}^\nu_I{} = \delta_\mu^\nu{},
\end{aligned}
\end{equation}
and $\hat {\embed t}$ is a tangential projection operator
\begin{equation}\label{eq:tangentialprojectionoperator}
\begin{aligned}
  & \hat {\embed t}_I{}_J{} \vcentcolon= \delta_I{}_J{} - \frac {X_I X_J} {|X|^2}, && \hat {\embed t}_I{}_J{} X^I = \hat {\embed t}_I{}_J{} X^J = 0, && \hat {\embed t}_I{}_J{} \, \hat {\embed t}^J_K = \hat {\embed t}_I{}_J{} \, \delta^J_K = \hat {\embed t}_I{}_K{}, && \hat {\embed t}_I^I = \DD - 1,
\end{aligned}
\end{equation}
which acts as the embedding space representation of $\delta_\mu{}_\nu{}$ ($\delta_\mu^\mu = \hat {\embed t}_I^I$) on $S^{d+1}$.

\subsection{Mellin transformed basis of fields}\label{sec:MellinTransformedBasisofFields}
\subsubsection{Restriction to the sphere}\label{sec:restrictiontothesphere}
There are many ways to define the restriction of a field $\embed T$ on $\RR^\DD$ to $T$ on $S^{d+1}$, all of which are technically equivalent. 
In other words, $T$ can be viewed to be $\embed T$ in a particular ``gauge''. 
One simplistic gauge, for example, is $\displaystyle T^{\cdots}_{\cdots}(w) = \embed T^{\cdots \neq t}_{\cdots \neq t}\Big|_{t = 0} = \int_{t} \embed T^{\cdots \neq t}_{\cdots \neq t} \, \delta(t)$. 
A more sound choice is to define $T$ as a radial average parameterised by $\Delta$:
\begin{equation}\label{eq:mellinisedFields}
\begin{aligned}
  \Phi_{[\Delta]} (\hat X) & \vcentcolon = \int_t e^{\Delta t} \, \embed \Phi (X)
  \\ A_{[\Delta]}{}_{\mu} & \vcentcolon = \int_t e^{(\Delta - 1) t} \, \embed {A}_{\mu}, && A_{[\Delta]}^{\mu} = \int_t e^{(\Delta + 1) t} \, \embed {A}^{\mu}
  \\ \chi_{[\Delta]} & \vcentcolon = \int_t e^{(\Delta - 1) t}\,  \embed {A}_{t} = \int_t e^{(\Delta + 1) t}\,  \embed {A}^{t}
  \\ T_{[\Delta]}{}_\mu{}_\nu{} & \vcentcolon = \int_t e^{(\Delta - 2) t} \, \embed {T}_{\mu \nu}, && T_{[\Delta]}^{\mu \nu} = \int_t e^{(\Delta + 2) t} \, \embed {T}^{\mu \nu}
  \\ \xi_{[\Delta]}{}_\mu{} & \vcentcolon = \int_t e^{(\Delta - 2) t} \, \embed {T}_{\mu t} = \int_t e^{\Delta t} \, \embed {T}_{\mu} {}^{t}{}, && \xi^{\mu}_{[\Delta]} = \int_t e^{(\Delta + 2) t} \, \embed {T}^{\mu t} = \int_t e^{\Delta t} \, \embed {T}^{\mu}{}_{t}{}
  \\ \chi_{[\Delta]}{} & \vcentcolon = \int_t e^{(\Delta - 2) t} \, \embed {T}_{t t} = \int_t e^{\Delta t} \, \embed {T}^t{}_{t}{} = && \int_t e^{(\Delta + 2) t} \, \embed {T}^{t t}  \qquad \cdots \text{ and so on},
\end{aligned}
\end{equation}
where $\chi$ and $\xi^\mu$ are additional scalars and vector fields defined in order to capture the entire (or if focus is restricted to the sphere, irrelevant) behaviour of the embedding space fields. 
Thus, when representing spin-$s$ fields on the sphere in embedding space, the additional fields of spin $< s$ can be viewed as parameterising gauge transformations of the spin-$s$ field. 
These transformations are separate from the {\it physical} gauge transformations, and instead are an additional mathematical redundancy stemming from the embedding space representation. 

\subsubsection{Reextension to embedding space}\label{sec:ReextensiontoEmbeddingSpace}
The embedding space field $\embed T$ can be recovered from the Mellin transformed basis of field $T$ by an inverse Mellin transform with respect to the radial component. 
Alternately, fields on the sphere can be extended to embedding space by assuming the tangentiality condition on the resultant embedding space field, $\embed T \cdot X = 0$. 
This is straightforward for scalar fields, $\embed \Phi$,
\begin{equation}
\begin{aligned}
  \embed \Phi (X) & 
  = \circint_{\Delta} |X|^{- \Delta} \, \Phi_{[\Delta]}(\hat X).
\end{aligned}
\end{equation}
Note that for these integrals to converge, a lower bound on $\Delta$ (and $\bdel$) is assumed: $\Real(\Delta) > 0$. 
A vector field $\embed A$ in spherical coordinates 
\begin{equation}
\begin{aligned}
  \embed {A}_{\mu} & = \circint_{\Delta} |X|^{- (\Delta - 1)} \,  A_{[\Delta]}{}_{\mu}, && \embed {A}^{\mu} = \circint_{\Delta} |X|^{- (\Delta + 1)} \,  A_{[\Delta]}{}^{\mu}
  \\ \embed {A}_{t} & = \circint_{\Delta} |X|^{- (\Delta - 1)} \,  \chi_{[\Delta]}{},  && \embed {A}^{t} = \circint_{\Delta} |X|^{- (\Delta + 1)} \,  \chi_{[\Delta]}{} 
\end{aligned}
\end{equation}
can be used to recover its flat space representation:
\begin{equation}
\begin{aligned}
  \embed A_{I} & = \circint_{\Delta} |X|^{- \Delta} \, \frac {X_I}{|X|} \chi_{[\Delta]}{} + \circint_{\Delta} |X|^{- \Delta} \, A_{[\Delta]}{}_I, && A_{[\Delta]}{}_I = \hat {e}^\mu_I \, A_{[\Delta]}{}_{\mu}.
\end{aligned}
\end{equation}
Note the change to $\Delta$-weight of $|X|$. 
The tangential projection operator, $\embed {\hat t}$, can be used to project out the radially dependent part of the field, $\chi$. 
This can be similarly extended to symmetric $2$-tensors, in spherical
\begin{equation}
\begin{aligned}
  \embed T_{\mu \nu} & = \circint_{\Delta} |X|^{- (\Delta - 2)} \, T_{[\Delta]}{}_\mu{}_\nu{}, \quad \embed T_{\mu t} = \circint_{\Delta} |X|^{- (\Delta - 2)} \, \xi_{[\Delta]}{}_\mu{}, \quad \embed T_{t t} = \circint_{\Delta} |X|^{- (\Delta - 2)} \, \chi_{[\Delta]}{}
\end{aligned}
\end{equation}
and flat coordinates
\begin{equation}
\begin{aligned}
  \embed T_{I J} & =  \circint_{\Delta} |X|^{- \Delta} \, \Bigg( \frac {X_I X_J} {|X|^2} \, \chi_{[\Delta]}{} + \frac {X_I} {|X|} \, \xi_{[\Delta]}{}_J{} + \frac {X_J}{|X|} \, \xi_{[\Delta]}{}_I{} 
  + T_{[\Delta]}{}_I{}_J{} \Bigg).
\end{aligned}
\end{equation}

\subsection{Eigenmodes of the Laplacian on the sphere in embedding space}\label{sec:EigenmodesoftheLaplacianontheSphereinEmbeddingSpace}
Spin-$s$ symmetric transverse traceless (STT) fields, forming the eigenmodes of the Laplacian on $S^{d+1}$, can be represented by radially normalised STT harmonics in embedding space $\RR^\DD$, as
\begin{equation}
\begin{aligned}
  & \Psi_{p,q}^{s,n}(X, \, U) = \frac {|p X|^{n}} {n! \, |X|^n} \frac {\big(|p X| \, |q U| - |p U| \, |q X|\big)^{s}} {s! \, |X|^s},
\end{aligned}
\end{equation}
symmetrised by defining the eigenmodes to be $\Psi_{p,q}^{s,n}{}_{a_1 \cdots a_s}(X) = \partial_{U^{a_1}} \cdots\partial_{U^{a_s}} \, \Psi(X, \, U)$. 
The required conditions of transversality, tracelessness, tangentiality and harmonicity of the underlying unnormalised eigenmode
\begin{equation}
\begin{aligned}
  & \text{transverse:} &&\partial_U \cdot \partial_X \, \Psi = 0, && \text{traceless:} && \partial_U^2 \, \Psi = 0
  \\ & \text{tangential:} && X\cdot \partial_U \, \Psi = 0,  && \text{harmonic:} && \partial_X^2 \, (|X|^{n+s} \, \Psi) = 0
\end{aligned}
\end{equation}
are satisfied when $\DD$-dimensional vectors $(p, \, q)$, $p \sim \CC^* \, p, \; q \sim\CC^* \, q$, are orthogonal and light-like $p^2 = q^2 = p \cdot q = 0$. 
Using a mode generating function $\embed \Psi$
\begin{equation}
\begin{aligned}
  \embed \Psi_{p,q}(X, \, U) =  \sum_{n}\sum_{s}\Psi_{p,q}(X, \, U) = e^{\frac {|p X|} {|X|}} \, e^{\frac {|p X| \, |q U| - |p U| \, |q X|} {|X|}}
\end{aligned}
\end{equation}
to compare the identically equal objects:
\begin{equation}
\begin{aligned}
  & \sum_{n,n',s,s'} \Xint{\times}_X^{\RR^\DD} \int_{U} e^{- U^2} \Psi^*_{p,q}(X, \, U) \, \Psi_{p',q'}(X, \, U) = \Xint{\times}_X^{\RR^\DD} \int_{U} e^{- U^2} \, \embed \Psi^*_{p,q}(X, \, U) \, \embed \Psi_{p',q'}(X, \, U), 
\end{aligned}
\end{equation}
the overlap of these eigenmodes is found to be:
\begin{equation}\label{eq:modeintegraloverlap}
\begin{aligned}
  \left\langle \Psi^{s,n}_{p, q}{} \, \Psi^{s',n'}_{p',q'}{} \right\rangle & = \delta_{s s'} \, \delta_{nn'} \, \frac { 2 \pi^{\frac{\DD}{2}} } {  \Gamma(\frac{\DD}{2} + n + s)} \, \frac { (n + s + 1)! \, s!} { 2^{n + s} \, (n+1)!\,  n! } \,  \Lambda^s \, (\bar p \cdot p')^n 
  \\ \Lambda &= (\bar q \cdot q') \, (\bar p \cdot p') - (\bar q \cdot p') \, (q' \cdot \bar p).
\end{aligned}
\end{equation}
Longitudinal spin-$s$ eigenmodes modes are derivatives of transverse fields and are given by
\begin{equation}
\begin{aligned}
  & \Upsilon_{p,q}^{s,n}(X, \, U) = |X|^{s - r} \, (U \cdot \partial_X)^{s - r} \, \Psi_{p,q}^{r,n}(X, \, U), \quad \forall \; r < s, \quad n \ge (s-r).
\end{aligned}
\end{equation}
For example, longitudinal vector eigenmodes are
\begin{equation}
\begin{aligned}
  & \Upsilon_{p}^{1,n}(X, \, U) = \frac {|p X|^{n-1}} {(n-1)! \, |X|^{n-1}} \, \left(|pU|  - \frac {|p X| \, |XU|} {|X|^{2}} \right).
\end{aligned}
\end{equation}
Using this, propagator expressions can be verified to satisfy
\begin{equation}\label{eq:propagatorexpressionsverifiedtosatisfy}
\begin{aligned}
  \left\langle\psi'(\hat X) \, G(\hat X, \, \hat Y) \, \psi(\hat Y) \right\rangle = \frac {1} {\lambda_{\psi} + m^2} \, \left\langle\psi'(\hat X) \, \psi(\hat X) \right\rangle, \quad G_{\cdots} = \mathcal G_{\cdots} \circ G
\end{aligned}
\end{equation}
for general eigenfunctions of the Laplacian, $\psi, \, \psi'$, where $\lambda_\psi$ is the eigenvalue of $\psi$. For future notational convenience:
\begin{equation}
\begin{aligned}
  \lambda^{(s)}_{n,\Delta} \equiv \lambda_{n+s,\Delta}  & \vcentcolon= (n  + s + \Delta) \, (n  + s + \bar \Delta) \cong (- \nabla^2 + m^2).
\end{aligned}
\end{equation}

\subsubsection{Verification of scalar propagator in embedding space formalism}\label{eq:ModeIntVerificationofScalarPropagatorinEmbeddingSpaceFormalism}
Integrals of eigenmodes with the scalar propagator can be used to verify \cref{eq:propandmodegiveinverseeigenvalue}. 
A sum over these eigenmode integrals can be modelled by
\begin{equation}
\begin{aligned}
  \Theta & \equiv \sum_{n,n'} \Gamma(\tfrac{\DD + n - \bar \Delta}{2}) \, \Gamma(\tfrac{\DD + n' - \Delta}{2}) \, \langle \bar\Psi^{(n)}_{p}(X)\, G(X, \, Y) \, \Psi^{(n)}_{p'}(Y)\rangle
  \\ & = \frac 1 {4 \pi^{\DD }}  \Xint{\times}^* \int_{W} \alpha_X^{\frac{\DD - \bar \Delta}{2}} \alpha_Y^{ \frac{\DD - \Delta}{2} } \, \tau \, \lambda^{\bar \Delta} \, \mu^{\Delta} e^{-f(W) }\, \sum_{n,n'} \frac {(\sqrt{\alpha_X} \, \bar p \cdot X)^{n}} { n!} \, \frac {(\sqrt{\alpha_Y} \, p' \cdot Y)^{n'}} { n'!}
  \\ & = 2 \pi^{\frac{\DD} {2}} \sum_{n} \frac {(\bar p \cdot p')^n} {2^{n+2} \, n! \, \Gamma(\frac{\DD}{2} + n)} \, \Gamma(\tfrac{\bar \Delta}{2} + n) \, \Gamma(\tfrac{\Delta}{2} + n),
\end{aligned}
\end{equation}
where
\begin{equation}
\begin{aligned}
  f(W \equiv \{P, \, X, \, Y\}) & = \alpha_X \, X^2 + \alpha_Y \,Y^2 + \tau \,P^2 + 2 \im \,P \, (\lambda \, X - \mu \,Y).
\end{aligned}
\end{equation}
Using \cref{eq:modeintegraloverlap} to normalise $\Theta$, it can be confirmed that
\begin{equation}
\begin{aligned}
  \langle \bar\Psi^{(n)}_{p}(X)\, G(X, \, Y) \, \Psi^{(n)}_{p'}(Y)\rangle & = \frac 1 {\lambda_{n,\Delta}} \,  \langle \bar\Psi^{(n)}_{p}(X)\, \Psi^{(n)}_{p'}(X)\rangle.
\end{aligned}
\end{equation}

\subsubsection{Verification of vector propagators in embedding space formalism}\label{eq:ModeIntVerificationofVectorPropagatorinEmbeddingSpaceFormalism}
Similarly, the massive and massless vector propagator eigenvalue equations, \cref{eq:vectorodesetup}, can be verified too, but with much greater tedium. 
Given some operator $\mathcal O^I{}^J{}$, the normalised inner product
\begin{equation}\label{eq:modeintegraloperatoreigenvalue}
\begin{aligned}
  \langle \mathcal O \rangle = \frac {\langle \bar\Psi^{(n)}_{p}{}_I{}(X)\, \mathcal O^I{}^J{} \circ G(X, \, Y) \, \Psi^{(n)}_{p'}{}_J{}(Y)\rangle } {\langle \bar\Psi^{(n)}_{p}{}_I{}(X)\,\Psi^{(n)}_{p'}{}^I{}(X)\rangle }
\end{aligned}
\end{equation}
with transverse $\langle \mathcal O \rangle_{\rm T}$ and longitudinal $\langle \mathcal O \rangle_{\rm L}$ vector eigenmodes is given by
\begin{equation}\label{eq:vectorpropagatormatrixelements}
\begin{aligned}
  \begin{array}{c | c | c}
\mathcal O^I{}^J{} & \langle \mathcal O \rangle_{\rm T}  & \langle \mathcal O \rangle_{\rm L}\\[1ex]\hline \rule{0pt}{2em}
\displaystyle\delta^I{}^J{} \cong \frac {4 \,\lambda \,\mu \,|PX| \,|PY|} {\Delta \,\bar \Delta} \delta^I{}^J{} & \displaystyle\frac {1} {\lambda_{n+1, \Delta}} & \displaystyle \frac {\lambda_{1,\Delta} +\lambda_{n,0}}  { \lambda_{n+1,\Delta} \, \lambda_{n-1,\Delta}  } \\\rule{0pt}{2em}
\displaystyle4 \,\lambda \, \mu \,|XY| \,P^I{} \,P^J{} &  \displaystyle\frac {1} {\lambda_{n+1, \Delta}} & \displaystyle\frac{ \lambda_{1,\Delta} + \lambda_{n,0} \, ( \lambda_{n,\Delta} - d - 2 ) }  {  \lambda_{n+1,\Delta} \, \lambda_{n-1,\Delta}  } \\\rule{0pt}{2em}
\displaystyle\frac {4 \,\lambda \, \mu \, |PX|} {\bar\Delta}  \, Y^I{} \,P^J{} = \frac {4 \, \lambda \, \mu \, |PY|} {\Delta}  \, X^J{}\, P^I{} &  \displaystyle\frac {1} {\lambda_{n+1, \Delta}} & \displaystyle\frac{   \lambda_{1,\Delta}  - \lambda_{n,0}   }  {  \lambda_{n+1,\Delta} \, \lambda_{n-1,\Delta}   }  \\\rule{0pt}{2em}
4 \, \lambda \,  \mu \, |X| \, |Y| \, P^I{} \, P^J{} & 0 & \displaystyle\frac {\lambda_{n,0}} {\lambda_{n,\Delta}} 
\\\rule{0pt}{2em} \displaystyle\frac {P^I{} \, P^J{}} {P^2} & 0 & \displaystyle\frac {\lambda_{n,0}} {\lambda_{n+1,\Delta} \, \lambda_{n-1,\Delta} }
  \end{array}
\end{aligned}.
\end{equation}
Using these results, the massive vector propagator is found to have the eigenvalues, $\frac {1} {\lambda_{n+1, \Delta}} = \frac {1} {\lambda^{(1)}_{n, \Delta}}$ for transverse modes and $\frac 1 {(\del-1) \, (\bdel-1)}$ for longitudinal modes, satisfying expectations.
The massless vector propagator has the eigenvalues $\frac {1}{(n + 2) \, (n+d)}$ and $\frac {d} {(d-2)} \, \frac {1}{n \, (n+d)}$ for transverse and longitudinal modes respectively, once again matching the requirements.

\section{Generalized Euler integrals and \texorpdfstring{$\mathcal A$}{A}-hypergeometric functions}\label{ch:Eulerintegrals}
Generalized Euler integrals form a class of integrals that appears conspicuously often in the study of quantum field theories perturbatively viewed as sums over Feynman integrals/diagrams/graphs, in the path integral formulation
\begin{equation} \label{eq:toocutetobetrue}
\mathcal Z = \int \de \Phi \, e^{- \, S_{\rm Eucl} [\Phi]},
\end{equation}
and in most applications of the principle of least action based on expansions around relevant physical saddle points. The term generalized Euler integrals was introduced in \cite{GELFAND1990255}, along with the algorithmically solvable system of equations (and their solution space) governing them, the $\mathcal A$-hypergeometric system (and functions). 

Famously appearing in the computation of periods determining the complex structure of Calabi-Yau manifolds as a means of deriving and solving Picard-Fuchs equations, the prototypical example being the construction of the mirror of the quintic Calabi-Yau manifold on $\CC \mathds P^4$, the application of this theory is far from new to physicists \cite{Greene:1990ud,Greene:1988ut,Greene:1998yz,Candelas:1990rm,Hosono:1994ax,Hosono:1995bm,Hori:2003ic,Bogner:2014mha}. 
Generalized Feynman integrals in flat space are Euler integrals considered in specific limits and over contours that are physically relevant. They satisfy systems of linear holonomic partial differential equations \cite{Kashiwara1976OnAC}, their Laurent series expansion coefficients correspond to their periods \cite{Bogner:2007mn}, and can be expressed as Mellin-Barnes integrals, \cref{eq:mellintransformdef}, which have representations as hypergeometric series \cite{Kalmykov:2008ofy}, with their correspondence to $\mathcal A$-hypergeometric systems discussed in more detail, and used as a means of studying their singular loci (Landau varieties) and explicit computation in \cite{1977387,KALMYKOV2012103,nasrollahpoursamami2016periods,Bitoun:2017nre,Klausen:2019hrg,delaCruz:2019skx,Klausen:2021yrt,Mizera:2021icv,Fevola:2023fzn}.

Given the prevalence of generalized Euler integrals, \cref{eq:toocutetobetrue2}, in physics, and specifically within the context of the current text, in the computation of Feynman integrals on a spherical background and by analytic continuation, on de Sitter space, application of the theory of $\mathcal A$-hypergeometric systems becomes indispensible, especially when seeking consistency in the ability to solve and evaluate, analytically and/or numerically, the aforementioned integrals. 

This appendix reviews the fundamental ideas and basic building blocks of the theory of Euler integrals and $\mathcal A$-hypergeometric systems, mainly based on \cite{GELFAND1990255,Gelfand:Discriminants,Saito:GrobnerBases}, with the inclusion of some further generalisions proven in \cite{schulze2012resonance}. Far from exhaustive, it can at best be considered an introduction or refresher to the topic and a means of establishing prerequisite understanding and notational conventions for \cref{sec:scalars,sec:VectorFields,ch:dSFeynmanIntegrals}, which heavily reference and utilize these ideas.

\Cref{sec:AHypergeometricSystem} begins with intiutive examples featuring some prevalent symmetry structures in physically relevant GKZ systems, then proceeds to introduce, motivate and explain the construction of $\mathcal A$-Hypergeometric ideal and its related terminology, and finally presents a simplified generic solution algorithm, all in manner to allow practical application, satisfying the aforementioned prerequisite. \Cref{sec:IdealstoSolutions} discusses ideals in Weyl algebras and their solution spaces in some greater generality to highlight the special properties of $\mathcal A$-Hypergeometric ideals. \Cref{sec:SolutionsofAHypergeometricSystems} shows some approaches to finding and constructing solutions to GKZ systems and is exposited through a running example, the physical significance of which becomes apparent in \cref{ch:dSFeynmanIntegrals}. 

For further reading and detailed discussions, also see \cite{Berkesch_2011,BEUKERS201130,schulze2012resonance,SAITO201331}. A recent review is \cite{reichelt2021algebraic}. GKZ ideals can be restructured and studied as Pfaffian systems, as shown in \cite{Chestnov:2022alh,Chestnov:2023kww}. 

\subsection{\texorpdfstring{$\mathcal A$}{A}-Hypergeometric system} \label{sec:AHypergeometricSystem}
Euler integrals can be symbolically written as
\begin{equation}\label{eq:toocutetobetrue2}
  I(z)  = \int_{\sigma} \, \frac {\de x} {x} \, \frac {x^{\beta}} {P(x; \, A ; \,z)^{\alpha}} .
\end{equation}
This compact notation, hiding within it many perturbative possibilities of \cref{eq:toocutetobetrue}
, is decompressed as follows:
\begin{enumerate}[left=0pt]
  \item The set of integration variables, $x_1, \, x_2, \cdots x_n \in \CC$, is denoted by $x$.
  $\sigma$ is a positively oriented closed contour within the domain of $x$ minus the zeroes of the polynomials, here $\CC^n \setminus \{P = 0\}$. $\alpha, \, \beta$ are vectors of complex parameters, which may be generic or fixed to some arbitrary value. The non-zero variables $z \in \CC\setminus \{0\}$ parameterise the coefficients of the polynomials in $x$, $P$.
  \item A multivariate monomial in $n$ variables $x_i$ with exponents $\beta_i$ in the multi-index notation is
  \begin{equation}
  x^{\beta} \equiv \prod_{i=1}^{n} \, x_i^{\beta_i} = \prod_{i=1}^{n} \, e^{\beta_i \, \log x_i}, \quad \beta \in \CC^n.
  \end{equation}
  \item A polynomial, $P$, in $n$ variables $x_i$ with $N$ monomial terms can be represented by an $n \times N$ matrix $A$ with non-negative integer entries, where each column consists of the exponents of the multivariate monomials, and an $N$-dimensional coefficient list/vector $z$. 
  \begin{equation}
  P(x\, ; \, A \, ; \; z)  = \sum_{k = 1}^{N} \, z_{k} \, x^{A_k}, \quad A = \left(\begin{smallmatrix} A_1 &  A_2 & \cdots & A_N \end{smallmatrix} \right), \quad A_i \in \NN_0^{n} \label{eq:polytoA}
  \end{equation}
  $P$ is a generic (usually sparse) polynomial and its associated matrix representation, $A$, is the support. 
  $P$ can be interpreted as a function of the coefficients $z$. Generic polynomials and their supports can be used interchangeably, with the former usually being preferred when the coefficients are fixed and the latter when they are variable. For example, 
  \begin{equation}
  A = \left(\begin{smallmatrix} 0 & 2 & 1 \\ 0 & 0 & 2 \end{smallmatrix} \right), \quad z = \{1, \, 2, \, 4\}, \quad P(x\, ; \, A \, ; \; z)  = 1 + 2 \, x_1^2 + 4 \, x_1 \, x_2^2 
  \end{equation}
  Within the context of \cref{eq:toocutetobetrue2}, $P(x; \, A ; \,z)^{\alpha}$ is understood to be in multi-index notation, i.e. a product of arbitrarily many polynomials raised to powers $\alpha$, each defined by some matrices $A$ and monomial coefficients $z$.
  \begin{equation}
    P(x; \, A ; \,z)^{\alpha} = \prod_{l=1}^{L} \, P(x; \, A^{(l)} ; \,z^{(l)})^{\alpha_{l}}
    \end{equation}
    \item Usage of such notation isn't limited to just variables but can be extended to operators and functions also, e.g.
    \begin{equation}
    \partial_x^{w} = \prod_{i=1}^{n} \, \partial_{x_i}^{w_i}, \quad \Gamma(w) = \prod_{i=1}^{n} \, \Gamma(w_i), \quad P( \partial_x \, ; \, \left(\begin{smallmatrix}2 & 3 \\ 0 & 1 \end{smallmatrix} \right) ; \; z) = z_1 \, \partial_{x_1}^2 + z_2 \, \partial_{x_1}^3 \, \partial_{x_2 }
    \end{equation}
    or as used in the case at hand $\tfrac {\de x} {x}  = \tfrac {\de x_1} {x_1} \, \tfrac {\de x_2} {x_2} \cdots \tfrac {\de x_n} {x_n} $.
  \item The Euler operator, $\theta_x$, is defined as
  \begin{equation}
  \theta_x \vcentcolon= x \, \partial_x = \partial_{\log x}, \quad \theta_x \, x^\alpha = \alpha \, x^\alpha, \quad \theta_x \, (\log x)^\alpha = \alpha \, (\log x)^{\alpha-1}
  \end{equation}
  The rising and falling Pochhammer symbols or factorials are
  \begin{equation}\label{eq:pochhammersymboldef}
  \begin{aligned}
    a^{(n)} & \vcentcolon= a \, (a+1)  \cdots (a+n-1) = \frac {\Gamma(a+n)} {\Gamma(a)}
    \\ a_{(n)} & \vcentcolon= a \, (a-1) \cdots (a-n+1) =  \frac {\Gamma(a)} {\Gamma(a-n)} && \text{resp.}
  \end{aligned}
  \end{equation}
  When it is obvious which variable is being referred to, operators like $\partial_{x_i}$, $\theta_{z_j}$ will be written as $\partial_{i}$, $\theta_{j}$ instead. Altogether, a falling form of the Euler operator that often graces discussions involving Euler integrals can be written in an extremely condensed and convenient form
  \begin{equation}
  \theta_{(m)}  = \prod_{i=1}^{n} \, \prod_{j=0}^{m-1} \, (\theta_{x_i} - j).\label{eq:fallingthetanotation}
  \end{equation}
\end{enumerate}
Interpreting these integrals as functions of the coefficients, $z$, of the polynomials in their integrand, which are allowed to be generically valued in their domain, is what makes the Euler integrals generalized. They satisfy a class of holonomic linear partial differential equations in $z$, often dubbed GKZ systems, that admit holomorphic solutions. The solution space, when represented in terms of generalized hypergeometric functions, is referred to as and by $\mathcal A$-hypergeometric functions, where $\mathcal A$ is a matrix describing the set of polynomials featured in the integral.

\subsubsection{Introductory examples}\label{sec:IntroductoryExamples}
Euler integrals are far from unfamiliar in their form, and the partial differential equations they satisfy can be found by intuitive arguments based on homogeneity and scaling symmetries. So it is instructive to first consider some examples which at least partially showcase the benefits of representing Feynman integrals in this more generalized and superficially superfluous format.

\begin{exmp}\label{sec:example_trivial}
A needlessly complicated representation of a bivariate monomial $I(z_0,\,z_1)$ is
\begin{equation}
I(z_0, \, z_1) = \int_{\sigma} \, \frac {\de x} {x} \, \frac {x^{\beta} } {(z_0 + z_1 \, x^2 )^{\alpha}}, \quad z_0,\,z_1 \in \CC \setminus \{0\}
\end{equation}
where $\sigma$ is a closed contour in $\CC$, and $\alpha, \, \beta$ are complex valued parameters. By rescaling\footnote{At $z_0 \, z_1 = 0$, the integral becomes singular and reducible, and doesn't merit discussion in the present context.} the variable $x$ by the factor $\sqrt {\tfrac  {z_0}{z_1}}$, $I(z_0, \, z_1)$ simplifies to
\begin{equation}
I(z_0, \, z_1) = z_0^{-\alpha + \frac{\beta}{2}} \, z_1^{-\frac{\beta}{2}} \int_{\sigma} \, \frac {\de x} {x} \, \frac {x^{\beta} } {( 1 + x^2 )^{\alpha}} = C_\sigma \, z_0^{-\alpha + \frac{\beta}{2}} \, z_1^{-\frac{\beta}{2}} \label{eq:trivial_sol_space}
\end{equation}
where $C_\sigma$ is a proportionality constant dependent on the integration contour. Identification of the scaling symmetries of this integral leads to the same conclusion.
\begin{equation}
I(\lambda \, z_0, \,  \lambda \, z_1) = \lambda^{-\alpha} \, I(z_0, \, z_1), \quad I(z_0, \, \lambda \, z_1) = \lambda^{-\frac{\beta}{2}} \, I(z_0, \, z_1)
\end{equation}
form a basis of the scaling symmetries of $I(z_0, \, z_1)$ and can equivalently be represented as
\begin{equation}
D_0 \circ I(z_0, \, z_1) = D_1 \circ I(z_0, \, z_1) = 0, \quad D_0 = \theta_{z_0} + \theta_{z_1} + \alpha, \quad D_1 = \theta_{z_1} + \tfrac{\beta}{2}.
\end{equation}
All scaling properties of $I(z_0, \, z_1)$ are spanned by linear combinations of $D_0,\, D_1$, better represented in the basis:
\begin{equation}
D_0' = \theta_{z_0} + \alpha - \tfrac{\beta}{2}, \quad D_1' = \theta_{z_1} + \tfrac{\beta}{2}.
\end{equation}
The simultaneous solution to these PDEs gives the exponents of $z_0, \, z_1$. Thus the solution space of the integral $I(z_0, \, z_1)$ is once again found to be \cref{eq:trivial_sol_space}. 

\par When the integral is convergent, some open integration intervals can also be considered. For example, when $\Real(\beta)>1$, $\Real(\alpha - \tfrac{\beta}{2})>0$, and $z_0, \, z_1 \in \RR^+$, some integration intervals that are more commonly seen in physics and their corresponding proportionality constants are
\begin{equation}
\begin{aligned}
\sigma_1 & = x \in (0, \, \infty ), && C_{\sigma_1} = \frac {\Gamma(\frac{\beta}{2}) \, \Gamma(\alpha - \frac{\beta}{2} )} {2 \, \Gamma(\alpha)}
\\ \sigma_2 & = x \in (-\infty, \, \infty ), && C_{\sigma_2} =  (1 - (-1)^{\beta}) \, C_{\sigma_1}  \qquad \text{etc.}
\end{aligned}
\end{equation}
\end{exmp}

\begin{exmp} \label{sec:example_easy} 
$I(z)$, $z \in \CC$, is an integral over a closed contour in $\CC$, $\sigma$:
\begin{equation}
I(z)  = \int_{\sigma} \, \frac {\de x} {x} \, \frac {x^{\beta}} {(z_1 + z_2 \, x^2)^{\alpha_1} \,  (z_3 + z_4 \, x^2)^{\alpha_2}}, \quad z_2 \neq z_4, \quad z_1,\,z_2,\,z_3, \, z_4 \in \CC \setminus \{0\}. \label{eq:unexpected2F1}
\end{equation}
The scaling behaviour of $I(z)$, also known as homogeneity conditions, are represented by
\begin{equation}\label{eq:exmpeasyhomogeneityconditions}
(\theta_1 + \theta_2 + \alpha_1) \circ I = (\theta_3 + \theta_4 + \alpha_2) \circ I =  (\theta_2 + \theta_4 + \tfrac{\beta}{2}) \circ I = 0
\end{equation}
which expectedly give rise to a set of linear equations between the exponents of $z$, allowing the relative dependence of $I(z)$ on any $3$ of $z_1,\,z_2,\,z_3,\,z_4$ to be fixed. A solution basis to \cref{eq:exmpeasyhomogeneityconditions} is
\begin{align}
I^{(1)} (z) & = z_2^{- \alpha_1} \, z_3^{\frac{\beta}{2} - \alpha_1 - \alpha_2} \, z_4^{\alpha_1 - \frac{\beta}{2}} \, \tilde I^{(1)}(\tfrac {z_1 \, z_4} {z_2 \, z_3})  \label{eq:easy_4rep1}
\\ I^{(2)} (z) & = 
z_1^{- \alpha_1} \, z_3^{\frac{\beta}{2} - \alpha_2} \, z_4^{-\frac{\beta}{2}} \, \tilde I^{(2)}(\tfrac  {z_1 \, z_4}{z_2 \, z_3}) \label{eq:easy_4rep2}
\\ I^{(3)} (z) & = 
z_1^{- \alpha_1 - \alpha_2 + \frac{\beta}{2}} \, z_2^{\alpha_2 -\frac{\beta}{2}} \, z_4^{-\alpha_2} \, \tilde I^{(3)}(\tfrac  {z_1 \, z_4} {z_2 \, z_3}) \label{eq:easy_4rep3}
\\ I^{(4)} (z) & = z_1^{\frac{\beta}{2} - \alpha_1}\,  z_2^{-\frac{\beta}{2}} \, z_3^{-\alpha_2} \, \tilde I^{(4)}(\tfrac {z_1 \, z_4} {z_2 \, z_3}) \label{eq:easy_4rep4}
\end{align}
for some as yet unknown set of functions $\tilde I$. Vectors consisting of the exponents of $z$ in the overall factors of these representations are called roots.  Roots are vectors in  $N$-dimensional affine space. In this case, there are four $4$ dimensional roots:
\begin{equation}
\begin{aligned}
r_1 & = \{0, \, - \alpha_1, \, \tfrac{\beta}{2} - \alpha_1 - \alpha_2,  \, \alpha_1 - \tfrac{\beta}{2}\}, && r_2 = \{- \alpha_1, \, 0,  \, \tfrac{\beta}{2} - \alpha_2, \,  -\tfrac{\beta}{2} \}, 
\\ r_3& = \{- \alpha_1 - \alpha_2 + \tfrac{\beta}{2}, \,  \alpha_2 -\tfrac{\beta}{2}, \, 0, \, -\alpha_2 \}, && r_4 = \{ \tfrac{\beta}{2} - \alpha_1, \, -\tfrac{\beta}{2}, \, -\alpha_2, \, 0  \}\label{eq:easy_root}
\end{aligned}
\end{equation}
with associated family of the solutions:
\begin{equation}
I^{(i)}(z)  = z^{r_i} \, \tilde I^{(i)}(z_t), \quad z_t \equiv \frac  {z_1 \, z_4} {z_2 \, z_3}.
\end{equation}
The dependence of $\tilde I^{(i)}$ on just $z_t$ and not the individual variables $z_1\,z_2,\,z_3,\,z_4$ can be confirmed by appropriately rescaling the variable $x$. For example, $x \rightarrow x \, \sqrt{ \tfrac { z_1   }  { z_2 }  }$ implies
\begin{equation}
I^{(4)}(z)  = z_1^{\frac{\beta}{2} - \alpha_1} \, z_2^{-\frac{\beta}{2}} \, z_3^{-\alpha_2} \, \tilde I^{(4)}(z_t), \quad  \tilde I^{(4)}(z_t) = \int_{\sigma} \, \frac {\de x} {x} \, \frac {x^{\beta}} {(1 + x^2)^{\alpha_1} \,  (1 + z_t \,  x^2)^{\alpha_2}}
\end{equation}
where $\tilde I^{(4)}(z_t)$ can be seen to remain unchanged under $T$, 
\begin{equation}
T: \, (z_1, \, z_2, \, z_3, \, z_4) \rightarrow (\lambda_1 \, z_1, \, \lambda_2 \, z_2, \, \lambda_3 \, z_3, \, \tfrac {\lambda_2 \, \lambda_3} {\lambda_1} \, z_4).
\end{equation}
Scale transformations like $T$ are called torus actions. $z_t$ represents the relative scaling symmetries of the variables $z$. It remains invariant under $T: \, z_t \rightarrow z_t$, and is expectedly called a toric invariant. Similarly, all $\tilde I$ can be verified to be functions of only the toric invariant $z_t$. $I$ is annihilated by the class of operators
\begin{equation}
\partial^n_T  \equiv \partial^{n}_{z_1} \, \partial^{n}_{z_4} - \partial^{n}_{z_2} \, \partial^{n}_{z_3}  \; \forall \; n \in \NN.
\end{equation}
This can be verified by writing the integral $I$ in its Schwinger parametric form.
\begin{equation}
\begin{aligned}
  I & = \int_0^\infty \tfrac {\de y_1 \, \de y_2 } {y_1 \, y_2} \,\tfrac {y_1^{\alpha_1}} {\Gamma(\alpha_1)} \,\tfrac {y_2^{\alpha_2}} {\Gamma(\alpha_2)} \int_{\sigma} \tfrac {\de x} {x} \,x^{\beta} \, e^{- y_1 \, (z_1 + z_2 \, x^2)} \, e^{- y_2 \, ( z_3 + z_4 \, x^2 )}
\\ \partial^n_T \circ I & = \int_0^\infty \tfrac {\de y_1 \, \de y_2 } {y_1 \, y_2} \,\tfrac {y_1^{\alpha_1}} {\Gamma(\alpha_1)} \,\tfrac {y_2^{\alpha_2}} {\Gamma(\alpha_2)}  \int_{\sigma} \Big(\partial^{n}_{z_1} \, \partial^{n}_{z_4} - \partial^{n}_{z_2} \, \partial^{n}_{z_3} \Big) \,\tfrac {\de x} {x} \,x^{\beta} \, e^{- y_1 \,  (z_1 + z_2 \, x^2)} \, e^{- y_2 \, ( z_3 + z_4 \, x^2 )}
\\ & = \int_0^\infty \tfrac {\de y_1 \, \de y_2 } {y_1 \, y_2} \,\tfrac {y_1^{\alpha_1}} {\Gamma(\alpha_1)} \,\tfrac {y_2^{\alpha_2}} {\Gamma(\alpha_2)} \int_{\sigma} \tfrac {\de x} {x} \,x^{\beta} \,e^{- y_1 \, (z_1 + z_2 \, x^2)} \,e^{- y_2 \, ( z_3 + z_4 \, x^2 )}
\\ & \; \; \times \Big( (- y_1)^{n} \, (-y_2 \, x^2)^{n} - (-y_1 \, x^2)^{n} (-y_2)^{n} \Big)   = 0.
\end{aligned}
\end{equation}
Assuming that $I(z)$ has a Laurent series expansion
\begin{equation}
I(z)  \equiv z^s \, \sum_{n \in \NN} \,c_n \, z_t^n = \sum \,c_{n} \, z_1^{s_1 + n} \, z_2^{s_2 - n} \, z_3^{s_3 - n} \, z_4^{s_4 + n} \label{eq:easy_lauren1}
\end{equation}
such that it satisfies the class of PDEs $\partial^n_T \circ I = 0$,
\begin{equation}
\partial_T \circ I  = \frac {z^{s}} {z_1 \, z_4} \, \sum_{n \in \NN} \,c_{n} \, (s_1 + n) \, (s_4 + n)\,  z_t^n - \frac {z^{s}} {z_2 \, z_3}  \, \sum_{n \in \NN} \,c_{n} \, (s_2 - n) \, (s_3 - n)  \, z_t^n = 0 \label{eq:easy_PDEsol1}
\end{equation}
the following relations need to be satisfied:
\begin{equation}\label{eq:easy_PDEsol2} 
\begin{aligned}
  & z^{s} \, \sum_{n \in \NN} \,\Big(  c_{n} \, (s_1 + n) \, (s_4 + n) - c_{n-1} \, (s_2 - n+1) \, (s_3 - n+1) \Big) \, z_t^n = 0 
\\ &  z^{s} \, c_{0} \, s_1 \, s_4 = 0. 
\end{aligned}
\end{equation}
Since $z_t$ is generic and $\not\equiv 0$, this implies either $s_1$ or $s_4 = 0$ and 
\begin{equation}
\frac {c_{n+1} } {c_{n}} = \frac {  (s_2 - n)\,  (s_3 - n)} {(s_1 + n+1) \, (s_4 + n+1)} \implies c_{n} = c_0 \, \frac {\Gamma(n - s_2) \, \Gamma(n-s_3)} {\Gamma(n+s_1 + 1) \, \Gamma(n + s_4 + 1)}. \label{eq:easy_PDEsol4}
\end{equation}
The representation of $I(z)$ corresponding to $s_1=0$ is \cref{eq:easy_4rep1} with the root $r_1$ given in \cref{eq:easy_root}. Thus, $I^{(1)}(z)$ is proportional to
\begin{equation}\label{eq:easy_PDEsol5}
\begin{aligned}
I^{(1)}(z) & = \; z_2^{- \alpha_1} \, z_3^{\frac{\beta}{2} - \alpha_1 - \alpha_2} \, z_4^{\alpha_1 - \frac{\beta}{2}} \, \sum_{n \in \NN} \,c_0 \, \frac {\Gamma(n + \alpha_1) \, \Gamma(n-\frac{\beta}{2} + \alpha_1 + \alpha_2)} {\Gamma(n + \alpha_1 - \frac{\beta}{2} + 1)} \, \frac{z_t^{n}} {n!} 
\\ & \propto \;  z_2^{- \alpha_1} \, z_3^{\frac{\beta}{2} - \alpha_1 - \alpha_2} \, z_4^{\alpha_1 - \frac{\beta}{2}} \,\,{}_2F_1(\alpha_1, \, \alpha_1 + \alpha_2-\tfrac{\beta}{2}; \, \alpha_1 - \tfrac{\beta}{2} + 1; \, z_t).
\end{aligned}
\end{equation}
Another representation of $I(z)$, \cref{eq:easy_4rep4}, with the root $r_4$ is similarly produced by the alternate solution $s_4 = 0$:
\begin{equation}
I^{(4)}(z)  \; \propto \;  z_1^{\frac{\beta}{2} - \alpha_1} \, z_2^{-\frac{\beta}{2}} \, z_3^{-\alpha_2} \,\,{}_2F_1(\tfrac{\beta}{2}, \, \alpha_2; \, \tfrac{\beta}{2} - \alpha_1 + 1; \, z_t).
\end{equation}
The series expansion of $I(z)$ presented in \cref{eq:easy_lauren1} is assumed to be in non-negative powers of $z_t$. An expansion in non-positive powers,
\begin{equation}
I(z)  \equiv z^s \, \sum_{n \in \NN}\, c_n \, z_t^{-n} = \sum_{n \in \NN}\, c_{n} \, z_1^{s_1 - n} \, z_2^{s_2 + n} \, z_3^{s_3 + n} \, z_4^{s_4 - n} \label{eq:easy_lauren2}
\end{equation}
that satisfies $\partial^n_T \circ I = 0$, analogously results in the conditions $s_2 \, s_3 = 0$ and
\begin{equation}
c_{n}  = c_0 \, \frac {\Gamma(n - s_1) \, \Gamma(n-s_4)} {\Gamma(n+s_2 + 1) \, \Gamma(n + s_3 + 1)}.
\end{equation}
Comparing \cref{eq:easy_4rep2,eq:easy_4rep3} and their corresponding roots $r_2,\,r_3$ to the cases $s_2 = 0$, $s_3=0$ respectively implies
\begin{equation}
\begin{aligned}
I^{(2)}(z) & \; \propto \; z_1^{- \alpha_1} \, z_3^{\frac{\beta}{2} - \alpha_2} \, z_4^{-\frac{\beta}{2}} \, {}_2F_1( \alpha_1, \, \tfrac{\beta}{2}; \, \tfrac{\beta}{2} - \alpha_2 + 1 ; \, z_t^{-1})
\\ I^{(3)}(z) & \; \propto \; z_1^{- \alpha_1 - \alpha_2 + \frac{\beta}{2}} \, z_2^{\alpha_2 -\frac{\beta}{2}} \, z_4^{-\alpha_2} \, {}_2F_1( \alpha_1 + \alpha_2 - \tfrac{\beta}{2}, \, \alpha_2 ; \, \alpha_2 -\tfrac{\beta}{2} +1; \, z_t^{-1}).
\end{aligned}
\end{equation}
Thus, the integral $I(z)$ takes the form of $4$ independent solutions of the Gauss hypergeometric ODE, around $z_t = 0, \, \infty$:
\begin{equation}
\begin{aligned}
I(z) & = C_1 \,\, z_2^{- \alpha_1} \, z_3^{\frac{\beta}{2} - \alpha_1 - \alpha_2} \, z_4^{\alpha_1 - \frac{\beta}{2}} \, \,{}_2F_1(\alpha_1, \, \alpha_1 + \alpha_2-\tfrac{\beta}{2}; \, \alpha_1 - \tfrac{\beta}{2} + 1; \, z_t) 
\\ & + C_2 \,\, z_1^{- \alpha_1} \, z_3^{\frac{\beta}{2} - \alpha_2} \, z_4^{-\frac{\beta}{2}} \, \,{}_2F_1( \alpha_1, \, \tfrac{\beta}{2}; \, \tfrac{\beta}{2} - \alpha_2 + 1 ; \, \tfrac {1} {z_t}) 
\\ & + C_3 \, \,z_1^{- \alpha_1 - \alpha_2 + \frac{\beta}{2}} \, z_2^{\alpha_2 -\frac{\beta}{2}} \, z_4^{-\alpha_2} \, \,{}_2F_1(  \alpha_2, \, \alpha_1 + \alpha_2 - \tfrac{\beta}{2} ; \, \alpha_2 -\tfrac{\beta}{2} +1; \, \tfrac {1} {z_t} ) 
\\ & + C_4 \, \,z_1^{\frac{\beta}{2} - \alpha_1} \, z_2^{-\frac{\beta}{2}} \, z_3^{-\alpha_2} \,\,{}_2F_1(\alpha_2, \, \tfrac{\beta}{2} ; \, \tfrac{\beta}{2} - \alpha_1 + 1; \, z_t)
\end{aligned}
\end{equation}
where $C_{1,\, 2,\, 3,\, 4}$ are proportionality constants that are dependent on the integration contour.
\end{exmp}

\begin{exmp}\label{sec:example_doable}
The simplest class of Euler integrals in $2$ variables with $2$ toric invariants is represented by
\begin{equation}\label{eq:example_doable}
I(z)  = \int_{\sigma} \, \frac {\de x_1\, \de x_2} {x_1\, x_2} \, \frac {x_1^{\beta_1} \, x_2^{\beta_2} } {(z_1 + z_2 \, x_1 \, x_2 + z_3 \, x_1^2 + z_4 \, x_2^2 + z_5\,  x_1^2 \, x_2^2)^{\alpha} }.
\end{equation}
The homogeneity conditions are
\begin{equation}
\begin{aligned}
& D_1 = \sum_{i=1}^{5} \,\theta_i + \alpha, \quad D_2 = \theta_2 + 2 \, \theta_3 + 2 \, \theta_5 + \beta_1, \quad D_3 = \theta_2 + 2 \, \theta_4 + 2 \, \theta_5 + \beta_2 \label{eq:doable_scalepde}
\\ & D_1 \circ I = D_2 \circ I = D_3 \circ I = 0.
\end{aligned}
\end{equation}
The toric symmetries can be found by representing this integral in its Schwinger parametric form,
\begin{equation}
I(z)  = \int_{\sigma} \, \frac {\de x_1\, \de x_2} {x_1\, x_2} \, x_1^{\beta_1} \, x_2^{\beta_2} \int_0^{\infty} \frac {\de y} {y} \frac {y^{\alpha}} {\Gamma(\alpha) } \, e^{- y \, (z_1 + z_2 \, x_1 \, x_2 + z_3 \, x_1^2 + z_4 \, x_2^2 + z_5 \, x_1^2 \, x_2^2)}
\end{equation}
and finding relations between its partial derivatives wrt the variables $z$:
\begin{equation}
\begin{aligned}
\partial_{z_1} I(z) & = (- y) \, I, \quad \partial_{z_2} I(z) = (- y \, x_1 \, x_2) \, I, \quad \partial_{z_3} I(z) = (- y \, x_1^2) \, I,
\\ \partial_{z_4} I(z) & = (- y \, x_2^2) \, I, \quad \partial_{z_5} I(z) = (- y  \, x_1^2 \, x_2^2) \, I
\end{aligned}
\end{equation}
where for the sake of notational brevity $I$ has been used instead of the entire integrand. All toric symmetries of this integral: 
\begin{equation}
\begin{aligned}
& \partial_{z_1} \, \partial_{z_5} \circ I = (-y)\, (-y  \, x_1^2 \, x_2^2) \, I = (- y \, x_1^2) \, (- y \, x_2^2) \, I = \partial_{z_3} \, \partial_{z_4} \circ I(z) 
\\ & \partial_{z_1} \, \partial_{z_5} \circ I = (-y)\, (-y  \, x_1^2 \, x_2^2) \, I = (- y \, x_1 x_2)^2\, I = \partial_{z_2}^2 \circ I(z) 
\end{aligned}
\end{equation}
can be represented by a basis of $2$ differential operators:
\begin{equation}
(\partial_{3} \, \partial_{4} - \partial_{1} \, \partial_{5}) \circ I = ( \partial_{1} \, \partial_{5} - \partial_2^{2}) \circ I =  0 \label{eq:toricpdedoable}
\end{equation}
corresponding to a working basis of toric invariants
\begin{equation}
t_1  = \frac {z_3 \, z_4} {z_1 \, z_5}, \quad t_2 = \frac {z_1 \, z_5} {z_2^2}.  \label{eq:doable_toric_inv}
\end{equation}
Alternate bases of PDEs and toric invariants are entirely equivalent choices as long as they commute with each other and span all toric symmetries of the integral, e.g.
\begin{align}
& ( \partial_{1} \, \partial_{5} - \partial_{3} \, \partial_{4}) \circ I = ( \partial_{3} \, \partial_{4} - \partial_2^{2}) \circ I =  0, \quad t'_1 = \frac {z_1 \, z_5} {z_3 \, z_4}  = t_1^{-1} , \quad t'_2 = \frac {z_3 \, z_4} {z_2^2} = t_1 \, t_2 \label{eq:doable_toric_inv2}
\\ &  (\partial_2^{2} - \partial_{3} \, \partial_{4}) \circ I =  (\partial_{1} \, \partial_{5} - \partial_2^{2}) \circ I = 0, \quad t''_1 = \frac {z_2^2}  {z_3 \, z_4} = (t_1 \, t_2)^{-1}, \quad  t''_2 = \frac {z_1 \, z_5} {z_2^2} = t_2. \label{eq:doable_toric_inv3}
\end{align}
The linear equations relating the exponents of the variables implied by \cref{eq:doable_scalepde} can be used to fix $3$ of $5$ exponents in $4$ (not $\binom{5}{3} = 10$) different ways. The choice of the dependent and independent variables isn't completely free in this scenario (unlike \cref{sec:example_easy}) and instead fixed to certain combinations. There are as many roots as the number of these allowed combinations and just as many linearly independent series solutions. Using \cref{eq:doable_toric_inv} as the basis of toric invariants,
\begin{equation}
\begin{aligned}
I_1(z) & = z_1^{\frac {\beta_1 + \beta_2} {2} - \alpha} \, z_2^{- \beta_2} \, z_3^{\frac {\beta_2 - \beta_1} {2}} \, \tilde I_1(z) , \quad \tilde I_1(z)  = \int_{\sigma} \tfrac {\de x} {x} \tfrac {x^{\beta} } {(1 + x_1 \, x_2 + x_1^2 + t_1 \, t_2 \, x_2^2 + t_2 \,  x_1^2 \, x_2^2)^{\alpha} } 
\\ I_2(z) & = z_1^{\frac {\beta_1 + \beta_2} {2} - \alpha} \, z_2^{- \beta_1} \, z_4^{\frac {\beta_1 - \beta_2} {2}} \, \tilde I_2(z) , \quad \tilde I_2(z)  = \int_{\sigma} \tfrac {\de x} {x} \tfrac {x^{\beta} } {(1 + x_1 \, x_2 + t_1 \, t_2 \,  x_1^2 +  x_2^2 + t_2 \,  x_1^2 \, x_2^2)^{\alpha} } 
\\ I_3(z) & = z_2^{\beta_1 - 2 \alpha} \, z_3^{\frac{\beta_2 - \beta_1}{2}} \, z_5^{\alpha - \frac{\beta_1 + \beta_2}{2}} \, \tilde I_3(z) , \quad \tilde I_3(z)  = \int_{\sigma} \tfrac {\de x} {x} \tfrac {x^{\beta} } {(t_2   +  x_1 \, x_2 +  x_1^2 + t_1 \, t_2 \,  x_2^2 +  x_1^2 \, x_2^2)^{\alpha} } 
\\ I_4(z) & = z_2^{\beta_2 - 2 \alpha} \, z_4^{\frac{\beta_1 - \beta_2}{2}} \, z_5^{\alpha - \frac{\beta_1 + \beta_2}{2}} \, \tilde I_4(z) , \quad \tilde I_4(z)  = \int_{\sigma} \tfrac {\de x} {x} \tfrac {x^{\beta} } {(t_2   +  x_1 \, x_2 +  t_1 \, t_2 \,  x_1^2 + x_2^2 +  x_1^2 \, x_2^2)^{\alpha} } 
\end{aligned}
\end{equation}
corresponding to the roots
\begin{equation}
\begin{aligned}
r_1 & = \{\tfrac {\beta_1 + \beta_2} {2} - \alpha, \, - \beta_2, \, \tfrac {\beta_2 - \beta_1} {2}, \, 0, \, 0\}, \quad r_2 = \{ \tfrac {\beta_1 + \beta_2} {2} - \alpha, \, - \beta_1, \, 0, \, \tfrac {\beta_1 - \beta_2} {2}, \, 0  \} 
\\ r_3 & = \{ 0, \, \beta_1 - 2 \alpha, \tfrac{\beta_2 - \beta_1}{2}, \, 0, \, \alpha - \tfrac{\beta_1 + \beta_2}{2}  \}, \quad r_4 = \{0, \, \beta_2 - 2 \alpha, \, 0, \, \tfrac{\beta_1 - \beta_2}{2} , \, \alpha - \tfrac{\beta_1 + \beta_2}{2} \}. \label{eq:doable_roots}
\end{aligned}
\end{equation}
All non-trivial behaviour is confined to $\tilde I(z)$ which is purely a function of $2$ toric invariants regardless of the representation. So it isn't unreasonable to expect $I$ to have a Laurent series expansion in powers of the toric invariants of the form
\begin{equation}
I_i (z)  \equiv z^{r_i} \, \sum \,c_{n} \, t^{n} = z^{r_i} \, \sum\, c_{n_1, \, n_2} \, t_1^{n_1} \, t_2^{n_2}.
\end{equation}
The above series representation of $I$ exhibits the toric symmetries in \cref{eq:toricpdedoable} when 
\begin{equation}\label{eq:doable_hypeseries}
\begin{aligned}
\tilde I_1(z)  & \,\propto\, F_4( \tfrac{\beta_2}{2} \, ;  \,\tfrac{\beta_2+1}{2} \, ;  \, \tfrac {\beta_2 - \beta_1} {2} + 1 \, ;  \,\tfrac {\beta_1 + \beta_2} {2} - \alpha + 1 \, ; \, 4 \, t_1 \, ;  \, 4 \,t_2  ) 
\\ \tilde I_2(z)  & \,\propto\, F_4( \tfrac{\beta_1}{2} \, ;  \,\tfrac{\beta_1+1}{2} \, ;  \, \tfrac {\beta_1 + \beta_2} {2}  + 1 \, ;  \, \tfrac {\beta_1 - \beta_2} {2} - \alpha + 1 \, ;  \, 4 \,t_1 \, ;  \, 4 \,t_2  ) 
\\ \tilde I_3(z)  & \,\propto\, F_4( \tfrac {2 \alpha - \beta_1} {2} \, ;  \, \tfrac {2 \alpha - \beta_1 + 1} {2}\, ; \, \tfrac {\beta_2 - \beta_1} {2} + 1 \, ;  \, \alpha - \tfrac {\beta_2 - \beta_1} {2} + 1 \, ; \, 4 \,t_1 \, ;  \,4 \,t_2 ) 
\\ \tilde I_4(z)  & \,\propto\, F_4( \tfrac {2 \alpha - \beta_2} {2} \, ;  \, \tfrac {2 \alpha - \beta_1 + 1} {2}\, ; \, \tfrac {\beta_1 - \beta_2} {2} + 1 \, ;  \, \alpha - \tfrac {\beta_1 - \beta_2} {2} + 1 \, ; \, 4 \, t_1 \, ;  \, 4 \,t_2 ) 
\end{aligned}
\end{equation}
where $F_4(\alpha_1 \, ; \, \alpha_2 \, ; \, \beta_1 \, ; \, \beta_2 \, ; \, t_1 \, ; \, t_2)$ is a Horn hypergeometric function defined as
\begin{equation}
\begin{aligned}
F_4(\alpha \, ; \, \beta \, ; \, t) & = \sum_{n_1, \, n_2 = 0}^{\infty} \, \frac { \alpha_1^{(n_1 + n_2)} \, \alpha_2^{(n_1 + n_2)} }  { \beta_1^{(n_1)} \, \beta_2^{(n_2)}  } \, \frac {t_1^{n_1}} {n_1!} \, \frac {t_2^{n_2}} {n_2!} 
\\ & = \frac  { \Gamma( \beta_1) \, \Gamma( \beta_2)  } { \Gamma(\alpha_1) \, \Gamma( \alpha_2) }   \sum_{n_1, \, n_2 = 0}^{\infty}  \,  \frac { \Gamma(\alpha_1 + n_1 + n_2) \, \Gamma( \alpha_2 + n_1 + n_2) }  { \Gamma( \beta_1 + n_1) \, \Gamma( \beta_2 + n_2)  } \, \frac {t_1^{n_1}} {n_1!}\,  \frac {t_2^{n_2}} {n_2!}
\end{aligned}
\end{equation}
with the domain of convergence $\sqrt{ t_1 } + \sqrt { t_2} < 1 $.
The linear independence of these $4$ representations is easily verified by checking their behaviours in particular limits.
\begin{enumerate}
  \item Assuming $\Real(\beta_1 - \beta_2),  \, \Real(2 \, \alpha - \beta_1 - \beta_2) > 0$, in the limit $z_4,\, z_5 \rightarrow 0$,  $ I_2 = I_3 = I_4 = 0$,
  \begin{equation}
  I_1(z)  = \int_{\sigma}\,  \frac {\de x} {x}\,  \frac {z_1^{\frac {\beta_1 + \beta_2} {2} - \alpha} \, z_2^{- \beta_2} \, z_3^{\frac {\beta_2 - \beta_1} {2}}  \, x^{\beta} } {(1 + x_1 \, x_2 + x_1^2)^{\alpha} } = C_1 \, z_1^{\frac {\beta_1 + \beta_2} {2} - \alpha} \, z_2^{- \beta_2} \, z_3^{\frac {\beta_2 - \beta_1} {2}}
  \end{equation}
  where $ z_1^{\frac {\beta_1 + \beta_2} {2} - \alpha} \, z_2^{- \beta_2} \, z_3^{\frac {\beta_2 - \beta_1} {2}}$ is the starting monomial of the series.
  \item Assuming $\Real(\beta_2 - \beta_1)>0,  \, \Real(2\, \alpha - \beta_1 - \beta_2)>0$, in the limit $z_3,\, z_5 \rightarrow 0$, $I_1 = I_3 = I_4 = 0$,
  \begin{equation}
  I_2(z)  =  \int_{\sigma} \, \frac {\de x} {x} \, \frac {z_1^{\frac {\beta_1 + \beta_2} {2} - \alpha} \, z_2^{- \beta_2} \, z_4^{\frac {\beta_1 - \beta_2} {2}} \, x^{\beta} } {(1 + x_1 \, x_2 +  x_2^2)^{\alpha} } = C_2 \, z_1^{\frac {\beta_1 + \beta_2} {2} - \alpha} \, z_2^{- \beta_2} \, z_4^{\frac {\beta_1 - \beta_2} {2}}.
  \end{equation}
  \item Assuming $\Real(\beta_1 - \beta_2), \, \Real(\beta_1 + \beta_2 - 2 \, \alpha) > 0$, in the limit $z_1,\, z_4 \rightarrow 0$, $I_1 = I_2 = I_4 = 0$,
  \begin{equation}
  I_3(z)  = \int_{\sigma}\,  \frac {\de x} {x} \, \frac {z_2^{\beta_1 - 2 \alpha} \, z_3^{\frac{\beta_2 - \beta_1}{2}} \, z_5^{\alpha - \frac{\beta_1 + \beta_2}{2}}  \, x^{\beta} } {(x_1 \, x_2 +  x_1^2 +  x_1^2 \, x_2^2)^{\alpha} } = C_3 \, z_2^{\beta_1 - 2 \alpha} \, z_3^{\frac{\beta_2 - \beta_1}{2}} \, z_5^{\alpha - \frac{\beta_1 + \beta_2}{2}}.
  \end{equation}
  \item Assuming $\Real(\beta_2 - \beta_1), \, \Real(\beta_1 + \beta_2 - 2 \, \alpha) > 0$, in the limit $z_1,\, z_3 \rightarrow 0$, $I_1 = I_2 = I_3 = 0$,
  \begin{equation}
  I_4(z)  =\int_{\sigma} \, \frac {\de x} {x} \, \frac { z_2^{\beta_2 - 2 \alpha} \, z_4^{\frac{\beta_1 - \beta_2}{2}} \, z_5^{\alpha - \frac{\beta_1 + \beta_2}{2}}  \, x^{\beta} } {( x_1 \, x_2 + x_2^2 +  x_1^2 \, x_2^2)^{\alpha} } = C_4 \, z_2^{\beta_2 - 2 \alpha} \, z_4^{\frac{\beta_1 - \beta_2}{2}} \, z_5^{\alpha - \frac{\beta_1 + \beta_2}{2}}.
  \end{equation}
\end{enumerate}
For generically valued $\{\alpha, \, \beta_1, \, \beta_2\}$, the $4$ starting monomials are independent. The proportionality constants $C_{1,\, 2,\, 3,\, 4}$ can be found comparatively easily in these specific limits and then carried over to the complete series solution:
\begin{equation}
I(z)  = \sum_{i = 1}^{4} \, C_i \, z^{r_i} \, \tilde I_i (z)
\end{equation}
where $r_i$ are the roots, \cref{eq:doable_roots}, and $\tilde I_i(z)$, \cref{eq:doable_hypeseries}, are Horn hypergeometric functions of the toric invariants $t_1 = \tfrac {z_3 \, z_4} {z_1 \, z_5}$, $t_2 = \tfrac {z_1 \, z_5} {z_2^2}$, parameterised by the roots.
\end{exmp}

\begin{exmp}\label{sec:example_tricky}
Euler integrals, \cref{eq:toocutetobetrue2}, defined at arbitrary choices of $\alpha$, $\beta$ and/or $z$ may have properties which deviate from the generalized versions. For example, the following integral $I$, over some contour $\sigma \in \CC^3$, is defined at a specific choice of $\beta_3$ and is to be evaluated at fixed $z$.
\begin{equation}
I  = \int_{\sigma} \de x_1 \, \de x_2 \, \de x_3 \, \mathscr I, \quad \mathscr I = \frac {x_1^{\beta_1} \, x_2^{\beta_2}} { g(x)^{\alpha}}
\end{equation}
where the polynomial $g(x)$ is
\begin{equation}
g(x)  = 1 - x_1^2 - x_2^2 + x_1^2  \, x_2^2 + \frac {1} {x_1 x_2} - \frac {x_1} {x_2}  - \frac {x_2} {x_1}  - x_3 - x_1 \,  x_2 \,  x_3.
\end{equation}
One way of evaluating this integral is to revert to the generalized format,
\begin{align}
  I & = \int_{\sigma} \de x \, \frac {x_1^{\beta_1 + \alpha} \, x_2^{\beta_2 + \alpha}\, x_3^{\beta_3}} { g(x; \, z)^{\alpha}} \Big|_{z = \bar z, \, \beta_3 = 0}, \quad \bar z = \{ 1, \,  - 1, \, - 1 , \, 1, \, - 1, \, -1, \, 1, \, - 1, \, - 1 \}
\\ g(x; z) & = z_1 + z_2 x_1^2 + z_3 x_2^2 + z_4 x_1 x_2 + z_5 x_1^3 x_2 + z_6 x_1 x_2^3 + z_7 x_1^3 x_2^3 + z_8 x_1 x_2 x_3 + z_9 x_1^2 x_2^2 x_3 \nonumber
\end{align}
and evaluating $I(z)$ in the limit $z \rightarrow \bar z$ and $\beta_3 \rightarrow 0$. However, this arbitrary choice of parameters allows another approach. The differential equations satisfied by the integrand $\mathscr I$ are
\begin{equation}
\partial_{1, \, 2} \, \mathscr I  = \left( \frac {\beta_{1, \, 2}} {x_{1, \, 2}} - \alpha \, \frac {\partial_{1, \,2} \; g(x)} {g(x)} \right) \, \mathscr I, \quad \partial_3 \, \mathscr I = - \alpha \, \frac {\partial_3 g(x)} {g(x)} \, \mathscr I.
\end{equation}
Upon eliminating the variable $x_3$ from them, $\mathscr I$ is found to satisfy
\begin{equation}
\partial_1 \left( \frac {x_1} {\beta_1 - \beta_2} \, \mathscr I\right) + \partial_2 \left( \frac {x_2} {\beta_2 - \beta_1}\, \mathscr I\right) + \partial_3 \left( \frac {2} {\beta_1 - \beta_2} \, \frac {x_2^2 - x_1^2} {x_1 x_2} \, \mathscr I\right) = \mathscr I.
\end{equation}
Thus, $I$ is a surface integral of a vector field:
\begin{equation}
\left( \frac {x_1} {\beta_1 - \beta_2} \, \mathscr I, \; \frac {x_2} {\beta_2 - \beta_1}\, \mathscr I, \; \frac {2} {\beta_1 - \beta_2} \, \frac {x_2^2 - x_1^2} {x_1 x_2} \, \mathscr I\right),
\end{equation}
and if $\sigma$ is a closed contour not enclosing any poles, it equals $0$. Setting $\sigma$ to an open contour, e.g. $x \in (\RR_+)^3$, and assuming  
$\Real(\alpha) > 0$ and $|\Real(\alpha)| > |\Real(\beta_1 + 1)|, \, |\Real(\beta_2 + 1)|$, $I$ reduces to an integral over $2$ variables:
\begin{equation}
I   = \tfrac {2} {\beta_1 - \beta_2} \int_{x_1, x_2}  \tfrac {x_1^2 - x_2^2} {x_1 x_2} \, \mathscr I \, \Big|_{x_3 = 0}
= \tfrac {2} {\beta_1 - \beta_2} \int_{x_1, x_2}  \Big( \tfrac {x_1} {x_2}  - \tfrac {x_2} {x_1} \Big)\, \tfrac {x_1^{\beta_1} \, x_2^{\beta_2}} { (1 - x_1^2 - x_2^2 + x_1^2 x_2^2 + \frac {1} {x_1 x_2} - \frac {x_1} {x_2}  - \frac {x_2} {x_1})^{\alpha}}.
\end{equation}
\end{exmp}

\begin{exmp}\label{sec:example_hard} 
${}_2F_1(a_1, \, a_2;  \, b ; \, x)$ is defined as a solution to the Gauss hypergeometric ODE
\begin{equation}\label{eq:gausshypergeometricODE}
{}_2D_1 = \Big( x  \,( 1 - x) \, \partial_x^2  + (b - x  \,(a_1 + a_2 + 1)) \, \partial_x  - a_1 \, a_2 \Big), \quad  {}_2D_1 \circ {}_2F_1(x) = 0.
\end{equation}
$I(a_1, \,a_2, \,b \,; \,p, \,q \,; \,x)$, a holomorphic function of $x$, is the analytic continuation of ${}_2F_1(x)$ and one of its integral representations, specifically the Euler integral representation, is
\begin{equation}
I(a_1, \,a_2, \,b \,; \,p, \,q \,; \,x)  = \int_{q}^p \de t \, f(t, \,x), \quad f(t, \,x) = \frac {1} {t  \,(1-t)} \left(\frac { t } {1 - t}\right)^{a_2}\frac {(1 - t)^{b}} {(1 - x \, t)^{a_1}} \label{eq:Eulerrep2F1}
\end{equation}
where $p, \,q \in \{0, \,1, \,\tfrac 1 x, \,\infty\}$.
The integrand $f(t,x)$ satisfies
\begin{equation}
\partial_t f  = \left(\frac {(a_2-1)} {t} - \frac {(b - a_2 - 1)} {1-t} + \frac {a_1\,  x } {1 - x \, t} \right)  \,f, \quad \partial_x f = \frac {a_1 \, t} {(1 - x \, t)}  \,f.
\end{equation}
Upon clearing the denominators to bring them into a commonly preferred form, operators $D_1$, $D_2$ are found to annihilate the integrand $f(t, \, x)$:
\begin{equation}
\begin{aligned}
D_1 & = t \, (1 - t) \, (1 - x \, t)\, \partial_t -  (a_2-1) \, (1 - t) \, (1 - x \, t)  
\\ & + (b - a_2 - 1) \, t \, (1 - x \, t) - a_1 \, x \, t \, (1 - t)
\\ D_2 & = (1 - x\,  t) \, \partial_x - a_1 \, t.
\end{aligned}
\end{equation}
All linear combinations of $D_1$ and $D_2$ also annihilate the integrand,
\begin{equation}
D_1 \oplus D_2  = d_1 \, D_1 + d_2 \, D_2, \quad D_1 \oplus D_2 \circ f (t, \, x) = 0
\end{equation}
where $d_1, \, d_2$ are any operators constructed from polynomial combinations of $x, \, \partial_x, \, t, \, \partial_t$.
A well intentioned choice of $d_1, \, d_2$ results in
\begin{equation}
D_{(1,2)} = (1 - a_1) \,\Big( x \,  {}_2D_1 + \partial_t  \,(a_1 \,  t + x  \,\partial_x - \partial_x) \Big) \label{eq:2f1pdelinearcombination}
\end{equation}
where all $t$ (and $\partial_t$) dependence is confined to a total derivative term wrt $t$ and the operator ${}_2D_1$ is the same as what is given in \cref{eq:gausshypergeometricODE}, defined only in terms of $x, \, \partial_x$. Integrating the identically null valued $D_{(1,2)} \circ f(t, \,x) = 0$ wrt $t$ over some contour in $\CC$, $\sigma$, gives
\begin{equation}
(1 - a_1) \, \left( \int_{\sigma} \de t \, x \,  {}_2D_1 \circ f(t, \,x) + \int_{\sigma} \de t \, \partial_t  \,(a_1 \,  t + x  \,\partial_x - \partial_x) \circ f(t, \,x)  \right)  = 0.
\end{equation}
Assuming $a_1 \neq 0, \, 1$ and $x \neq 0$, this simplifies to
\begin{equation}
\begin{aligned}
{}_2D_1 \circ \int_{\sigma} \de t \, f(t, \,x) & = \int_{\sigma} \de t \, \partial_t  \,\left( \frac 1 {x}  \, \partial_x - \partial_x - \frac {a_1 \,  t } {x}  \right) \circ f(t, \,x) \\ & = - a_1 \, \int_{\sigma} \de t \, \partial_t  \, F(t, \, x),\qquad \qquad  \! \! F(t, \, x) = \frac {t^{b} \, (1 - t)^{b - a_2}} { (1 - x \, t)^{a_1 + 1}}.
\end{aligned}
\end{equation}
When $\Real(b),  \, \Real(b-a_2),  \, \Real(a_1 + 1)$ and $\Real(2 b - a_1 - a_2 - 1) > 0$, $F(t, \, x)$ tends to $0$ at the singular points of $f(t,\,x)$, $\Sigma_0 = \{0, \, 1, \, \tfrac 1 {x}, \, \infty\}$. If $\sigma$ is a closed contour not enclosing $\Sigma_0$, or an open interval $(p, \, q)$, where $\{p, \, q\} \in \Sigma_0$, the RHS equals $0$, thus proving the original claim that \cref{eq:Eulerrep2F1} is an integral representation of ${}_2F_1(x)$.

The dimension of the solution space of a degree $n$ ODE
\begin{equation}
\sum_{i=0}^{n}  \, f_{i}(x) \, \partial_x^i \circ I(x) = 0  
\end{equation}
on $U \subset \CC$ equals $n$ for $f_n(U) \neq 0$ (and/or including any regular singularities) and $f_0(x)$ is the indicial polynomial. Since the ODE in the present context is second order, ${}_2F_1(x)$, or rather $I(x)$, has 2 linearly independent non-degenerate Laurent series solutions for $x \, (1 - x) \neq 0$. The series expansion of $I(x)$ around $x=0$ such that ${}_2D_1 \circ I (x) = 0$ is
\begin{equation}
I (x) = x^{\alpha} \, \sum_{n=0}^{\infty}  \, c_{n} \, x^{n}, \quad b_n = \frac {c_{n+1}} {c_n} = \frac {(\alpha + n + a_1) \, (\alpha + n + a_2)} {(\alpha + n + 1) \, (\alpha + n + b)}, \quad \alpha \, (\alpha + b - 1) = 0.
\end{equation}
Such a rational relation between the coefficients $c_n$, $b_n$, is called the Bernstein - Sato polynomial. The equation satisfied by the roots $\alpha$ is called the indicial equation and it expectedly has $2$ solutions, each giving rise to a linearly independent series, together spanning the entire solution space:
\begin{equation}
\begin{aligned}
I (x) & \; \propto \; 
\sum_{n=0}^{\infty}  \, \tfrac {\Gamma(a_1 + n) \, \Gamma(a_2 + n) } {\Gamma(b+n)} \, \tfrac {x^{n}} {n!} \, \oplus \, 
x^{1-b} \, \sum_{n=0}^{\infty}  \, \tfrac {\Gamma(a_1 + 1 - b + n) \, \Gamma(a_2 + 1-b + n) } {\Gamma(2 - b + n)} \, \tfrac {x^{n}} {n!} 
\\ & =  \,{}_2 F_1 ( a_1, \, a_2 \, ;\,  b \, ; \, x ) \, \oplus \, x^{1-b} \,  \, {}_2F_1 (a_1 + 1 - b, \, a_2 + 1 - b \, ; \, 2 - b; \, x). \label{eq:2F1solspace}
\end{aligned}
\end{equation}
It is no coincidence that the superficially different looking integrals \cref{eq:unexpected2F1,eq:Eulerrep2F1} describe the same class of functions. As was shown in \cite{GELFAND1990255} over $3$ decades ago, Euler integrals are classified by their reduced form and \cref{eq:unexpected2F1} does reduce to \cref{eq:Eulerrep2F1} under a change of variables that is obvious in hindsight.
\end{exmp}

\begin{exmp}\label{sec:example_intractable}
The most general $2$\textsuperscript{nd} order polynomial in $n$ variables $x$ is
\begin{equation}
P(x \, ; z)  = \sum_{i=1}^{n}  \, z_i \, x_i^{2} + \sum_{i < j}^{n}  \, z_{i,\,j} \, x_i \, x_j.
\end{equation}
The class of Euler integrals featuring this polynomial,
\begin{equation}
I (z)  = \int_{\sigma} \, \frac {\de x} {x} \, \frac {x^{\beta}} {(z_0 + P)^{\alpha}}
\end{equation}
satisfies $n+1$ homogeneity conditions that can be found by considering the scale transformations $z \rightarrow \rho \, z$, and $x_i \rightarrow \rho \, x_i$, $i \in [n]$:
\begin{equation}
D_{0} =  \theta_0 + \alpha - \tfrac 1 2 \, \sum  \, \beta_i, \quad D_{i} = 2 \,\theta_i + \sum_{j=1}^{i-1}  \, \theta_{j,\,i} + \sum_{j=i+1}^{n}  \, \theta_{i,\,j} + \beta_i. \label{eq:example_intractacle_Hideal}
\end{equation}
Thus, the integral can be represented as the product of a monomial in $z$, with trivial dependence on $z_0$, and a hypergeometric function, $\tilde I$, of $\tfrac {n\,(n-1)} {2}$ toric invariants, $t$,
\begin{equation}
I (z)  = z_0^{\frac 1 2 \sum \beta - \alpha} \, z^{r} \, \tilde I(t).
\end{equation}
The toric symmetries are spanned by $3$ types of PDEs:
\begin{equation}\label{eq:example_intractable_toric_PDEs}
D_{i[2]}  = \partial_{i_1} \, \partial_{i_2} - \partial_{i_1,\,i_2}^2 , \quad D_{i[3]} = \partial_{i_1, \,i_2} \, \partial_{i_2, \,i_3}  - \partial_{i_1, \,i_3} \, \partial_{i_2}, \quad D_{i[4]} = \partial_{i_1,\, i_2} \, \partial_{i_{3}, \,i_4} - \partial_{i_1,\, i_4} \, \partial_{i_2, \,i_3}.
\end{equation}
Akin to \cref{sec:example_doable}, each PDE has a representative toric invariant. It is immediately apparent that there are more toric PDEs than independent toric invariants. However, unlike \cref{sec:example_doable}, which has an intuitive basis of toric invariants, \cref{eq:doable_toric_inv}, it is not immediately apparent which of the over-complete set of toric invariants suggested by \cref{eq:example_intractable_toric_PDEs} form an appropriate linearly independent basis. For the sake of concrete illustration, let $n=3$, $z = \{z_1, \, z_2, \, z_3, \, z_{1,\,2}, \, z_{1,\,3}, \, z_{2,\,3}\}$. The homogeneity conditions are simple:
\begin{equation}
\begin{aligned}
D_{0}& = \theta_0 + \alpha -  \tfrac 1 2 \sum  \, \beta, \quad D_{1} = 2 \,\theta_1 +  \theta_{1,\,2} +  \theta_{1,\,3} + \beta_1 
\\ D_{2} & = 2 \,\theta_2 +  \theta_{1,\,2} +  \theta_{2,\,3} + \beta_2, \quad D_{3} = 2 \,\theta_3 +  \theta_{1,\,3} +  \theta_{2,\,3} + \beta_3.
\end{aligned}
\end{equation}
The toric PDEs are spanned by $6$ differential operators
\begin{equation}
\begin{aligned}
D_{1,\,2} & = \partial_{1} \, \partial_{2} - \partial_{1,\,2}^2, \quad D_{1,\,3} = \partial_{1} \, \partial_{3} - \partial_{1,\,3}^2, \quad D_{2,\,3} = \partial_{2} \, \partial_{3} - \partial_{2,\,3}^2 
\\ D_{1,\,2,\,3} & = \partial_{1, \,2} \, \partial_{2, \,3}  - \partial_{1, \,3} \, \partial_{2}, \quad D_{2,\, 3,\, 1} = \partial_{1, \,3} \, \partial_{2,\, 3}  - \partial_{1,\, 2} \, \partial_{3}, \quad D_{3,\,1,\,2} = \partial_{1,\, 3} \, \partial_{1, \,2}  - \partial_{2, \,3} \, \partial_{1}
\end{aligned}
\end{equation}
even though the basis of toric invariants consists of only $3$ elements. One such basis is
\begin{equation}
t_1  = \frac {z_1 \, z_{2,\,3}} {z_{1,\,2} \, z_{1,\,3}} , \quad t_2 = \frac {z_2 \, z_{1,\,3}} {z_{1,\,2} \, z_{2,\,3}}, \quad t_3 = \frac {z_3 \, z_{1,\,2}} {z_{1,\,3} \, z_{2,\,3}}. \label{eq:example_intractable_orig_toricinvbasis}
\end{equation}
Operating in this basis, the solution space of $D \circ I(z) = 0$ will consist of Laurent series solutions in both positive and negative powers of each toric invariant, of which obviously only one can be convergent. Further, since all possible combinations get explored, there will always be at least one convergent series solution at any value of $t$ for generically valued roots, $r$:
\begin{equation}
I (z)  = z_0^{\frac 1 2 \sum \beta - \alpha} \, z^{r} \,  \sum_{n \in \mathbb Z}  \, c_{n_1, \, n_2, \, n_3} \, t_1^{n_1} \, t_2^{n_2} \, t_3^{n_3}. \label{eq:exampleintractable_laurent}
\end{equation}
There are $4$ independent roots, $r$, to this system (exponent of $z_0$ is dropped in the following)
\begin{equation}\label{eq:example_intractable_rootlist}
\begin{aligned}
r_{(1)} & = \{0, 0, 0, \tfrac {\beta_3-\beta_1 - \beta_2} {2} , \tfrac {\beta_2 - \beta_1 - \beta_3} {2} , \tfrac {\beta_1 - \beta_2 - \beta_3} {2}\}, && r_{(2)} = \{0, 0, \tfrac { \beta_1 + \beta_2 - \beta_3} {2}, 0, -\beta_1 , -\beta_2 \},
\\ r_{(3)} & = \{0, \tfrac {\beta_1  + \beta_3 - \beta_2} {2} , 0, -\beta_1,0 ,  -\beta_3 \}, && r_{(4)} = \{\tfrac { \beta_2 + \beta_3 -\beta_1} {2}  , 0, 0, -\beta_2, -\beta_3, 0\}.
\end{aligned}
\end{equation}
The entire solution space isn't represented in terms of the same basis of toric invariants. The series originating from the first root is indeed in terms of the originally suggested basis, \cref{eq:example_intractable_orig_toricinvbasis},
\begin{equation}
\begin{aligned}
I_1(z) & = z_{1,\,2}^{ \frac {\beta_3-\beta_1 - \beta_2} {2}} \, z_{1,\,3}^{\frac {\beta_2 -\beta_1  - \beta_3} {2}} \, z_{2,\,3}^{\frac {\beta_1 - \beta_2 - \beta_3} {2}} \,\sum_{n \in \NN_0}  \, \tfrac {(t_1)^{n_1}} {n_1!} \; \tfrac {(t_2)^{n_2}} {n_2!} \; \tfrac {(t_3)^{n_3}} {n_3!} 
\\  &  \times  \tfrac {\Gamma(\frac {\beta_3-\beta_1 - \beta_2} {2} + 1)} {\Gamma(\frac {\beta_3-\beta_1 - \beta_2} {2} + n_3-n_1-n_2 + 1)} \; \tfrac {\Gamma(\frac {\beta_2 -\beta_1  - \beta_3} {2} + 1)} {\Gamma(\frac {\beta_2 -\beta_1  - \beta_3} {2} + n_2-n_1-n_3 + 1)} \; \tfrac {\Gamma(\frac {\beta_1 - \beta_2 - \beta_3} {2} + 1)} {\Gamma(\frac {\beta_1 - \beta_2 - \beta_3} {2} + n_1-n_2-n_3 + 1)}
\end{aligned}
\end{equation}
but that is not the case for the other $3$, which after shifting the summation range to $\NN_0$ are
\begin{equation}
\begin{aligned}
I_2(z) & = \tfrac{ z_{3}^{ \frac { \beta_1 + \beta_2 - \beta_3} {2} } }{ z_{1,3}^{\beta_1} \, z_{2,3}^{\beta_2}  } \, 
\sum_{n \in \NN_0} \, \tfrac {\Gamma( \beta_1 + 2 n_1 + n_3) \, \Gamma(\beta_2  + 2 n_2 + n_3)} {\Gamma(\frac { \beta_1 + \beta_2 - \beta_3} {2} + n_1 + n_2 + n_3 + 1)} \,  \tfrac {(t_1  \, t_3)^{n_1}} {n_1!} \; \tfrac {(t_2 \, t_3)^{n_2}} {n_2!} \; \tfrac {(t_3)^{n_3}} {n_3!} 
\\ 
I_3(z) & =
\tfrac{ z_{2}^{ \frac { \beta_1 + \beta_3 - \beta_2} {2} } }{ z_{1,2}^{\beta_1} \, z_{2,3}^{\beta_3}  }  \, \sum_{n \in \NN_0}  \, \tfrac {\Gamma( \beta_1 + 2 n_1 + n_2) \, \Gamma(\beta_3  + 2 n_3 + n_2)} {\Gamma(\frac { \beta_1 + \beta_3 - \beta_2} {2} + n_1 + n_2 + n_3 + 1)} \,  \tfrac {(t_1 \, t_2)^{n_1}} {n_1!} \; \tfrac {(t_2)^{n_2}} {n_2!} \; \tfrac {(t_2  \, t_3)^{n_3}} {n_3!} 
\\ 
I_4(z) & =
\tfrac{ z_{1}^{ \frac { \beta_2 + \beta_3 - \beta_1} {2} } }{ z_{1,2}^{\beta_2} \, z_{1,3}^{\beta_3}  }   \, \sum_{n \in \NN_0}  \, \tfrac {\Gamma( \beta_2 + 2 n_2 + n_1) \, \Gamma(\beta_3  + 2 n_3 + n_1)} {\Gamma(\frac { \beta_1 + \beta_3 - \beta_2} {2} + n_1 + n_2 + n_3 + 1)} \,  \tfrac {(t_1)^{n_1}} {n_1!} \; \tfrac {(t_1  \, t_2)^{n_2}} {n_2!} \; \tfrac {(t_1 \, t_3)^{n_3}} {n_3!}.
\end{aligned}
\end{equation}
The list of roots, \cref{eq:example_intractable_rootlist}, isn't exhaustive. Shifting it by any element of the kernel $\mathcal K$ leaves the hypergeometric system unchanged
\begin{equation}
\mathcal K  = \mathbb Z \, \{1,0,0,-1,-1,1\}  \, \oplus  \, \mathbb Z \, \{0,1,0,-1,1,-1\}  \, \oplus  \, \mathbb Z \,  \{0,0,1,1,-1,-1\}.
\end{equation}
Using these linear combinations, all possible toric invariants and roots can be found. Each root corresponds to a singular point of the integral (zero of the polynomial), encoding the zero locus of the polynomial within the root system of the Euler integral. Thus, depending on the contour $\sigma$, different roots and their associated solutions dominate. 
\end{exmp}

\begin{exmp} \label{sec:example_intractable_2} 
Similar in construction but far more simplistic than the famous mirror quintic is the family of elliptic curves $P_\psi = x_1^3 + x_2^3 + x_3^3 - 3 \, \psi \, x_1 \, x_2 \, x_3$, parameterised by $\psi$, with period $\Omega$ invariant under linear scale transformations $(x_1, \, x_2, \, x_3) \rightarrow \lambda \, (x_1, \, x_2, \, x_3)$, and hence well-defined on $\mathbb P^2$ as
\begin{equation}
\begin{aligned}
  \Omega_{\psi} & = \int_{\gamma_P} \frac {- x_1 \wedge \de x_2 \wedge \de x_3 + \de x_1 \wedge x_2 \wedge \de x_3 - \de x_1 \wedge \de x_2 \wedge x_3} {P_\psi}
\end{aligned}
\end{equation}
where $\gamma_P$ is a small loop around the surface $P_\psi = 0$ \cite{Hori:2003ic}. Considering the polynomial in its generalized form $P_{a, \, \psi} = a_1 \, x_1^3 + a_2 \, x_2^3 + a_3 \, x_3^3 - 3  \, \psi \, x_1 \, x_2 \, x_3$, $\Omega_{a, \, \psi}$ satisfies
\begin{equation}
\begin{aligned}
  & (\theta_{a_1} + \theta_{a_2} + \theta_{a_3} + \theta_{\psi} + 1) \,\Omega_{a, \, \psi} = 0, && (3 \,\theta_{a} + \theta_{\psi} + 1) \, \Omega_{a, \, \psi} = 0
  \\ & \left( \partial_{a_1} \, \partial_{a_2} \, \partial_{a_3} + (\tfrac 1 {3} \, \partial_{\psi})^3 \right) \, \Omega_{a, \, \psi} = 0.
\end{aligned}
\end{equation}
The scaling equations require $\Omega_{a, \, \psi}$ to take the form $\tfrac 1 {3 \, \psi} \, \bar\Omega(\varphi)$, where $\varphi = \tfrac {a_1 \, a_2 \, a_3}{(3 \psi)^3}$ and the toric PDE simplifies to an ODE in $\varphi$
\begin{equation}
\begin{aligned}
  & (\tfrac 1 3 \, \theta_{\varphi})^3 \, \bar\Omega(\varphi) = \left(\theta^2_\varphi - \theta_\varphi + \tfrac {2} {9} \right) \, \theta_\varphi \, \varphi \, \bar\Omega(\varphi).
\end{aligned}
\end{equation}
Series solutions of this ODE around $\varphi = 0$ can be iteratively built upon a linearly independent basis of solutions of $(\tfrac 1 3 \, \theta_{\varphi})^3 \, \bar\Omega(\varphi) = 0$ that serve as the starting monomials.

A basis of $n$ solutions of $\theta_z^n \, f(z) = 0$ are $f_k(z) = (\tfrac 1 {2 \pi \im} \, \log z)^{k}$, $k \in [0, \cdots, n-1]$. Rotating the argument by an angle of $2 \pi$ relates the solutions as $f_n(e^{2 \pi \im} \, z) = (\tfrac 1 {2 \pi \im} \, \log z + 1)^{n} = \sum\limits_{k = 0}^n \binom{n}{k} f_k (z) $. Thus, the columns of the monodromy matrices, $M_{n}$, are binomial coefficients:
\begin{equation}
\begin{aligned}
  M_2 & = \left( \begin{smallmatrix}
        1 & 1
        \\ 0 & 1
      \end{smallmatrix} \right), && M_3 = \left( \begin{smallmatrix}
        1 & 1 & 1
        \\ 0 & 1 & 2
        \\ 0 & 0 & 1
      \end{smallmatrix} \right), && M_4 = \left( \begin{smallmatrix}
        1 & 1 & 1 & 1
        \\ 0 & 1 & 2 & 3
        \\ 0 & 0 & 1 & 3
        \\ 0 & 0 & 0 & 1
      \end{smallmatrix} \right), && M_5 = \left( \begin{smallmatrix}
        1 & 1 & 1 & 1 & 1
        \\ 0 & 1 & 2 & 3 & 4
        \\ 0 & 0 & 1 & 3 & 6
        \\ 0 & 0 & 0 & 1 & 4
        \\ 0 & 0 & 0 & 0 & 1
      \end{smallmatrix} \right) \text{ and so on.}
\end{aligned}
\end{equation}
In this case, the starting monomials are hence $\bar\Omega_0(\varphi) = \{1, \, \tfrac 1 {2 \pi \im} \, \log \varphi, \, (\tfrac 1 {2 \pi \im} \, \log \varphi)^2\}$. The most general zeroth and first order terms of the series are
\begin{equation}
\begin{aligned}
  & \bar\Omega_0(\varphi) = c_{0,0} + c_{0,1} \, \tfrac 1 {2 \pi \im} \, \log \varphi +  c_{0,2} \, (\tfrac 1 {2 \pi \im} \, \log \varphi)^2
  \\ & \bar\Omega_1(\varphi) =  \varphi \, (c_{1,0} + c_{1,1} \, \tfrac 1 {2 \pi \im} \,\log \varphi +  c_{1,2} \, (\tfrac 1 {2 \pi \im} \, \log \varphi)^2).
\end{aligned}
\end{equation}
Requiring $\left(\theta^2_\varphi - \theta_\varphi + \tfrac {2} {9} \right) \, \theta_\varphi \, \varphi \, \bar\Omega_0(\varphi) = (\tfrac 1 3 \, \theta_{\varphi})^3 \, \bar\Omega_1(\varphi) $ fixes the constants $c_{1}$ to
\begin{equation}
\begin{aligned}
   & \{c_{1,0}, \,  c_{1,1}, \, c_{1,2}\} = \{3 \, (2 \, c_{0,0} + 5 \, c_{0,1} - 6 \, c_{0,2}) , \, 6 \, (c_{0,1} + 5 \, c_{0,2}) , \, 6 \, c_{0,2}\}.
\end{aligned}
\end{equation}
Subsequent terms in the series can be iteratively found in the same way.
\end{exmp}

\subsubsection{\texorpdfstring{$\mathcal A$}{A}-Hypergeometric ideal}\label{sec:AHypergeometricIdeal}
Any Euler integral can be resolved into a set of linear PDEs with series solutions that are in one-to-one correspondence to a set of vectors, known as roots. Finding them is equivalent to solving the integral, and as previously advertised and presented in the following, the procedure to find them is not only algorithmic, but also better visualised in the appropriate mathematical language. 

A generalized Euler integral, $I_{\sigma} [\alpha, \, \beta; P](z)$, is an integral over $n$ integration variables $x \in \CC^n$ over a closed contour $\sigma$, with the integrand $\mathscr I(\alpha, \, \beta, \, P)$ defined by generic complex valued vectors $\alpha \in \CC^{m}$, $\beta \in \CC^{n}$ and $m$ polynomials $P$, interpreted as function of $N$ variables $z \in \CC \setminus \{0\}$.
\begin{equation}
I[\alpha, \, \beta; P](z)   = \int_{\sigma} \,  \mathscr I(\alpha, \, \beta; \, P), \quad \mathscr I(\alpha, \, \beta; \, P) = \prod_{i=1}^{n} \,\frac {\de x_i} {x_i}\, x_i^{\beta_i} \,\prod_{j=1}^{m} \, P_j(x ; \,A^{(j)} ; \,z^{(j)})^{-\alpha_j} \label{eq:genEulerIntegral}
\end{equation}
By definition, this induces the shift relations:
\begin{equation}\label{eq:Eulerintshiftrelation}
I[\alpha, \, \beta; \, x^{\omega} \, P](z)   = I[\alpha, \, \beta + \alpha \, \omega ;\, P](z), \quad  x^{\omega} \, P = \prod_{i=1}^{m} \, x^{\omega_i} \, P_i
\end{equation}
allowing infinitely many trivially equivalent representations of $I$, at least one of which has the polynomials in the form $P_i = 1 + P'_i(x,\,z)$. Equivalent shift relations given in \cref{eq:generalderivativeactiononeulerint,eq:thetaonI,eq:authentication1,eq:authentication2} are also induced by derivatives wrt $z$. The support of a polynomial $P(x\,; \,z)$ is $\{ x \in \CC^{n} \; | \; P(x\,;\,z) \neq 0 \}$. Extrapolating to the integrand, its support, $\Sigma(P)$ is
\begin{equation}
\Sigma(P)= (\CC^*)^{n} \; \setminus \; \underset{i \in [m]}{\bigcup} \{ P_i(x\,;\,z) \neq 0\}.
\end{equation}
The contour $\sigma$ is an $n$-cycle defined as a formal sum of maps from the standard $n$-simplex to $\Sigma(P)$:
\begin{equation}
\sigma^{(n)} = \sum \, \sigma^{(n)}_i : {\bf \Delta}^{n} \rightarrow \Sigma(P), \quad \de \sigma = 0, \quad {\bf \Delta}^{n} = \{v \, \Big| \sum_{i=0}^{n}  \,v_i = 1, \quad \, v_i \in \RR_+\}, \label{eq:generalcontour}
\end{equation}
schematically pictured in \cref{fig:simplextocontour} (when $q = n$). 
\begin{figure}[H]
\centering
\begin{tikzpicture}[scale=0.75]
\draw [lightgray, ->] (0,0) -- (2,0);
\draw [lightgray] (1,0.1) -- (1,-0.1);
\draw [lightgray, ->] (0,0) -- (0,2);
\draw [lightgray] (0.1,1) -- (-0.1,1);
\draw [lightgray, ->] (0,0) -- (-1.5,-1.5);
\draw [lightgray] (-0.75,-0.65) -- (-0.65,-0.75);
\draw [black] (1,0) -- (0,1) -- (-0.7,-0.7) -- cycle;
\node at (0,-2.5){\tiny $\RR_+^q$};
\draw [lightgray, xshift=6cm, <->] (-2,0) -- (2,0);
\draw [lightgray, xshift=6cm, <->] (0,-2) -- (0,2);
\draw [lightgray, xshift=6cm, <->] (-1.5,-1.5) -- (1.5,1.5);
\node at (6,-2.5){\tiny $\Sigma(P) \subset (\CC^*)^n$};
\fill[red, xshift=6cm] (-0.5,0.5) circle (1.5pt);
\fill[red, xshift=6cm] (0.5,-0.5) circle (1.5pt);
\fill[red, xshift=6cm] (0.3,0.7) circle (1.5pt);
\fill[red, xshift=6cm] (-0.3,-0.7) circle (1.5pt);
\draw [black, xshift = 6cm] plot [smooth cycle] coordinates {(1.5,1) (-1,1) (-1,-0.3) (-0.7,-1.5) (1,-0.5)};
\draw [black, dashed, xshift = 6cm] plot [smooth cycle] coordinates {(1,1.5) (-0.5,1) (-1.4,-0.8) (-0.3,-1) (1.3,-0.4)};
\draw [black, ->] plot [smooth, tension=1] coordinates{ (0.6,0.6) (3,1.5) (4.8,1)};
\draw [black, ->] plot [smooth, tension=1] coordinates{ (0.6,-0.4) (2,-1) (4.4,-0.8)};
\node at (3,1.5) [anchor=south] {$\sigma^{(q)}_1$};
\node at (2,-1) [anchor=north] {$\sigma^{(q)}_2$};
\draw [lightgray, xshift=12cm, <->] (-2,0) -- (2,0);
\draw [lightgray, xshift=12cm, <->] (0,-2) -- (0,2);
\node at (12,-2.5){\tiny $\CC^N$};
\draw [black, xshift=12cm ] plot [smooth, tension=1] coordinates{ (-1.5,0.5) (0.1,1.5) (1.5,0.5) (2.5,1)};
\draw [black, xshift=12cm, dashed ] plot [smooth, tension=1] coordinates{ (-1.5,1.5) (0.1,0.5) (1.5,1) (2.5,0.5)};
\draw [black, ->] plot [smooth, tension=1] coordinates{ (7.7,1) (9,1.2) (10.6,0.8)};
\node at (9,1.2) [anchor=south] {$P_1$};
\draw [black, ->] plot [smooth, tension=1] coordinates{ (7.6,-0.4) (9,0) (10.3,1.5)};
\node at (9,0) [anchor=north] {$P_2$};
\end{tikzpicture}
\caption{ $\mathcal L ({\bf \Delta}^{q} \times \CC^N)$: contour $\sigma^{(q)} = \sigma^{(q)}_1 + \sigma^{(q)}_2$ from simplex ${\bf \Delta}^{q}$ to $\Sigma(P)$ to $\CC^N$}\label{fig:simplextocontour}
\begin{flushleft}\singlespacing \vspace{-0.5\baselineskip} 
Red points in $ (\CC^*)^n$ are indicative of the zeroes of the polynomials $P$ and are not included in $\Sigma(P)$. $\sigma_1 \circ P_1, \, \sigma_2 \circ P_2$ are complex valued functions on the simplex ${\bf \Delta}^{q}$.
\end{flushleft}
\end{figure}
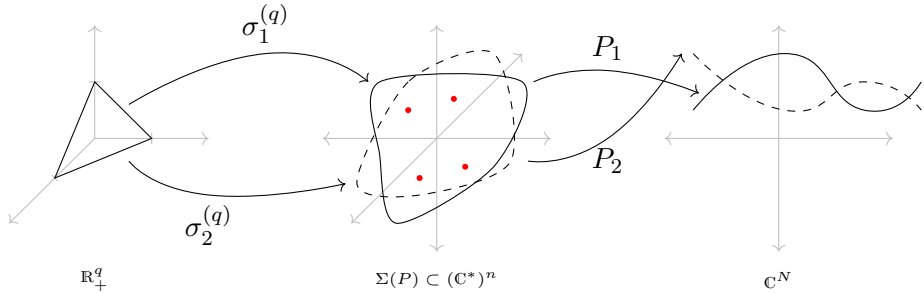\noindent
The space of integrals $I[\alpha, \, \beta;  \, P]$ is isomorphic to the solution space of the left ideal, $\mathcal I_{\mathcal A} \subseteq D_{N}$, where $ D_{N} = \CC [ z, \, \partial_z ]$ is the $N$-dimensional Weyl algebra. $\mathcal I_{\mathcal A}$, often dubbed the $\mA$ hypergeometric ideal, GKZ hypergeometric ideal, or GKZ differential equations, can be represented as a $D_N$-linear combination of the left ideals $H_{\mathcal A}$ and $J_{\mathcal A}$,
\begin{equation}
\mathcal I_{\mathcal A} = D_{N} \circ J_{\mathcal A} + D_{N} \circ H_{\mathcal A} = \langle \mathcal G_{\mathcal A} \rangle, \quad \mathcal I_{\mathcal A}  \circ I(z) = 0. \label{eq:IdealJ+H}
\end{equation}
There exists no non-trivial element in $D_N \setminus \, \mathcal I_{\mathcal A}$ which annihilates $I[\alpha, \, \beta ; \, P]$, making $\mathcal I_{\mathcal A}$ a maximal ideal in $D_N$. $J_{\mathcal A} \subset \CC\langle \partial \rangle$ is the toric ideal and $H_{\mathcal A} \subset \CC\langle \theta \rangle$ is torus fixed, respectively generated by
\begin{align}
J_{\mathcal A} & = \left\langle \partial^{u^+} - \partial^{u^-} \; \Big| \; \mathcal A \,u^+ = \mathcal A \, u^- , \; \{u^+, \, u^-\} \in \NN_0^{N} \right\rangle \label{eq:Jideal}
\\ H_{\mathcal A} & = 
\left\langle \mathcal A \, \theta + \gamma \right\rangle, \quad \gamma = \{ \alpha_1, \, \cdots \, \alpha_m, \, \beta_1, \cdots \, \beta_n \} 
\label{eq:Hideal}
\end{align}
where the vector $\gamma$ is non-resonant, and $\mathcal A$ is an $(n+m) \times N$ matrix formed of the support matrices $A$ of the polynomials $P$
\begin{equation}\label{eq:genAmatrixform}
  \mathcal A = \left( \begin{smallmatrix}
  \mathds{1}_1 & 0 & \cdots & 0\\
  0 & \mathds{1}_2 & \cdots & 0\\
  \cdots & \cdots & \cdots & \cdots\\
  0 & 0 & \cdots & \mathds{1}_m\\
  A^{(1)} & A^{(2)} & \cdots & A^{(m)}
  \end{smallmatrix} \right), \quad \mathds{1}_j \equiv \{ 1 , \,\cdots, \,1 \}.
\end{equation}
There are an infinitely many ways to represent the same Euler integral.
\begin{equation*}
\begin{aligned}
  I(z) & = \int \tfrac {\de x_1 \, \de x_2} {x_1 \, x_2}  \tfrac { x_1^{\beta_1} \, x_2^{\beta_2} } { (z_1 + z_2 x_1 + z_3 x_2^2)^{\alpha_1} (z_4  + z_5 x_2 + z_6 x_1^2)^{\alpha_2} }  = \int \tfrac {\de x_1 \, \de x_2} {x_1 \, x_2}  \tfrac { x_1^{\beta_1 + \alpha_1} \, x_2^{\beta_2} } { (z_1 x_1 + z_2 x_1^2 + z_3 x_1 \, x_2^2)^{\alpha_1} (z_4  + z_5 x_2 + z_6 x_1^2)^{\alpha_2} } 
\\ & = \int \tfrac {\de x_1 \, \de x_2} {x_1 \, x_2}  \tfrac { 2 \, x_1^{2 \, \beta_1} \, x_2^{\beta_2} } { (z_1 + z_2 x_1^{2} + z_3 x_2^2)^{\alpha_1} (z_4  + z_5 x_2 + z_6 x_1^4)^{\alpha_2} } = \cdots 
\end{aligned}
\end{equation*}
Following along with the above procedure, each of these technically equal representations produce different possible $\mathcal A$ matrices,
\begin{equation}\label{eq:einttoAmatexample}
  \mathcal A^{(1)} = \left( \begin{smallmatrix}
  1 & 1 & 1 & 0 & 0 & 0 \\ 
  0 & 0 & 0 & 1 & 1 & 1 \\ 
  0 & 1 & 0 & 0 & 2 & 0 \\
  0 & 0 & 2 & 0 & 0 & 1 
  \end{smallmatrix} \right),
  \quad 
  \mathcal A^{(2)} = \left( \begin{smallmatrix}
  1 & 1 & 1 & 0 & 0 & 0 \\ 
  0 & 0 & 0 & 1 & 1 & 1 \\ 
  1 & 2 & 1 & 0 & 2 & 0 \\
  0 & 0 & 2 & 0 & 0 & 1 
  \end{smallmatrix} \right),
  \quad
  \mathcal A^{(3)} = \left( \begin{smallmatrix}
  1 & 1 & 1 & 0 & 0 & 0 \\ 
  0 & 0 & 0 & 1 & 1 & 1 \\ 
  0 & 2 & 0 & 0 & 4 & 0 \\
  0 & 0 & 2 & 0 & 0 & 1 
  \end{smallmatrix} \right).
\end{equation}
\paragraph{Irreducibility conditions on \texorpdfstring{$\mA$}{A}}\label{sec:AIdealCons}
For any given Euler integral, there always exists a class of reduced form representations, such that its associated $\mathcal A$ matrix satisfies the following conditions.
\begin{enumerate}[left=0pt]
  \item The union of the columns of each $A_i$ generates $\mathbb Z^{n}$ and there is at least one column which is $0$ (i.e. there exists a non-zero constant term in the polynomial). These conditions together imply that union of the columns of $\mathcal A_i$ generate $\mathbb Z^{n+m}$.
  \begin{equation}
  \begin{aligned}
  & \exists \; c_i \in \mathbb Z \; | \; \xi = \sum  \, c_i \, A_i \; \forall \; \xi \in \mathbb Z^{n}, \quad \exists \; j \; | \; A_i{}_j{} = 0 
  \\ \implies & \exists \; c_i \in \mathbb Z \; | \; \xi = \sum  \, c_i \, \mathcal A_i \; \forall \; \xi \in \mathbb Z^{n+m}\label{eq:AmustgenerateZ}
  \end{aligned}
  \end{equation}

  \item There exists a group homomorphism $h: \mathbb Z^{n+m} \rightarrow \mathbb Z$ such that all $\mathcal A_i \mapsto 1$. One row of $\mathcal A$ can always be brought to a constant (which is by default chosen to be $1$ in \cref{eq:genAmatrixform}) under appropriate unimodular row transformations.
\end{enumerate}
The converse is also true, i.e. given a finite subset $\mathcal A$ of $\mathbb Z^n$ satisfying the above conditions, the solution space of the system of holonomic differential equations \cref{eq:IdealJ+H} is spanned by Euler integrals of the form \cref{eq:genEulerIntegral} defined over generically valued vectors $\alpha, \, \beta$. Further, the solution space associated with $\mathcal A$ is isomorphic to that of all $G  \, \mathcal A$, where $G$ is an invertible $n \times n$ matrix, with the parameter vector $\gamma$ similarly transforming to $G \, \gamma$. 

Thus, it is worth noting that even if $\mathcal A$ doesn't satisfy \cref{eq:AmustgenerateZ}, it is enough for it to be of rank $n+m$, i.e. full rank, as shown in \cite{schulze2012resonance}, thereby relaxing the constraints on $\mA$ matrices in given \cite{GELFAND1990255}. Returning to the example in \cref{eq:einttoAmatexample}, even though only $\mathcal A^{(1)} $ satisfies these conditions, the GKZ ideal, \cref{eq:IdealJ+H}, of all three  $\mathcal A^{(1, \, 2, \, 3)}$ is the same, generated by:
\begin{equation}
\begin{aligned}
  & \theta_1 + \theta_2 + \theta_3 + \alpha_1, && \theta_4 + \theta_5 + \theta_6 + \alpha_2, && \theta_2 + 2 \, \theta_5 + \beta_1, && 2  \,\theta_3 + \theta_6 + \beta_2
  \\ & \partial_1^2 \, \partial_5 - \partial_2^2\, \partial_4, && \partial_3 \,\partial_4^2 - \partial_1 \,\partial_6^2.
\end{aligned}
\end{equation}
If $\mathcal A$ doesn't satisfy the homogeneity condition, i.e. cannot be brought into a form with a row of $\mathds 1$, then the ideal $\mathcal I_{\mathcal A}$ is instead found to annihilate a confluent hypergeometric integral, which includes exponentials of polynomials:
\begin{equation}
I[\beta;\, P](z)  = \int_{\sigma}  \, \frac {\de x} {x} \, x^{\beta} \,\prod_{j=1}^{m} \,e^{P_{j}(x ; \,A^{(j)} ; \,z^{(j)}) }. \label{eq:genConfluentEulerIntegral}
\end{equation}
However, the confluent integrals $I[\beta;\, P]$ may not span the entire solution space of their associated GKZ system of equations.
\begin{exmp}
The confluent integral $\int_0^{\infty}  \, \tfrac {\de t} {t} \, t^{\Delta} \, e^{z_1 \,  t + z_2  \, t^2}$ is a solution of $\langle \left(\begin{smallmatrix} 1 & 2 \end{smallmatrix} \right) \left(\begin{smallmatrix} \theta_1 \\ \theta_2 \end{smallmatrix} \right) + \Delta, \; \partial_1^2 - \partial_2  \rangle$. This can be confirmed by eliminating $t$ (like in \cref{sec:example_hard}) from
\begin{equation}
\begin{aligned}
  & \partial_1 - t, \quad \partial_2 - t^2, \quad \partial_t - \frac {\Delta - 1} {t} - z_1 - 2 \, z_2 \, t \cong \partial_t  \, t - (\Delta + z_1  \, \partial_1 + 2 \, z_2 \, \partial_1^2)
  \\ & (\Delta + z_1  \, \partial_1 + 2 \, z_2 \, \partial_1^2) \int_0^{\infty}\tfrac {\de t} {t} \, t^{\Delta} \, e^{z_1  \, t + z_2  \, t^2} = \int_0^{\infty} \partial_t  \, t  \,  \tfrac {\de t} {t} \, t^{\Delta} \, e^{z_1  \, t + z_2  \, t^2} = 0, \quad \Real(z_2) < 0
  \\ & (\Delta + z_1  \, \partial_1 + 2 \, z_2 \, \partial_1^2) = (\Delta + z_1  \, \partial_1 + 2 \, z_2 \, \partial_2) = (\Delta + \theta_1 + 2  \, \theta_2).
\end{aligned}
\end{equation}
The solution space expectedly consists of ${}_1F_1$ functions
\begin{equation}
\begin{aligned}
  z_2^{- \frac {\Delta} {2}} \sum_n  \frac {\Gamma(n + \frac {\Delta} {2} + 1)} {\Gamma(n + 1)  \, \Gamma(n + \frac{1}{2}) } \left(-\tfrac {z_1^{2}} {4 \,  z_2}\right)^n \; \oplus \; \tfrac {1} {4}  \,  z_1  \, z_2^{- \frac {1 + \Delta} {2}} \sum_n \frac {\Gamma(n + \frac {1 + \Delta} {2} + 1)} {\Gamma(n + 1)  \, \Gamma(n + \frac {3} {2} ) } \left(-\frac {z_1^{2}} {4 z_2}\right)^n.
\end{aligned}
\end{equation}
\end{exmp}

\paragraph{Non-resonance condition}\label{sec:nonresonancecondition}
The vector $\gamma$ is non-resonant when $\gamma \notin \mathbb Z^{n+m} + \CC$-linear span of the codimension-$1$ faces (i.e. $n+m-1$ dimensional faces/facets) of the convex cone of $\mathcal A$. An example representation of the projection of resonant values of $\gamma$ onto the real plane is shown by the dashed arrows in \cref{fig:NonResonanceCon}. 
\begin{figure}[htpb]
\centering
  \begin{tikzpicture}[scale=0.75]
    \draw[thin,lightgray,->] (0,0) -- (5.5,0);
    \draw[thin,lightgray,->] (0,0) -- (0,5.5);
    \foreach \x in {1,2,3,4,5} \draw[thin, lightgray] (\x,1pt) -- (\x,-1pt);
    \foreach \y in {1,2,3,4,5} \draw[thin, lightgray] (1pt,\y) -- (-1pt,\y);
    \filldraw[fill=gray!20!white, draw=black,thick] (0,0) -- (2,1) -- (1,2) -- cycle;
    \draw[thin,black,->] (0,0) -- (4,2);
    \draw[thin,black,->] (0,0) -- (2,4);
    \fill[black] (0,0) circle (1.5pt) node[anchor=north]{$\mA_1$};
    \fill[black] (2,1) circle (1.5pt) node[anchor=north]{$\mA_2$};
    \fill[black] (1,2) circle (1.5pt) node[anchor=east]{$\mA_3$};
    \draw[thin,black,dashed,->] (1,0) -- (5,2);
    \draw[thin,black,dashed,->] (2,0) -- (6,2);
    \draw[thin,black,dashed,<->] (0,1) -- (2,5);
    \draw[thin,black,dashed,<->] (0,2) -- (2,6);
    \draw[thin,black,dashed,<->] (3,0) -- (0,3);
    \draw[thin,black,dashed,<->] (4,0) -- (0,4);
    \end{tikzpicture}
    \caption{Non-resonance Condition}\label{fig:NonResonanceCon}
    \begin{flushleft}
    \singlespacing \vspace{-0.5\baselineskip} 
    The shaded region is the convex hull of $\mathcal A$ with vertices $\{\mA_1, \mA_2, \cdots \mA_N\}$. The interior of the arrows originating from $\mA_1$ is the $\RR$ linear span thereof, i.e. the convex cone of $\mA$, with the arrows also forming the codimension-$1$ faces (facets) of the convex cone. Dashed arrows are representative of the projection onto the real plane of the facets shifted by all $\mathbb Z^{n+m}$. 
  \end{flushleft}
\end{figure}
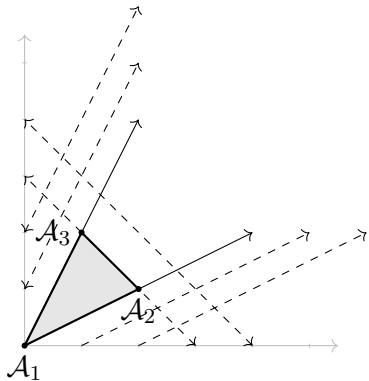\noindent

\subsubsection{Equivalence of Euler integrals and solution space of the GKZ ideal}
It can be verified explicitly that all the elements of the left ideal $\mathcal I_{\mathcal A}$ annihilate the integral $I(z)$. The action of a general element in $\CC\langle \partial_z \rangle$, $\partial^{u_+}$, on the integral $I(z)$ is
\begin{equation}\label{eq:generalderivativeactiononeulerint}
\partial^{u_+} \circ I (z) = \int_{\sigma}\tfrac {\de x} {x}  \, x^{\beta} \, \prod_{j=1}^{m} (-\alpha_j)_{(\sum u_j{}_+{})} \, x^{A^{(j)} u_{j_+}}  \, P_{j}(x ;  \, A^{(j)} ;  \, z^{(j)})^{-\alpha-\sum u_j{}_+{}}.
\end{equation}
This implies $\partial^{u_+} \circ I (z) = \partial^{u_-} \circ I (z)$ iff
\begin{equation}
\sum u_j{}_+{}  = \sum u_j{}_-{} = \text{constant}, \quad A^{(j)}  \, u_{j_+} = A^{(j)}  \, u_{j_-} \quad \forall \; j \in [m],
\end{equation}
a set of conditions which are better represented as $J_{\mathcal A}$ as given in \cref{eq:Jideal}, with the option of replacing any $\mathds 1$ with any positive integer multiple thereof in \cref{eq:genAmatrixform} to produce equivalent representations of $\mathcal A$. The action of the generators of $\CC\left\langle \theta \right\rangle$, $\theta_{z^{(j)}_i{}}$, on $I(z)$ is
\begin{align}
\theta_{z^{(j)}_i{}} \circ I (z) & = \int_{\sigma} \tfrac {\de x} {x} \, x^{\beta} \, \tfrac {- \alpha_{j} \, z^{(j)}_i{} \, x^{A^{(j)}_i}} {P_j(x ; \,A^{(j)} ;\, z^{(j)})} \, \prod_{j=1}^{m} \, P_{j}(x ;\, A^{(j)} ; \,z^{(j)})^{-\alpha_j} \label{eq:thetaonI}
\\ & = \int_{\sigma} \prod_{j_1 \neq k} \tfrac {\de x_{j_1}} {x_{j_1}}\, x_{j_1}^{\beta_{j_1}} \, \prod_{j_2 \neq j}^{m}  P_{j_2}(x ; \,A^{(j_2)} ; \,z^{(j_2)})^{-\alpha_{j_2}} \nonumber
\\ &  \times \de x_{k} \, x_k^{\beta_k} \, \tfrac {(- \alpha_{j}) } {P_{j}(x ; \,A^{(j)} ; \,z^{(j)})^{\alpha_j}} \,\tfrac {z^{(j)}_i{} \, \partial_{x_k} \, x^{A^{(j)}_i}} {P_j(x ; \,A^{(j)} ; \,z^{(j)})} \, \tfrac {1} {A^{(j)}{}_i{}_k{}} \quad \forall \; k \in [n]. \label{eq:thetaonIpartialx}
\end{align}
One straightforward way of eliminating all $x$ from these $N$ PDEs when presented in the form of \cref{eq:thetaonI} is to simply sum them up per polynomial,
\begin{equation}
\sum_{i} \theta_{z^{(j)}_i{}} \circ I (z)  = - \alpha_{j} \, I (z)  \quad \forall \; j \in [m]\label{eq:authentication1}
\end{equation}
producing $m$ homogeneity relations. Summing rescaled versions of all $N$ PDEs presented in the form of \cref{eq:thetaonIpartialx} per each variable $x$ gives
\begin{equation}
\sum_{j} \sum_{i} A^{(j)}{}_{i}{}_{k}{} \, \theta_{z^{(j)}_i{}} \circ I (z)  = \int_{\sigma} \prod_{j_1 \neq k} \tfrac {\de x_{j_1}} {x_{j_1}} \, x_{j_1}^{\beta_{j_1}} \, \de x_{k} \, x_k^{\beta_k} \, \partial_{x_k} \,\prod_{j_2 \neq j}^{m}  P_{j_2}(x ; \,A^{(j_2)} ; \,z^{(j_2)})^{-\alpha_{j_2}}. 
\end{equation}
Since $\sigma$ is a $n$-cycle within the support $\Sigma$, $\mathscr I$ is smooth, which together imply
\begin{equation}
\begin{aligned}
  \sum_{j} \sum_{i} A^{(j)}{}_{i}{}_{k}{} \, \theta_{z^{(j)}_i{}} \circ I (z) &= - \int_{\sigma} \prod_{j_1 \neq k} \tfrac {\de x_{j_1}} {x_{j_1}}  \, x_{j_1}^{\beta_{j_1}} \, \de x_{k} \, (\partial_{x_k} x_k^{\beta_k}) \, \prod_{j_2 \neq j}^{m}  P_{j_2}(x ; A^{(j_2)} ; z^{(j_2)})^{-\alpha_{j_2}}
  \\ & = - \beta_k \, I(z) \quad \forall \; k \in [n]\label{eq:authentication2}
\end{aligned}
\end{equation}
thus producing $n$ more homogeneity relations. Altogether these relations are concisely represented by $H_{\mathcal A}$ as given in \cref{eq:Hideal}. This verifies that \cref{eq:IdealJ+H} does annihilate \cref{eq:genEulerIntegral}, or equivalently, proves that the solution space of $\mathcal I_{\mathcal A}$ contains $I[\alpha, \, \beta ; \, P]$. 

The solution space of the ideal, $\mathcal I_{\mathcal A}$, 
is spanned by Euler integrals, $ I[\alpha, \, \beta ; \, P]$, \cref{eq:genEulerIntegral}, whilst assuming the vector $\gamma \equiv \{\alpha, \, \beta\}$ satisfies the \nameref{sec:nonresonancecondition}. This statement and a stronger generalisation thereof are the core result of \cite{GELFAND1990255}. 

\paragraph{Monodromy exponents:} $\mathcal D_{V}$ is the sheaf of rings of differential operators with holomorphic 
coefficients on $V \equiv \CC^{N} \ni z$. The geometric Fourier transform, $\mathscr F$, of $V$, $V^*$, induces the coordinate transformations:
\begin{equation}\label{eq:geometricfouriertransform}
z \mapsto - \partial^*, \quad \partial \mapsto z^*, \quad \theta \mapsto - (\theta^* + 1).
\end{equation}
Thus, $\mI_\mA$ defined on $V$ (see \cref{eq:IdealJ+H} for notation) transforms to $\mI^*_\mA$ defined on $V^*$,
\begin{equation}
\mI_\mA = \langle \partial^{u+} - \partial^{u-} \rangle + \langle \mathcal A \theta + \gamma \rangle \mapsto \mI^*_\mA = \langle z^*{}^{u+} - z^*{}^{u-} \rangle + \langle  - \mathcal A  \, (\theta^* + 1) + \gamma \rangle.
\end{equation}
The quotient module $\mD_\mA = \mD_V / \mD_V \circ \mI_\mA$ similarly transforms to $\mD^*_\mA = \mD_{V^*} / \mD_{V^*} \circ \mI^*_\mA$.

The first set of generators of $\mI^*_\mA$ are algebraic, to wit: $J^* = \langle z^*{}^{u+} - z^*{}^{u-} \rangle$ are homogeneous binomials by design, and so $\mI^*_\mA \circ f(z^*) = 0$ cannot have any analytic functions as solutions. This means that $J^*_\mA$ must be identically $0$, which is the case on the loci of $z^*$ defined by the zero set of the aforementioned binomials, and so $\mD^*_\mA$ is not supported at these values. This subspace of $V^*$ is the characteristic variety, ch$(V^*) \equiv \bar V^*$. Its dual is $\bar V$ and the complement of $\bar V$ in $V$, $V_0 = V \setminus \bar V$, is the proper domain of definition of $z$ with non-singular solutions.

The orbit of the point $(\mathds 1_{n+m})$ by the torus $(x_1, \cdots x_{n+m}) \in (\CC^*)^{n+m}$ is represented in $V^*$ as $x^{\mathcal A} \equiv (x^{\mathcal A_1}, \cdots x^{\mathcal A_{N}}) \in \bar V_0^* \subset \bar V^*$, with its embedding in $V^*$ being $\hat J : \bar V_0^* \rightarrow V^*$.  The torus action of $x \in (\CC^*)^{n+m}$ on $V^* \ni z^*$ is $ ( x^{\mathcal A_1} \, z^*_1, \cdots x^{\mathcal A_{N}} \, z^*_N ) $, and it obviously leaves ch$(V^*)$ unchanged. Note that the orbit $\bar V_0^*$ is open. It doesn't include $z^*{}^{u+} = z^*{}^{u-} = 0$ including but not limited to the trivially singular $z^* = 0$.
\begin{exmp}\label{sec:exampledemystify}
For the sake of some demystification, \cref{sec:example_doable} is recast in this language (see \cref{examplequick} for preliminary setup). $z \in (\CC)^5 = V$, $z^* \in V^*$, and $\mD_V$, $\mD_{V^*}$ consist of all possible differential operators that are polynomial in $\partial_z, \, \partial_{z^*}$ and rational in $z, \, z^*$ respectively. The characteristic variety is the zero set of $J^*$, $\bar V^* = \{z^*_1  \, z^*_5 = z^*_2{}^2 = z^*_3  \, z^*_4 \}$. It can be explicitly checked that $\bar V^*$ remains invariant under the torus action $T$ parameterised by $\lambda \in (\CC^*)^3$
\begin{equation}
z \rightarrow T z, \quad z^{*} \rightarrow T^{-1} z^*, \quad T =  (\lambda_1, \, \lambda_1  \, \lambda_2  \, \lambda_3, \, \lambda_1  \, \lambda_2^2, \, \lambda_1 \,  \lambda_3^2, \, \lambda_1  \, \lambda_2^2 \,  \lambda_3^2) \label{eq:exampleofopenorbitparamaterisation}
\end{equation}
and the integral itself exhibits the expected homogeneity relation $I(Tz) = \lambda_1^{-\alpha}  \, \lambda_2^{-\beta_1}  \, \lambda_3^{-\beta_3} I(z)$. Thus, the orbit of $\mathds 1_{3}$ in $V^*$, $\bar V_0^*$, as parameterised by $\lambda \in (\CC^*)^3$, doesn't include the hyperplane $z_1^*  \, z_5^*  = z_2^* = z_3^*  \, z_4^* = 0$, and its Zariski closure is indeed $\bar V^*$. 
\end{exmp}
Another quotient module is defined on $\bar V_0^*$, $\mD^*_{H^*}{} = \mD_{\bar V_0^*} / \mD_{\bar V_0^*} \circ H^*$, $H^* = \langle  - \mathcal A  \, (\theta^* + 1) + \gamma \rangle$, which is locally isomorphic to $\mD^*_\mA$ by definition. Its extension to $V^*$ by $\hat J$, denoted $ \hat J \mD^*_{H^*}{} $, is irreducible and isomorphic to $\mD^*_\mA$. Since $\mD^*_\mA$ is irreducible, so is its Fourier transform $\mD_\mA$, the original $D$-module of interest. The solution space of $\mD^*_{H^*}{}$ is locally isomorphic to a system represented in $\bar V_0^*$ by branches of functions of the form $x^\gamma$, i.e. with monodromy exponents $\gamma$ around $x = 0$. 
\begin{continueexample}{sec:exampledemystify}
Requiring $H^* \circ \bar I(\bar z^*)= 0$ on the open orbit $\bar V_0^* \ni \bar z^*$ as defined in \cref{eq:exampleofopenorbitparamaterisation} implies that $\bar I(\bar z^*) = \tfrac  { \lambda_1^{\alpha} \,  \lambda_2^{\beta_1}  \, \lambda_3^{\beta_2}} {(\lambda_1 \,  \lambda_2  \, \lambda_3)^5} $, i.e. with the monodromy exponents $\gamma = \{\alpha,\, \beta_1, \, \beta_2\}$. 
\end{continueexample}

\subsubsection{Summary of generic solution algorithm}\label{sec:quickstartguide}
The following algorithm finds the $\mathcal A$-hypergeometric series describing the solution space of any GKZ ideal granting complete genericity of the parameters. If followed blindly, it may not yield the most efficient or elucidating results, but yield it will.
\begin{exmp}\label{examplequick}
The setup of \cref{sec:example_doable} is reused to illustrate.
\begin{equation*}\tag{\ref{eq:example_doable}}
I(z) = \int_{\sigma} \frac {\de x_1\, \de x_2} {x_1\, x_2}  \, \frac {x_1^{\beta_1} \, x_2^{\beta_2} } {(z_1 + z_2 \, x_1 \, x_2 + z_3 \, x_1^2 + z_4 \, x_2^2 + z_5\,  x_1^2 \, x_2^2)^{\alpha} }
\end{equation*}
\end{exmp}
\begin{enumerate}[left=0pt]
\item Given a generalized Euler integral, that may be converted into a form satisfying the \nameref{sec:AIdealCons}, 
\begin{equation}
  I(z) = \int_{\sigma} \frac {\de x_1\, \de x_2} {x_1\, x_2}  \, \frac {x_1^{\frac{\beta_1 + \beta_2}{2}} \, x_2^{\beta_2} } {(z_1 + z_2 x_1 \,x_2 + z_3 \, x_1 + z_4 x_1 \, x_2^2 + z_5\, x_1^2 \,  x_2^2)^{\alpha} }
\end{equation}
its $\mathcal A$ matrix and $\gamma$ vector are found to be
\begin{equation}
\mA = \left(\begin{smallmatrix}
1 & 1 & 1 & 1 & 1\\ 0 & 1 & 1 & 1 & 2 \\ 0 & 1 & 0 & 2 & 2
\end{smallmatrix} \right), \quad \gamma = \{\alpha,  \, \tfrac{\beta_1 + \beta_2}{2} , \,  \beta_2\}.
\end{equation}
If confirmed to be convergent, an Euler integral over an open contour or a Feynman integral (usually with arbitrary fixed values of $z$ and $\gamma$, and a contour over $\RR / \RR_+$) can be treated in the same manner. They are first assumed to be generalized, and then their appropriate limiting behaviours are considered.

\item $\mathcal K$ is a reduced basis representation in $\mathbb Z^{N}$ of the kernel of $\mathcal A$. The generators of the toric ideal $J$ are $\partial^{|\mathcal K_+|} - \partial^{|\mathcal K_-|} $, where $\mathcal K_{\pm}$ refer to the sets of positive/negative integer valued elements in $\mathcal K$. 
\begin{equation}
\mathcal K = \left( \begin{smallmatrix}
-1 & 0 & 1 & 1 & -1 \\ 1 & - 2 & 0 & 0 & 1
\end{smallmatrix} \right), \quad J = \langle \partial_{3} \, \partial_{4} - \partial_{1} \, \partial_{5},  \; \partial_{1} \, \partial_{5} - \partial_2^{2} \rangle
\end{equation}
Putting it together with \cref{eq:Hideal} gives the $\mA$ hypergeometric ideal.
\begin{equation}
\mI  = J + H, \quad H = \langle \sum_{i=1}^{5} \theta_i + \alpha, \; \theta_2 + \theta_3 + \theta_4 + 2 \, \theta_5 + \tfrac{\beta_1 + \beta_2}{2}, \; \theta_2 + 2 \, \theta_4 + 2 \, \theta_5 + \beta_2 \rangle
\end{equation}

\item A Gröbner basis of $J$ can be found by using Macaulay 2 \cite{M2} (with many useful packages and their usage described in \cite{MacManual}) or any other relevant math software. $M$ is a monomial ideal of $J$, formed by only retaining a single monomial from all the binomials that generate $J$. The choice of monomial to be retained can be made by either picking the term with the higher weight wrt {\it any} weight vector $w \in \RR^N$ or greater precedence wrt any order $\prec$. For example, the monomial ideal wrt $w = \{0, \, 1, \, 1, \, 0, \, 0\}$ is
\begin{equation}\label{eq:Jidealquick}
J\Big|_{(w)} = \langle \partial_{3} \, \partial_{4}\Big|_{(1)} - \partial_{1} \, \partial_{5}\Big|_{(0)},  \; \partial_{1} \, \partial_{5}\Big|_{(0)} - \partial_2^{2}\Big|_{(2)} \rangle \implies M = \langle \partial_{3} \, \partial_{4}, \; \partial_2^2 \rangle.
\end{equation}

\item Given a monomial ideal $M = \langle \partial^{a} \rangle$, its distraction is $\tilde M = \langle \theta_{(a)} \rangle$ (see \cref{eq:fallingthetanotation} for notation). 
\begin{equation}
  \tilde M  = \langle \theta_{3} \, \theta_{4}, \; \theta_2 \,  (\theta_2 - 1) \rangle 
\end{equation}
The roots, $\{s\}$, are the simultaneous solution to $\tilde M + H$.
\begin{equation}
\begin{aligned}
  s & = \{\tfrac {\beta_2} {2} - \alpha  , \, 0 , \,  0, \, \tfrac {\beta_1 - \beta_2} {2} , \, - \tfrac {\beta_1} {2}\}, 
 && \{ \tfrac {\beta_1} {2} - \alpha , \, 0 ,  \, \tfrac {\beta_2 - \beta_1} {2} ,  \, 0,  \, - \tfrac {\beta_2} {2}\}, 
\\ & 
\quad \{\tfrac {\beta_2 - 1} {2} - \alpha  , \, 1 ,  \, 0, \, \tfrac {\beta_1 - \beta_2} {2} , \, - \tfrac {\beta_1 + 1} {2}\},
&& \{ \tfrac {\beta_1 - 1} {2} - \alpha , \, 1 ,  \, \tfrac {\beta_2 - \beta_1} {2} ,  \, 0,  \, - \tfrac {\beta_2 + 1} {2}\} 
\end{aligned}
\end{equation}
These roots are not the same as the ones presented in \cref{sec:example_doable}, which correspond to a different monomial ideal $\langle \partial_{3} \, \partial_{4}, \;  \partial_{1} \, \partial_{5} \rangle$. Another possible monomial ideal is $\langle \partial_{1} \, \partial_{5}, \; \partial_2^2 \rangle$ with another set of associated roots. Though the choice of monomial ideal and hence set of roots is non-unique, the solution spaces spanned by each set are identical.

\item The integral $I$ is spanned by a linear combination of the series:
\begin{equation}
I = \sum_{s} \; N_s \; z^s \; \sum_{t \in \mathbb Z}  \, \frac {\Gamma(s + 1)} {\Gamma(s + t_1 \, \mathcal K_1 + t_2 \, \mathcal K_2 + 1)} \, z^{t_1 \, \mathcal K_1 + t_2 \, \mathcal K_2}, \quad N_s \in \CC.
\end{equation}
Considering limits of the original integral with $z \rightarrow 0$ matching the zeroes in the roots can be used to find the normalisation constants, $N_s$. 

\end{enumerate}

\subsection{Ideals to solutions}\label{sec:IdealstoSolutions}
Given the isomorphism between Euler integrals and $\mathcal A$-hypergeometric systems' solution space, the next step is to find the latter and to strip away the magic from \cref{sec:quickstartguide}.

An $N$-dimensional Weyl algebra,  $D_N$, is a free associative non-commutative algebra defined over a field of characteristic zero, here $\CC$, $D_N = \CC \langle x_1, \, x_2, \, \cdots x_N, \, \partial_1, \, \partial_2, \, \cdots \partial_N \rangle $ modulo the commutation relations $[x_{i}, \, x_j] = [\partial_{i}, \, \partial_j] = 0$ and $[\partial_{i}, \, x_j] = \delta_i{}_j{}$. The ambiguity of multiple equivalent representations of the same elements in $D_N$ is fixed by demanding that every element be normal ordered by commuting all $\partial$ to the left, 
\begin{equation}
d = \sum_{(\alpha, \,\beta) \in E} \,  c_{\alpha, \,\beta} \, x^\alpha \, \partial^\beta, \quad c_{\alpha, \,\beta} \neq 0 \; \forall \; (\alpha, \,\beta) \in E, \quad d \in D_N \label{eq:general_weylalgebraelement}
\end{equation}
making $D_N$ the space of differential operators $\CC[\partial_x]$ with coefficients in the ring of polynomials $\CC[x] \, : \CC^N \rightarrow \CC$ with a vector space isomorphism $\Psi : \CC[x, \,\partial_x] \rightarrow \CC[x, \,\xi]$, where $\CC[x, \,\xi]$ is a $2N$ dimensional commutative ring. Products and Poisson brackets on $D_N$ are pushed forward by $\Psi$ to 
\begin{equation}
\Psi(f (x,\,\xi) )\, \Psi(g(x,\,\xi)) = \sum_{k\in \NN_0^{n}} \tfrac 1 {k_1! \cdots k_n!} \,\Psi( \tfrac {  \partial^k f} { \partial \xi^k } \,\tfrac {  \partial^k g} { \partial x^k } ), \quad [f, \,g]_{\text{PB}} = \sum_{i=1}^{n} \; \tfrac {\partial f} {\partial \xi_i} \,\tfrac {\partial g} {\partial x_i} - \tfrac {\partial g} {\partial \xi_i} \,\tfrac {\partial f} {\partial x_i}. \label{eq:weyl_poisson_def}
\end{equation}
$D_N$ is a subset of the ring of differential operators with coefficients in rational functions, $\mathcal D_N$, by definition. $\CC[x]$, $\CC[\partial]$, $\CC[\theta]$ are commutative sub-rings in $D_N$, the relations between which can be concretised by using identities like
\begin{equation}\label{eq:restrictiontoCtheta}
\begin{aligned}
  &x^{\alpha} \,\partial^{\alpha} = \theta_{(\alpha)},  && \partial^{\alpha} \,x^{\alpha} = \theta^{(\alpha)}, && \theta_i \vcentcolon= x_i \,\partial_i
  \\ & f(\theta) \, x^{\alpha} = x^{\alpha} \,f(\theta + \alpha), && x^{\alpha} \,f(\theta) \, \partial^{\beta} \rightarrow \theta^{(\alpha)}  \, f(\theta - b)  \, \theta_{(\beta)}, && f \in \CC[\theta].
\end{aligned}
\end{equation}
The Mellin and inverse Mellin transforms are defined as 
\begin{equation}\label{eq:mellintransformdef}
\begin{aligned}
  \mathcal {M} \circ f\,  (s)& = \int_0^{\infty}  \, \frac {\de x} {x} \, x^{s} \, f(x), && \mathcal {M}^{-1} \circ f\, (x) = \circint_{s} \, x^{-s} \, f(s), && \circint_{s} \equiv \int_{\delta - i \infty}^{\delta + i \infty}  \, \frac {\de s} {2 \pi \im} ,
\end{aligned}
\end{equation}
where $\delta \in \RR$ is such that the integral is absolutely convergent with $f(s)$ being analytic over the line integral and uniformly tending to $0$ at the end points. The Mellin transform induces an isomorphism from $\CC[x, \, \theta]$ to the ring $\langle s, \, \delta_s \rangle$, where $\delta_s$ is the difference operator satisfying the commutation relation $\delta_s \, s = (s + 1) \, \delta_s$, via the transformations $s + 1 \mapsto - \theta, \; \delta_s \mapsto x$, i.e. matching the geometric Fourier transform defined in \cref{eq:geometricfouriertransform}. As is to be expected, Euler integrals and GKZ systems can also be represented in the form of Mellin-Barnes integrals \cite{nilsson2013mellin}.

\subsubsection{Varieties of ideals}
A left ideal or $D$-ideal, $\mathcal I$, is a subset of the Weyl algebra, $D_{N}$, satisfying the properties: 
\begin{equation}
(1) \quad 0 \in \mI, \quad (2) \quad f_1, \, f_2 \in \mathcal I \Rightarrow f_1 + f_2 \in \mathcal I, \quad (3) \quad f \in I, \; d \in D_N \Rightarrow d \circ f \in \mathcal I.
\end{equation}
A right ideal, $\mathcal I_{\rm R}$, is equivalently defined with the third condition changed to $f \circ d \in \mathcal I_{\rm R}$. The $\mathcal A$-hypergeometric system, $\mathcal I_{\mathcal A}$, forms a left ideal in $D_N$ and its solution space corresponds to the $D$-module $\mathcal D_N / \mathcal I_{\mathcal A}$. 

A {variety}, $V$, defined by a set of polynomials, $\{f(x)\}$, is their zero locus/solution space in affine $\CC^{N}$. This implies intersections and unions of varieties (which can be defined by the union and products of the sets of defining polynomials respectively) are also affine varieties. This can be extended to include rational functions $g_i = \tfrac 1 {f_i}$ by introducing dummy variables $y_i$, replacing the rational functions with $g'_i = y_i \,f_i - 1$ and subsequently eliminating $y_i$ \cite{Cox:AlgebraicGeometry}. The ideal of a variety, $\mI(V)$, is the set of functions which have the variety as the solution space, i.e.
\begin{equation}
\mI(V) = \{f \; | \; f(v) = 0 \; \forall \; v \in V\}.
\end{equation}
The {Zariski closure} of some subset $U \subseteq \CC^{N}$ is the smallest variety containing $U$, i.e. if there exists a variety containing $U$, it contains the Zariski closure. The variety of an ideal, $V(\mI)$, is the Zariski closure of the zero locus of the polynomials generating the ideal. The {radical} of an ideal, $\sqrt {\mI}$, is a superset of $\mI$ that contains of all elements such that
\begin{equation}
f^{n} \in \mI, \; n \in \NN  \implies f \in \sqrt {\mI}.
\end{equation}
The radical, $\sqrt {\mI}$, equals $\mI ( V (\mI) )$ when the field is algebraically closed. A trivial example: the variety of the ideal $\mI_0 = \langle x^2 \rangle \in \CC[x]$ is the origin $V(\mI_0) =  \{0\}$. However, the ideal of $\{0\}$ is $\mI(V(\mI_0)) = \langle x \rangle$, which isn't equal to the original ideal but includes it and is its radical $\sqrt{\mI_0}$.

An ideal is {prime} if for every $(f_1 \, f_2) \in \mI$, either $f_1 \in \mI$ or $f_2 \in \mI$, but not both.
An ideal $\mI$ is {primary} if there exists an $m\in \NN$ such that $g^{m} \in Q$ for all $f \notin Q$, $f  \, g \in Q$. Any proper ideal has an irreducible primary ideal decomposition, i.e. it can be represented as an intersection of finitely many primary ideals $Q_i$ such that
\begin{equation}
\mI  = \bigcap_{i} Q_i, \quad Q_k \not\subseteq Q_{i} \bigcap Q_{j}, \quad \sqrt{Q_i} \neq \sqrt{Q_j}, \quad  i \neq j \neq k. \label{eq:primary_ideal_decomposition}
\end{equation}
The set of $\sqrt{Q}$ are the {associated prime} ideals. $V(\mI)$ is the union of the irreducible pieces $V(\sqrt{Q})$, and $V(\mI (V)) = V$. An {elimination} ideal, $\mI_{x'}$, is defined as $\mI \bigcap \CC[x']$, i.e. it consists of all elements of $\mI$ from which the variables $x \setminus x'$ are eliminated. 
The integrand of an Euler integral belongs to the Weyl algebra $\CC\langle x, \, \partial_x, \, z, \, \partial_z \rangle$. Given $\mI$ that annihilates the integrand, the ideal annihilating the Euler integral is the elimination ideal $\mI_{z, \, \partial_z}$ (see \cref{sec:example_hard}).

\par An ideal in a commutative ring is freely generated by a finite basis $\mI = \langle f_1, \, f_2, \, \cdots f_\omega \rangle$, $\omega < \infty$. $D$-ideals in $D_N$ have finite bases too, as can be intuited given the isomorphism $\Psi$. The associated graded ring of $D_N$, gr${}_{(u, \, v)} (D_N)$, wrt the {weight vector} $(u, \, v) \in \RR^{2N}$, $u_i + v_i \ge 0$ is generated by
\begin{equation}
\{x\} \bigcup_{} \, \{ \partial_{i} \, | \, u_i + v_i > 0\} \, \bigcup_{} \, \{ \xi_{i} \, | \, u_i + v_i = 0\}
\end{equation}
with the {initial form} of a general element $d$, being its restriction to gr${}_{(u, \, v)} (D_N)$,
\begin{equation}
\text{in}_{(u, \, v)} (d) = \sum_{u  \, \alpha + v  \, \beta = \max E_{(u, \, v)}}  \, c_{\alpha, \, \beta} \; x^{\alpha} \prod_{u_i + v_i = 0}  \, \partial_i^{\beta_i}  \, \prod_{u_i + v_i > 0}  \, \xi_i^{\beta_i} \in \text{gr}_{(u, \, v)} (D)
\end{equation}
and the {initial ideal} of a left ideal $\mI$ forming a left ideal in gr${}_{(u, \, v)} (D_N)$
\begin{equation}
\text{in}_{(u, \, v)} (\mathcal I) = \{ \text{in}_{(u, \, v)} (f) \; | \; \forall \; f \in \mI \}.
\end{equation}
Weight vectors that produce the same initial ideal are equivalent. An {ordering prescription}, $\prec$, attaches a symbolic weight to the generators of $D_N$, e.g. $\cdots x_2 \prec x_1 \prec \cdots \partial_2 \prec \partial_1$ (lexicographic), $\partial \prec x$ (reverse lexicographic), etc. $\prec$ is called a term order when it satisfies the property:
\begin{equation}
1  \prec x^\alpha  \, \partial^\beta \prec x^{\alpha + s}  \, \partial^{\beta + t} , \quad \alpha + \beta \ge 1,  \quad \forall \; (s, \, t) \in \NN^{2 n}, \quad 1 \prec \log x.
\end{equation}
An induced order, $\prec_{(w)}$, uses $\prec$ as a tie breaker between terms of equal weight, for example the graded lexicographic order $\prec_{(\mathds 1)}$ induces the order $x_2 \prec x_1 \prec x_1  \, x_2^2 \prec x_1^2  \, x_2$ in $D_2$, where the last 2 terms have the same total degree so the degree of the lexicographically `highest' ordered variable, $x_1$, is used for tie-breaking. An {initial monomial} of a general element $d$, in${}_{\prec}(d)$, according to the order $\prec$ is the ``largest'' term in $d$ such that
\begin{equation}
\text{in}_{\prec} (d)  = c_{\bar\alpha,  \, \bar\beta} \, x^{\bar\alpha} \, \partial^{\bar \beta},  \quad c_{\bar\alpha,  \, \bar\beta} \, x^{\bar\alpha} \, \partial^{\bar \beta} \succeq c_{\alpha,  \, \beta} \, x^{\alpha} \, \partial^{\beta} \; \forall \; \{\alpha,  \, \beta\} \in E.
\end{equation}
When the coefficient of in${}_{\prec}(d)$ is $1$, $d$ is said to be in {monic} form. The {monomial ideal} of $\mI$ wrt $\prec$, in${}_{\prec}(\mI)$, consists of the initial monomials of all the elements in $\mI$. Ordering prescriptions allow the equivalent of subtraction (S-pair) and division (Normal form) operations to be defined in $D_N$. The S-pair of $f, \, g$ wrt $\prec$, with in${}_{\prec}(f) = f_{\alpha, \, \beta} \, x^{\alpha}  \, \xi^{\beta}$, in${}_{\prec}(g) \equiv g_{a, \, b} \, x^{a} \,  \xi^{b}$, is
\begin{equation}\label{eq:spair_def}
\text{Sp}_{\prec}(f, \, g) = x^{\alpha'}  \, \partial^{\beta'}  \, f - \frac {f_{\alpha, \, \beta}} {g_{a, \, b}}  \, x^{a'}  \, \partial^{b'} g, \quad \alpha' = \max(a, \, \alpha) - \alpha, \cdots.
\end{equation}
For example the S-pair of $f = x_1 + 2  \, x_2  \, \partial_1$, $g = x_2 + x_1^2  \, \partial_2$ wrt $x \prec \partial$ is
\begin{equation}
\begin{aligned}
  \text{in}_{\prec} (f)  & = 2  \, x_2  \, \partial_1, \quad \text{in}_{\prec} (g) = x_1^2  \, \partial_2, 
  \\ \text{Sp}_{\prec}(f, \, g)&  =  \, x_1^2  \, \partial_2  \, f - 2  \, x_2  \, \partial_1  \, g = 2 \,  x_1^2  \, \partial_1 - 4   \, x_1  \, x_2  \, \partial_2.
\end{aligned}
\end{equation}
The Normal form, NF$_{\prec} (d) $ by $G \subset D_N$, equals the result of the recursive algorithm: Start with NF$_{\prec} (d) = 0$, 
\begin{enumerate*}[label=(\roman*)]
  \item While $\exists \, g \in G$ such that in${}_\prec (d) = x^{p} \, $in${}_\prec (g) \, \partial^{q}$, $p,q \in \NN_0$, $d = $ Sp${}_{\prec} (d,g)$.
  \item NF${}_{\prec} (d) = $ in${}_\prec(d) + $ NF${}_{\prec}(d - $in${}_\prec(d))$. When NF$_{\prec} (d) $ by $G \subset D_N$ equals $0$, $d$ has a representation in $G$.
\end{enumerate*} 
For example NF${}_{\prec}(x_1^4  \, \partial_2^3 + x_2^2  \, \partial_1^2)$ by $G = \{ x_1  \, \partial_2, \; x_2  \, \partial_1 \}$ wrt $x \prec \partial$ is expectedly $0$.
\begin{equation}
\begin{aligned}
& \text{in}_{\prec}(d) = x_1^4  \, \partial_2^3 \in D \circ g_1 
\\ & \implies d \rightarrow d' = \text{Sp}_{\prec}(d,  \, g_1) = x_1^4  \, \partial_2^3 + x_2^2  \, \partial_1^2 - x_1^3  \, \partial_2^2  \, g_1 = x_2^2  \, \partial_1^2
\\ & \text{in}_{\prec}(d') = x_2^2 \,\partial_1^2 \in D \circ f_2 
\\ & \implies d' \rightarrow d'' = \text{Sp}_{\prec}(d, \,g_2) = x_2^2 \,\partial_1^2 - x_2 \,\partial_1 \,g_2 = 0 
\end{aligned}
\end{equation}
A Gröbner basis, $\mathcal G$, of $\mI$ wrt the weight $(u, \,v)$ is a finite basis of $\mI$ that respects the grading gr${}_{(u,\,v)} (D_N)$, i.e. the initial forms of $\mathcal G_{(u,\,v)}$, in${}_{(u,\,v)}(\mathcal G_{(u,\,v)})$, form a basis of the initial ideal in${}_{(u,\,v)}(\mI)$ in gr${}_{(u,\,v)} (D_N)$. Similarly, the initial monomials of a Gröbner basis, $\mathcal G_{\prec}$, of $\mI$ wrt an order $\prec$ form a basis of the monomial ideal in${}_{\prec}(\mI)$.
\begin{equation}
\mI = \langle \mathcal G_{*} \rangle, \quad \text{in}_{*}(\mI) = \big\langle \text{in}_{*} (\mathcal G_{*}) \big\rangle, \quad * = (u,v) , \; \; \prec \label{eq:GbasisCondition}
\end{equation}
Given any finite subset of $D_N$ and an order $\prec$ on it, there exists a non-negative integer weight $(u,\,v)$ such that the initial forms wrt $(u,\,v)$ equal the initial monomials wrt $\prec$. A Gröbner basis $\mathcal G$ wrt $(u,\,v)$ or $\prec$ admits a {standard representation} of all elements of $\mI$ such that
\begin{equation}
f = \sum \,d_i \, g_i, \quad f \in \mI , \quad d_i \in D_N, \quad g_i \in \mathcal G_*, \quad \text{in}_* (f) \succeq  \text{in}_* (d_i \, g_i) \quad  \forall \; i.
\end{equation}
The {homogenised Weyl algebra}, $D^{(h)}_N$, is generated by the elements of the Weyl algebra $D_N$ and an additional homogenisation variable $h$. $h$ commutes with all the generators of $D_N$ and acts as a generalisation of $1 = x^{0} \,\partial^{0} \in D_N$, such that the commutation relation $[\partial_i, \, x_j] = \delta_i{}_j{}$ in $D_N$ changes to $[\partial_i, \, x_j] = h^2 \,\delta_i{}_j{}$ in $D^{(h)}_N$. A weight vector, $w = (t,\, u, \, v)$, in $\RR^{2N+1}$, with $u + v \ge 2 \,t$ grades $D^{(h)}_N$, where $t$ is the weight of $h$. The homogenisation of an element $d$, $H(d)$, makes the total degree of every term in $d$ equal,
\begin{equation}
H(d)  = \sum_{(\alpha, \, \beta) \in E} \,c_\alpha{}_\beta{} \, h^{\deg(d) - |\alpha| - |\beta|} \, x^{\alpha} \,\partial^{\beta}, \quad \deg(d) = \max_{(\alpha, \, \beta) \in E}(|\alpha| + |\beta|).
\end{equation}
So $H(d)$ equals its initial form wrt the weight $w = (\mathds{1})$. The maximum number of monomials in any $H(d)$ is set by deg$(d)$, formally precluding any infinitely long elements from a finite dimensional $D^{(h)}$. The {homogenised ideal}, $H(\mI)$, is expectedly defined as $\{H(f) \; \forall \; f \in \mI\}$. Setting $h=1$ dehomogenises $D^{(h)}_N$ to $D_N$ and can be viewed as a restriction of $D^{(h)}_N$ to a slice in affine $\RR^{2N+1}$.  

There always exists a {universal Gröbner basis} for any ideal in $D^{(h)}_N$ that satisfies \cref{eq:GbasisCondition} for generic weights and orders. An irredundant and unique form thereof is the {reduced Gröbner basis}, $\mathcal G^{(h)}$, such that for any pair of distinct elements $g, \,g' \in \mathcal G^{(h)}$, no term of $\Psi(g')$ is divisible by in$_{*}(g)$.\footnote{Divisibility is defined exactly as expected: $g_1$ is divisible by $g_2$ in $D$ if $ \exists \, d \in D$ such that  $d \circ g_2 = g_1$.} Every element of the ideal has a unique standard representation in $\mathcal G^{(h)}$. These properties are also carried over to $D_N$ by the dehomogenisation of $\mathcal G^{(h)}$, i.e. by the sequence of operations: $\mI \in D_N \rightarrow H(\mI) \in D^{(h)}_N \rightarrow \mathcal G^{(h)} \in D^{(h)}_N \rightarrow \mathcal G^{(h)}\big|_{h=1} = \mathcal G \in D_N$.

\subsubsection{Newton polytopes to Gröbner cones}
An $n$-dimensional polytope, $Q$, has vertices on the lattice $\mathbb Z^n$ and has normalised volume such that a regular simplex, $\Sigma_n$, with vertices given by  $(0,\,e_1,\, \cdots\,e_n)$ has volume $1$, where $e_i$ are the basis vectors of the lattice.
\begin{equation}
\Sigma_n = \{v \in \RR_+^{n} \; | \; \sum_{i=1}^{n} \,v_i = 1\}, \quad \Vol(\Sigma_n) = 1, \quad \Vol(Q) = n! \, \Vol_{\rm Euclidean} (Q)
\end{equation}
The Newton polytope of an element $d$ as defined in \cref{eq:general_weylalgebraelement} is the convex hull of (i.e. smallest convex space enclosing) the points $(\alpha, \,\beta) \in E$ in affine $\RR^{2N}$ space. It is represented as an intersection of half planes defined by vectors $w_i$ and constants $\chi_i$,
\begin{equation}
\text{NP}(d) =  \bigcap_{i}\, \{ v \in \RR^{2N} \; | \; v \cdot w_i \ge \chi_i \}.
\end{equation}
For example, the Newton polytope of $d = 1 + x + \partial + x \,\partial \in D_1$ is a unit square in $\RR^{2}$ is
\begin{equation}
\text{NP}_{1^2} = \{ v \in \RR^{2} \; | \;  v \cdot (1,\,0) \ge 0, \; v \cdot (0,\,1) \ge 0, \; v \cdot (-1,\,0) \ge -1, \; v \cdot  (0,\,-1) \ge -1 \}
\end{equation}
which is a rather long form representation of $\{ (x,\,y) \in \RR^{2} \; | \; 1 \ge x \ge 0, \; 1 \ge y \ge 0\}$ with normalised volume $2$. NP${}_{1^2}$ is also the Newton polytope of the polynomial $1 + x_1 + x_2 + x_1 \, x_2$. 

The NP associated with the $\mA$-hypergeometric ideal $\mI_{\mathcal A}$ is the convex hull of the points with coordinates given by the columns vectors of $\mA$. Any translations in affine space correspond to multiplying the polynomials in an Euler integral by some monomial, both being expectedly redundant operations.

The face, $\text{F}_{w}(Q)$, of a polytope, $Q$, wrt a weight $w$ is the section of the polytope with the maximum $w$-weight:
\begin{equation}
\text{F}_{w}(Q)  = \{ v \in Q \; | \; v \cdot w \ge v' \cdot w \; \forall \; v' \in Q \}.
\end{equation}
A facet is an $n-1$ dimensional face of $n$ dimensional $Q$. An edge is a $1$ dimensional face. 
The edges (also facets in this case) and vertices of the 2D unit square, NP${}_{1^2}$, are its faces wrt weight vectors $(0, \,\pm 1)$, $(\pm 1, \,0)$ and $(\pm 1, \,\pm 1)$, $(\pm 1, \, \mp 1)$ respectively. If $Q$ is the Newton polytope of $d$, then every face of $Q$ corresponds to an initial form of $d$. 

The equivalence class of weights that generate the same face of a polytope is called a normal cone. The sum of the dimensions of a face and its normal cone equals the dimension of the polytope. A cone, $\mathcal C$, in affine space can be described by some basis vectors $v_i$ as $\mathcal C = \{\sum \,w^i \cdot v_i \}$, $w^i \in \RR_+$. It is strongly convex when $\mathcal C \bigcap - \mathcal C = \varnothing$. The polar/dual cone $\mathcal C^*$ is $\{v \; | \; v \cdot \mathcal C \ge 0\}$.
\begin{figure}[htpb]
\centering
\begin{tikzpicture}[scale=0.75]
    \draw[thin,lightgray,->] (0,0) -- (3.5,0);
    \draw[thin,lightgray,->] (0,0) -- (0,3.5);
    \draw[thin,lightgray,->] (0,0) -- (-1.8,-1.8);
    \foreach \x in {1,2,3} \draw[thin, lightgray] (\x,1pt) -- (\x,-1pt);
    \foreach \y in {1,2,3} \draw[thin, lightgray] (1pt,\y) -- (-1pt,\y);
    \foreach \y in {-0.7,-1.4}\draw[thin, lightgray] (\y-0.05,\y) -- (\y+0.05,\y); 
    \draw[thick,black,->] (0.5,0.5) -- (-1.5,0.5);
    \draw[thick,black,->] (0.5,0.5) -- (0.5,-1.5);
    \draw[thick,black,->] (0.5,0.5) -- (1.9,1.9); 
    \filldraw[fill=gray!20!white, draw=black,thick] (0,0) -- (0,3) -- (1,2) -- (2,0) -- cycle;
    \filldraw[fill=gray!20!white, draw=black,thick] (0,0) -- (0,3) -- (-0.7,-0.7) -- cycle;
    \filldraw[fill=gray!20!white, draw=black,thick] (0,0) -- (2,0) -- (1.3,-0.7) -- (-0.7,-0.7) -- cycle;
    \fill[black] (0.5,0.5) circle (1.5pt);
    \draw[thick,black] (0.5,0.5) -- (-0.3,0.5);
    \draw[thick,black] (0.5,0.5) -- (0.5,-0.4);
    \draw[thick,black] (0.5,0.5) -- (0.9,0.9);
    \fill[black] (0,0) circle (1.5pt);
    \fill[black] (0,3) circle (1.5pt);
    \fill[black] (2,0) circle (1.5pt);
    \fill[black] (1,2) circle (1.5pt);
    \fill[black] (-0.7,-0.7) circle (1.5pt);
    \fill[black] (1.3,-0.7) circle (1.5pt);
    \fill[black] (0.2,0.3) circle (1.5pt);
    \draw[thin,black,dashed] (0,3) -- (0.2,0.3) -- (-0.7,-0.7);
    \draw[thin,black,dashed] (0.2,0.3) -- (1.3,-0.7);
    \end{tikzpicture} \qquad
    \begin{tikzpicture}[scale=0.75]
    \fill[gray!20!white,opacity = 0.5] (0,3) -- (1,2) -- (2,0)  -- (1.3,-0.7) -- (0.2,0.3) -- cycle;
    \fill[gray!20!white,opacity = 0.5] (-0.7,-0.7)  -- (1.3,-0.7) -- (0.2,0.3) -- cycle;
    \fill[gray!20!white,opacity = 0.5] (-0.7,-0.7)  -- (0.2,0.3) -- (0,3) -- cycle;
    \draw[thin,lightgray,->] (0,0) -- (3.5,0);
    \draw[thin,lightgray,->] (0,0) -- (0,3.5);
    \draw[thin,lightgray,->] (0,0) -- (-1.8,-1.8);
    \draw[thin,gray,dotted] (0,0) -- (2,0);
    \draw[thin,gray,dotted] (0,0) -- (0,3);
    \draw[thin,gray,dotted] (0,0) -- (-0.7,-0.7);
    \foreach \x in {1,2,3} \draw[thin, lightgray] (\x,1pt) -- (\x,-1pt);
    \foreach \y in {1,2,3} \draw[thin, lightgray] (1pt,\y) -- (-1pt,\y);
    \foreach \y in {-0.7,-1.4}\draw[thin, lightgray] (\y-0.05,\y) -- (\y+0.05,\y); 
    \fill[black] (0,0) circle (1.5pt);
    \draw[thick,gray!80!white,dashed] (0.5,0.5) -- (0,-0.38);
    \draw[thick,gray!80!white,dashed] (0.5,0.5) -- (-0.2,0.77);
    \draw[thick,gray!80!white,dashed] (0.5,0.5) -- (1,0.95);
    \fill[gray!80!white] (0.5,0.5) circle (1.5pt);
    \draw[thick,black,->] (0,-0.38) -- (-0.5,-1.2);
    \draw[thick,black,->] (-0.2,0.77) -- (-1,1.1);
    \draw[thick,black,->] (1,0.95) -- (2.2,2);
    \draw[black,thick] (0,3) -- (1,2) -- (2,0)  -- (1.3,-0.7) -- (0.2,0.3) -- cycle;
    \draw[black,thick] (-0.7,-0.7)  -- (1.3,-0.7) -- (0.2,0.3) -- cycle;
    \draw[black,thick] (-0.7,-0.7)  -- (0.2,0.3) -- (0,3) -- cycle; 
    \fill[black] (0,3) circle (1.5pt);
    \fill[black] (2,0) circle (1.5pt);
    \fill[black] (1,2) circle (1.5pt);
    \fill[black] (-0.7,-0.7) circle (1.5pt);
    \fill[black] (1.3,-0.7) circle (1.5pt);
    \fill[black] (0.2,0.3) circle (1.5pt);
    \end{tikzpicture}
  \caption{Newton polytope, Facets, and Normal Cones}
  \begin{flushleft}\singlespacing \vspace{-0.5\baselineskip}
    The Newton polytope of the polynomial $1 + x + y^2 + z^3 + y\,z^2 + x \,y^2 +  x \,y \,z$ is pictured. Its facets/codimension-$1$ faces are shaded in and the arrows are representative vectors of the facets' normal cones.
  \end{flushleft}
  \end{figure}
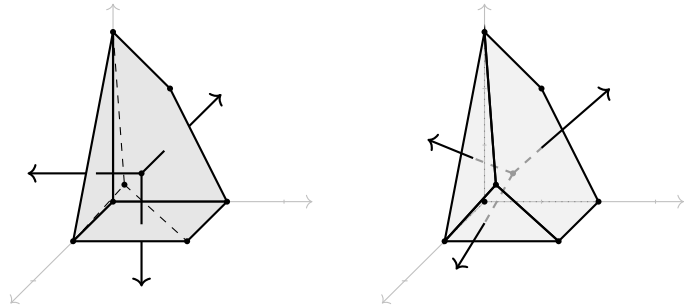\noindent
  Weight vectors within the same equivalency class sweep out a cone in $\RR^{2N}$ and form a {Gröbner cone}, with each cone corresponding to an initial ideal. A fan can be roughly described as a collection of cones. The {small Gröbner fan} consists of the all Gröbner cones of weight vectors of the form $(-w,\,w)$, with the fan covering all in${}_{(-w,\,w)}(\mI)$. The {Gröbner fan} consists of all possible Gröbner cones and so effectively enumerates all possible initial ideals. It is necessarily finite implying the existence of a finite universal Gröbner basis. The Gröbner fan of $\mI \subset D$ is the $t=0$ slice of the Gröbner fan of $H(\mI) \subset D^{(h)}$. 
  \begin{figure}[htpb]
  \centering
  \includegraphics[width=0.5\textwidth]{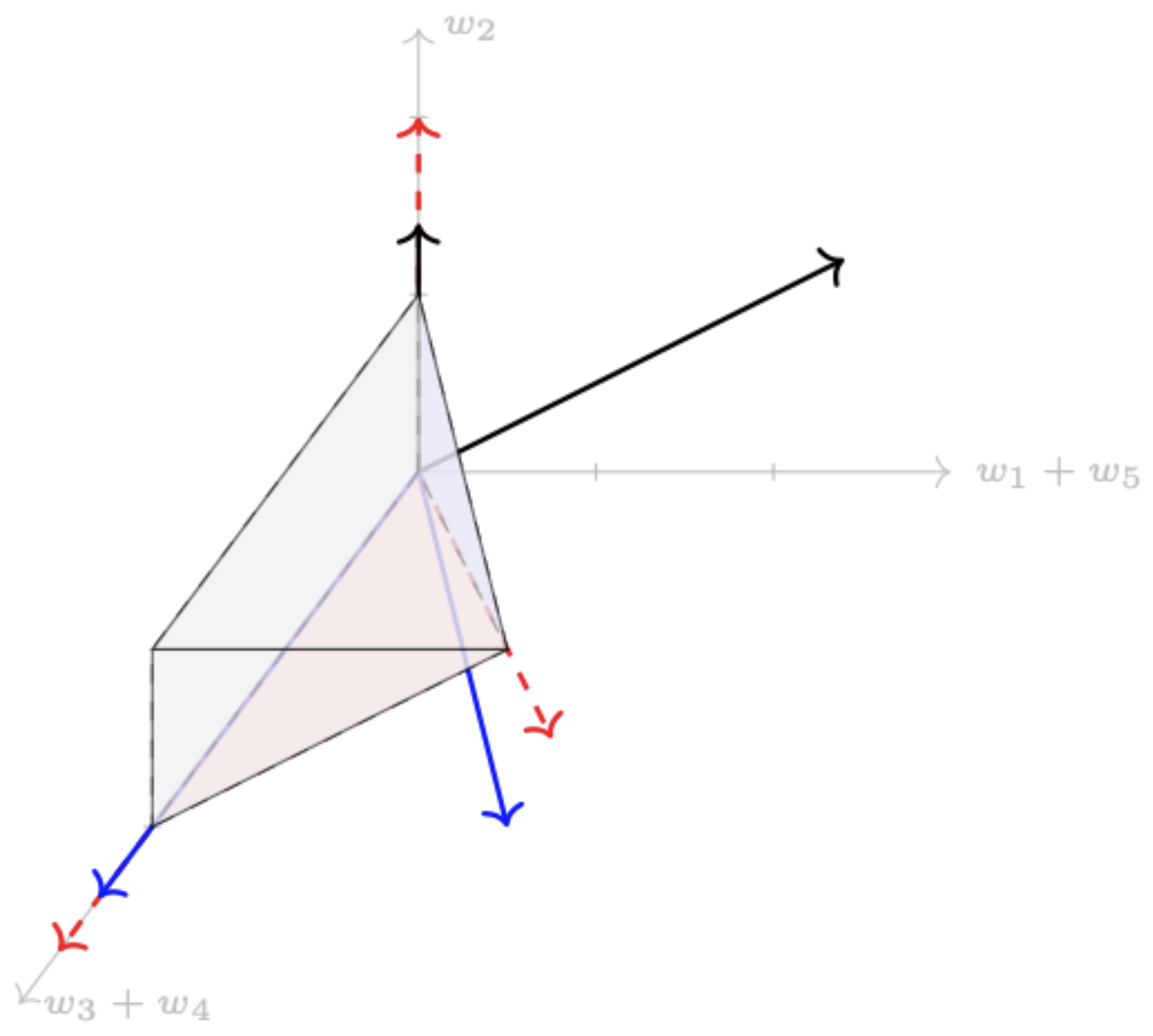}
    \caption{Gröbner cone of Toric ideal : \cref{eq:Jidealquick} in \cref{examplequick}}
    \begin{flushleft}\singlespacing \vspace{-0.5\baselineskip}
    The choice of weight vector in \cref{eq:Jidealquick} is obviously non-unique and any weight vector of the form $w =\{w_i\}$ such that $w_3 + w_4 > w_1 + w_5$ (interior of blue arrows) and $2 \,w_2 > w_1 + w_5$ (interior of black arrows) belongs to the equivalence class of $\{0,\,1,\, 1, \, 0,\, 0\}$. The dashed red arrows serve as a basis for the Gröbner cone, enclosing it.
    \end{flushleft}
    \end{figure}\noindent
    \vspace{-\baselineskip}
\subsubsection{Holonomic characteristics}
    The {characteristic ideal} is the initial ideal with respect to the weight $(0,\,\mathds 1)$, in${}_{(0,\,\mathds 1)}(\mI)$, defined in the commutative ring $\CC[x,\, \xi]$. The zero set of the characteristic ideal in affine $\CC^{2N}$ space is the {characteristic variety}, ch$(\mI)$, equalling the characteristic variety of the $D$-module $\mathcal D / \mI$.

    For a commutative ideal, $\mathcal J$, the number of monomials of total degree $\le k$ which aren't divisible by in$_{\prec}(\mathcal J)$ is polynomial for sufficiently large $k$. When $\mathcal J = $ in${}_{(0,\,\mathds 1)}(\mI)$, it is called the Hilbert polynomial, $p(k)$, of the ideal $\mI$. It is known to have the form $p(k) = \tfrac {m} {N!} \,k^N + \cdots$.

    The degree of the Hilbert polynomial, $N$, is the (Krull) dimension of $\mI$. A $0$-dimensional ideal is called Artinian and has a finite zero set i.e. points, corresponding to the roots. When $\mI$ is homogeneous, $m$ is called the degree of $\mI$. A $D$-ideal, $\mI \subset D_N$, is {holonomic} if its characteristic ideal has dimension $N$.

    An ideal is integrable if it is closed under Poisson brackets, a characteristic ideal being so trivially. If an ideal is integrable, so is its characteristic ideal and the radical of its characteristic ideal. Thus, the characteristic variety, which is the zero set of the characteristic ideal, is also integrable (see \cref{eq:primary_ideal_decomposition}). A linear space in $\CC^{2N}$ ceases to be integrable if both $x_i = \xi_i = 0$ on it. Thus, it must have a dimension of at least $N$. The characteristic ideal of a proper $D$-ideal also has dimension $\ge N$. A proper $D$-ideal $\mI$ is holonomic iff the dimension of in${}_{(u,\,v)}(I)$ is $N$ for generic $(u, \,v) \in \RR^{2N}_+$ such that $u+v > 0$.

The dimension of the complex vector space of holomorphic solutions to $\mI \circ f = 0$ on a simply connected domain in $\CC^{n} \backslash$ S$(\mI)$ is rank$(\mI)$, i.e the number of $\CC$-linearly independent holomorphic solutions outside the singular locus. The singular locus, S$(\mI)$, is the Zariski closure of the image of ch$(\mI) \backslash \{ \xi = 0 \}$ under the coordinate projection $\CC^{2 N} \rightarrow \CC^{N} : (x, \,\xi) \mapsto x$, i.e. it is the smallest variety enclosing the zero set of the projected ideal, which is found by eliminating $\xi$ from the characteristic ideal. The rank of holonomic ideals is finite and defined as the vector space dimension over the field $\CC$:
\begin{equation}
\text{rank}(\mI) \vcentcolon= \dim ( \CC(x) \, [\xi] / \CC(x) \, [\xi] \circ \text{in}_{(0,\,\mathds 1)}(\mI) ).
\end{equation}
If $M \subset \CC[\xi]$ is a monomial ideal wrt the order $\prec$ of the restriction of the characteristic ideal in$_{(0,\,\mathds 1)}(\mI)$ by $x \mapsto 1$ to $\CC[\xi]$, the rank is also given by
\begin{equation}
\text{rank}(\mI) = \dim (\CC[\xi]/M) = \# \{ \xi^\alpha \notin M, \; \alpha \in \NN_0^{N} \}.\label{eq:holoranksimpleprocess}
\end{equation}
\begin{exmp}
Given a left ideal $\mI = \left\langle x_1 \,\partial_2, \; x_2 \,\partial_1 \right\rangle$ in $D_2$, the characteristic ideal, $\text{in}_{\prec_{(0,\,\mathds 1)} } (\mI)$ is $\left\langle x_1 \,\xi_2, \; x_2 \,\xi_1 \right\rangle$. The monomial ideal, $M$, upon projection $x \mapsto 1$ is $\left\langle \xi_1, \,\xi_2 \right\rangle$. Obviously $M$ can generate every element all of $\CC[\xi_1, \, \xi_2]$ apart from $1$. Thus, rank$(\mI) = 1$.
\end{exmp}

\subsubsection{Torus fixed ideals}
The action of the $N$-dimensional algebraic torus $T = (\CC^*)^N$ on $D_N$ is
\begin{equation}
T \times D_N \rightarrow D_N \; : \; (t_i, \,\partial_i) \mapsto t_i \,\partial_i \; : \; (t_i,\, x_i) \mapsto t_i^{-1}\, x_i.
\end{equation}
A $D$-ideal which remains invariant under the torus action, $T \circ \mathcal I = \mI$, is {torus fixed/invariant}. $\CC[\theta]$ represents all torus fixed elements of $D_N$, with a general $D$-ideal $\mI$ being torus fixed iff it consists of elements of the form $x^{\alpha} \, f (\theta) \, \partial^\beta$, $\alpha, \,\beta \in \NN$, $f(\theta) \in \CC[\theta]$ or equivalently iff in${}_{(-w,\,w)}(\mI) = \mI$ for all non-negative weights $w \in \RR_+^{N}$. For sufficiently generic $w$, in${}_{(-w,\,w)}(\mI)$ is torus fixed by default. The {distraction}, $\tilde{\mI}$, of a torus fixed $\mI$ is 
\begin{equation}
\tilde{\mI}  = \CC[x][\partial] \circ \mI \bigcap \CC[\theta] = \langle \theta_{(b)} \, f(\theta - b) \rangle, \quad \mI = \langle x^{\alpha} \, f (\theta) \, \partial^\beta \rangle. \label{eq:distractiondef}
\end{equation}
$\mI$ is a {Frobenius} ideal if $\mI \subseteq \CC[\theta]$. It is {Artinian} if $\CC[\theta]/ \mI$ is finite dimensional, with the dimension equalling the rank of $\mI$. The {indicial ideal} is a generalisation of a Frobenius ideal, defined as
\begin{equation}
\text{ind}_{(w)}(\mI)  = \CC[x][\partial] \circ \text{in}_{(-w,\,w)}(\mI) \bigcap \CC[\theta] , \quad w \in \RR^N.
\end{equation}
It is a holonomic Frobenius ideal with rank equal to the rank of the initial ideal in${}_{(-w,\,w)}(\mI)$. The zeros of the indicial ideal in affine $\CC^N$ are called {exponents}. When counted with multiplicity, there are rank$(\mI)$ exponents. 
In short, for a holonomic ideal $\mI$:
\begin{equation}
\text{rank}(\mI)  \underset{= \text{ for regular}}{\ge} \text{rank}(\text{in}_{(-w,\,w)}(\mI))  = \text{rank}(\text{ind}_{(w)}(\mI)) = \# \text{Exponents}.
\end{equation}
The $b$-function, $b(\omega)$, or indicial polynomial of an ideal $\mI$ can be algorithmically computed by eliminating $\{\theta\}$ from the intersection of $\CC\{\omega - w \, \theta\}$ and in${}_{(-w, \, w)} \mI \bigcap \,\CC[\theta]$, found by applying \cref{eq:restrictiontoCtheta} on the Gröbner basis of $\mI$ wrt weight $(-w, \, w)$.

\subsubsection{Solutions of holonomic \texorpdfstring{$D$}{D}-ideals}
Holonomic $D$-ideals have convergent series solutions of the form $f$ around the point $c$,
\begin{equation}
\mI \circ f = 0, \quad f = \CC[(x - c), \,\log (x - c)] = \sum \,c_\alpha{}_\beta{} \; x^\alpha \,(\log x)^\beta, \quad \alpha \in \CC, \quad \beta \in \NN_0
\end{equation}
if the singular locus, S$(\mI)$, is a normal crossing divisor at $c$, i.e. it can be locally represented by a polynomial of the form
\begin{equation}
{\rm S}(\mI) \Big|_{|x - c| \ll 1}  \cong (x-c)^s \cong x^{s} \,(1 + \sum_{u>0} \, c_u \, x^u ). \label{eq:convexhullfromsing}
\end{equation}
The Newton polytope of S$(\mI)$, NP$(\mI)$, is the convex hull of the points $A = \{A_i\}$ in $\RR^n$ such that 
\begin{equation}
{\rm S}(\mI) = \sum \,c_i \, x^{A_i}, \quad c_i \in \CC^*.
\end{equation}
A cone, $\mathcal C_{A_i}$, originating from a vertex $A_i$ enclosing NP$(\mI)$ is unimodular if it can be represented by $n$ integer vectors $u_i$ such that
\begin{equation}
\mathcal C_{A_i} = \sum_{i=1}^{n} \,c_i \, u_i, \quad  c_i \in  \RR_+, \quad u_i \in \mathbb Z^n \quad U = \{u_1, \, u_2, \cdots u_n\}, \quad |\det U|^2 = 1. \label{eq:unimodularconeofSingNP}
\end{equation}
Its polar/dual cone, $\mathcal C^*$, is generated by the vectors $(U^{-1})^{\rm Tr} \equiv U^*$.
\begin{exmp}\label{sec:exmp_AppellF2}
Appell's $F_2(x,\,y)$ is defined as the solution to the differential equations generated by $\langle \theta_x^2 - x \,(\theta_x + \theta_y + a) \,(\theta_x + b_1)$, $\theta_y^2 - y \,(\theta_x + \theta_y + a) \,(\theta_y + b_2) \rangle$. This representation is a Gröbner basis wrt any weight $(-w,\,w)$, $w>0$. Its singular locus is $x\, y \,(1 - x) \,(1 - y) \,(1 - x - y)$ \cite{Saito:GrobnerBases}.
\begin{figure}[htpb]
\centering
  \begin{tikzpicture}
    \draw[thin,lightgray,->] (0,0) -- (3.5,0);
    \draw[thin,lightgray,->] (0,0) -- (0,3.5);
    \foreach \x in {1,2,3} \draw[thin, lightgray] (\x,1pt) -- (\x,-1pt);
    \foreach \y in {1,2,3} \draw[thin, lightgray] (1pt,\y) -- (-1pt,\y);
    \coordinate (1) at (1,1);
    \coordinate (2) at (2,1);
    \coordinate (3) at (3,1);
    \coordinate (4) at (3,2);
    \coordinate (5) at (2,3);
    \coordinate (6) at (1,3);
    \coordinate (7) at (1,2);
    \coordinate (8) at (2,2);
    \draw[thick,blue,dashed,->] (4) -- (2.1,1.1);
    \draw[thick,blue,dashed,->] (4) -- (2.1,2);
    \draw[thick,blue,->] (4) -- (3,0.7);
    \draw[thick,blue,->] (4) -- (2.5,2.5);
    \draw[thick,blue,dashed,->] (5) -- (1.1,2.1);
    \draw[thick,blue,dashed,->] (5) -- (2,2.1);
    \draw[thick,blue,->] (5) -- (2.5,2.5);
    \draw[thick,blue,->] (5) -- (0.7,3);
    \draw[thick,red,->] (1) -- (1.8,1);
    \draw[thick,red,->] (1) -- (1,1.8);
    \draw[thick,red,->] (6) -- (1.8,3);
    \draw[thick,red,->] (6) -- (1,2.2);
    \draw[thick,red,->] (3) -- (3,1.8);
    \draw[thick,red,->] (3) -- (2.2,1);
    \foreach \c in {1,2,3,6,7,8} \fill[black] (\c) circle (1.5pt) node[anchor=north east]{\tiny$\c$};
    \foreach \c in {4,5} \fill[black] (\c) circle (1.5pt) node[anchor=south west]{\tiny$\c$};
    \end{tikzpicture}
    \caption{Newton Polytope of S$(\mI)$, Unimodular Cone, Polar Cones of Appell's $F_2$}\label{fig:appellsf2}
    \end{figure}
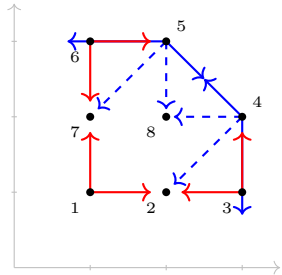
\vspace{-\baselineskip}

\noindent The interior of the black dots in \cref{fig:appellsf2} is the newton polytope. The red arrows indicate the generators of the unimodular cones that coincide with their dual cones originating from the vertices $\{1,\,3,\,6\}$. The blue and dashed blue arrows encapsulate the unimodular and polar cones respectively at vertices $\{4,\,5\}$. 
\begin{equation}
\begin{aligned}
  \mathcal C_{1} & = \left( \begin{smallmatrix} 1 & 0 \\ 0 & 1\end{smallmatrix} \right), && \mathcal C_{3} = \left( \begin{smallmatrix} -1 & 0 \\ 0 & 1\end{smallmatrix} \right), && \mathcal C_{4} = \left( \begin{smallmatrix} -1 & 1 \\ 0 & -1\end{smallmatrix} \right), && \mathcal C_{5} = \left( \begin{smallmatrix} -1 & 0 \\ 1 & -1\end{smallmatrix} \right), && \mathcal C_{6} = \left( \begin{smallmatrix} 1 & 0 \\ 0 & -1\end{smallmatrix} \right)
  \\ \mathcal C^*_{1} & = \mathcal C_{1}, && \mathcal C^*_{3} = \mathcal C_{3}, && \mathcal C^*_{4}  = \left( \begin{smallmatrix} -1 & 0 \\ -1 & -1\end{smallmatrix} \right), && \mathcal C^*_{5} = \left( \begin{smallmatrix} -1 & -1 \\ 0 & -1\end{smallmatrix} \right), && \mathcal C^*_{6} = \mathcal C_{6}
\end{aligned}
\end{equation}
Vertices $\{2,\,7,\,8\}$ cannot serve as origins of convex unimodular cones enclosing the NP.

\end{exmp}\noindent
    The degree/$w$-weight of a monomial including logarithmic terms, $x^\alpha \,\log x^\beta$, wrt some non-negative weight vector $w \in \RR^N$ is defined as Re$(w \cdot \alpha)$. Min${}_w(f)$ is the minimum of the $w$-weights of all the terms in $f$. The sum of the terms which have this minimum $w$-weight form the {initial series}, in${}_w(f)$. The initial series can be further ordered wrt some prescription $\prec$ and the initial monomial of the initial series is the {starting monomial} of the series.

    Given a series solution $f$ of $\mI$, the initial series, in${}_{(w)}(f)$, is a solution to the initial ideal in${}_{(-w,\,w)}(\mI)$, which is also referred to as a {Gröbner deformation} of $\mI$. Obviously,
    \begin{equation}
    \text{rank} \big( \text{in}_{(-w,\,w)} (\mI) \big) \le \text{rank} \big(\mI\big).
    \end{equation}
    If the $D$-ideal $\mI$ is regular holonomic, the equality holds. A basis of $\CC$-linearly independent series solutions, $N(\mI) = \{f_i\}$, of regular holonomic ideal $\mI$
    \begin{equation}
    f = \sum_{ N(\mI)} \,c_i \; f_i, \quad \forall \; \mI \circ f = 0,  \quad  \sum_{N(\mI)}\, c_i \; f_i = 0 \implies c_i = 0, \quad c_i \in \CC
    \end{equation}
    are called {canonical solutions} wrt $\prec_w$, if every basis solution $f_i \in N(\mI)$ has a {\it unique} starting monomial, which doesn't appear in any $f_j \in N(\mI)$, $j \neq i$. There are rank$(\mI)$ elements in $N(\mI)$, and so just as many $\CC$-linearly independent {\it finite} initial series and unique starting monomials. The initial monomial wrt $\prec_{w}$ of any series solution of $\mI$ falls within the set of starting monomials. The canonical series are convergent when $0 < | x^{u_i} | \ll 1$ which is the same condition as $ U \cdot (- \log | x |) \gg 0$, where $U$ is as described in \cref{eq:unimodularconeofSingNP}. This means, the series converges when
    \begin{equation}\label{eq:weightchoice}
    \begin{aligned}
      (- \log |x|) \in w + \mathcal C^*
    \end{aligned}
    \end{equation}
    for some point $w$, which influences the starting monomials. The degree of logarithmic terms in these solutions is at most rank$(\mI) - 1$. Given a set of starting monomials $\{\tilde f_i\}$, the canonical series are of the form
    \begin{equation}
    f = \tilde f_i \, \sum_{t_i \in \mathbb Z} \; \sum_{|s| \in Q}  \,c_{t,\, s} \; x^{t_i \, u^*_i} \; (\log x)^{s}  , \quad Q = \{0, \,1, \,2 , \cdots \text{rank}(\mI) - 1\}
    \end{equation}
    where $|s| = \sum s_i$, $\{u^*_1, \, \cdots u^*_n \}$ is the basis of $\mathcal C^*$, and the constants $c_{t,\,s}$ can be solved for inductively by requiring the series to satisfy $\mathcal I \circ f = 0$ order by order. 
    \begin{continueexample}{sec:exmp_AppellF2}
    Since this is a system of rank $4$, the maximum degree of the logarithmic terms will be $\le 3$.  For any given weight $(-w,\,w)$, $w \in \RR_+^2$, the initial ideal is $\left\langle\theta_x^2, \, \theta_y^2 \right\rangle$, yielding the starting monomials $\tilde f = \{1, \, \log x, \, \log y, \, \log x \, \log y\}$. Using the basis of $\mathcal C^*_1$, the series are represented by
    \begin{equation}
    \begin{aligned}
      f = \tilde f_i  \,\sum_{t_i \in \mathbb Z}  \,\sum_{s_i \in \{0, \, 1\}} \, c_{t,\, s} \; x^{t_1} \, y^{t_2} \; (\log x)^{s_1} \, (\log y)^{s_2} 
    \end{aligned}
    \end{equation}
    The action of the rest of the ideal $\langle - x \,(\theta_x + \theta_y + a) \,(\theta_x + b_1), \, - y \,(\theta_x + \theta_y + a) \,(\theta_y + b_2) \rangle$ on the starting monomials is
    \begin{equation}
    \begin{aligned}
      \mathcal I_1 \,\tilde f_1 & = - a \, b_1 \, x, && \mathcal I_2 \,\tilde f_1 = - a \, b_2 \, y
      \\ \mathcal I_1 \,\tilde f_2 & = - ( a  + b_1 + a \, b_1 \, \log x) \, x , && \mathcal I_2 \,\tilde f_2 = - (a \, b_2 \, \log x  + b_2 ) \, y
      \\ \mathcal I_1 \,\tilde f_3 & = - ( b_1 + a \, b_1 \, \log y ) \, x, && \mathcal I_2 \,\tilde f_3 = - ( a   + b_2 + a \, b_2 \, \log y) \, y
      \\ \mathcal I_1 \,\tilde f_4 & = - \big( 1 + b_1 \, \log x  + (a + b_1) \, \log y  && \mathcal I_2 \,\tilde f_4 = - \big( 1 + b_2 \, \log y  + (a + b_2) \, \log x
      \\ & \phantom{= - } + a \, b_1 \, \log x \, \log y \big) \, x, &&  \phantom{\mathcal I_2 \, \tilde f_4 = - } + a \, b_2 \, \log x \, \log y \big) \, y.
    \end{aligned}
    \end{equation}
    The series solution can be found inductively, like in \cref{sec:example_intractable_2}, to get
    \begin{equation}
    \begin{aligned}
      c_{1,\,0,\,0,\,0} & = \tfrac {a \, b_1 \, ( a \, (2 \, b_1 - 1) - (b_1 + 1))} {(1 + a + b_1 - 2 \, a \, b_1)^2} \, c_{0,\,0,\,0,\,0} - \tfrac {b_1 \, (b_1 + 1)} {(1 + a + b_1 - 2\,  a \, b_1)^2}  \, c_{0,\,0,\,0,\,1}
      \\ c_{1,\,0,\,1,\,0} & = - \tfrac {a^2 \,b_1^2 \,(a \,(2 \,b_1 - 1)  - (b_1 + 1))} {(1 + a + b_1 - 2 \,a \,b_1)^2} \, c_{0,\,0,\,0,\,0} - \tfrac {a \,b_1^2 \,( a \,(2 \,b_1 - 1) - 2 \,(b_1 + 1))} {(1 + a + b_1 - 2 \,a \,b_1)^2} \, c_{0,\,0,\,0,\,1}
      \\ c_{1,\,0,\,0,\,1}  & = -\tfrac {a \, b_1}{1 + a + b_1 - 2 \, a \, b_1} \, c_{0,\,0,\,0,\,1}
      \\ c_{1,\,0,\,1,\,1} & = \tfrac {a^2 \, b_1^2}{1 + a + b_1 - 2 \, a \, b_1} \, c_{0,\,0,\,0,\,1}
      \\ c_{0,\,1,\,0,\,0} & = \tfrac {a \, b_2 \, ( a \, (2 \, b_2 - 1) - (b_2 + 1))} {(1 + a + b_2 - 2 \, a \, b_2)^2} \, c_{0,\,0,\,0,\,0} - \tfrac {b_2 \, (b_2 + 1)} {(1 + a + b_2 - 2\,  a \, b_2)^2}  \, c_{0,\,0,\,1,\,0}
      \\ c_{0,\,1,\,0,\,1} & = - \tfrac {a^2 \,b_2^2 \,(a \,(2 \,b_2 - 1)  - (b_2 + 1))} {(1 + a + b_2 - 2 \,a \,b_2)^2} \, c_{0,\,0,\,0,\,0} - \tfrac {a \,b_2^2 \,( a \,(2 \,b_2 - 1) - 2 \,(b_2 + 1))} {(1 + a + b_2 - 2 \,a \,b_2)^2} \, c_{0,\,0,\,1,\,0}
      \\ c_{0,\,1,\,1,\,0}  & = -\tfrac {a \, b_2}{1 + a + b_2 - 2 \, a \, b_2} \, c_{0,\,0,\,1,\,0}
      \\ c_{0,\,1,\,1,\,1} & = \tfrac {a^2 \, b_2^2}{1 + a + b_2 - 2 \, a \, b_2} \, c_{0,\,0,\,1,\,0}
    \end{aligned}
    \end{equation}
    and so on. However, induction quickly starts to become rather cumbersome even at $2$\textsuperscript{nd} order, and hence isn't particularly suitable for analytic evaluation of solutions in more than one variable.
    \end{continueexample}
    When the roots of the indicial ideal (i.e. exponents), $\beta$, form the starting monomials, $x^\beta$, the canonical series solutions are called the {Nilsson ring}, $N_{w}(\mI)$. In this case, the unimodular cone of relevance becomes the Gröbner cone wrt the weight $w$, $\mathcal C_w(\mI)$, satisfying in${}_{(-w,\,w)}(\mI) = $ in${}_{(-w',\,w')}(\mI)$ for all $w' \in \mathcal C_w(\mI)$, which can in turn be used to find the polar cone $\mathcal C^*_w(\mI)$ and its basis vectors $u^*$. However, there is apparently a slightly easier procedure to find $u^*$. Given a Gröbner basis, $\mathcal G_w$, a cone $\mathcal C_w(\mathcal G)$, defined as 
    \begin{equation}
    \text{in}_{(-w,\,w)}(\mathcal G_w) = \text{in}_{(-w',\,w')}(\mathcal G_w), \quad \; \forall \; w' \in \mathcal C_w(\mathcal G) 
    \end{equation}
    includes the proper Gröbner cone $\mathcal C_w(\mI)$ and can instead be used to compute the basis $u^*$.

\subsection{Solutions of \texorpdfstring{$\mA$}{A}-Hypergeometric systems}\label{sec:SolutionsofAHypergeometricSystems}
The $\mA$-hypergeometric system of equations, $\mI_{\mA}$, with $\mA$ as defined in \cref{eq:genAmatrixform} i.e. with $\mathds {1}$ as one of the rows, is a regular holonomic ideal. This automatically makes the generating set of the toric ideal $J_{\mA}$ in \cref{eq:Jideal} consist of homogenous binomials, also represented as a toric variety of dimension $N-1$ in $\mathbb P^{N-1}$.
    \begin{exmp}\label{exmp:runningexample}
    A running example throughout this section is based on the integral
    \begin{equation}\label{eq:runningexampleint}
    \begin{aligned}
      I_{1}  & = \int \tfrac {\de x} {x} \,\tfrac{ x^{\beta} } { ( z_1 + z_2 \,x_1^2 + z_3 \,x_2^2 + z_4 \,x_3^2 + z_5 \,x_4^2 + z_6 \,x_1^2 \,x_4^2 + z_7\, x_2^2 \,x_3^2 + z_8 \,x_1 \,x_2 \,x_3\, x_4)^{\beta_0} } 
      \\ & = \int \tfrac {\de x} {8 \, x} \,\tfrac{ x_1^{\frac {\beta_1 + \beta_4} 2} \,x_2^{\frac {\beta_2 + \beta_3} 2} \,x_3^{\beta_3} \,x_4^{\frac{\beta_3 + \beta_4}{2}}} { ( z_1 + z_2 \,x_1 + z_3\, x_2 + z_4 \,x_2 \,x_3^2 \,x_4 +z_5 \,x_1 \,x_4 + z_6 \,x_1^2 \,x_4 + z_7 \,x_2^2 \,x_3^2 \,x_4 + z_8\, x_1 \,x_2\, x_3 \,x_4)^{\beta_0} } \equiv I'_{1},
    \end{aligned}
    \end{equation}
    as such chosen because it will come to have physical significance (see \cref{sec:2melonfeynmanint}). Its associated $\mA$ system with generic parameters:
    \begin{equation}\label{eq:runningexamplegenericgammas}
    \begin{aligned}
      \gamma_1 & = \{ \beta_0, \, \beta_1, \,  \beta_2, \,  \beta_3, \,  \beta_4  \}, && \gamma'_1 = \{ \beta_0, \, \tfrac {\beta_1 + \beta_4} 2, \,  \tfrac {\beta_2 + \beta_3} 2, \,  \beta_3, \,  \tfrac{\beta_3 + \beta_4}{2}  \}, 
    \end{aligned}
    \end{equation}
    is given by
    \begin{equation}\label{eq:runningexample}
    \begin{aligned}
      \mathcal A_{1}  & = \left( \begin{smallmatrix}
    1 & 1 & 1 & 1 &1 & 1 & 1 & 1\\
    0 & 2 & 0 & 0 &0 & 2 & 0 & 1 \\
    0 & 0 & 2 & 0 &0 & 0 & 2 & 1 \\
    0 & 0 & 0 & 2 &0 & 0 & 2 & 1 \\
    0 & 0 & 0 & 0 &2 & 2 & 0 & 1 \\ \end{smallmatrix} \right), \quad \mathcal A'_{1}  = \left( \begin{smallmatrix}
    1 & 1 & 1 & 1 &1 & 1 & 1 & 1\\
    0 & 1 & 0 & 0 &1 & 2 & 0 & 1 \\
    0 & 0 & 1 & 1 &0 & 0 & 2 & 1 \\
    0 & 0 & 0 & 2 &0 & 0 & 2 & 1 \\
    0 & 0 & 0 & 1 &1 & 1 & 1 & 1 \\ \end{smallmatrix} \right)
    \\ \mathcal J_{1} & = \left\langle\partial_2 \,\partial_5 - \partial_1 \,\partial_6, \; \partial_3 \,\partial_4 -  \partial_1 \,\partial_7 , \; \partial_6 \,\partial_7 - \partial_8^2 \right\rangle
    \end{aligned}
    \end{equation}
    where both $\mA$ and $\mA'$ have the same kernel and irreducible toric ideal $\mathcal J_1$. 
    The $\mA$ hypergeometric ideal is $\langle \mathcal G_{1} \rangle + \mathcal H_1, \;   \mathcal H_1 = \langle \mA_1 \,\theta + \gamma_1\rangle$, where $\mathcal G_{1}$ is the universal Gröbner basis of $\mathcal J_1$.
    \begin{equation} \label{eq:runningexampleJid}
    \begin{aligned}
    \mathcal G_{1} & = \mathcal J_{1} + \langle \partial_1 \,\partial_8^2 - \partial_2 \,\partial_5 \,\partial_7, \; \partial_1 \,\partial_8^2 - \partial_3 \,\partial_4 \,\partial_6, \; \partial_2 \,\partial_5 \,\partial_7 - \partial_3 \,\partial_4 \,\partial_6, 
    \\ & \phantom{= J_{1} + \langle  } \partial_3 \,\partial_4 \,\partial_8^2 - \partial_2 \,\partial_5 \,\partial_7^2, \; \partial_2 \,\partial_5 \,\partial_8^2 - \partial_3 \,\partial_4 \,\partial_6^2 \rangle
    \end{aligned}
    \end{equation}
    with zero set of $\mathcal J_{1}$ embedded in $\mathbb P^{7}$ as $\{1 : \,p_2^2 : \,p_3^2 : \,p_4^2 : \,p_5^2 : \,p_2^2 \,p_5^2 : \,p_3^2 \,p_4^2 : \,p_2 \,p_3 \,p_4 \, p_5\}$.
    \end{exmp}
    When $\mA$ is an $N \times N$ matrix, the volume of the Newton polytope, $\Vol(\mA)$, is simply $\det \mA$. If the dimension of the newton polytope of $\mA$ is $< n$, it is degenerated, else it is full dimensional. However, since $\mA$ is required to satisfy the \nameref{sec:AIdealCons} in the current context, it will always produce a full dimensional polytope. 
\begin{figure}[htpb]
\centering
\begin{subfigure}[t]{0.48\textwidth}
\centering
\includegraphics[width=\textwidth]{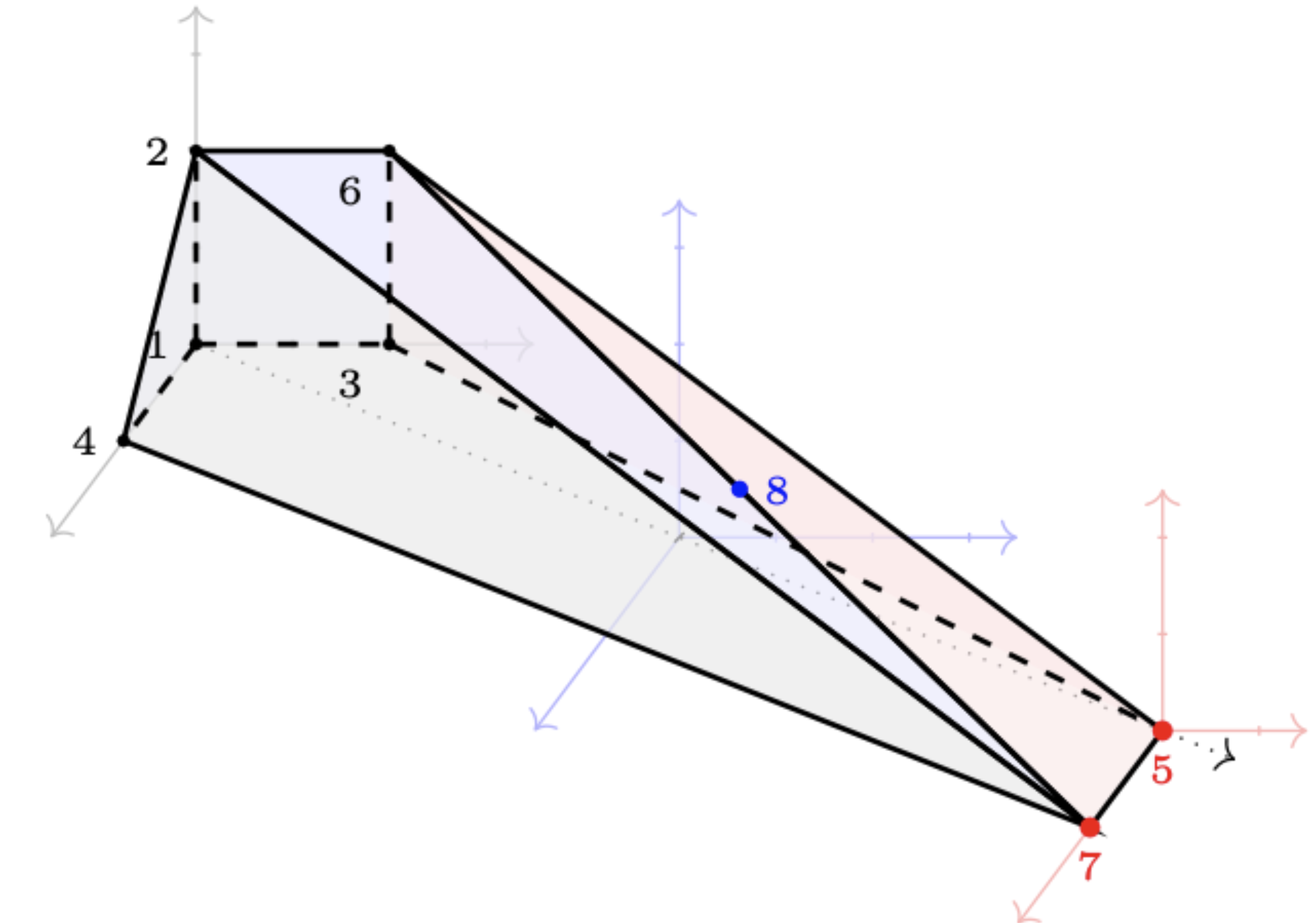}
\caption{NP${}_1$: Newton Polytope of $\mA_1$}\label{fig:4DNPfirst}
\begin{flushleft}\singlespacing \vspace{-0.5\baselineskip} 
\hspace{0.5\baselineskip} Vertex $8$ falls on the surface of the convex hull evidencing its degeneracy.
\end{flushleft}
    \end{subfigure}%
    \begin{subfigure}[t]{0.48\textwidth}
        \centering
  \includegraphics[width=0.9\textwidth]{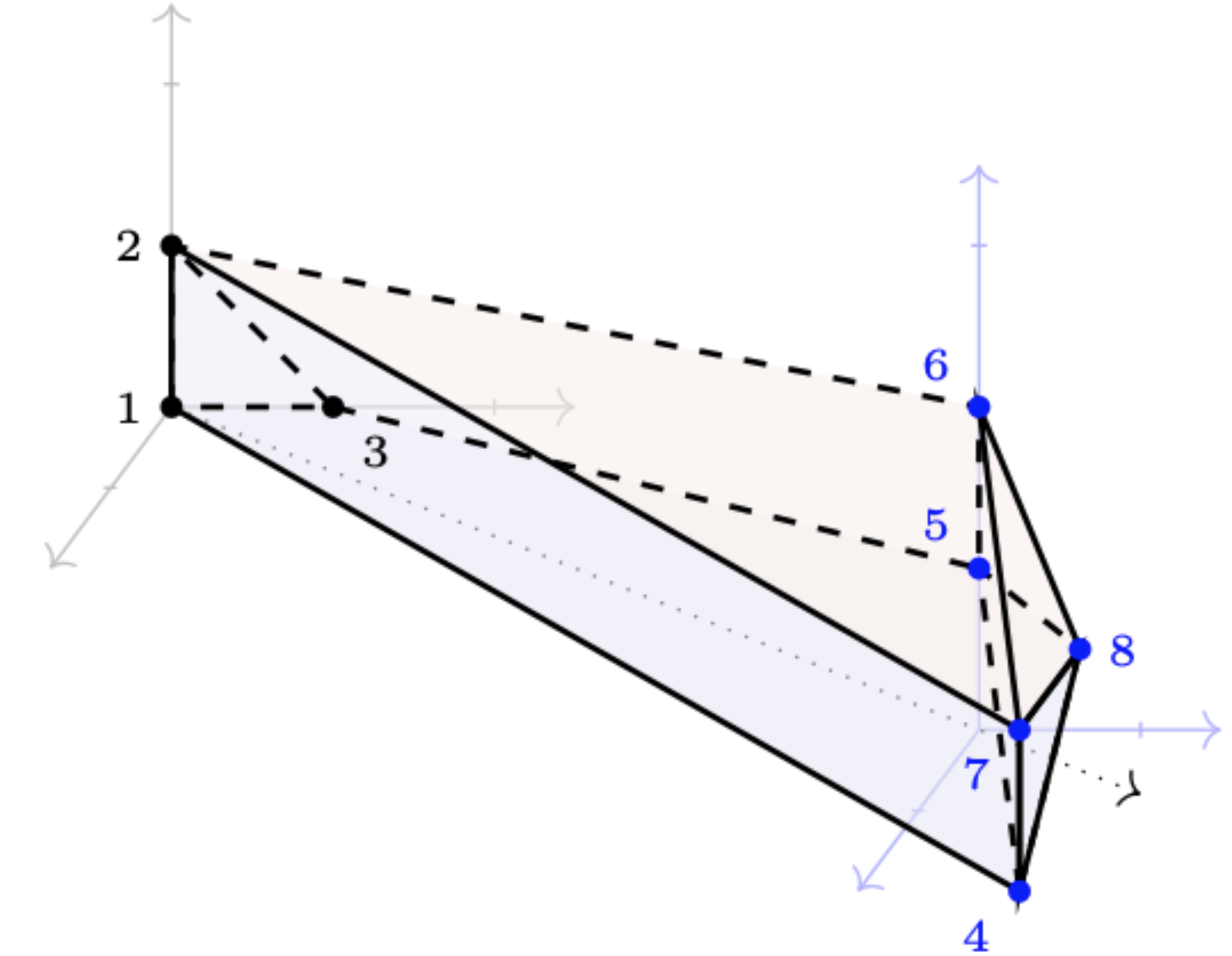}
        \caption{NP$'{}_1$: Newton Polytope of $\mA'_1$}\label{fig:4DNPsecond}
\begin{flushleft}\singlespacing \vspace{-0.5\baselineskip} 
\hspace{0.5\baselineskip} The irreducibility conditions imposed on $\mA'_1$ ensure that there are no degenerate points.
  \end{flushleft}{}
    \end{subfigure}%
\caption{Newton Polytopes of $\mA_1$, $\mA'_1$ matrices in \cref{eq:runningexample}}\label{fig:4DNP}
  \begin{flushleft}\singlespacing \vspace{-0.5\baselineskip} 
  The black, blue and red $3$D axes are slices at $0, \, 1$ and $2$ resp. in the $4$\textsuperscript{th} dimension, represented by the dotted line. Solid dots are colour coded accordingly and label the vertices of the polytopes corresponding to the columns of $\mA$.
  \end{flushleft}   
  \end{figure}

\vspace{-\baselineskip}
\noindent The newton polytopes in \cref{fig:4DNP} are $4$-dimensional and can be represented as an intersection of half spaces. For example, NP${}_1$ is defined by $\{ 2 \ge \eta_i \ge 0, \; \eta_1 + \eta_4 \le 4, \; \eta_2 + \eta_3 \le 4, \; \sum \eta \le 4  \}$. 

The degree of a toric ideal, $J_\mA$, is equal to $\Vol(\mA)$. In general, the degree of $\mI_{\mA} \ge \Vol(\mA)$ but equality exists if $\gamma$ is sufficiently generic and/or $J_\mA$ is Cohen-Macaulay, i.e. it has a square-free monomial ideal.
\begin{equation}
  \text{rank}(\mI_{\mA}) \ge \text{degree} (J_{\mA}) = \Vol(\mA) 
\end{equation}
The singular locus of $\mI_{\mA}$ is the zero set of the {principal} $\mA$-determinant, S${}_{\mA}$, of the polynomials in $n$-variables, $\{P_1, \, P_2 \cdots P_m\}(x ; \,\mA ; \,z)$, defined by $\mA$. It is defined as the zero locus of the product of the polynomials from which the variables $x$ have been eliminated. Using the Cayley trick, which at the level of the polynomials boils down to the Feynman parameterisation, this can also be written in terms of one polynomial, $P_{\mA} = \sum_{i=1}^{m} y_i \, P_i\, (x, \,A_i ,\, z)$, in variables $x$ and $y$,
  \begin{equation}
  {\rm S}_\mA = \bigcup_{i=1}^{m}\{P_i \,(x,\, A_i,\,z) = 0\} \; \bigcap \; \CC \langle z \rangle = \{P_{\mA}\,(x, \,y, \,\mA , \,z) = 0\}  \bigcap \; \CC \langle z \rangle.
  \end{equation}
  This makes $S_\mA$ the resultant of $n+m$ functions $\{ \theta_x \,P_{\mA}, \, \theta_y \,P_{\mA} \}$ (i.e. the intersection of the zero loci of the polynomials and their derivatives). 
   \cite{Gelfand:Discriminants} is the prototypical textbook on this topic. 

  \subsubsection{Nilsson series}
  In order to construct series solutions of the Nilsson ring wrt $w$, the initial and indicial ideals, in${}_{(-w,\,w)}(\mI_\mA)$ and ind${}_{(w)}(\mI_\mA)$, need to be computed. This is computation has a huge time complexity in general, however subsets of these ideals, dubbed {fake initial} and {fake indicial ideals} respectively are easier to compute. $M$ is a monomial ideal wrt $\prec_{w}$ of the toric ideal $J_\mA$ and $\tilde M$ is its distraction.
  \begin{equation} \label{eq:fakeideals}
  \begin{aligned}
    M & = \text{in}_{(-w,\,w)}(J_{\mathcal A}),  && \text{in}^{\rm fake}_{(-w,\,w)} (\mI_{\mA}) = M + H_{\mA} \subseteq  \text{in}_{(-w,\,w)} (\mI_{\mA})
  \\ \tilde M  & = \text{ind}_{(w)} (J_{\mA}),  && \text{ind}^{\rm fake}_{(w)} (\mI_{\mA}) = \tilde M + H_{\mA}  \subseteq \text{ind}_{(w)} (\mI_{\mA})
  \end{aligned}
  \end{equation}
  When the vector $\gamma$ is held to be sufficiently generic, equality holds, making the roots of the fake indicial ideal, called {fake exponents}, equal to the exponents of the original indicial ideal. The monomial ideal, $M_1$, and its distraction, $\tilde M_1$, of $J_{1}$ in \cref{eq:runningexampleJid} wrt $w_1 = (0,1,1,0,0,0,0,1)$ are
  \begin{equation}
  M_1 = \left\langle\partial_2 \,\partial_5,  \, \partial_3 \,\partial_4, \,\partial_8^2 \right\rangle, \quad \tilde M_1 = \left\langle\theta_2 \,\theta_5,  \,\theta_3 \,\theta_4, \,(\theta_8)_{(2)} \right\rangle. \label{eq:fakeexample}
  \end{equation}
  A different choice of weight, for instance, $(0,1,1,1,1,1,1,0)$, would produce a different monomial ideal, $\left\langle\partial_2 \,\partial_5,  \, \partial_3 \,\partial_4,\, \partial_6 \,\partial_7 \right\rangle$ and distraction $\left\langle\theta_2 \,\theta_5,\,  \theta_3 \,\theta_4, \,\theta_6 \,\theta_7 \right\rangle$. However, the nature of the generic solution space will not change.

  When considering specific ranges of variables, it is useful to choose a weight with \cref{eq:weightchoice} in mind to ensure convergence. For example, granting some foreknowledge of the physically relevant values of $z$ \cref{eq:runningexmpphysicallyrelevantz} in \cref{exmp:runningexample}, a weight belonging to the class $w_2 = \{0,0,0,0,0,1,1,1\}$ can be considered, in line with \cref{eq:weightchoice}, resulting in
  \begin{equation}
  \begin{aligned}
    M_2 & = \langle \partial_1 \,\partial_7,  \;  \partial_1 \,\partial_6,  \; \partial_3 \,\partial_4 \,\partial_6, \; \partial_6 \,\partial_7, \; \partial_1 \,\partial_8^2, \; \partial_2 \,\partial_5 \,\partial_7^2 \rangle
    \\ \tilde M_2 & = \langle \partial_1\, \partial_6, \;  \partial_1\,\partial_7, \; \partial_3 \,\partial_4 \,\partial_6, \; \partial_6 \,\partial_7, \; \partial_2 \,\partial_5 \,\partial_7 \,(\partial_7 - 1) , \; \partial_1 \,\partial_8 \,(\partial_8 -1 )\rangle.
  \end{aligned}
  \end{equation}

  \subsubsection{Horn hypergeometric functions}
  The elements of the kernel of $\mathcal A$, $u \in \mathcal K \subset \mathbb Z^N$, satisfy $\mA \, u = 0$. Given a complete basis $\{u_i\}$, the span of $\mathcal K$ is represented as $\{t^i \, u_i\}$, $t_i \in \mathbb Z$. For notational ease, every element of the kernel, $u$, can be separated into two non-negative integer vectors $u_+, \, u_-$ such that $u = u_+ - u_-$ paralleling \cref{eq:Jideal}. A formal series solution, which isn't holomorphic in general, associated with a starting monomial $x^s$, with $s$ satisfying $\mA \, s + \gamma = 0$ is
  \begin{equation}
  I^{(s)} = \sum_{u \in \mathcal K} \, \frac {(s)_{(u_-)}} {(s + u)_{(u_+)}} \, x^{s + u}. \label{eq:general_formal_sol}
  \end{equation}
Formal solutions of $\mI_\mA$ take the form of Horn hypergeometric functions, hence the name. A multidimensional Horn hypergeometric series is a Laurent series $\sum \,c(\omega) \; x^{\omega}$, \; $\omega \in \mathbb Z^{N}, \; c(\omega) \in \CC^*$, such that there exist some non-zero rational functions $b_i : \,\mathbb Z^N \rightarrow \CC^*$, called $b$-function(s):
\begin{equation}
b_i(\omega)  = \frac {c (\omega + e_i)} {c (\omega)} \; \forall \; i \in [N]
\end{equation}
with $e_i$ being the standard basis vectors of $\mathbb Z^N$ and $b_i$ satisfying the consistency condition:
\begin{equation}
\tfrac {c (\omega + e_i + e_j)} {c (\omega)} = \tfrac {c (\omega + e_i)} {c (\omega)} \,\tfrac {c (\omega + e_i + e_j)} {c (\omega + e_i)} = \tfrac {c (\omega + e_j)} {c (\omega)} \,\tfrac {c (\omega + e_i + e_j)} {c (\omega + e_j)} \implies \tfrac {b_j(\omega + e_i)} { b_j(\omega) } = \tfrac {b_i(\omega + e_j)} {  b_i(\omega) }.  
\end{equation}
$b_i$ form a 1-cocycle from $\mathbb Z^N$ to $\CC^*$ where the group action is additive in abelian $\mathbb Z^N$ and multiplicative in $\CC^*$ (specifically a subset of $\CC^*$ spanning ratios of $c(\omega)$).

\subsubsection{Exponents and log-free series solutions}
A Gale transform, $\mathcal K \subset  \mathbb Z^N$, of $\mathcal A$ is a reduced basis of the kernel of $\mathcal A$ such that
\begin{equation}
\begin{aligned}
  & \mathcal K = \left( \begin{smallmatrix} u_1 & \cdots &  u_i & \cdots & u_K \end{smallmatrix} \right), \quad  \mA \, u_i = 0, \quad u_i{} \in \mathbb Z^N,
  \\ & \text{if} \; \mA \, v = 0, \quad \exists \, t \in \mathbb Z^{N} \; {\rm s.t.} \; v = \sum_{i=1}^{K}\,  t_i \,u_i.
\end{aligned}
\end{equation}
It is obviously non-unique. A Gale transform of $\mA_1, \, \mA'_1$ in \cref{eq:runningexample} is
\begin{equation}
\mathcal K_1 = (u_1 \; u_2 \; u_3) = \left( 
\begin{smallmatrix}
-1 & 1 & 0 & 0 & 1 & -1 & 0 & 0  \\ 
-1 & 0 & 1 & 1 & 0 & 0 & -1 & 0 \\
0 & 0 & 0 & 0 & 0 & -1 & -1 & 2
\end{smallmatrix} \right)^{\rm Tr}. \label{eq:runningexampleKernel}
\end{equation}
Thus a general term of $\mathcal K_1$ can represented as
\begin{equation}
v \equiv \{-(t_1 + t_2), \; t_1 , \; t_2 , \; t_2 , \; t_1 , \; -(t_1 + t_3) , \; - (t_2 + t_3) , \; 2 \, t_3\}, \quad t \in \mathbb Z^3.\label{eq:runningexampleKernelgeneralterm}
\end{equation}
A standard pair $(\partial^a, \, \sigma)$, $ a \in \NN_0^N$, $ \sigma \subseteq [N] $, of a monomial ideal $M = \langle \partial^a \rangle$, is defined such that
\begin{enumerate*}[label=(\roman*)]
  \item $a_i = 0$ $\forall \;  i \in \sigma$
  \item $\forall \; b_j \in \NN_{0}$, $ j \in \sigma$, $\partial^a{} \, \partial^b{} \notin M$
  \item $\forall \; l \notin \sigma$, there exists $b_j{} \in \NN_{0}$ such that $\partial^a \,\prod\limits_{l \in \sigma'} \,\partial_l^{b_l}{} \,\prod\limits_{j \in \sigma} \,\partial^{b_j} \in M$, where the sets of $\sigma$ are triangulations of the Newton polytope. 
\end{enumerate*} 
$\mathcal T(M)$ is the set of all standard pairs of a monomial ideal $M$. $M_1$ in \cref{eq:fakeexample} has the standard pairs
\begin{equation}\label{eq:runningexampletriangulationTM}
\begin{aligned}
\mathcal T(M_1) & = \{ 1, \{1,2,3,6,7\} \}, \; \{ 1, \{1,2,4,6,7\} \}, \; \{ 1, \{1,3,5,6,7\} \}, \; \{ 1,\{1,4,5,6,7\} \}, 
\\ &  \{ \partial_8, \{1,2,3,6,7\}  \}, \; \{ \partial_8, \{1,2,4,6,7\} \}, \; \{ \partial_8,  \{1,3,5,6,7\}  \}, \; \{ \partial_8, \{1,4,5,6,7\} \},
\end{aligned}
\end{equation}
and the associated triangulation of its Newton polytope is $\{1,2,3,6,7\}$, $\{1,2,4,6,7\}$, $\{1,3,5,6,7\}$, $\{1,4,5,6,7\}$, each with a volume of $16$. There are $16$ possible triangulations, each corresponding to a different monomial ideal. However, only $5$ of these monomial ideals are generated by just the representative terms in $J_1$. Although there do exist weights for which the initial forms of the Gröbner basis $\mathcal G_1$ are restricted to the initial forms of $J_1$, it is not the case in general.
\begin{figure}[htpb]
\centering
\includegraphics[width=\textwidth]{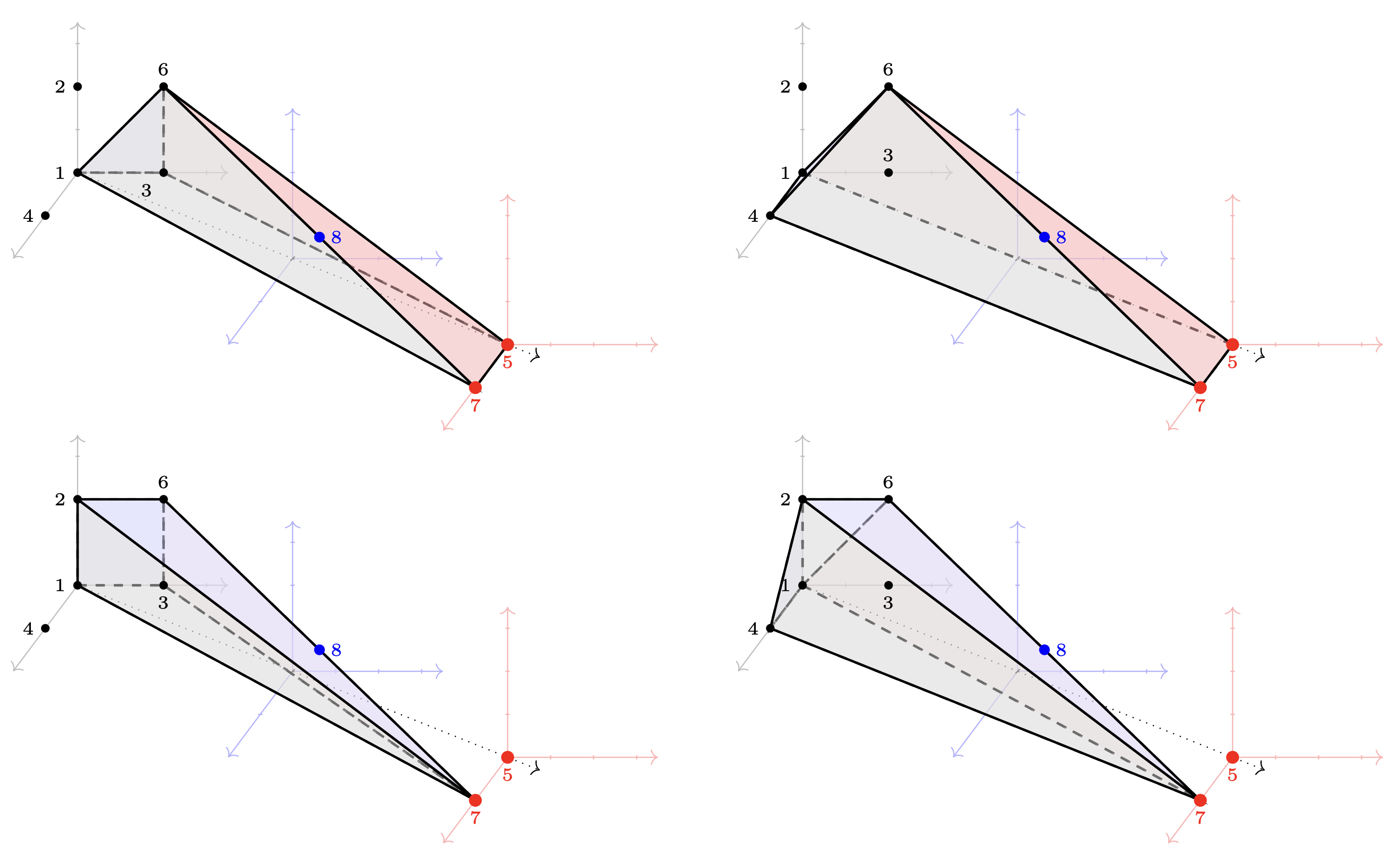}
  \caption{Triangulations of the Newton Polytope in \cref{fig:4DNPfirst}}
  \begin{flushleft}\singlespacing\vspace{-0.5\baselineskip}   
  The triangulations correspond to \cref{eq:runningexampletriangulationTM}. Each triangulation has vertex $8$ on one of its edges. 
  \end{flushleft}
  \end{figure}\noindent
  If the $n+m \times n+m$ sized matrices formed by columns of $\mA$, $\mA_{\sigma}$, that constitute a triangulation $(\partial^a, \, \sigma) \in \mathcal T(M)$ are non-singular (i.e. have non-zero determinants), the triangulation is {regular}. This can be confirmed for the case at hand. 
  When this determinant, $|\mA_{\sigma}| = 1$, the triangulation is called unimodular. When in$_{(w)}(J)$ is radical for generic weight $w$, i.e. square-free for generic weight vector $w$, the regular triangulation, $\sigma \in \Sigma_{w}$, of the Newton polytope is unimodular.
  \begin{equation}
  \text{in}_{(w)}(J) = \left\langle\partial_{i} \; | \; i \notin \Sigma_{w} \right\rangle = \bigcap_{\sigma \in \Sigma_w} \left\langle\partial_j \; | \; j \notin \sigma \right\rangle
  \end{equation}
  For non-resonant $\gamma$, the number of standard pairs equals the number of exponents and hence the rank of the system. So \cref{eq:runningexample} will have a solution basis of $8$ series. The primary ideal decomposition of $M$ in terms of $\mathcal T(M)$ is
  \begin{equation}
  M = \bigcap_{(\partial^a, \,\sigma) \in \mathcal T(M)} \langle \partial_i^{a_i + 1} \; | \; i \notin \sigma \rangle.
  \end{equation}
  The indicial ideal is based upon the decomposition of the monomial ideal and is hence equal to
  \begin{equation}
  \text{ind}_{w}(\mI) = \bigcap_{(\partial^a,\,\sigma) \in \mathcal T(M)} \langle (\theta_i - a_i ) \; | \; i \notin \sigma \rangle, \quad M = \text{in}_{(-w,\,w)}(\mI). \label{eq:indicialidealoftoricidealfromstandardpairs}
  \end{equation}
  So the decomposition of $M_1$ is
  \begin{equation}
  \begin{aligned}
  M_1 = \bigcap \; & \langle \partial_4,\,\partial_5,\,\partial_8 \rangle, \; \langle \partial_3,\,\partial_5,\,\partial_8 \rangle, \; \langle \partial_2,\,\partial_4,\,\partial_8 \rangle, \; \langle \partial_2,\,\partial_3,\,\partial_8 \rangle, 
  \\ &  \langle \partial_4,\,\partial_5, \,\partial_8^2 \rangle, \; \langle \partial_3,\,\partial_5,\,\partial_8^2 \rangle, \; \langle \partial_2,\,\partial_4,\,\partial_8^2 \rangle, \; \langle \partial_2,\,\partial_3,\,\partial_8^2 \rangle
  \end{aligned}
  \end{equation}
  and indicial ideal $\tilde M_1$ can be verified to be $\left\langle\theta_2\, \theta_5, \, \theta_3 \,\theta_4, \,(\theta_8)_2 \right\rangle$. The root/exponent, $s_{(\partial^a ,\, \sigma)}$, associated with a standard pair, $(\partial^a , \,\sigma)$ is defined as 
\begin{equation}
s_{(\partial^a , \,\sigma)}  \equiv \tilde s, \quad \mathcal A \cdot \tilde s + \gamma = 0, \quad \tilde s_i = a_i, \quad i \notin \sigma.
\end{equation}
Alternately, the (fake) indicial equation ind${}_{(w)}(\mI) = 0$ can be solved to find the (fake) exponents. 
For example, the indicial ideal in \cref{eq:fakeideals,eq:fakeexample} results in the exponents
\begin{equation}\label{eq:runningexampleRoots}
\left[\begin{smallmatrix}
s_1 =& - \beta_0 + \tfrac{1}{2} ({\beta_1}+{\beta_2} )  & \tfrac{{\beta_4}-{\beta_1}}{2} & \tfrac{{\beta_3}-{\beta_2}}{2} & 0 & 0
& -\tfrac{{\beta_4}}{2} & -\tfrac{{\beta_3}}{2} & 0 \\
s_2 =& - \beta_0 + \tfrac{1}{2} ({\beta_1}+{\beta_3}) & \tfrac{{\beta_4}-{\beta_1}}{2} & 0 & \tfrac{{\beta_2}-{\beta_3}}{2} & 0 &
-\tfrac{{\beta_4}}{2} & -\tfrac{{\beta_2}}{2} & 0 \\
s_3 =& - \beta_0 + \tfrac{1}{2} ({\beta_2}+{\beta_4}) & 0 & \tfrac{{\beta_3}-{\beta_2}}{2} & 0 & \tfrac{{\beta_1}-{\beta_4}}{2} &
-\tfrac{{\beta_1}}{2} & -\tfrac{{\beta_3}}{2} & 0 \\
s_4 =&  - \beta_0 + \tfrac{1}{2} ({\beta_3}+{\beta_4}) & 0 & 0 & \tfrac{{\beta_2}-{\beta_3}}{2} & \tfrac{{\beta_1}-{\beta_4}}{2} &
-\tfrac{{\beta_1}}{2} & -\tfrac{{\beta_2}}{2} & 0 \\
s_5 =&  -\beta_0 + \tfrac{1}{2} ({\beta_1}+{\beta_2}) & \tfrac{{\beta_4}-{\beta_1}}{2} & \tfrac{{\beta_3}-{\beta_2}}{2} & 0 & 0
& - \tfrac{ {\beta_4}+1 }{2}  & - \tfrac{{\beta_3}+1}{2}  & 1 \\
s_6 =&  - \beta_0 + \tfrac{1}{2} ({\beta_1}+{\beta_3}) & \tfrac{{\beta_4}-{\beta_1}}{2} & 0 & \tfrac{{\beta_2}-{\beta_3}}{2} & 0 &
-\tfrac{{\beta_4}+ 1}{2}  & - \tfrac{ {\beta_2} + 1}{2}  & 1 \\
s_7 =& - \beta_0 + \tfrac{1}{2} ({\beta_2}+{\beta_4}) & 0 & \tfrac{{\beta_3}-{\beta_2}}{2} & 0 & \tfrac{{\beta_1}-{\beta_4}}{2} &
-\tfrac{{\beta_1}+1}{2}  & - \tfrac{ {\beta_3}+1 }{2}  & 1 \\
s_8 =& - \beta_0 + \tfrac{1}{2} ({\beta_3}+{\beta_4}) & 0 & 0 & \tfrac{{\beta_2}-{\beta_3}}{2} & \tfrac{{\beta_1}-{\beta_4}}{2} &
-\tfrac{{\beta_1}+1}{2}  & -\tfrac{{\beta_2}+1}{2}  & 1 \\
\end{smallmatrix}\right].
\end{equation}
Each triangulation of the Newton polytope, or alternately, each possible monomial ideal will lead to a different {\it set} of exponents, however, some standard pairs and, hence, exponents may overlap. Thus, the set of all possible standard pairs $\bigcup \mathcal T(M)$ produces all possible pertubative expansions of the solution around the origin. Here, there are $54$ standard pairs in $\bigcup \mathcal T(M)$ and hence, just as many exponents. When the considered triangulation, $\sigma \in \Sigma_{(w)}$, wrt some weight $w$ is unimodular, the fake exponents are simply
\begin{equation}
s_{\sigma} = - \mA_{\sigma}^{-1} \,\gamma.
\end{equation}
The $b$-function for generic $\gamma$ and weight $w$ is 
\begin{equation}
b_{(w)}(\omega) = \prod_{s \in \alpha} (\omega - w \, s), \quad \alpha = \{s_{(\partial^{a}, \, \sigma)} \; | \quad (\partial^{a}, \, \sigma) \in \mathcal T(M\}
\end{equation}
where $\alpha$ is the set of (fake) exponents associated with the standard pairs in $\mathcal T(M)$. The parametric $b$-functions supply information regarding the singular hyperplanes of $\gamma$:
\begin{equation}\label{eq:bfunctionforcreationannihilationoperators}
\begin{aligned}
  b_i(\omega) = \prod_{s \in \alpha} \,(\omega - s_i), \quad \alpha = \{s_{(\partial^{a}, \, \sigma)} \; | \quad (\partial^{a}, \, \sigma) \in \bigcup \mathcal T(M) \}.
\end{aligned}
\end{equation}
A log-free series solution with the starting monomial $z^s$ is
\begin{equation}
   I^{(s)}(z) 
   =  \sum_{u \in \mathcal K} \, \frac {z^{s + u}} {(s + u)_{(u)}}  = z^{s} \, \sum_{u \in \mathcal K} \, z^{u} \, \left( \frac {\Gamma(s+1)} {\Gamma(u + s + 1)} \quad \text{or} \quad \frac{(-1)^{u} \, \Gamma(- u - s) }{ \Gamma(-s)} \right)\label{eq:general_logfree_sol}
   \end{equation}
   where the summation range of $u$ can be set by requiring the $\Gamma$ functions in the denominator to have arguments in $\CC^* \backslash \mathbb Z^-$.

   Since the number of standard pairs equals the rank, the set of log-free series solutions $I^{(\alpha)} = \{I^{(s)} \; | \; s \in \alpha\}$ form the Nilsson ring for {\it generic} $\gamma$, with each starting monomial being different by definition making the generated series linearly independent. e.g.
   \begin{equation}
   I_1^{(k)} = \sum_{t_1,\, t_2, \, t_3 \in \mathbb Z} \, \prod_{i=1}^{8} \,  \frac {\Gamma(s_{k}{}_i{}+1)} {\Gamma(u_1{}_i{} \, t_1 + u_2{}_i{} \, t_2 + u_3{}_i{} \, t_3 + s_{k}{}_i{} + 1)} \, z_i^{s_k{}_i{} + u_1{}_i{} \, t_1 + u_2{}_i{} \, t_2 + u_3{}_i{} \, t_3}\label{eq:runningexampleGenericSol}
   \end{equation}
   which can be concisely denoted as
   \begin{equation}
   I_1^{(k)} = z^s \,\sum_{t\in \mathbb Z^3} \frac {\Gamma(s_k+1)} {\Gamma(u \cdot t + s_{k}{} + 1)} \, \omega^{t}, \quad \omega \equiv \{z^{u_1}, \, z^{u_2}, \, z^{u_3} \} = (\tfrac {z_2 \, z_5} {z_1\, z_6}, \, \tfrac {z_3 \, z_4} {z_1\, z_7}, \,  \tfrac {z_8^2} {z_6\, z_7} )\label{eq:runningexampleToricInvariants}
   \end{equation}
   where $\omega$ are the toric invariants.

Specific $\gamma$ vectors may have special features, which may prevent the direct translation of the generic solution. If an exponent $s \in \NN^{N}$ for some arbitrary $\gamma$, the associated canonical series is finite, i.e. polynomial. 

A fake exponent that isn't an exponent has at least one negative integer component for some given $\gamma$. The negative support of a fake exponent is defined as
\begin{equation}
\text{nsupp} (s)  = \{ i \in \{1, \, \cdots \, N\} \; | \; s_i \in \mathbb Z^{-}  \}.
\end{equation}
$s$ has {\it minimal} negative support iff there is no $u \in \mathcal K$ such that nsupp$(s + u) \subset$ nsupp$(u)$. The series solution for roots with non-empty minimal negative supports are not directly defined by \cref{eq:general_logfree_sol}, since the coefficients would hit poles of the $\Gamma$ functions and become ill-defined, but instead by \cref{eq:general_formal_sol} with the summation range of $u$ restricted to $\mathcal K^{(s)} \subset \mathcal K$:
\begin{equation}
\mathcal K^{(s)} = \{ u \in \mathcal K \; | \; \text{nsupp} (s) = \text{nsupp} (s + u ) \}.
\end{equation}
The log free series solution consists of the series $\{I^{(s)}\}$, where $s$ belongs to the set of roots with minimal negative support. For example, the arbitrary (and physically relevant) choice of
\begin{equation}
\gamma_1 = \{\tfrac {d} {2}+1, \, \bar \Delta_1 = \tfrac {d} {2} - \im \nu_1, \, \bar \Delta_2 = \tfrac {d} {2} - \im \nu_2, \, \Delta_1 =  \tfrac {d} {2} + \im \nu_1, \, \Delta_2 = \tfrac {d} {2} + \im \nu_2 \}, \; \; \nu_+ \equiv \tfrac {\nu_1 + \nu_2} {2}  \label{eq:2pexampleGammaVec}
\end{equation}
in \cref{eq:runningexamplegenericgammas} produces fake exponents with negative integer terms:
\begin{equation}
\left[\begin{smallmatrix}
\alpha_1 = & -1 - \im \nu_{+} & \im \nu_+ & \im \nu_+ & 0 & 0 & -\tfrac{{\Delta_2}}{2} & -\tfrac{{\Delta_1}}{2} & 0 \\
\alpha_2 =& -1 & \im \nu_+ & 0 & -\im \nu_+ & 0 & -\tfrac{{\Delta_2}}{2} & -\tfrac{{\bar \Delta_2}}{2} & 0 \\
\alpha_3 =& -1 & 0 & \im \nu_+ & 0 & -\im \nu_+ & -\tfrac{{\bar \Delta_1}}{2} & -\tfrac{{\Delta_1}}{2} & 0 \\
\alpha_4 =& -1 + \im \nu_{+} & 0 & 0 &-\im \nu_+ & -\im \nu_+ & -\tfrac{{\bar \Delta_1}}{2} & -\tfrac{{\bar \Delta_2}}{2} & 0 \\
\alpha_5 =& -1 - \im \nu_{+} & \im \nu_+ & \im \nu_+ & 0 & 0 & - \tfrac{ {\Delta_2}+1 }{2}  & - \tfrac{{\Delta_1}+1}{2}  & 1 \\
\alpha_6 =& -1 & \im \nu_+ & 0 & -\im \nu_+ & 0 & -\tfrac{{\Delta_2}+ 1}{2}  & - \tfrac{ {\bar \Delta_2} + 1}{2}  & 1 \\
\alpha_7 =& -1 & 0 & \im \nu_+ & 0 & -\im \nu_+ & -\tfrac{{\bar \Delta_1}+1}{2}  & - \tfrac{ {\Delta_1}+1 }{2}  & 1 \\
\alpha_8 =& -1 + \im \nu_{+} & 0 & 0 & -\im \nu_+ & -\im \nu_+ & -\tfrac{{\bar \Delta_1}+1}{2}  & -\tfrac{{\bar \Delta_2}+1}{2}  & 1 \\
\end{smallmatrix}\right].\label{eq:2parbitraryroots}
\end{equation}
Assuming $d, \, \nu_{1,2}$ are generically valued, four of the log-free solutions, $\{ I^{(1,4,5,8)} \}$, are exponents. Fake exponents $\alpha_{(2,3,6,7)}$ have the same negative support $\{1\}$. It is minimal because no element of $\mathcal K_1$, \cref{eq:runningexampleKernelgeneralterm}, can shift the roots such that the size of their negative support set will reduce, i.e. there exists no $u' \in \mathcal K_1$ such that nsupp$(\alpha_{(2,3,6,7)} + u' ) \neq \emptyset$, the only possible subset of $\{1\}$.

Requiring the negative support to remain the same restricts the kernel to \cref{eq:runningexampleKernelgeneralterm} with $t \in \NN_0^{3}$. Thus, the series solution corresponding to $\alpha_{(2)}$ is
\begin{equation}
\begin{aligned}
  I_1^{(2)}& = z^{\alpha_2} \sum_{t \in \NN_0^{3}} \tfrac {(t_1 + t_2)!} {t_1! \, t_2! \, 2 \, t_3!} \tfrac {\Gamma(1 + \im \nu_+)} {\Gamma(t_1 + 1 + \im \nu_+)} \tfrac {\Gamma(1 - \im \nu_+)} {\Gamma(t_2 + 1 - \im \nu_+)} \tfrac {\Gamma(t_1 + t_3 + \frac{{\Delta_2}}{2})} {\Gamma(\frac{{\Delta_2}}{2})} \tfrac {\Gamma(t_2 + t_3 + \frac{{\bar\Delta_2}}{2})} {\Gamma(\frac{{\bar\Delta_2}}{2})} \, \omega^{t}.
\end{aligned}
\end{equation}
In order to construct the complete Nilsson ring if there are fake exponents without negative support or overlapping exponents, a generic deformation to $\gamma$ in the form of $\gamma + \epsilon \,\gamma'$ can be considered and the initial terms/monomials expanded in orders of small $\epsilon$. Appropriate linear combinations of these terms give distinct initial terms, with the terms of lowest order in $\epsilon$ serving as starting monomials to linearly independent solutions in the Nilsson ring. 
This is implemented in the example by giving a small generic deformation to $\gamma_1$ in \cref{eq:runningexamplegenericgammas,eq:2pexampleGammaVec}, say $\beta_0 \rightarrow \beta_0 + \epsilon$, to forego any considerations regarding fake exponents.

\subsubsection{Normalisation constants}
Non-degenerate limiting values of the Euler integral can be used to find the normalisation constants of the series solutions $I^{(s)}$. These normalisation constants take the place of boundary conditions that typically appear in solutions to differential equations. 

A direct approach to finding them is to consider $z \rightarrow 0$ limits matching the zeroes of the exponents, assuming the integrals in the limiting cases are convergent on the relevant contour.
\begin{equation}
\lim_{z_j \rightarrow 0} I = \lim_{z_j \rightarrow 0} \, N_i \, I^{(s_i)}, \quad (s_i)_{j} = 0 \; \forall \; j
\end{equation}
For example, root $s_1$ suggests the limit $\{ z_4, \, z_5, \, z_8 \} \rightarrow  0$. In this limit, the toric invariants $\omega = 0$, so all the series collapse to just their initial terms $z^{s}$. The {\it generalized} Euler integral (i.e. in terms of the remaining variables $z$) in this limit can be uniquely identified with the initial series of $I_1^{(1)}$ on the basis of their monodromy exponents around $\{z=0\}$. Consequently, they are indicative of the singular hyperplanes of $\gamma_1$ at which the particular series will diverge.

With the integration range chosen to be $x \in (\RR_+)^4$, the normalisation constant $N_1$ of the series $I_1^{(1)}$ turns out to be
\begin{equation}
N_1 = \frac{\Gamma \left(\frac{\beta_3}{2}\right)\,  \Gamma \left(\frac{\beta_4}{2}\right) \, \Gamma \left(\frac{\beta_1- \beta_4}{2}\right) \, \Gamma \left(\frac{\beta_2 -\beta_3}{2}\right) \, \Gamma \left( \gamma_0 - \frac{\beta_1+\beta_2}{2} \right)}{16 \, \Gamma \left(\beta_0 \right)}.
\end{equation}
Similarly, as suggested by the exponent $s_5$, the $\{z_4, \, z_5 \} \rightarrow  0$ limit of the integral can be evaluated and upon comparison of the monodromy exponents with those of the initial series, the integral is found to be dependent only on $I_1^{(1)} \oplus I_1^{(5)}$. Given the previous piece of information, $N_5$ is then
\begin{equation}
N_5 = - \frac{\Gamma \left(\frac{\beta_3+1}{2}\right) \, \Gamma \left(\frac{\beta_4+1}{2}\right) \, \Gamma \left(\frac{\beta_1- \beta_4}{2}\right) \, \Gamma \left(\frac{\beta_2 -\beta_3}{2}\right) \, \Gamma \left( \gamma_0 - \frac{\beta_1+\beta_2}{2} \right)}{16 \, \Gamma \left(\beta_0 \right)}.
\end{equation}
This example is relatively simple since the only non-radical term in the generators of in${}_{w}(J_1)$ is $\partial_8^2$, which constrains the inductive process of finding these normalisation constants to a tolerable depth of $1$. This can be bypassed by requiring $\partial_{z_8} I_1 = \sum\,  N_i \, \partial_{z_8}  \, I_1^{(i)}$ in the limit $\{z_4, \, z_5, \, z_8\} \rightarrow 0$, which will once again allow unique identification of the LHS with the starting monomial of $I_1^{(5)}$. In general, the minimal system of linear equations supplying the normalisation constants is:
\begin{equation}
\lim_{z_j \rightarrow 0} \partial_{z_j}^{(s_i{})_j} \, I = \lim_{z_j \rightarrow 0} \, N_i \, \partial_{z_j}^{(s_i{})_j} \, I^{(s_i)} = \lim_{z_j \rightarrow 0} \, N_i \, (s_i{})_j! \, I^{(s_i)}, \quad (s_i)_{j} \in \NN_0 \; \forall \; j.
\end{equation}
The symmetry structure of the exponents in \cref{eq:runningexampleRoots} can be used to deduce the remaining:
\begin{equation}
\begin{aligned}
  & N_{2/6} = N_{1/5}\Big|_{\beta_2 \leftrightarrow \beta_3}, && N_{3/7} = N_{1/5}\Big|_{\beta_1 \leftrightarrow \beta_4}, && N_{4/8} = N_{1/5} \Big|_{\beta_2 \leftrightarrow \beta_3, \; \beta_1 \leftrightarrow \beta_4}.
\end{aligned}
\end{equation}

\subsubsection{Restriction of solution spaces}\label{sec:Restriction_of_solution_spaces}
Although solution spaces of $\mI \subset D_N \equiv \CC\langle z, \, \partial_z \rangle$ are defined in terms of generic variables $z$, more often than not, interest is limited to only specific values thereof. This implies that the modules, $\mathcal D = \mathcal D_N / \mI$, need to be restricted to particular slices of $z$, say $\{z_k = \bar z_k\}$. The corresponding restriction ideal $\mI'$ is the intersection of $(\mI + {\rm R}_z)$, ${\rm R}_z \equiv \CC\langle z_k - \bar z_k \rangle$, with $D_{N - k}$, and has a finite Gröbner basis. If $\mI$ is holonomic, so is $\mI'$, and this property of holonomy presents itself in $\mathcal D$ and its restriction by ${\rm R}_z$, $\mathcal D'$, too. \cite{OAKU199761,oaku2001algorithms} are referred to for an algorithm to restrict by $\{z_k = 0\}$. Restriction of Pfaffian systems of Feynman integrals, specifically by $\{z_k = 1\}$, is discussed in \cite{Chestnov:2022alh,Chestnov:2023kww} as means of reducing the rank of the solution space. 

The naive approach of restricting the generic solution $I^{(s)}(z)$ of $\mI$ to the subspace $\{z_k = \bar z_k\}$ remains valid. Granting that the Euler integral at $I(\bar z_k)$ is convergent, even if individual series $I^{(s)}(\bar z_k)$ diverge, their sum will be necessarily converging (see \cref{sec:limitingvalsofhypfns}). For example, \cref{eq:runningexampleint} in a physically relevant (see \cref{sec:2melonfeynmanint}) limit is to be restricted to
\begin{equation} \label{eq:runningexmpphysicallyrelevantz}
\begin{aligned}
  z = \{1,\,1,\,1,\,1,\,1,\,1,\,1,\,-2\}, \quad \omega = \{1, \,1,\, 4\}.
\end{aligned}
\end{equation}
In order to find a $\log$-free solution, the integral is evaluated by considering the entire generic solution space as follows: A general term of the kernel $\mathcal K_1$, \cref{eq:runningexampleKernelgeneralterm}, is symmetric under the exchanges $\upsilon_3 \leftrightarrow \upsilon_4$ and $\upsilon_2 \leftrightarrow \upsilon_5$, translating to $\beta_2 \leftrightarrow \beta_3$ and $\beta_1 \leftrightarrow \beta_4$ respectively in terms of the roots in \cref{eq:runningexampleRoots}. Relabeling the parameters for notational convenience as $\rho \equiv \{ \beta_1, \, \beta_4 \}$, $\sigma \equiv \{ \beta_2, \, \beta_3 \}$, $\bar \sigma \equiv \sigma_1 - \sigma_2$, $\bar \rho \equiv \rho_1 - \rho_2$, a general root can be represented as
\begin{equation}
s_{j,\,k,\,\Theta} = \{ \tfrac{\rho_j + \sigma_k}{2} - \beta_0  , \; (-1)^j \tfrac{{\bar\rho}}{2} , \; (-1)^k  \tfrac{{\bar\sigma }}{2} , \; 0 , \; 0 , \; -\tfrac{{\rho_j + (-1)^j \bar\rho + \Theta}}{2} , \; -\tfrac{{\sigma_k + (-1)^k \bar\sigma + \Theta}}{2}, \; \Theta \},
\end{equation} 
where $j,\,k \in \{1,\,2\}$ and $\Theta = \{0,\,1\}$. The initial term corresponding to a root labelled $s_{j,\,k,\,\Theta}$ is $(-2)^{\Theta}$. To ensure genericity, $\gamma_0$ is given a small deformation $\beta_0 \rightarrow \tfrac {d+2} {2} + \epsilon$, shifting the negative integers appearing in \cref{eq:2parbitraryroots}.

Summing over $t_1$ and setting $\{ z_1, \,z_2, \,z_5, \,z_6 \} = 1$ in the generic solution reduces the holonomic rank of the characteristic variety of $\mI_1$, ch$(\mI_1)$, by $4$ to $4$ (picking a partial tie-breaking order $z_1, \,z_5, \,z_6 \succ z_4,\, z_7$ and $z_2 \succ z_3, \,z_8$ and applying \cref{eq:holoranksimpleprocess} makes this clear). In particular, the series corresponding to roots $s_{1,\,k,\,\Theta}$ and $s_{2,\,k,\,\Theta}$ no longer remain independent for each $\{ k, \, \Theta \}$, with the divergent parts cancelling against each other order-by-order. It is useful to note at this stage that the series become well-defined and non-degenerate in the $\epsilon \rightarrow 0$ limit, thus allowing straightforward subsequent evalution (\cref{app:2propmelonExtraDetails}), resulting in
\begin{equation}\label{eq:2Pexamplefinalresultforarbitraryparameters}
I_{1} \prop  (\sin (\pi \, \Delta_1) + \sin (\pi \, \bar \Delta_1)) \,\Gamma (\Delta_1)\, \Gamma (\bar\Delta_1 ) - (\sin (\pi \, \Delta_2) + \sin (\pi \, \bar\Delta_2)) \,\Gamma (\Delta_2 ) \,\Gamma ( \bar\Delta_2 ).  
\end{equation}

\subsubsection{Limiting values of hypergeometric functions}\label{sec:limitingvalsofhypfns}
Convergence of an Euler integral, $I(z)$, must obviously also be accompanied by the convergence of the series solution representing it, $\sum N_s \, I^{(s)}(z)$. Thus, even if individual series $I^{(s)}(\bar z)$ may appear to diverge when analytically evaluated at specific limiting values of the variables $z = \bar z$, these divergences precisely cancel out against each other. This can be observed by studying the divergent parts of each series $I^{(s)}(\bar z)$ order by order.

For example, at the simplest order, this can be observed by using the Euler transformation of the classical Gauss hypergeometric function ${}_2F_1$:
\begin{equation}\label{eq:2f1eulertransformation}
\begin{aligned}
  {}_2F_1(a_1, \, a_2 \, ; b \, ; x) & = \tfrac {\Gamma(b) \,\Gamma(b - a_1 - a_2)} {\Gamma(b - a_1)\, \Gamma(b - a_2)} \, {}_2F_1(a_1, \, a_2 \, ; a_1 + a_2 + 1 - b \, ; 1 - x) 
  \\ & + \tfrac {\Gamma(b) \,\Gamma(a_1 + a_2 - b)} {\Gamma(a_1)\, \Gamma(a_2) \, (1 - x)^{a_1 + a_2 - b}} \,  \, {}_2F_1 (b - a_1, \, b - a_2 \, ; 1 + b - a_1 - a_2 \, ; 1 - x) .
\end{aligned}
\end{equation}
When $\Real(b - a_1 - a_2) > 0$, the series has a well known limiting value at $x = 1$. However, even when such is not the case i.e. $\Real(b - a_1 - a_2) \le  0$, the series representation can be separated into convergent and divergent parts. A general univariate hypergeometric function ${}_pF_q(a;\,  b; \, x)$ is 
\begin{equation}
{}_pF_q(a_1, \, a_2, \cdots a_p; b_1, \,  b_2, \cdots b_q; x) \vcentcolon= \sum_{n=0}^{\infty} \tfrac {a^{(n)}} { b^{(n)} } \, \tfrac {x^n} {n!} = \sum_{n=0}^{\infty} \tfrac {a_1^{(n)} \, a_2^{(n)} \cdots \, a_p^{(n)}} { b_1^{(n)} \, b_2^{(n)} \cdots \, b_q^{(n)} } \, \tfrac {x^n} {n!},
\end{equation}
normalised to equal $1$ at $x = 0$. If $\Real(\sum b - \sum a) \le 0$, $\;{}_pF_q(a;\,  b; \, x)$ diverges as $x \rightarrow 1$. As suggested in \cite{buehring2003partial}, using the limiting value of ${}_2F_1$ at $x = 1$, $\;{}_{p+1}F_p(a;\,  b; \, x)$ can be rewritten as
\begin{equation}
\begin{aligned}
  {}_{p+1}F_{p}(a; b; x) & = \sum_{n=0}^{\infty} \tfrac {a_1^{(n)} \, \cdots \, a_{p}^{(n)}} { b_1^{(n)} \cdots \, b_{p-2}^{(n)} } \,\tfrac {\Gamma(b_{p})\,\Gamma(b_{p-1})} {\Gamma(a_{p+1})} \, \tfrac {{}_2F_1(b_{p} - a_{p+1}, \; b_{p-1} - a_{p+1}, \; b_{p} + b_{p-1} - a_{p+1} + n, \; 1)} {\Gamma(b_{p} + b_{p-1} - a_{p+1}  + n )}  \, \tfrac {x^n} {n!}
  \\ & = \sum_{n=0}^{\infty} \tfrac {a_1^{(n)} \, \cdots \, a_{p}^{(n)}} { b_1^{(n)} \cdots \, b_{p-2}^{(n)} } \,\tfrac {\Gamma(b_{p}) \,\Gamma(b_{p-1})} {\Gamma(a_{p+1})} \, \tfrac {{}_2F_1(a_{p+1} - b_{p}, \; a_{p+1} - b_{p-1}, \; a_{p+1} + n, \; 1)} {\Gamma(b_{p} + b_{p-1} - a_{p+1} + n)}  \, \tfrac {x^n} {n!}.
\end{aligned}
\end{equation}
and the ${}_2F_1$ function reexpanded to find a recursion relation
\begin{equation}
\begin{aligned}
  {}_{p+1}F_{p}(a; b; x) & = \sum_{m = 0}^{\infty} \tfrac { \Gamma(b_{p})\,\Gamma(b_{p-1})\, \Gamma(b_{p} - a_{p+1} + m) \,\Gamma(b_{p-1} - a_{p+1} + m) } {\Gamma(a_{p+1})  \,\Gamma(b_{p-1} - a_{p+1}) \,\Gamma(b_{p} - a_{p+1}) \,\Gamma(b_{p} + b_{p-1} - a_{p+1} + m) \, m!}   
  \\ & \phantom{\sum_{m = 0}^{\infty} } \times {}_{p}F_{p-1}\left( \begin{smallmatrix}
       a_1, \, \cdots \, a_{p} \\ b_1, \cdots \, b_{p-2}, \,  (b_{p} + b_{p-1} - a_{p+1} + m) 
      \end{smallmatrix} \Big| x \right)
\end{aligned}
\end{equation}
that can be inductively carried on to ${}_2F_1$, allowing the divergences of ${}_{p+1}F_{q}$ functions at unit argument to be separated as series in $(1 - x)$ with initial terms $\{1, \, (1 - x)^{\sum b - \sum a} \}$, and when $(\sum b - \sum a)$ is an integer, also $\log(1 - x)$. It is expected that this generalisation can be used find the aforementioned cancellations of divergences. For example, see \cref{app:2propmelonExtraDetails}.

\subsubsection{Creation and annihilation operators}
Appropriate combinations of the shift relations induced by derivatives $\partial_z$ form the toric ideal and annihilate the integral, with each individually acting as raising operators on the vector $\gamma$, specifically a non-negative integer vector $u$ in \cref{eq:generalderivativeactiononeulerint} induces shifts of the form:
\begin{equation}
\begin{aligned}
  & \partial_{z^{(j)}_k{}} \circ I (z) : && \{\alpha, \, \beta\}  \rightarrow \{\alpha + 1, \, \beta + A^{(j)}_{k}\}
  \\ & \partial^{u} \circ I (z): && \{\alpha, \, \beta\}  \rightarrow \{\alpha + \sum u_j{}, \, \beta + \sum_{j = 1}^{m} A^{(j)} \cdot u_{j}\}.
\end{aligned}
\end{equation}
The homogeneity relations \cref{eq:Hideal} can be viewed as products of these raising operators and their inverses, i.e. operators that induce downward shifts in the $\gamma$ vector, called `creation' operators \cite{10.2748/tmj/1178227247,SAITO_STURMFELS_TAKAYAMA_1999}. Since the GKZ ideal is the maximal ideal in $D_N$, all relations satisfied by the Euler integral can and will be encoded within it, and so these creation operators can be found from the GKZ system. Recently reviewed in \cite{Caloro:2023cep}, the fundamental idea behind the construction of creation operators is to find a relation of the form
\begin{equation}
\begin{aligned}
  \mathcal C_i \, \partial_i \, I & = b_i(\gamma) \, I = b_i(\theta) \, I 
\end{aligned}
\end{equation}
where the $b_i(\gamma)$, \cref{eq:bfunctionforcreationannihilationoperators}, describes singular hyperplanes of the parameter vector $\gamma$ away from the vertex $\mathcal A_i$ of the Newton polytope. $b_i(\theta)$ is found by using the homogeneity relations, \cref{eq:Hideal}, and then creative use of the toric ideal, \cref{eq:Jideal}, results in a form divisible by $\partial_i$, thus allowing identification of $\mathcal C_i$.

For example, consider the integral with integrand, $\mathscr I$, and it associated GKZ equations:
\begin{equation}
\begin{aligned}
  \mathscr I & = \frac {\lambda^{\Delta} } {(z_0 + z_1  \,\lambda_1 + z_2  \,\lambda_2 + z_3  \,\lambda_1  \,\lambda_2)^{\alpha}}
  \\ \text{GKZ ideal} & = \sum \theta + \alpha, \quad \theta_1 + \theta_3 + \Delta_1, \quad \theta_2 + \theta_3 + \Delta_2, \quad \partial_1 \partial_2 - \partial_0 \partial_3.
\end{aligned}
\end{equation}
Taking any one of the scaling symmetry equations to start, say $(z_1 \,\partial_1 + z_3 \,\partial_3 ) \,\mathscr I (\Delta_1, \,\Delta_2, \,\alpha) = - \Delta_1 \,\mathscr I (\Delta_1, \,\Delta_2, \,\alpha)$, the intention is to convert it into the form: $\mathcal C_i \, \partial_i \mathscr I (\gamma) = f(\gamma, \,\gamma') \,\mathscr I (\gamma')$, enabling the identification of an operator $\mathcal C$ that shifts the $\gamma$ vectors instead of holding it constant like the GKZ equations are designed to. Applying $\partial_0$ on both sides of the chosen equation:
\begin{equation}
\begin{aligned}
  & \partial_0 \,(z_1 \,\partial_1 + z_3 \,\partial_3 )\, \mathscr I (\Delta_1, \,\Delta_2, \,\alpha) 
  = - \Delta_1 \,(-\alpha) \,\mathscr I (\Delta_1, \,\Delta_2, \,\alpha + 1)
  \\ \implies & (z_1  \,\partial_0 + z_3 \, \partial_2 ) \,\partial_1 \mathscr I (\Delta_1, \,\Delta_2, \,\alpha) = (z_1 \, \partial_0 + z_3  \,\partial_2 )\, (-\alpha)\,\mathscr I (\Delta_1+1, \,\Delta_2, \,\alpha+1) 
  \\ & = - \Delta_1 \,(-\alpha) \,\mathscr I (\Delta_1,\, \Delta_2, \,\alpha + 1).
\end{aligned}
\end{equation}
Following this process, the creation operators can be read off from:
\begin{equation}
\begin{aligned}
  & (z_1  \,\partial_0 + z_3 \, \partial_2 )\, \mathscr I (\Delta_1+1) = - \Delta_1 \,\mathscr I (\Delta_1)
  \\ & (z_2  \,\partial_0 + z_3  \,\partial_1 ) \,\mathscr I (\Delta_2+1) = - \Delta_2 \,\mathscr I (\Delta_2)
  \\ & (z_1  \,\partial_0 + z_3  \,\partial_2 ) \,(z_0\, \partial_1 +  z_2 \,\partial_3 ) \,\mathscr I (\alpha+1) = - \Delta_1 \,(\Delta_1 - \alpha) \,\mathscr I (\alpha).
\end{aligned}
\end{equation}
A simple way to envision the singular hyperplaces of $\gamma$ is via the poles of the normalisation constants, though the information found in each one if far come complete. For example, some variable rescalings change \cref{eq:runningexampleint} to 
\begin{equation}
\begin{aligned}
  & I_{1}  = \int \tfrac {\de x} {x} \tfrac{ z_1^{- \beta_0 + \frac {\beta_1 + \beta_2 + \beta_3 + \beta_4} {2}} \, z_2^{- \frac {\beta_1} 2} \, z_3^{- \frac {\beta_2} 2} \, z_4^{- \frac {\beta_3} 2} \, z_5^{- \frac {\beta_4} 2} \; x^{\beta} } { ( 1 +  x_1^2 + x_2^2 + x_3^2 + x_4^2 + q_1 \,x_1^2 \,x_4^2 + q_2 \, x_2^2 \,x_3^2 + q_3 \,x_1 \,x_2 \,x_3\, x_4)^{\beta_0} }, 
\end{aligned}
\end{equation}
where $ \{q_1, \, q_2, \, q_3 \} \equiv \{\tfrac {z_1\, z_6}{z_2\, z_5}, \, \tfrac {z_1\, z_7}{z_3\, z_4}, \, \tfrac {z_1 \, z_8}{\sqrt{z_2\, z_3\, z_4\, z_5}} \}$. This makes some relevant singular surfaces of the $\gamma$ vector evident, namely $(\beta_0 - \frac {\sum_{1}^{4} \beta}{2}), \, \frac {\beta_1}{2}, \, \frac {\beta_2}{2}, \, \frac {\beta_3}{2}, \, \frac {\beta_4}{2} = 0$. Considering the exponents in the rescaled integral, the homogeneous ideal, $\mathcal H_1$, simply becomes $\langle \theta_6 - \theta_7, \, 2 \, \theta_6 + \theta_8 \rangle$, i.e. the integral is purely a function of $\frac {z_8^2}{z_6 \, z_7} = \frac {q_3^2}{q_1 \, q_2}$, with the overall normalisation $\frac {\Gamma(\beta_0 - \frac {\sum_{1}^{4} \beta}{2})\, \Gamma(\frac {\beta_1}{2}) \, \Gamma(\frac {\beta_2}{2}) \, \Gamma(\frac {\beta_3}{2}) \, \Gamma(\frac {\beta_4}{2})} {16 \, \Gamma(\beta_0)}$, found in the $\{q_1, \, q_2, \, q_3 \} \rightarrow 0$ limit. It can be deduced from the arguments of the $\Gamma$ functions in the normalisation that the integral has singularities not only at the aforementioned surfaces but at all negative integer shifts thereof.

\subsection{Series solutions to some Feynman integrals}

\subsubsection{\texorpdfstring{$2$}{2}-Melon: series}\label{app:2propmelonExtraDetails}
Summing over $t_1$ of the series solutions results in:
\tiny
\begin{equation*}
\begin{aligned}
\phi_1 & = \tfrac{\Gamma \left(\frac{\bar\Delta_1}{2}-\frac{\Delta_2}{2}\right) \Gamma \left(\frac{\bar\Delta_2}{2}-\frac{\Delta_1}{2}\right) \Gamma \left(-\frac{\bar\Delta_2}{2}+\frac{\Delta_1}{2}+1\right)  \Gamma \left(t_3+\frac{\Delta_2}{2}\right) \Gamma \left(t_2+t_3+\frac{\Delta_1}{2}\right)  \Gamma \left(t_2+\frac{\mathbb{D}}{2}-\frac{\bar\Delta_1}{2}-\frac{\bar\Delta_2}{2}\right)}{16 \Gamma \left(\frac{\mathbb{D}}{2}\right) \Gamma (t_2+1) \Gamma (2 t_3+1) \Gamma \left(t_2-\frac{\bar\Delta_2}{2}+\frac{\Delta_1}{2}+1\right)}
 \\ & \times z_3^{-\frac{\bar\Delta_2}{2}+\frac{\Delta_1}{2}+t_2}  z_4^{t_2} z_7^{-\frac{\Delta_1}{2}-t_2-t_3} z_8^{2 t_3} \omega_1^{\frac{\Delta_2}{2}+t_3} \,_2F_1\left(t_2+\tfrac{\mathbb{D}}{2}-\tfrac{\bar\Delta_1}{2}-\tfrac{\bar\Delta_2}{2},t_3+\tfrac{\Delta_2}{2};-\tfrac{\bar\Delta_1}{2}+\tfrac{\Delta_2}{2}+1;\omega_1\right)
 \\ \phi_2 & = \tfrac{ \Gamma \left(\frac{\bar\Delta_1}{2}-\frac{\Delta_2}{2}\right) \Gamma \left(\frac{\bar\Delta_2}{2}-\frac{\Delta_1}{2}+1\right) \Gamma \left(\frac{\Delta_1}{2}-\frac{\bar\Delta_2}{2}\right) \Gamma \left(t_3+\frac{\Delta_2}{2}\right) \Gamma \left(t_2+t_3+\frac{\bar\Delta_2}{2}\right) \Gamma \left(t_2+\frac{\mathbb{D}}{2}-\frac{\bar\Delta_1}{2}-\frac{\Delta_1}{2}\right)}{16 \Gamma \left(\frac{\mathbb{D}}{2}\right) \Gamma (t_2+1) \Gamma (2 t_3+1) \Gamma \left(t_2+\frac{\bar\Delta_2}{2}-\frac{\Delta_1}{2}+1\right)}
 \\ & \times z_3^{t_2} z_4^{\frac{\bar\Delta_2}{2}-\frac{\Delta_1}{2}+t_2} z_7^{-\frac{\bar\Delta_2}{2}-t_2-t_3}  z_8^{2 t_3}   \omega_1^{\frac{\Delta_2}{2}+t_3} \,_2F_1\left(t_2+\tfrac{\mathbb{D}}{2}-\tfrac{\bar\Delta_1}{2}-\tfrac{\Delta_1}{2},t_3+\tfrac{\Delta_2}{2};-\tfrac{\bar\Delta_1}{2}+\tfrac{\Delta_2}{2}+1;\omega_1\right)
 \\ \phi_3 & = \tfrac{ \Gamma \left(\frac{\Delta_2}{2}-\frac{\bar\Delta_1}{2}\right) \Gamma \left(\frac{\bar\Delta_2}{2}-\frac{\Delta_1}{2}\right) \Gamma \left(-\frac{\bar\Delta_2}{2}+\frac{\Delta_1}{2}+1\right) \Gamma \left(t_3+\frac{\bar\Delta_1}{2}\right) \Gamma \left(t_2+t_3+\frac{\Delta_1}{2}\right)  \Gamma \left(t_2+\frac{\mathbb{D}}{2}-\frac{\bar\Delta_2}{2}-\frac{\Delta_2}{2}\right) }{16 \Gamma \left(\frac{\mathbb{D}}{2}\right) \Gamma (t_2+1) \Gamma (2 t_3+1) \Gamma \left(t_2-\frac{\bar\Delta_2}{2}+\frac{\Delta_1}{2}+1\right)}
 \\ & \times  z_3^{-\frac{\bar\Delta_2}{2}+\frac{\Delta_1}{2}+t_2}  z_4^{t_2} z_7^{-\frac{\Delta_1}{2}-t_2-t_3}z_8^{2 t_3} \omega_1^{\frac{\bar\Delta_1}{2}+t_3}  \,_2F_1\left(t_3+\tfrac{\bar\Delta_1}{2},t_2+\tfrac{\mathbb{D}}{2}-\tfrac{\bar\Delta_2}{2}-\tfrac{\Delta_2}{2};\tfrac{\bar\Delta_1}{2}-\tfrac{\Delta_2}{2}+1;\omega_1\right)
 \\ \phi_4 & = \tfrac{ \Gamma \left(\frac{\Delta_2}{2}-\frac{\bar\Delta_1}{2}\right) \Gamma \left(\frac{\bar\Delta_2}{2}-\frac{\Delta_1}{2}+1\right) \Gamma \left(\frac{\Delta_1}{2}-\frac{\bar\Delta_2}{2}\right) \Gamma \left(t_3+\frac{\bar\Delta_1}{2}\right) \Gamma \left(t_2+t_3+\frac{\bar\Delta_2}{2}\right) \Gamma \left(t_2+\frac{\mathbb{D}}{2}-\frac{\Delta_1}{2}-\frac{\Delta_2}{2}\right) }{16 \Gamma \left(\frac{\mathbb{D}}{2}\right) \Gamma (t_2+1) \Gamma (2 t_3+1) \Gamma \left(t_2+\frac{\bar\Delta_2}{2}-\frac{\Delta_1}{2}+1\right)}
 \\ & \times z_3^{t_2} z_4^{\frac{\bar\Delta_2}{2}-\frac{\Delta_1}{2}+t_2} z_7^{-\frac{\bar\Delta_2}{2}-t_2-t_3}  z_8^{2 t_3}  \omega_1^{\frac{\bar\Delta_1}{2}+t_3} \, _2F_1\left(t_3+\tfrac{\bar\Delta_1}{2},t_2+\tfrac{\mathbb{D}}{2}-\tfrac{\Delta_1}{2}-\tfrac{\Delta_2}{2};\tfrac{\bar\Delta_1}{2}-\tfrac{\Delta_2}{2}+1;\omega_1\right)
\\ \phi_5 & = -\tfrac{ \Gamma \left(\frac{\bar\Delta_1}{2}-\frac{\Delta_2}{2}\right) \Gamma \left(\frac{\bar\Delta_2}{2}-\frac{\Delta_1}{2}\right) \Gamma \left(-\frac{\bar\Delta_2}{2}+\frac{\Delta_1}{2}+1\right)  \Gamma \left(t_3+\frac{\Delta_2}{2}+\frac{1}{2}\right) \Gamma \left(t_2+t_3+\frac{\Delta_1}{2}+\frac{1}{2}\right)  \Gamma \left(t_2+\frac{\mathbb{D}}{2}-\frac{\bar\Delta_1}{2}-\frac{\bar\Delta_2}{2}\right) }{16 \Gamma \left(\frac{\mathbb{D}}{2}\right) \Gamma (t_2+1) \Gamma (2 t_3+2) \Gamma \left(t_2-\frac{\bar\Delta_2}{2}+\frac{\Delta_1}{2}+1\right)}
 \\ & \times z_3^{-\frac{\bar\Delta_2}{2}+\frac{\Delta_1}{2}+t_2} z_4^{t_2}  z_7^{-\frac{\Delta_1}{2}-t_2-t_3-\frac{1}{2}}  z_8^{2 t_3+1} \omega_1^{\frac{\Delta_2}{2}+t_3+\frac{1}{2}}  \, _2F_1\left(t_2+\tfrac{\mathbb{D}}{2}-\tfrac{\bar\Delta_1}{2}-\tfrac{\bar\Delta_2}{2},t_3+\tfrac{\Delta_2}{2}+\tfrac{1}{2};-\tfrac{\bar\Delta_1}{2}+\tfrac{\Delta_2}{2}+1;\omega_1\right)
 \\ \phi_6 & = -\tfrac{ \Gamma \left(\frac{\bar\Delta_1}{2}-\frac{\Delta_2}{2}\right) \Gamma \left(\frac{\bar\Delta_2}{2}-\frac{\Delta_1}{2}+1\right) \Gamma \left(\frac{\Delta_1}{2}-\frac{\bar\Delta_2}{2}\right) \Gamma \left(t_3+\frac{\Delta_2}{2}+\frac{1}{2}\right) \Gamma \left(t_2+t_3+\frac{\bar\Delta_2}{2}+\frac{1}{2}\right)  \Gamma \left(t_2+\frac{\mathbb{D}}{2}-\frac{\bar\Delta_1}{2}-\frac{\Delta_1}{2}\right) }{16 \Gamma \left(\frac{\mathbb{D}}{2}\right) \Gamma (t_2+1) \Gamma (2 t_3+2) \Gamma \left(t_2+\frac{\bar\Delta_2}{2}-\frac{\Delta_1}{2}+1\right)}
 \\ & \times z_3^{t_2} z_4^{\frac{\bar\Delta_2}{2}-\frac{\Delta_1}{2}+t_2} z_7^{-\frac{\bar\Delta_2}{2}-t_2-t_3-\frac{1}{2}} z_8^{2 t_3+1}  \omega_1^{\frac{\Delta_2}{2}+t_3+\frac{1}{2}} \, _2F_1\left(t_2+\tfrac{\mathbb{D}}{2}-\tfrac{\bar\Delta_1}{2}-\tfrac{\Delta_1}{2},t_3+\tfrac{\Delta_2}{2}+\tfrac{1}{2};-\tfrac{\bar\Delta_1}{2}+\tfrac{\Delta_2}{2}+1;\omega_1\right)
 \\ \phi_7 & = -\tfrac{ \Gamma \left(\frac{\Delta_2}{2}-\frac{\bar\Delta_1}{2}\right) \Gamma \left(\frac{\bar\Delta_2}{2}-\frac{\Delta_1}{2}\right) \Gamma \left(-\frac{\bar\Delta_2}{2}+\frac{\Delta_1}{2}+1\right) \Gamma \left(t_3+\frac{\bar\Delta_1}{2}+\frac{1}{2}\right) \Gamma \left(t_2+t_3+\frac{\Delta_1}{2}+\frac{1}{2}\right)  \Gamma \left(t_2+\frac{\mathbb{D}}{2}-\frac{\bar\Delta_2}{2}-\frac{\Delta_2}{2}\right) }{16 \Gamma \left(\frac{\mathbb{D}}{2}\right) \Gamma (t_2+1) \Gamma (2 t_3+2) \Gamma \left(t_2-\frac{\bar\Delta_2}{2}+\frac{\Delta_1}{2}+1\right)}
 \\ & \times z_3^{-\frac{\bar\Delta_2}{2}+\frac{\Delta_1}{2}+t_2}  z_4^{t_2} z_7^{-\frac{\Delta_1}{2}-t_2-t_3-\frac{1}{2}} z_8^{2 t_3+1} \omega_1^{\frac{\bar\Delta_1}{2}+t_3+\frac{1}{2}}  \, _2F_1\left(t_3+\tfrac{\bar\Delta_1}{2}+\tfrac{1}{2},t_2+\tfrac{\mathbb{D}}{2}-\tfrac{\bar\Delta_2}{2}-\tfrac{\Delta_2}{2};\tfrac{\bar\Delta_1}{2}-\tfrac{\Delta_2}{2}+1;\omega_1\right)
 \\ \phi_8 & = -\tfrac{ \Gamma \left(\frac{\Delta_2}{2}-\frac{\bar\Delta_1}{2}\right) \Gamma \left(\frac{\bar\Delta_2}{2}-\frac{\Delta_1}{2}+1\right) \Gamma \left(\frac{\Delta_1}{2}-\frac{\bar\Delta_2}{2}\right) \Gamma \left(t_3+\frac{\bar\Delta_1}{2}+\frac{1}{2}\right) \Gamma \left(t_2+t_3+\frac{\bar\Delta_2}{2}+\frac{1}{2}\right)  \Gamma \left(t_2+\frac{\mathbb{D}}{2}-\frac{\Delta_1}{2}-\frac{\Delta_2}{2}\right) }{16 \Gamma \left(\frac{\mathbb{D}}{2}\right) \Gamma (t_2+1) \Gamma (2 t_3+2) \Gamma \left(t_2+\frac{\bar\Delta_2}{2}-\frac{\Delta_1}{2}+1\right)}
 \\ & \times z_3^{t_2} z_4^{\frac{\bar\Delta_2}{2}-\frac{\Delta_1}{2}+t_2} z_7^{-\frac{\bar\Delta_2}{2}-t_2-t_3-\frac{1}{2}} z_8^{2 t_3+1}  \omega_1^{\frac{\bar\Delta_1}{2}+t_3+\frac{1}{2}} \, _2F_1\left(t_3+\tfrac{\bar\Delta_1}{2}+\tfrac{1}{2},t_2+\tfrac{\mathbb{D}}{2}-\tfrac{\Delta_1}{2}-\tfrac{\Delta_2}{2};\tfrac{\bar\Delta_1}{2}-\tfrac{\Delta_2}{2}+1;\omega_1\right).
\end{aligned}
\end{equation*}\normalsize
Setting $\omega_1 = 1$ and expanding the ${}_2F_1$ functions into convergent and divergent parts, the divergent parts cancel against each other, reducing to
\tiny
\begin{equation*}
\begin{aligned}
\phi_{1+3} & = z_3^{\frac{i \nu_1}{2}+\frac{i \nu_2}{2}+t_2} z_4^{t_2} z_7^{-\frac{d}{4}-\frac{i \nu_1}{2}-t_2-t_3} z_8^{2 t_3} \tfrac{\Gamma \left(-\frac{i \nu_1}{2}-\frac{i \nu_2}{2}\right) \Gamma \left(\frac{i \nu_1}{2}+\frac{i \nu_2}{2}+1\right) \Gamma \left(\frac{d}{4}+t_3-\frac{i \nu_1}{2}\right) \Gamma \left(\frac{d}{4}+t_3+\frac{i \nu_2}{2}\right)  \Gamma \left(\frac{d}{4}+t_2+t_3+\frac{i \nu_1}{2}\right) }{16 \Gamma \left(\frac{d}{2}+1\right) \Gamma (2 t_3+1) \Gamma \left(\frac{d}{4}+t_2+t_3+\frac{i \nu_2}{2}+1\right)} 
\\ \phi_{2+4} & = z_3^{t_2} z_4^{-\frac{i \nu_1}{2}-\frac{i \nu_2}{2}+t_2} z_7^{-\frac{d}{4}+\frac{i \nu_2}{2}-t_2-t_3} z_8^{2 t_3}  \tfrac{\Gamma \left(-\frac{i \nu_1}{2}-\frac{i \nu_2}{2}+1\right) \Gamma \left(\frac{i \nu_1}{2}+\frac{i \nu_2}{2}\right) \Gamma \left(\frac{d}{4}+t_3-\frac{i \nu_1}{2}\right) \Gamma \left(\frac{d}{4}+t_3+\frac{i \nu_2}{2}\right)  \Gamma \left(\frac{d}{4}+t_2+t_3-\frac{i \nu_2}{2}\right) }{16 \Gamma \left(\frac{d}{2}+1\right) \Gamma (2 t_3+1) \Gamma \left(\frac{d}{4}+t_2+t_3-\frac{i \nu_1}{2}+1\right)} 
\\ \phi_{5+7} & = - z_3^{\frac{i \nu_1}{2}+\frac{i \nu_2}{2}+t_2} z_4^{t_2} z_7^{-\frac{d}{4}-\frac{i \nu_1}{2}-t_2-t_3-\frac{1}{2}} z_8^{2 t_3+1} \tfrac{\Gamma \left(-\frac{i \nu_1}{2}-\frac{i \nu_2}{2}\right) \Gamma \left(\frac{i \nu_1}{2}+\frac{i \nu_2}{2}+1\right) \Gamma \left(\frac{d}{4}+t_3-\frac{i \nu_1}{2}+\frac{1}{2}\right) \Gamma \left(\frac{d}{4}+t_3+\frac{i \nu_2}{2}+\frac{1}{2}\right)  \Gamma \left(\frac{d}{4}+t_2+t_3+\frac{i \nu_1}{2}+\frac{1}{2}\right) }{16 \Gamma \left(\frac{d}{2}+1\right) \Gamma (2 t_3+2) \Gamma \left(\frac{d}{4}+t_2+t_3+\frac{i \nu_2}{2}+\frac{3}{2}\right)} 
\\ \phi_{6+8} & = - z_3^{t_2} z_4^{-\frac{i \nu_1}{2}-\frac{i \nu_2}{2}+t_2} z_7^{-\frac{d}{4}+\frac{i \nu_2}{2}-t_2-t_3-\frac{1}{2}} z_8^{2 t_3+1}\tfrac{\Gamma \left(-\frac{i \nu_1}{2}-\frac{i \nu_2}{2}+1\right) \Gamma \left(\frac{i \nu_1}{2}+\frac{i \nu_2}{2}\right) \Gamma \left(\frac{d}{4}+t_3-\frac{i \nu_1}{2}+\frac{1}{2}\right) \Gamma \left(\frac{d}{4}+t_3+\frac{i \nu_2}{2}+\frac{1}{2}\right)  \Gamma \left(\frac{d}{4}+t_2+t_3-\frac{i \nu_2}{2}+\frac{1}{2}\right) }{16 \Gamma \left(\frac{d}{2}+1\right) \Gamma (2 t_3+2) \Gamma \left(\frac{d}{4}+t_2+t_3-\frac{i \nu_1}{2}+\frac{3}{2}\right)}.
\end{aligned}
\end{equation*}\normalsize
Then summing over $t_2$ and once again following the same process, the series reduce to
\small
\begin{equation}
\begin{aligned}
  \phi_{1+2+3+4} & = \tfrac {\pi \csch ( \pi \nu_+ ) z_8^{2 t_3}} {8 \Gamma \left(\frac{d}{2}+1\right) (\nu_1-\nu_2)  }  \tfrac {\Gamma \left(\frac{d}{4}+t_3-\frac{i \nu_2}{2}\right) \Gamma \left(\frac{d}{4}+t_3+\frac{i \nu_2}{2}\right) -  \Gamma \left(\frac{d}{4}+t_3-\frac{i \nu_1}{2}\right) \Gamma \left(\frac{d}{4}+t_3+\frac{i \nu_1}{2}\right)  } {\Gamma(2 t_3 + 1)} 
\\ \phi_{5+6+7+8} & = \tfrac {\pi \csch ( \pi \nu_+ ) z_8^{2 t_3 + 1}} {8 \Gamma \left(\frac{d}{2}+1\right) (\nu_1-\nu_2)  }  \tfrac{ \Gamma \left( \frac{d}{4} + t_3- \frac{i \nu_1 }{2} + \frac{1}{2} \right) \Gamma \left( \frac{d}{4} + t_3 + \frac{i \nu_1 }{2} + \frac{1}{2}  \right) - \Gamma \left( \frac{d}{4} + t_3- \frac{i \nu_2 }{2} + \frac{1}{2}  \right) \Gamma \left( \frac{d}{4} + t_3 + \frac{i \nu_2 }{2} + \frac{1}{2}  \right)}{ \Gamma (2 t_3+2)}
\end{aligned}
\end{equation}\normalsize
Finally summing over $t_3$ results in the stated solution:
\begin{equation}
\begin{aligned}
  \mathcal I_{2P} & = \tfrac {\Gamma (-d) \left( (\sin \pi\Delta_1 + \sin \pi\bar\Delta_1) \Gamma (\Delta_1) \Gamma (\bar\Delta_1 ) - (\sin \pi\Delta_2 + \sin \pi\bar\Delta_2) \Gamma (\Delta_2 ) \Gamma ( \bar\Delta_2 ) \right)}{ \pi (\nu_1^2-\nu_2^2) }.
\end{aligned}
\end{equation}

\subsubsection{Propagator with \texorpdfstring{$1$}{1} vertex insertion: root system}\label{app:Propwith1vertexExtraDetails}
The roots used to construct this series solution are
\begin{equation}
\begin{aligned}
  \left(
\begin{smallmatrix}
 \frac{d}{2}-\frac{\Delta_1}{2}-\frac{\Delta_2}{2}-1 & 0 & 0 & -\frac{d}{2}+\frac{\Delta_1}{2}+\frac{\Delta_2}{2} & -\frac{d}{2}+\frac{\Delta_1}{2}+\frac{\Delta_2}{2} &
   -\frac{\Delta_1}{2} & -\frac{\Delta_2}{2} & 0 \\
 -1 & 0 & \frac{d}{2}-\frac{\Delta_1}{2}-\frac{\Delta_2}{2} & 0 & -\frac{d}{2}+\frac{\Delta_1}{2}+\frac{\Delta_2}{2} & -\frac{\Delta_1}{2} & \frac{\Delta_1}{2}-\frac{d}{2} & 0 \\
 -1 & \frac{d}{2}-\frac{\Delta_1}{2}-\frac{\Delta_2}{2} & 0 & -\frac{d}{2}+\frac{\Delta_1}{2}+\frac{\Delta_2}{2} & 0 & \frac{\Delta_2}{2}-\frac{d}{2} & -\frac{\Delta_2}{2} & 0 \\
 -\frac{d}{2}+\frac{\Delta_1}{2}+\frac{\Delta_2}{2}-1 & \frac{d}{2}-\frac{\Delta_1}{2}-\frac{\Delta_2}{2} & \frac{d}{2}-\frac{\Delta_1}{2}-\frac{\Delta_2}{2} & 0 & 0 &
   \frac{\Delta_2}{2}-\frac{d}{2} & \frac{\Delta_1}{2}-\frac{d}{2} & 0 \\
 \frac{d}{2}-\frac{\Delta_1}{2}-\frac{\Delta_2}{2}-1 & 0 & 0 & -\frac{d}{2}+\frac{\Delta_1}{2}+\frac{\Delta_2}{2} & -\frac{d}{2}+\frac{\Delta_1}{2}+\frac{\Delta_2}{2} &
   -\frac{\Delta_1}{2}-\frac{1}{2} & -\frac{\Delta_2}{2}-\frac{1}{2} & 1 \\
 -1 & 0 & \frac{d}{2}-\frac{\Delta_1}{2}-\frac{\Delta_2}{2} & 0 & -\frac{d}{2}+\frac{\Delta_1}{2}+\frac{\Delta_2}{2} & -\frac{\Delta_1}{2}-\frac{1}{2} & -\frac{d}{2}+\frac{\Delta_1}{2}-\frac{1}{2} & 1 \\
 -1 & \frac{d}{2}-\frac{\Delta_1}{2}-\frac{\Delta_2}{2} & 0 & -\frac{d}{2}+\frac{\Delta_1}{2}+\frac{\Delta_2}{2} & 0 & -\frac{d}{2}+\frac{\Delta_2}{2}-\frac{1}{2} & -\frac{\Delta_2}{2}-\frac{1}{2} & 1 \\
 -\frac{d}{2}+\frac{\Delta_1}{2}+\frac{\Delta_2}{2}-1 & \frac{d}{2}-\frac{\Delta_1}{2}-\frac{\Delta_2}{2} & \frac{d}{2}-\frac{\Delta_1}{2}-\frac{\Delta_2}{2} & 0 & 0 &
   -\frac{d}{2}+\frac{\Delta_2}{2}-\frac{1}{2} & -\frac{d}{2}+\frac{\Delta_1}{2}-\frac{1}{2} & 1 \\
\end{smallmatrix}
\right).
\end{aligned}
\end{equation}
They were initially assumed to be generic by setting $\DD = d + 2 + \epsilon$ and the series were evaluated in the limit $\epsilon \rightarrow 0$, similar to \cref{app:2propmelonExtraDetails}.

\section{Notation and useful formulae}

\begin{table}[H]
\caption{Notation}\label{tb:notation}
\begin{center}
\begin{tabular}{| l | l |}
\hline
 $\DD - 1 = D = d + 1$ & dimensions \\
\hline
${\rm d}S_{d+1}$ & $(d+1)$-dimensional de Sitter space \\
$S^{\DD-1} = S^{D} = S^{d+1}$ & $(d+1)$-dimensional sphere of unit radius\\
\hline
  $\RR^{n} : X = (X^1, \cdots X^n)$   & Flat euclidean space with metric $\de s^2 = \delta_i{}_j{} \de X^i \de X^j{}$\\
  $\RR^\DD \cong \RR^* \times S^{d+1}$  & $\DD$ dimensional euclidean space serving as embedding space for $S^{d+1}$ \\
\hline
$\Lambda, \; \ell$ & Cosmological constant, Length scale \\
\hline
$\Delta, \; \bar \Delta$ & Mass parameters related by $d = \Delta + \bar \Delta$ \\
\hline
\rule{0pt}{2em}$\displaystyle\int^*_\sigma, \; \dashint^*_\sigma, \; \Xint{\times}, \; \XintAdd{\times}{*}{\sigma}$ & $\displaystyle\int_0^\infty \frac {\de \sigma} {\sigma}, \; \int_0^1 \frac {\de \sigma} {\sigma \, (1-\sigma)}, \; \frac 1 {\volR} \int, \; \frac 1 {\volR} \int^\infty_0 \frac {\de \sigma} {\sigma}$  \\
\rule{0pt}{2em}$\displaystyle\int^*_{\lambda} =  \dashint^*_s$ & $\displaystyle s = \frac {\lambda} {\lambda + 1} \Leftrightarrow \lambda = \frac {s}{1 - s} $  \\
\rule{0pt}{2em}$\displaystyle\int_{t} $ & $\displaystyle\int_{-\infty}^{\infty} \de t$ when the range of integration is obvious in context\\[2ex]
\rule{0pt}{2em}$\displaystyle\ointctrclockwise_{\Delta}$ & $\displaystyle\int_{\delta  - i \infty}^{\delta  + i \infty} \frac { \de \Delta} {2 \pi i} $, where $\displaystyle \delta \in \RR$ such that the integral is convergent. \\[2ex]
\hline
\rule{0pt}{2em}$\displaystyle \xdotY, \; \xdotY_{1,2}, \;  \theta, \; \geodissq, \; \geodis$ & $\displaystyle\hat X \cdot \hat Y, \; \hat X_1 \cdot \hat X_2, \;  \cos^{-1} (\hat X \cdot \hat Y), \;  \tfrac {1 + \hat X \cdot \hat Y} {2}, \; \sqrt{\tfrac {1 + \hat X \cdot \hat Y} {2}} $ \\[2ex]
\hline
\end{tabular}
\end{center}
\end{table}

\begin{table}[H]
\caption{Parameterisations and some relevant formulae}\label{tb:Parameterisations}
\begin{center}
\begin{tabular}{| l | l |}
\hline
\rule{0pt}{2em}Schwinger 
  parameterisation & \(\displaystyle \frac 1 {f^{s}} = \int^\infty_0 \frac {\de z} {z} \, \frac {z^{s}} {\Gamma(s)} \, e^{- z \, f}\) \\[2ex]
\hline
\rule{0pt}{2em}Feynman parameterisation & \( \displaystyle \tfrac 1 {f_1^{s_1} \, \cdots \, f_n^{s_n}} = \int^1_0 \tfrac {\de y_1} {y_1} \cdots  \tfrac {\de y_n} {y_n} \, \tfrac {y_1^{s_1}} {\Gamma(s_1)} \, \cdots \, \tfrac {y_n^{s_n}} {\Gamma(s_n)} \, \tfrac {\Gamma(\sum s_i) \, \delta(1 - \sum y_i)} {(\sum y_i \, f_i)^{\sum s_i}} \) \\[2ex]
\rule{0pt}{2em} & \( \displaystyle \phantom{\tfrac 1 {f_1^{s_1} \, \cdots \, f_n^{s_n}} }= \tfrac {\Gamma(\sum s)} {\Gamma(s_1) \cdots \Gamma(s_n)} \, \int_0^\infty \tfrac {\de y_2} {y_2} \cdots  \tfrac {\de y_n} {y_n} \, \tfrac {y_2^{s_2} \cdots y_n^{s_n}} {(f_1 + y_2 \, f_2 \cdots + y_n \, f_n)^{\sum s_i}} \) \\[2ex]
\hline
\rule{0pt}{2em}Mellin transform & $\displaystyle \mathcal M \circ f\, (\Delta) =  \int_{r}^* \, r^{\Delta} \, f(r)$\\[2ex]
  \rule{0pt}{2em}Inverse Mellin transform & $\displaystyle f(r) = \ointctrclockwise_{\Delta} \, r^{-\Delta} \, \mathcal M \circ f\, (\Delta), \quad \ointctrclockwise_{\Delta} = \int_{\delta - i \infty}^{\delta + i \infty} \, \frac {\de \Delta} {2 \pi i}$\\[2ex]
  \hline
 \rule{0pt}{2em}Mellin representation of $\Gamma$ function& $\displaystyle\Gamma(z) = \int^* s^{z} \, e^{- s} = \mathcal M \circ e^{- x}$\\[2ex]
  \hline
 \rule{0pt}{2em}Euler's reflection & $\displaystyle\Gamma(1 - z) \, \Gamma(z) = \pi \, \csc (\pi \, z)$ \\[2ex]
  \hline
  \rule{0pt}{2em}Legendre duplication & $\displaystyle \Gamma(z) \, \Gamma(z + \tfrac 1 2) = \frac {2\, \sqrt \pi} {2^{2 z}} \, \Gamma(2 z)$ \\[2ex]
  \hline
\rule{0pt}{2em}Limiting value of ${}_2F_1$ & $\displaystyle {}_2F_1(a,\,b;\,c;\,1) = \frac {\Gamma(c)\,\Gamma(c-a-b) }{\Gamma(c-a) \, \Gamma(c-b)}, \quad \Real(c - a - b) > 0$
\\[2ex]
  \hline
\end{tabular}
\end{center}
\end{table}\noindent

\begin{table}[H]
\caption{Scalar propagators on $S^{\rm odd}$ at geodesic distance $\theta$ with mass parameter $\Delta = \frac {d} {2} + \im \nu$}\label{tb:oddSprop}
  \begin{center}
  \begin{tabular}{ |c| c | c|} 
   \hline
  \rule{0pt}{1em} $\displaystyle S^{2\,n+1}$  & Scalar Propagator  \\ \hline

  \rule{0pt}{2em}$\displaystyle S^{1}$ &  $\displaystyle - \frac {\cos (\Delta (\pi - \theta))} {4 \, \Delta \, \sin (\pi \Delta)}  = \frac {\cosh((\pi - \theta) \, \nu)} {4 \, \nu \, \sinh( \pi \nu) } $\\[2ex] \hline

   \rule{0pt}{2em}$\displaystyle S^{3}$  &  $\displaystyle \frac{\sin (\Delta(\pi - \theta) + \theta)} {4 \pi  \, \sin (\pi \Delta) \, \sin \theta} = \frac {\sinh((\pi - \theta) \, \nu)}{4 \pi \, \sinh( \pi \nu) \, \sin \theta}$ \\[2ex] \hline

  \rule{0pt}{2em} $\displaystyle S^{5}$ &$\displaystyle \frac{3 \big( (\Delta-1) \sin ( \Delta (\pi-\theta )+ 3 \theta ) - ( \Delta - 3) \sin ( \Delta (\pi -\theta )+\theta ) \big)}{8 \pi^{2} \,  \sin (\pi  \Delta) \, \sin^3 \theta }$  \\[2ex] 
  \rule{0pt}{2em}  & $\displaystyle = \frac{3 \big( \nu \, \cosh((\pi - \theta) \, \nu) \, \sin \theta + \sinh((\pi - \theta)\, \nu) \, \cos\theta \big)}{4 \pi^{2} \,  \sinh (\pi  \nu) \, \sin^3 \theta }$  \\[2ex] \hline

  \rule{0pt}{2em} $\displaystyle S^{7}$ & $\frac{15 \big((\Delta - 1)\, (\Delta - 2) \, \cos 4 \theta - 3 \, (\Delta -1)\, (\Delta - 4) \, \cos 2 \theta + 2 (\Delta - 2) \, (\Delta- 4) \big)}{8 \pi^{3}  \,  \sin (\pi  \Delta) \, \sin^5 \theta} \, \sin (\Delta \, (\pi - \theta)) \, \cos \theta$ \\[2ex] 
  \rule{0pt}{2em}  & $ + \frac{15 \big( (\Delta - 1) \, (\del - 2) \, \cos 4 \theta - (\del - 1) \, (\del - 8) \, \cos 2\theta + 6 \big)}{8 \pi^{3}  \,  \sin (\pi  \Delta) \, \sin^5 \theta} \, \cos (\Delta \, (\pi - \theta)) \, \sin \theta $ \\[2ex]
  \rule{0pt}{2em}  & $ \displaystyle= \frac{15\Big(  3 \, \nu \, \sin 2 \theta\, \cosh((\pi - \theta) \,\nu)  + ((2 - \nu^2) \, \cos 2 \theta + \nu^2 + 4) \, \sinh((\pi - \theta) \,\nu) \, \Big)}{8 \pi^{3}  \,  \sinh (\pi  \nu) \, \sin^5 \theta}  $ \\[2ex] \hline
  \end{tabular}
  \end{center}
\end{table}

\begin{table}[H]
\caption{Scalar propagators on $S^{\rm even}$ at geodesic distance $\theta$ at some mass parameters $\Delta$}\label{tb:evenSprop}
  \begin{center}
  \begin{tabular}{ |c| c | c|} 
   \hline
  \rule{0pt}{1em}  $\displaystyle S^{2 \, n}$ & $\Delta$ & Scalar Propagator  \\ \hline

  \rule{0pt}{2em} $\displaystyle S^{2}$ & $\frac 1 2$, $\frac 3 2$   & $\displaystyle\frac{K(\geodissq)}{2 \pi^{\frac {3}{2}} }$, $\displaystyle \frac{K(\geodissq)-2 \, E(\geodissq)}{2 \pi ^{\frac 3 2}} $ \\[2ex]  \hline
  \rule{0pt}{2em} $\displaystyle S^{4}$ & $\frac 1 2$, $\frac 3 2$   & $\displaystyle \frac { \tan \frac{\theta }{2} \,  K(\geodissq)+ 2 \, \cot \theta \,  E(\geodissq)} {2 \pi^{\frac 5 2} \, \sin \theta}$, $\displaystyle \frac { E(\geodissq) -\sin^2 \frac {\theta}{2} \, K(\geodissq)} {\pi^{\frac 5 2} \, \sin^2 \theta}$ \\[2ex]  \hline
  \rule{0pt}{2em} $\displaystyle S^{6}$ & $\frac 3 2$   & $\displaystyle \frac {4 \,\big( 2 \, (1  - \geodissq \, (1 - \geodissq))\, E(\geodissq) - (2 - \geodissq) \, (1 - \geodissq)\, K(\geodissq) \big)} {\pi^{\frac 7 2} \, \sin^4\theta}$ \\[2ex] 
  \rule{0pt}{2em}  & $\frac 5 2$   & $\displaystyle \frac {4  \,\big( (2 - 3 \, \geodissq) \, (1 - \geodissq)\, K(\geodissq) - 2 \, (1 - 2 \, \geodissq)\, E(\geodissq) \big)} {\pi^{\frac 7 2} \, \sin^4\theta}$ \\[2ex]  \hline
  \end{tabular}
  \end{center}
\end{table}\noindent
Euler transformations of the scalar propagator $G(\hat X, \, \hat Y)$: 
\begin{equation}\label{eq:scpropEulertransform}
\begin{aligned}
 {\rm Euler} : \; G \rightarrow & \; \frac {\Gamma(\Delta) \Gamma(d - \Delta)} {(4 \pi)^{\frac{d+1}{2}} \Gamma(\frac {d+1} {2})} \, \frac 1 {(1 -\geodissq)^{\frac {d - 1} {2}} } \, {}_2 F_1 (\tfrac {\bar\Delta - \Delta +1} {2} , \, \tfrac {\Delta - \bar\Delta + 1} {2} ; \, \tfrac {d+1} {2}; \, \geodissq) 
  \\   G  \rightarrow & \;  \frac {\Gamma(\Delta) \Gamma(d - \Delta) \Gamma(\frac {1 - d} {2}) } {(4 \pi)^{\frac{d+1}{2}}\, \Gamma(\frac {\bar \Delta - \Delta + 1 } {2}) \, \Gamma(\frac {\Delta - \bar \Delta + 1 } {2})}  \, {}_2 F_1 (\Delta, \, d - \Delta; \, \tfrac {d + 1} {2}; \, 1 - \geodissq) 
  \\  & + \frac {\Gamma(\frac {d - 1} {2})} {(4 \pi)^{\frac{d+1}{2}}} \, \frac {1}{ (1 - \geodissq)^{\frac {d - 1} {2}}} {}_2 F_1 (\tfrac {\bar \Delta - \Delta + 1 } {2}, \, \tfrac {\Delta - \bar  \Delta + 1 } {2} ; \, \tfrac {3 - d} {2}; \, 1 - \geodissq). 
\end{aligned}
\end{equation}
Pfaff transformation of the scalar propagator $G(\hat X, \, \hat Y)$: 
\begin{equation}\label{eq:scpropPfafftransform}
\begin{aligned}
  {\rm Pfaff} : \;  G &  \rightarrow  \frac {\Gamma(\Delta) \Gamma(d - \Delta)} {(4 \pi)^{\frac{d+1}{2}} \Gamma(\frac {d+1} {2})} \, \frac {1}{(1 - \geodissq)^{\Delta} } \,{}_2 F_1 (\Delta, \, \tfrac {\Delta - \bar \Delta + 1} {2}; \, \tfrac {d+1} {2}; \, \tfrac {\xdotY + 1} { \xdotY - 1}).
\end{aligned}
\end{equation}

\nocite{Henneaux:1989jq,Henneaux:1992ig,Cox:AlgebraicGeometry,tantau:2013a}
\begin{singlespace}
\printbibliography[title={References}]
\end{singlespace}

\end{document}